\documentclass[twocolumn,showpacs,amsmath,amssymb,nofootinbib,%
               longbibliography]{revtex4-1}

\usepackage{graphicx,amsmath,amssymb,amsfonts,mathrsfs,bm,mathtools,multirow}
\usepackage{dcolumn}
\usepackage[usenames,dvipsnames]{color}
\usepackage{xspace}
\usepackage{braket}
\usepackage{slashed}
\usepackage{xargs}

\usepackage[colorlinks,linkcolor=blue,anchorcolor=blue,citecolor=blue]{hyperref}

\newcommand{\elemA}[2]{\ensuremath{{}^{#1}}\textrm{#2}}

\newcommand{\red}[1]{\textcolor{red}{#1}}

\newcommandx{\yyy}[2][1=xxx,2=yyy]{\red{[#1,#2]}\xspace}
\newcommandx{\zzz}[3][1=xxx,2=yyy,3=zzz]{\red{[#1,#2,#3]}\xspace}
\newcommand{\bbz}{$0\nu\beta\beta$\xspace}
\newcommand{\bbt}{$2\nu\beta\beta$\xspace}
\newcommand{\bb}{$\beta\beta$\xspace}

\begin{document}
    
\title{Status and Future of Nuclear Matrix Elements for Neutrinoless Double-Beta
Decay: A Review}
 \author{Jonathan Engel}
\email{engelj@physics.unc.edu}
\affiliation{Department of Physics and Astronomy, University of North Carolina, Chapel 
Hill, NC 27516-3255, USA}
\author{Javier Men\'endez}
\email{menendez@nt.phys.s.u-tokyo.ac.jp}
\affiliation{Department of Physics, University of Tokyo, Hongo, Tokyo 113-0033, Japan}

\date{\today}

\begin{abstract}
The nuclear matrix elements that govern the rate of neutrinoless double beta
decay must be accurately calculated if experiments are to reach their full
potential.  Theorists have been working on the problem for a long time but have
recently stepped up their efforts as ton-scale experiments have begun to look
feasible.  Here we review past and recent work on the matrix elements in a wide
variety of nuclear models and discuss work that will be done in the near future.
\textit{Ab initio} nuclear-structure theory, which is developing rapidly,
holds out hope of more accurate matrix elements with quantifiable error bars. 

\end{abstract}

\maketitle

\tableofcontents

\section{Introduction}
\label{s:intro}

Neutrinos are the only neutral fermions we know to exist.  They are thus the
only known particles that may be Majorana fermions, that is, their own
antiparticles.  Because neutrinos are so light, the difference in behavior
between Majorana neutrinos and Dirac neutrinos, which would be distinct from
their antiparticles, is slight.  The easiest way to determine which of the two
possibilities nature has chosen --- and it is far from easy --- is to see
whether certain nuclei undergo neutrinoless double-beta (\bbz) decay, a
second-order weak-interaction process in which the parent nucleus decays into
its daughter with two fewer neutrons and two more protons, while emitting two
electrons but, crucially, no (anti)neutrinos.

Experiments to observe \bbz decay are becoming more and more sensitive, and
international teams are trying to push the sensitivity to the point at which 
they can identify a few decay events per year in a ton of material
\cite{Henning16,
DellOro16, Cremonesi14, GomezCadenas12}.  The hope is to be sensitive enough to
detect \bbz decay if neutrinos are indeed Majorana particles and their masses
are arranged in a pattern known as the ``inverted hierarchy'' (discussed in
Sec.~\ref{ss:hier}).  Because the decay takes place inside nuclei, the amount
of material required to fully cover the inverted-hierarchy region depends not
only on the masses of the three kinds of neutrinos, but also on the nuclear
matrix element (or elements, since present and planned \bbz decay
experiments~\cite{KamLAND-Zen16,EXO14,Next16,GERDA13,MAJORANA14,
CUORE15,SNO+16,SUperNEMO16,LUCIFER15,CANDLES16,AMoRE16,COBRA16,TINTIN15,ZICOS16}
may consider about a dozen different nuclei) of a subtle two-nucleon
operator between the ground states of the decaying nucleus and its decay
product.  Since \bbz decay involves not only nuclear physics but also unknown
neutrino properties, such as the neutrino mass scale, the matrix elements
cannot be measured; they must be calculated.  And at present they are not
calculated with much accuracy.  We need to know them better.

Fortunately, nuclear-structure theory has made rapid progress in the last
decade and the community is now in a position to improve calculated matrix
elements materially.  This review describes work that has already been carried
out, from early pioneering studies to more recent and sophisticated efforts,
and discusses what is needed to do significantly better.  We are optimistic
that recent progress in the use of chiral effective field theory ($\chi$EFT) to
understand nuclear
interactions~\cite{chiral,Machleidt11,Hammer13,Machleidt:2016rvv}, and of
nonperturbative methods to efficiently solve the nuclear many-body problem from
first principles (with controlled
errors)~\cite{carlson15,Hebeler15,hagen14,Hergert16,Soma14,Meissner16} will
produce reliable matrix elements with quantified uncertainties over the next
five or so years.  We will outline the ways in which that might happen.

This review is structured as follows:  Section \ref{s:signigicance} discusses
the significance of \bbz decay and the nuclear matrix elements that govern it.
Section \ref{s:present} reviews calculations of the matrix elements and
indicates where we stand at present.  Section \ref{s:ga} is a slight detour
into a more general problem, the ``renormalization of the axial vector coupling
$g_A$," that has important consequences for \bbz nuclear matrix
elements.  Section \ref{s:improving} is about ways in which matrix-element
calculations should improve in the next few years, and ways in which the
uncertainty in new calculations can be assessed.  Section \ref{s:conclusion} is
a conclusion.

\section{Significance of Double-Beta Decay}
\label{s:signigicance}

\subsection{Neutrino Masses and Hierarchy}
\label{ss:hier}

Before turning to nuclear-structure theory, we very briefly review the neutrino
physics that makes it necessary.  References\ \cite{avi08} and \cite{ver12}
contain pedagogical reviews of both the neutrino physics and the nuclear matrix
elements that are relevant for \bb decay.

Flavor oscillations of neutrinos from the atmosphere~\cite{SuperKamiokande98},
from the sun~\cite{SNO02}, and from nuclear reactors~\cite{KamLAND03} have
revealed neutrino properties that were unknown a few decades ago.  Neutrinos
have mass, but the three kinds of neutrino with well-defined masses are linear
combinations of the kinds with definite flavor that interact in weak
processes.  We know with reasonable accuracy the differences in squared mass
among the three mass eigenstates, with one smaller difference $\Delta
m^2_{\text{sun}}\simeq 75\ \text{meV}^2$~\cite{pdg14} coming mainly from
solar-neutrino experiments and one larger difference $\Delta
m^2_{\text{atm}}\simeq 2400\ \text{meV}^2$~\cite{pdg14} coming mainly from
atmospheric-neutrino experiments.  We also know, with comparable accuracy, the
mixing angles that specify which linear combinations of flavor eigenstates have
definite mass~\cite{GonzalezGarcia14}.

\begin{figure}[t]
\begin{center}
\includegraphics[width=0.5\textwidth,clip=]{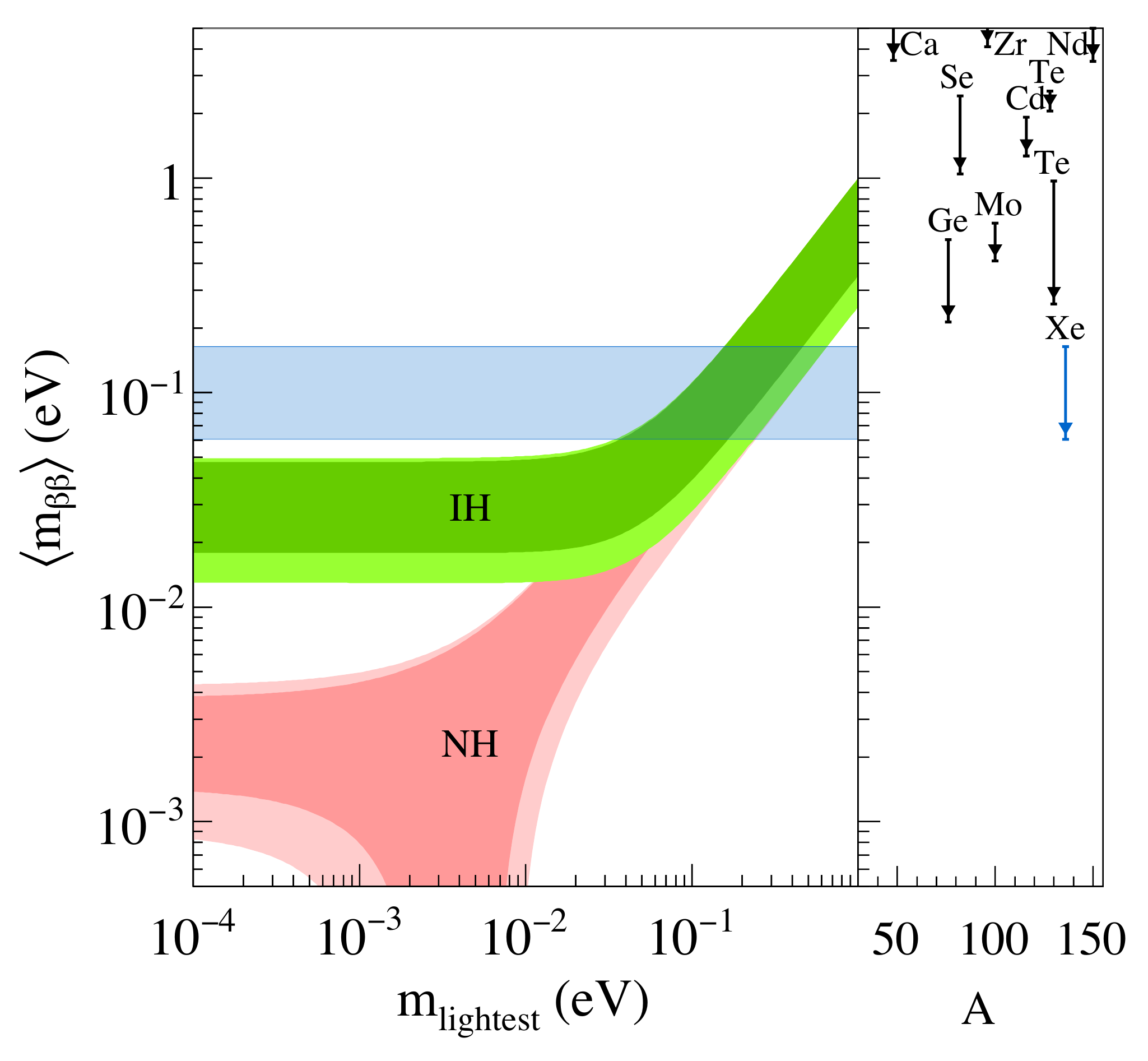}
\end{center}
\caption{Left panel: Bands for the value of the parameter $m_{\beta\beta}$ as a
function of the mass of the lightest neutrino, for the case of normal (NH, red
band) and inverted (IH, green band) neutrino-mass hierarchies. The present best
experimental upper limits on $m_{\beta\beta}$ are shown in the blue band.
Right panel: Present best upper limits, with uncertainty bars, on
$m_{\beta\beta}$ from experiments performed on each \bb emitter, as a function
of mass number $A$.  The uncertainty bands and bars include experimental
uncertainties and ranges of calculated nuclear matrix elements. Figure adapted
from Ref.~\cite{KamLAND-Zen16}, courtesy of the KamLAND-Zen collaboration.}
\label{f:m_bb}
\end{figure}

The arrangement of the masses, called the ``hierarchy," is still unknown,
however.  There are two possibilities:  either the two mass eigenstates
that mix most strongly with electron flavor are lighter than the third (the
``normal hierarchy," because it is similar to the hierarchy of quark mass 
eigenstates) or they are heavier (the ``inverted hierarchy'').  Long baseline
neutrino-oscillation experiments can eventually determine the hierarchy with a
confidence level corresponding to four standard deviations or more, but for now
they show just a two-$\sigma$ preference for the normal
hierarchy~\cite{T2K15,NOvA16}. Figure~\ref{f:m_bb} shows the present
experimental \bbz decay limits on the combination of neutrino masses
$m_{\beta\beta}$ [defined by Eq.~\eqref{eq:mass} in Sec.~\ref{sss:LNE}],
together with the regions corresponding to the normal and inverted hierarchies,
as a function of the mass of the lightest neutrino.  If the hierarchy is normal
and the lightest neutrino is lighter than about 10 meV, then a detection of
\bbz decay is out of reach for the coming generation of experiments unless the
decay is driven by the exchange of a heavy particle, the existence of which we
have not yet discovered, or some other new physics (see Sec.~\ref{sss:bsm}).
If the hierarchy is inverted, the experiments to take place in the next decade
have a good chance to see the decay, provided they have enough material.
Indeed, Fig.~\ref{f:m_bb} shows that the current experimental limit almost
touches the upper part of the inverted-hierarchy region.

How much material will be needed to completely cover the region, so that we can
conclude in the absence of a \bbz signal that either the neutrino hierarchy is
normal or neutrinos are Dirac particles?  And in the event of a signal, how
will we tell whether the exchange of light neutrinos or some other mechanism is
responsible?  If it is the latter, what is the underlying new physics?  To
answer any of these questions, we need accurate nuclear matrix elements.

\subsection{Neutrinoless Double-Beta Decay}
\label{ss:0vdb}

\subsubsection{Light-neutrino Exchange}
\label{sss:LNE}

The beginning of this section closely follows Ref.\ \cite{avi08}, which itself
is informed by Ref.\ \cite{bil03}. More detailed derivations of the \bb
transition rates can be found in Refs.~\cite{doi85,hax84,tom91}.

\begin{figure}[t]
\begin{center}
\includegraphics[width=0.35\textwidth,clip=]{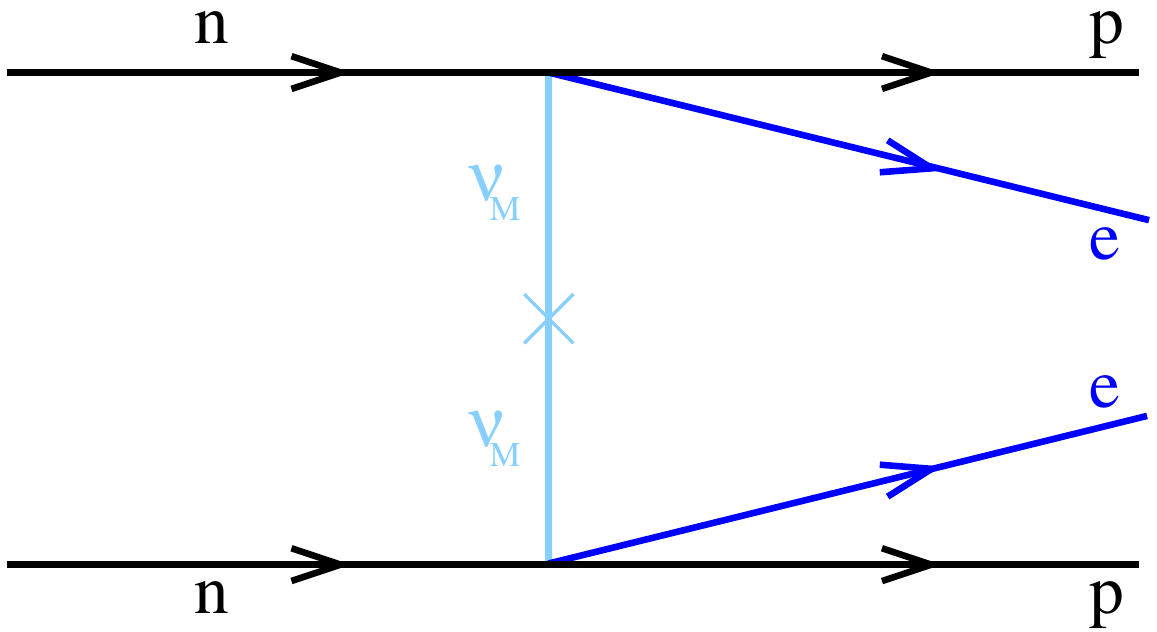}
\end{center}
\caption{Feynman diagram for \bbz decay mediated by light-neutrino exchange.
Two neutrons (n) decay into two protons (p), emitting two
electrons (e). No neutrinos are emitted, implying that they are Majorana
particles ($\nu_M$).}
\label{f:0nu_light_diagram}
\end{figure}

The rate for \bbz decay, if we assume that it is mediated by the exchange of
the three light Majorana neutrinos and the Standard Model weak interaction as
represented in Fig.\ \ref{f:0nu_light_diagram}, is
\begin{equation}
\label{eq:rate} [T^{0\nu}_{1/2}]^{-1}=\sum_{\textrm{spins}} \int |Z_{0\nu}|^2
\delta(E_{e1}+E_{e2}+E_f-E_i) \frac{d^3\bm{p}_1}{2\pi^3}\frac{d^3\bm{p}_2}
{2\pi^3}~,
\end{equation}
where $E_{e1}$, $E_{e2}$ and $\bm{p}_1$, $\bm{p}_2$ are the energies and
momenta of the two emitted electrons, $E_{i}$ and $E_{f}$ are the energies of
the initial and final nuclear states, and $Z_{0\nu}$ is an amplitude
proportional to an $S$-matrix element up to delta functions that enforce energy
and momentum conservation. The $S$ matrix depends on the product of leptonic
and hadronic currents in the effective low-energy semi-leptonic Lagrangian
density: 
\begin{equation}
\label{eq:eff-semilep-la}
\mathcal{L}(x)= G_F/\sqrt{2}\{
\overline{e}(x) \gamma_{\mu} (1-\gamma_5) \nu_e(x) J_L^{\mu}(x)\} + \text{h.c.}
\,,
\end{equation}
with $J_L^{\mu}$ the left-handed charge-changing hadronic current density.
Because $Z_{0\nu}$ is second order in the weak-interaction Lagrangian, it
contains a lepton part that depends on two space-time positions $x$ and $y$,
which are contracted and ultimately integrated over: 
\begin{equation}
\label{eq:amplitude}
\begin{split}
\sum_k & \overline{e}(x) \gamma_{\mu}
(1-\gamma_5) U_{ek} \nu_k\!\underbracket{(x) \  \overline{e}(y)
\gamma_{\nu} (1-\gamma_5) U_{ek} \nu_k}(y)    \\
 =-\sum_k &
\overline{e}(x) \gamma_{\mu} (1-\gamma_5) U_{ek}
\nu_k\!\underbracket{(x) \ \overline{\nu_k^c}}(y) \gamma_{\nu}
(1+\gamma_5) U_{ek} e^c(y) \,.
\end{split}
\end{equation}
Here $\nu_k$ is the Majorana mass eigenstate with mass $m_k$ and $U_{ek}$ is
the element of the neutrino mixing matrix that connects electron flavor with
mass eigenstate $k$.  We denote the charge conjugate of a field $\psi$ by
$\psi^c \equiv i \gamma^2\psi^*$ (in the Pauli-Dirac representation), and
because $\nu_k$ are Majorana states we can take $\nu_k^c = \nu_k$.  

The contraction of $\nu_k$ with $\overline{\nu_k^c}$ turns out to be the usual
fermion propagator, so that the lepton part above becomes
\begin{align}
\label{eq:nu-prop}
-\frac{i}{4} \int &\sum_k \frac{d^4q}{(2\pi)^4}e^{-i
q\cdot(x-y)} \overline{u}(p_1) \gamma_{\mu}(1-\gamma_5) e^{-i(p_1 \cdot x +
p_2 \cdot y)}\nonumber \\
&\times \frac{\slashed{q}+m_k}{q^2-m_k^2} \gamma_{\nu}(1+\gamma_5)
 u^c(p_2)~U_{ek}^2 \,,
\end{align}
where $q$ is the 4-momentum of the virtual neutrino.  The term with
$\slashed{q}$ vanishes because the two currents are left handed and if we
neglect the very small neutrino masses in the denominator, the decay amplitude
becomes proportional to
\begin{align}
\label{eq:mass}
   & m_{\beta\beta} \equiv \left| \sum_k m_k U^2_{ek}\right| \\
&= \big| m_1
    |U_{e1}|^2 
    + m_2|U_{e2}|^2 e^{i(\alpha_2-\alpha_1)} + m_3 |U_{e3}|^2 
    e^{i (-\alpha_1-2\delta)} \big| \,. \nonumber
 \end{align}
Here $\delta$ is the so-called Dirac phase, and $\alpha_1, \alpha_2$ are
Majorana phases that vanish if neutrinos are Dirac particles.  We have inserted
the absolute value in Eq.\ \eqref{eq:mass} consistently with the amplitude in
Eq.~\eqref{eq:rate}, because the expression inside can be complex.

To obtain the full amplitude $Z_{0\nu}$, one must multiply the lepton part
above by the nuclear matrix element of two time-ordered hadronic currents and
integrate the product over $x$ and $y$.  Because $J_L^\mu(x)= e^{iH x_0}
J_L^{\mu}(\bm{x})e^{-iH x_0}$ ($H$ is the hadronic Hamiltonian and the current
on the right-hand side is evaluated at time $x_0=0$), one can write the matrix
element of an ordinary product of hadronic currents between initial (i) and
final (f) nuclear states as
\begin{align}
\label{eq:hadron}
\bra{f} J_L^{\mu}(x) J_L^{\nu}(y) \ket{i}
= \sum_n & \bra{f}  J_L^{\mu}(\bm{x}) \ket{n} 
\bra{n} J_L^{\nu}(\bm{y}) \ket{i} \\
&\times e^{-i(E_n-E_f) x_0}e^{-i(E_i-E_n) y_0} \,, \nonumber
\end{align}
where the $|n\rangle$'s are a complete set of intermediate nuclear states, with
corresponding energies $E_n$.  Time ordering the product of currents and
combining the phases in Eq.~\eqref{eq:hadron} with similar factors from the
lepton currents yields the following amplitude, after integration first over
$x_0$, $y_0$, and $q^0$, then over $\bm{x}$ and $\bm{y}$:
\begin{align}
\label{eq:preclosure} 
 \sum_n &
\left[\frac{\langle f |J_L^{\mu}(\bm{q})| n \rangle \langle n |
J_L^{\nu}(-\bm{q})|i\rangle}  {|\bm{q}|(E_n+|\bm{q}|+E_{e2}-E_i)} +
\frac{\langle f |J_L^{\nu}(\bm{q})| n \rangle \langle n |
J_L^{\mu}(-\bm{q})|i\rangle}  {|\bm{q}|(E_n+|\bm{q}|+E_{e1}-E_i)}\right] \nonumber  \\
& \times 2\pi \delta(E_f+E_{e1}+E_{e2}-E_i)  \,,
\end{align}
where the tiny neutrino masses in the denominator of Eq.~\eqref{eq:nu-prop} and
the electron momenta $|\bm{p}_1|$ and $|\bm{p}_2|$ have been neglected because
they are much smaller than a typical momentum transfer $|\bm{q}|$.  The
energy-conservation condition comes from the definition of $Z_{0\nu}$.  

To go further one needs to know the nuclear current operators.  At this point,
most authors make two important approximations.  The first is the ``impulse
approximation,'' i.e.\ the use of the current operator for a collection of free
nucleons.  The operator is then specified by its one-body matrix elements:
\begin{align}
\label{eq:one-b-me}
\langle p | J_L^{\mu}(x)| p^\prime \rangle = & e^{iqx} 
\overline{u}(p)\Big( g_V(q^2)\gamma^{\mu} - g_A(q^2)
\gamma_5 \gamma^{\mu} \\ 
& - i g_M(q^2) \frac{\sigma^{\mu\nu}}{2m_N} q_{\nu}
+ g_P(q^2) \gamma_5 q^{\mu} \Big){u}(p^{\prime}) \,, \nonumber
\end{align}
where $q=p^\prime-p$, the conservation of the vector current tells us that $g_V
\equiv g_V(0)=1$, and $g_M(q^2)= g_M g_V(q^2)$ with $g_M \equiv g_M(0) \simeq
4.70 g_V$ (as given by the proton and neutron anomalous magnetic
moments~\cite{pdg14}), $g_A=g_A(0)\simeq 1.27$~\cite{pdg14} (from neutron
$\beta$-decay measurements~\cite{UCNA13}), and the Goldberger-Treiman relation
$g_P(q^2)=2m_N g_A(q^2)/(\bm{q}^2+m_\pi^2)$, with $m_N$ and $m_\pi$ the nucleon
and pion masses, connects the pseudoscalar and axial terms and is accurate
enough for our purposes.  The momentum-transfer dependence of the axial and
vector terms can be parameterized in several ways by fitting experimental
data~\cite{Dumbrajs83,Bernard01}.  A non-relativistic reduction of the matrix
elements in Eq.\ \eqref{eq:one-b-me} leads to the form $J_L(\bm{q}) = \sum_a
\text{exp}(-i\bm{q}\cdot \bm{x}_a) \hat{O}(\bm{x}_a) \tau_a^+$, where the
operator $\hat{O}(\bm{x}_a)$ acts on space and spin variables of the
$a^\mathrm{th}$ nucleon and the isospin-raising $\tau_a^+$ operator makes the
nucleon a proton if it is initially a neutron. 

The second approximation, known as closure, begins with the observation that to
contribute significantly to the amplitude, the momentum transfer must be on the
order of an average inverse spacing between nucleons, about $100$ MeV.  The
closure approximation is to neglect the intermediate-state-dependent quantity
$E_n-E_i$ (which is generally small compared to $|\bm{q}|$) in the denominator
of Eq.~\eqref{eq:preclosure}, so that $E_n$ can be replaced by a
state-independent average value $\bar{E}$ and the contributions of intermediate
states can be summed implicitly in Eq.\ \eqref{eq:preclosure}.  This
approximation avoids the explicit calculation of excited states of the
intermediate odd-odd nucleus up to high energies, a nuclear structure
calculation that is computationally much more involved than obtaining the
initial and final states in the decay.  Because the momentum transfer in \bbt
decay (limited by the Q-value of the transition) is of the same order of
magnitude as $E_n-E_i$, the closure approximation cannot be used there.  For
that reason, some methods that focus on low-lying states or even-even nuclei can
be applied to \bbz decay but not to \bbt decay.  Approaches that do allow an
evaluation of the contributions of each intermediate state suggest that a
sensible choice of $\bar{E}$ can allow the closure approximation to reproduce
the unapproximated \bbz matrix element to within 10\%
\cite{mut94,sim11,sen13,sen14,sen14b}.  It is worth noting, however, that tests
of the closure approximation have not included states above 10's of MeV.
Since higher-energy/shorter-range dynamics could be important,
future closure tests should include them.

Assuming the closure approximation is accurate, and neglecting terms associated
with the emission of $p$-wave electrons (which are expected to be a few percent
of those associated with $s$-wave electrons) and the small electron energies
$(E_{e1}-E_{e2})/2$ in the denominator of Eq.\ \eqref{eq:preclosure}, one has
the expression
\begin{equation}
\label{eq:finalrate}
[T^{0\nu}_{1/2}]^{-1}=G_{0\nu}(Q,Z)
  \left|M_{0\nu}\right|^2  m_{\beta\beta}^2~,
\end{equation}
where $Q\equiv E_i-E_f$, $Z$ is the proton number, and $G_{0\nu}(Q,Z)$ comes
from the phase-space integral and has recently been re-evaluated with improved
precision \cite{kot12,Stoica13}.  The ``nuclear matrix element'' $M_{0\nu}$
\cite{sim99,rod06,hor10} is given by
\begin{equation}
\label{eq:monu}
M_{0\nu} = M_{0\nu}^{GT}-\frac{g_V^2}{g_A^2}M_{0\nu}^{F}+M_{0\nu}^T \,,
\end{equation}
with, in the approximations mentioned above,
\begin{widetext}
\begin{equation}
\label{eq:nme_explicit}
\begin{split}
M_{0\nu}^{GT}&  = 
\frac{2R}{\pi g_A^2} \int_0^\infty \!\!\! |\bm{q}| \, d|\bm{q}|
\bra{f} \sum_{a,b}\frac{
j_0(|\bm{q}|r_{ab}) h_{GT}(|\bm{q}|) \bm{\sigma}_a \cdot \bm{\sigma}_b}
{|\bm{q}|+\overline{E}-(E_i+E_f)/2} \tau^+_a \tau^+_b\ket{i}\,, \\
M_{0\nu}^F &=
\frac{2R}{\pi g_A^2} \int_0^\infty \!\!\!|\bm{q}| \, d|\bm{q}| \bra{f}
\sum_{a,b} \frac{
j_0(|\bm{q}|r_{ab}) h_{F}(|\bm{q}|)}{|\bm{q}|+\overline{E}-(E_i+E_f)/2} 
\tau^+_a \tau^+_b\ket{i}\,, \\
M_{0\nu}^T &=
\frac{2R}{\pi g_A^2} \int_0^\infty \!\!\! |\bm{q}| \, d|\bm{q}| \bra{f}
\sum_{a,b}\frac{
j_2(|\bm{q}|r_{ab}) h_T(|\bm{q}|)\left[3\bm{\sigma}_j \cdot \hat{\bm{r}}_{ab}
\bm{\sigma}_k \cdot \hat{\bm{r}}_{ab}- \bm{\sigma}_a \cdot \bm{\sigma}_b\right]}
{|\bm{q}|+\overline{E}-(E_i+E_f)/2} \tau^+_a \tau^+_b\ket{i}  \,.
\end{split}
\end{equation}
\end{widetext}
Here the nucleon coordinates are all operators that, like spin and isospin
operators, act on nuclear states. The nuclear radius, $R$, is inserted by
convention to make the matrix element dimensionless, with a compensating factor
in $G_{0\nu}$ in Eq.\ \eqref{eq:finalrate}.  The quantity
$r_{ab}=|\bm{x}_a-\bm{x}_b|$ is the magnitude of the inter-nucleon position
vector, and $\hat{r}_{ab} =(\bm{x}_a-\bm{x}_b)/r_{ab}$ is the corresponding unit
vector.  The objects $j_0$ and $j_2$ denote spherical Bessel functions, and the
$h$'s, called neutrino potentials, are defined in momentum space by
\begin{align}
h_{GT}(|\bm{q}|)&\equiv g_A^2(\bm{q}^2) -\frac{g_A(\bm{q}^2) g_P(\bm{q}^2) \bm{q}^2}{3m_N} \nonumber\\
&\hspace*{2cm} + \frac{g_P^2(\bm{q}^2)\bm{q}^4}{12m_N^2}+
\frac{g_M^2(\bm{q}^2)\bm{q}^2}{6m_N^2}\,,\nonumber\\
h_F(|\bm{q}|) &\equiv \frac{g_V^2(\bm{q}^2)}{g_A^2}\,,\\
h_T(|\bm{q}|)&\equiv \frac{g_A(\bm{q}^2) g_P(\bm{q}^2)
\bm{q}^2}{3m_N}-\frac{g_P^2(\bm{q}^2)\bm{q}^4}{12m_N^2}+\frac{g_M^2(\bm{q}^2)\bm{q}^2}{12m_N^2} \,,
\nonumber
\end{align}
where terms of higher order in $1/m_N$, coming from the nonrelativistic
expansion of Eq.~\eqref{eq:one-b-me}, have been neglected.

Sometimes the operators inside the matrix elements of
Eq.~\eqref{eq:nme_explicit} are multiplied by a radial function $f(r_{ab})$,
designed to take into account short-range correlations that are omitted by
Hilbert-space truncation in most many-body calculations.  Several
parameterizations of $f$ have been proposed; they are based either based on a
Jastrow ansatz~\cite{miller76}, the unitary correlator operator
method~\cite{rot05}, Brueckner-Goldstone
calculations~\cite{Brueckner55,Muther00} or nuclear matter correlation
functions~\cite{Benhar14}.  Even though the prescriptions differ from one
another, those that preserve isospin symmetry (the Jastrow ansatz does
not~\cite{engel11}) have small effects on \bbz matrix elements when the
momentum dependence of the transition operator is taken fully into
account~\cite{kor07,sim09}.

\begin{figure}[t]
\begin{center}
\includegraphics[width=0.35\textwidth,clip=]{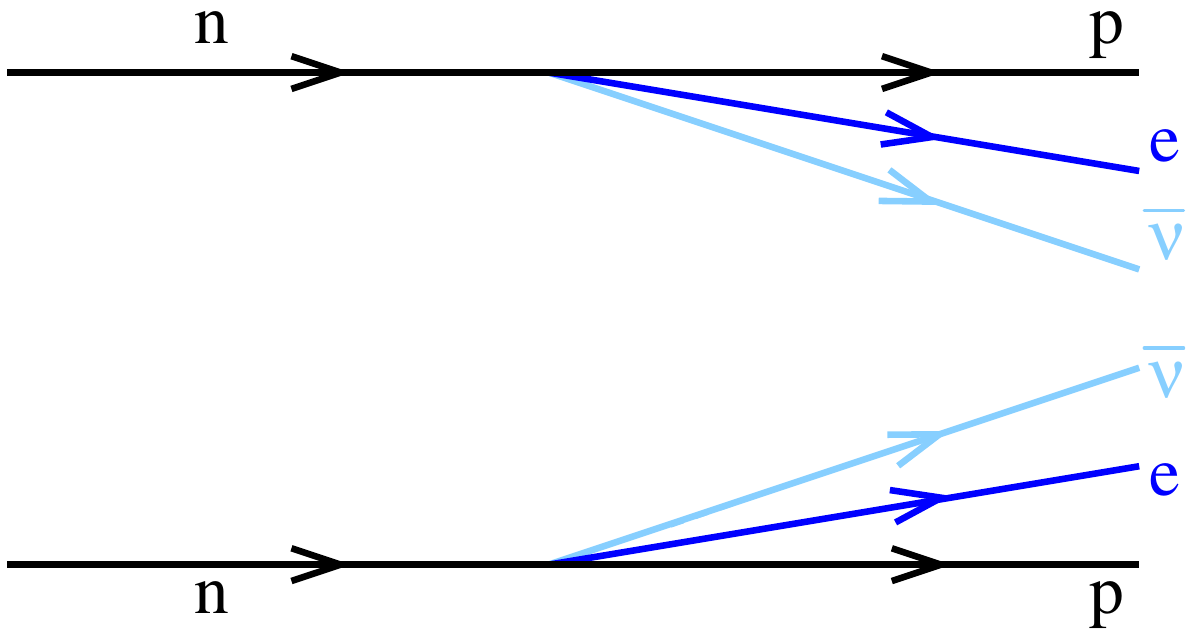}
\end{center}
\caption{Feynman diagram for \bbt decay.}
\label{f:2nu_diagram}
\end{figure}

For completeness, we write down the decay rate for \bbt decay, which is
permitted in the Standard Model and therefore does not depend on neutrino mass
or charge-conjugation properties.  This process is sketched in
Fig.~\ref{f:2nu_diagram} and the decay rate can be derived in a similar way as
for \bbz decay, with the result that 
\begin{equation}
\label{eq:tworate}
[T^{2\nu}_{1/2}]^{-1}=G_{2\nu}(Q,Z)
|M^{2\nu}_{GT}-\frac{g_V^2}{g_A^2}M^{2\nu}_F|^2 \,,
\end{equation}
where $G_{2\nu}(Q,Z)$ is the corresponding phase-space factor (also calculated
to high precision in Refs.\ \cite{kot12,Stoica13}) and
\begin{equation}
\label{eq:two-nu-me}
\begin{split}
M^{2\nu}_F&= \sum_n \frac{\langle f |\sum_a \tau^+_a |n\rangle \langle n|
\sum_b \tau^+_b |i\rangle}{E_n-(M_i+M_f)/2}\,, \\
M^{2\nu}_{GT} &=\sum_n \frac{\langle f |\sum_a \bm{\sigma}_a \tau^+_a
|n\rangle \langle n|\sum_b
\bm{\sigma}_b \tau^+_b |i\rangle}{E_n-(M_i+M_f)/2} \,.
\end{split}
\end{equation}
Isospin symmetry forces all the Fermi strength to lie in the isobar analog state
in the daughter nucleus, so that $M^{2\nu}_F$ for the transition to the daughter
ground state is negligibly small.

\subsubsection{New Physics Mechanisms}
\label{sss:bsm}

It is not just light-neutrino exchange that can contribute to \bbz decay.
Although the occurrence of \bbz decay immediately implies that neutrinos are
Majorana particles~\cite{sch82} --- once a lepton-number violating operator
appears in the Lagrangian, all possible effective operators violating the
symmetry are generated --- some other lepton-number violating mechanism could
be the dominant cause of the decay~\cite{Rodejohann11,Deppisch12}.  If that
were the case, the detection of \bbz decay would not give us information about
the absolute neutrino mass, but could be used as a low-energy test of new
physics that would complement high-energy searches at accelerators such as the
Large Hadron Collider~\cite{Gouvea13,Helo13,Pas15,Peng16}.

Several mechanisms for \bbz decay have been proposed.  Besides the exchange of
sterile neutrinos via left-handed currents~\cite{Blennow10}, the most popular
are the exchange of light or heavy neutrinos in left-right symmetric
models~\cite{moh80,moh81}, the exchange of supersymmetric
particles~\cite{moh86,ver87}, and the emission of Majorons (bosons that appear
in theories with spontaneous breaking of the global baryon-lepton number
symmetry $B-L$ ~\cite{chi81,gel81,geo81}).  Combining the contributions from all
possible mechanisms, one finds that the \bbz decay rate takes the form
\begin{equation}
\label{eq:bsmrate}
[T^{0\nu}_{1/2}]^{-1}=\sum_i G_{0\nu}^i(Q,Z) |M^{0\nu}_i|^2 \eta_i^2\,,
\end{equation}
with new-physics parameters $\eta_i$ that are distinct for every mechanism and
mode: a combination of the light neutrino masses ($m_{\beta\beta}$) for the
usual mechanism and combinations of heavy-neutrino masses, the
right-handed $W_R$ boson mass, the left- and right-handed $W_L-W_R$ boson mixing
angles, supersymmetric couplings, couplings of Majorons to neutrinos, etc. for
nonstandard mechanisms.  It is these parameters that \bbz decay experiments can
constrain, provided that the associated nuclear matrix elements $M^{0\nu}_i$ are
known. (Updated phase-space factor calculations can be found in
Ref.~\cite{Stefanik15}.) Detailed treatments of the matrix elements governing
these new-physics \bbz decay modes appear in
Refs.~\cite{ver12,doi85,hax84,tom91}.  Matrix elements calculations for the
sterile-neutrino exchange~\cite{Blennow10,fae14,bar15a,Hyvarinen15,hor16},
left-right symmetric models~\cite{Retamosa95,cau96,hor16a,Stefanik15}, and the
exchange of supersymmetric particles~\cite{Hirsch1995,hir96,hor13b,Meroni13} are
common in the literature.

\begin{figure}[t]
\begin{center}
\includegraphics[width=0.35\textwidth,clip=]{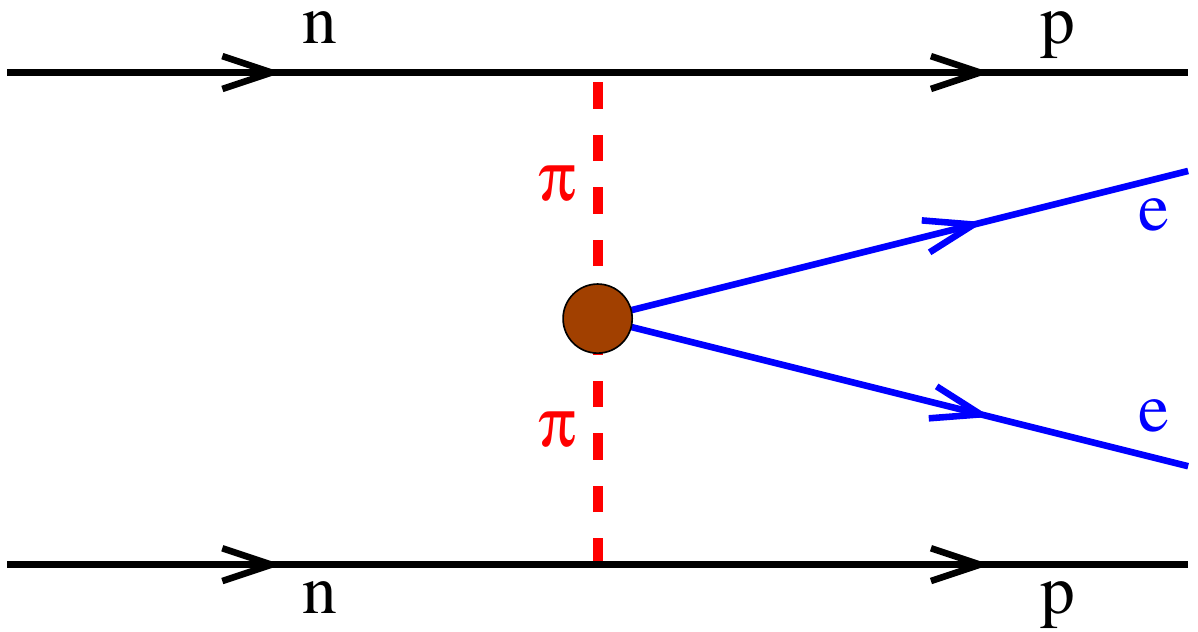}
\includegraphics[width=0.35\textwidth,clip=]{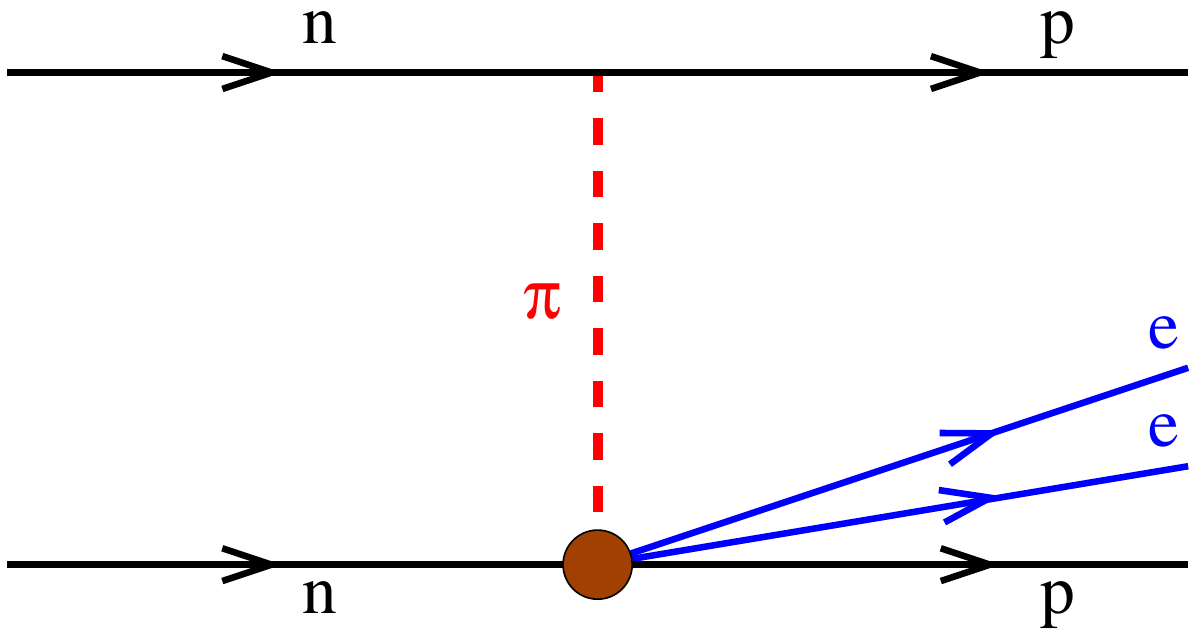}
\includegraphics[width=0.35\textwidth,clip=]{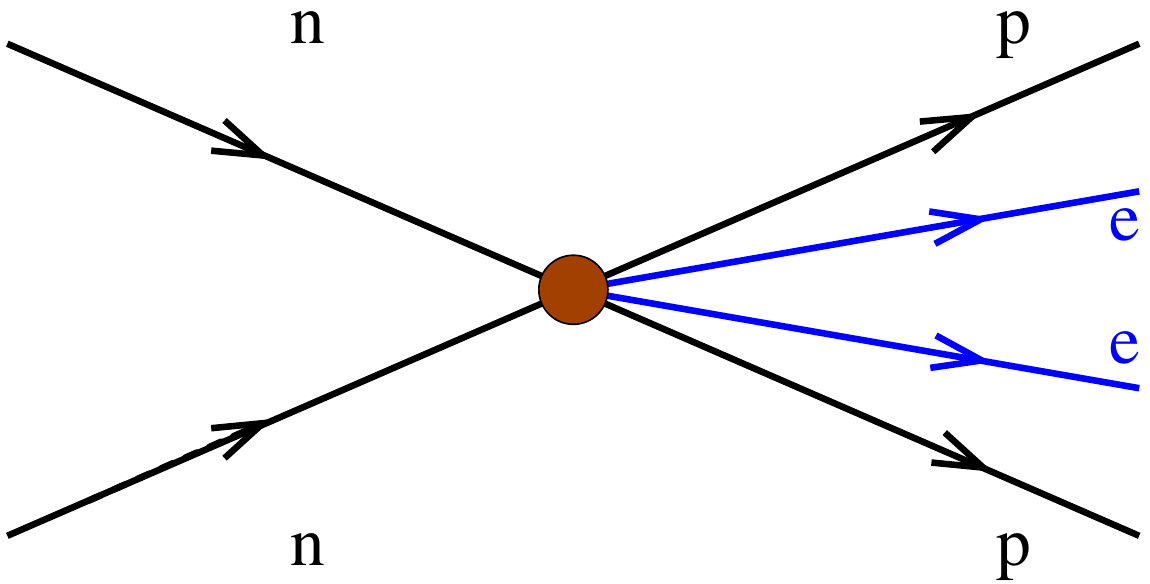}
\end{center}
\caption{Diagrams for the two-pion-exchange (top), one-pion-exchange (middle)
and contact (bottom) modes of \bbz decay caused by lepton-number violation
associated with the exchange of a heavy particle.}
\label{f:0nu_heavy_diagrams}
\end{figure}

Most of the new-physics mechanisms involve the exchange of heavy particles.
However, the direct exchange between nucleons, represented by the contact
operator in the bottom diagram in Fig.~\ref{f:0nu_heavy_diagrams} in the
heavy-particle limit, occurs less often in most models than exchange between
pions or between a pion and a nucleon, shown in the top and middle diagrams of
the figure. In $\chi$EFT each pion propagator carries a factor
$\Lambda_b^2/m_\pi^2$, where $\Lambda_b \sim 500$ MeV$ - 1$ GeV is the
chiral-symmetry breaking scale, at which the effective theory breaks down. Each
ordinary two-nucleon--pion ($NN\pi$) vertex comes with a derivative, which
results in a factor of $p/\Lambda_b$ or $m_\pi/\Lambda_b$, where $p$ is a
typical momentum. Because the contact interaction has no derivatives in most
models, pion mediation enhances the amplitude~\cite{pre03}. 
The two-pion mode at the top of the figure is thus generally the dominant one.
The one-pion graph in the middle is nominally smaller by a factor of
$\Lambda_b/m_\pi$ and the four-nucleon graph at the bottom is smaller by another
factor of the same quantity. The leading one-pion-exchange contribution to $0^+
\rightarrow 0^+$ \bbz decay is forbidden by parity symmetry, however, and so the
middle graph ends up contributing at the same order as the contact
term \cite{pre03}. 
The counting is different for nuclear forces, where the contact and one-pion
exchange interactions both appear at leading order~\cite{chiral,Machleidt11}.
The usual one-pion exchange interaction diagram contains a derivative at each
vertex; the derivatives counteract the pion propagator, placing the diagram at
the same chiral order as the four-nucleon contact diagram.  Two-pion exchange
occurs at higher order.  Computations of matrix elements in supersymmetric
models, even when they do not rely explicitly on $\chi$EFT, support the
statement that pion-exchange modes are the most important
\cite{fae97,fae98a,fae08a}.

The $\chi$EFT counting should be confirmed by explicit calculations, as
additional suppression or enhancement may occur~\cite{Savage99}.  Lattice QCD
studies that explicitly incorporate hadronic degrees of freedom are
underway~\cite{Nicholson16}, and will provide accurate input for the effective
field theory treatment of these decay modes. 

The four-nucleon contact vertex represented at the bottom of Fig.\
\ref{f:0nu_heavy_diagrams} is further affected by short-range physics.  In the
light-neutrino exchange \bbz decay mode, typical internucleon distances are of
the order of few femtometers.  The exchange of heavy particles, with mass
$m_H\gtrsim 100$~GeV~\cite{pre03}, requires nucleons to be closer to each other
and will thus be suppressed. Pions have a mass of $m_{\pi}\simeq 138$~MeV
$\approx 1.4$ fm$^{-1}$, a distance comparable to the average internucleon
spacing, and so the graphs with pions propagating between nucleons will not be
suppressed.  This behavior is apparent in potentials associated with the three
modes of heavy-particle exchange.  In momentum space, they have the form
\begin{equation}
\label{eq:heavy-part-pot}
\begin{aligned}
h_{NN-GT/NN-T}(|\bm{q}|)&\sim \bm{q}^2\,, \\
h_{\pi-GT/\pi-T}(|\bm{q}|)&\sim \frac{\bm{q}^2}{\bm{q}^2+m_\pi^2}\,, \\
h_{\pi\pi-GT/\pi\pi-T}(|\bm{q}|)&
\sim \frac{\bm{q}^2}{\left(\bm{q}^2+m_\pi^2\right)^2} \,.
\end{aligned}
\end{equation}
The first of these is clearly more strongly affected at high momentum transfer
than the two pion-exchange modes\footnote{The induced
pseudoscalar term discussed in Sec.\ \ref{sss:LNE} also
involves pion-exchange, in combination with the usual exchange of a light
neutrino.  There the pion brings no enhancement because the light-neutrino
exchange is already long range.}.
Additional powers of the momentum transfer can be present
in subleading contributions to each diagram.

The effects of short-range physics must be treated with care
when working in $\chi$EFT, from which that physics has effectively been
integrated out, or in many-body approaches with severe truncation.  The contact
coupling constant must then be renormalized from its naive value by an amount
that can be computed with the similarity renormalization group
(SRG)~\cite{bog10,Furnstahl13} or related schemes.  Until such methods are
perfected (we discuss progress in Sec.\ \ref{s:improving}) the strength of the
contact term will carry some uncertainty.  No such uncertainty plagues the
strong contact interaction in the $\chi$EFT Hamiltonian, the parameters of which
are fit directly to data.  It is fortunate that the \bb contact operators are
less important than those involving pion exchange.

Within specific models, heavy-particle exchange with $m_H$ of the order of one
TeV can compete with the light-neutrino exchange~\cite{pre06,cir04}.  Once \bbz
decay has been observed, therefore, one must cope with the question of its
cause. The easiest process to distinguish from light-neutrino exchange is decay
with the emission of a Majoron; the additional emitted particle causes the
energy spectrum of the electron pairs to be spread out rather than concentrated
at a single energy.  More challenging for experiments is exploiting the angular
correlation between the two electrons, which is different if one of them is
emitted in a $p$-wave.  The emission in $p$-waves is suppressed in
light-neutrino exchange (where $s$-waves are always more likely), but turns out
to be important in models with right-handed lepton
currents~\cite{doi85,hor16a}, where the parity-odd $q^{\rho}\gamma_{\rho}$ term
in Eq.\ (\ref{eq:nu-prop}) contributes to the amplitude instead of the term
containing the neutrino masses.  Some upcoming \bbz decay experiments should be
able to measure angular distributions, given enough events
~\cite{arno10,Next16}.

Unfortunately, other kinds of new physics can produce the same angular electron
correlations as does light neutrino exchange~\cite{ali06}.  In those cases,
however, one might be able to determine the physics responsible for the decay by
combining measurements in different isotopes (such measurements will be required
in any event to confirm detection). The matrix elements governing
heavy-particle-mediated and light-neutrino-mediated decay can depend differently
on the nuclear species in which the decay occurs
~\cite{dep06,geh07,sim10,Meroni13,hor16a}.  The task will be difficult, however,
if the dependence is similar, as Refs.~\cite{fae11,Lisi15} suggest.  Another
possibility is to compare the decay rates to the ground state and the first
excited $0^+$ state of the daughter nucleus; the ratio of these quantities can
also depend on the decay mode~\cite{sim01a}.  Unfortunately, it is difficult to
observe the decay to an excited state, which is suppressed by the small
$Q$-value associated with the transition ~\cite{kot12}. The suppression is too
great to be compensated for by differences in nuclear matrix
elements~\cite{sim01a,men09,bar15,hor16,Hyvarinen16}.  

\subsection{Importance of Nuclear Matrix Elements for Experiments}
\label{ss:importance}

The next generation of \bbz decay experiments will aim to fully cover the region
$m_{\beta\beta}\gtrsim 10$~meV in Fig.\ \ref{f:m_bb}, so that a signal will be
detected if neutrinos are Majorana particles and the mass hierarchy is inverted.
As Eq.~\eqref{eq:finalrate} shows, the \bbz decay lifetime depends on the square
of the nuclear matrix element that we need to calculate.  This sensitivity makes
matrix elements calculations important in a number of ways.

First, if an experiment is completely background free, the amount of material
needed to be sensitive to any particular neutrino mass $m_{\beta\beta}$ in a
given time is proportional to the lifetime and thus, from Eq.\
\eqref{eq:finalrate}, the inverse square of the matrix element. An uncertainty
of a factor of three in the matrix element thus corresponds to nearly an order
of magnitude uncertainty in the amount of material required, e.g., to cover the
parameter space corresponding to the inverted hierarchy.  If the experiment is
background-limited, the uncertainty is even larger \cite{GomezCadenas11}.  An
informed decision about how much material to use in an expensive experiment will
require a more accurate matrix element.  

Second, the uncertainty affects the choice of material to be used in \bbz decay
searches, a choice that is a compromise between experimental advantages and the
matrix element value.  Figure~\ref{f:NMEcompare} (top) shows nuclear matrix
elements calculated in different approaches, and because of the spread of the
results (roughly the factor of three above) we can conclude only that the
matrix element of $^{48}$Ca is smaller than those of the other \bbz decay
candidates.  And the differences in the expected rate, a product of the nuclear
matrix elements and phase-space factors, are even more similar (see Fig.
\ref{f:NMEcompare} bottom, and Eq.~\eqref{eq:finalrate}) \cite{hamish13}.
Better calculations would make it easier to select an optimal isotope. 

Finally, and perhaps most obviously, we need matrix elements to obtain
information about the absolute neutrino masses once a \bbz decay lifetime is
known.  Reducing the uncertainty in the matrix element calculations will be
crucial if we wish to fully exploit an eventual measurement of the decay
half-life.  Even the interpretation of limits is hindered by matrix-element
uncertainty.  The blue band in Figure \ref{f:m_bb} represents the upper limit
of $m_{\beta\beta}<61-165$~meV from the KamLAND-Zen experiment
\cite{KamLAND-Zen16}.  The uncertainty, again a factor of about three, is due
almost entirely to the matrix element.  And the real theoretical uncertainty,
at this point, must be taken to be larger; the ``$g_A$ problem," which we
discuss in Sec.~\ref{s:ga}, has been ignored in this analysis.  We really
need better calculations.  Fortunately, we are now finally in a position to
undertake them.

\section{Nuclear Matrix Elements At Present}
\label{s:present}

\begin{figure}[t]
\begin{center}
\includegraphics[width=0.48\textwidth,clip=]{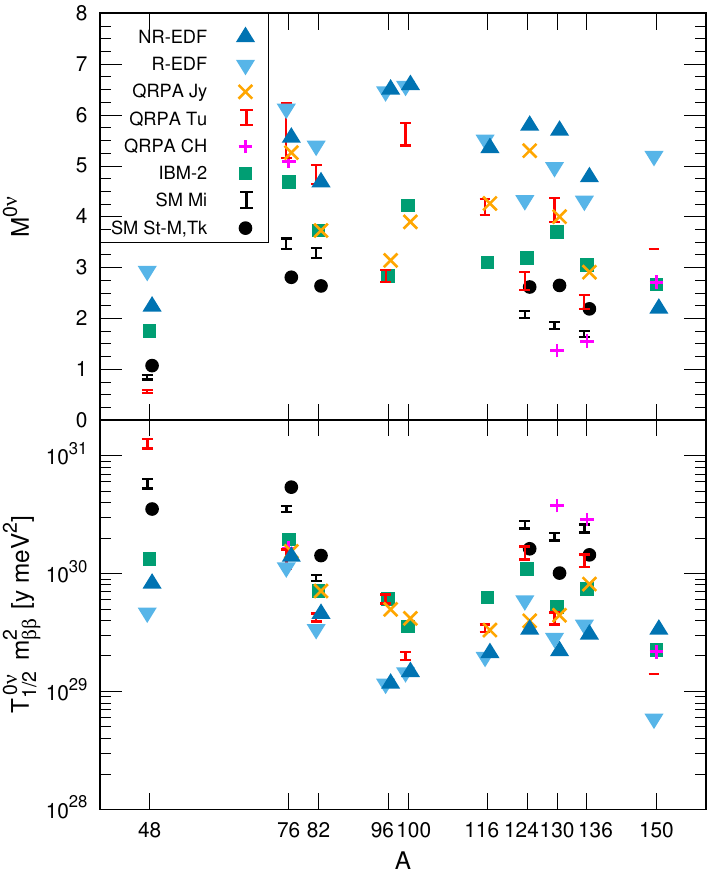}
\end{center}
\caption{Top panel: Nuclear matrix elements ($M^{0\nu}$) for \bbz decay
candidates as a function of mass number $A$.  All the plotted results are
obtained with the assumption that the axial coupling constant $g_A$ is
unquenched and are from different nuclear models: the shell model (SM) from the
Strasbourg-Madrid (black circles)~\cite{men08}, Tokyo (black circle in
\elemA{48}Ca)~\cite{Iwata16}, and Michigan (black bars)~\cite{hor16} groups;
the interacting boson model (IBM-2, green squares)~\cite{bar15}; different
versions of the quasiparticle random-phase approximation (QRPA) from the
T\"ubingen (red bars)~\cite{sim13,fang15}, Jyv\"askyl\"a (orange times
signs)~\cite{Hyvarinen15}, and Chapel Hill (magenta crosses)~\cite{Mustonen13}
groups; and energy density functional theory (EDF), relativistic (downside cyan
triangles)~\cite{yao15,yao16} and non-relativistic (blue
triangles)~\cite{vaq13}. QRPA error bars result from the use of two realistic
nuclear interactions, while shell model error bars result from the use of
several different treatments of short range correlations.  Bottom panel:
Associated \bbz decay half-lives, scaled by the square of the unknown parameter
$m_{\beta\beta}$.}
\label{f:NMEcompare}
\end{figure}

As we have noted, calculated matrix elements at present carry large
uncertainties.  Matrix elements obtained with different nuclear-structure
approaches differ by factors of two or three.  Figure~\ref{f:NMEcompare}
compares matrix elements produced by the shell
model~\cite{men08,Iwata16,hor16}, different variants of the quasiparticle
random phase approximation (QRPA)~\cite{sim13,fang15,Hyvarinen15,Mustonen13},
the interacting boson model (IBM)~\cite{bar15}, and energy density functional
(EDF) theory~\cite{yao15,yao16,vaq13}.  The strengths and weaknesses of each
calculation are discussed in detail later in this Section.

Some of these methods can be used to compute single-$\beta$ and \bbt decay
lifetimes.  It is disconcerting to find that predicted lifetimes for these
processes are almost always shorter than measured lifetimes, i.e.\ computed
single Gamow-Teller and \bbt matrix elements are too
large~\cite{cau05,bar13,Pirinen15}.  The problems are usually ``cured'' by
reducing the strength of the spin-isospin Gamow-Teller operator ${\bm
\sigma}{\bm \tau}$, which is equivalent to using an effective value of the
axial coupling constant that multiplies this operator in place of its ``bare''
value of $g_A\simeq1.27$.  This phenomenological modification is sometimes
referred to as the ``quenching'' or ``renormalization'' of $g_A$.  In Sec.\
\ref{s:ga} we review possible sources of the renormalization, none of which has
yet been shown to fully explain the effect, and their consequences for \bbz
matrix elements. 

\subsection{Shell Model}
\label{ss:sm}

The nuclear shell model is a well-established many-body method, routinely used
to describe the properties of medium-mass and heavy
nuclei~\cite{cau05,bro01,ots01}, including candidates for \bb-decay
experiments.  The model, also called the ``configuration interaction method''
(particularly in quantum chemistry~\cite{carsky98,Sherrill99}), is based on the
idea that the nucleons near the Fermi level are the most important for
low-energy nuclear properties, and that all the correlations between these
nucleons are relevant.  Thus, instead of solving the Schr\"odinger equation for
the full nuclear interaction in the complete many-body Hilbert space, one
restricts the dynamics to a limited configuration space (sometimes called the
valence space) containing only a subset of the system's nucleons.  In the
configuration space one uses an effective nuclear interaction $H_{\text{eff}}$,
defined (ideally) so that the observables of the full-space calculation are
reproduced, e.g.\
\begin{equation}
\label{eq:sm_diag}
H\ket{\Phi_i}=E_i\ket{\Phi_i} 
\rightarrow H_{\text{eff}}\ket{\bar{\Phi}_i}=E_i\ket{\bar{\Phi}_i}
\,.
\end{equation}
The states $\ket{\Phi_i}$ and $\ket{\bar{\Phi}_i}$ are defined in the full
space and the configuration space, respectively, and have associated energy
$E_i$.  

The configuration space usually comprises only a relatively small number of
``active'' nucleons outside a core of nucleons that are frozen in the
lowest-energy orbitals and not included in the calculation.  The active
nucleons can occupy only a limited set of single-particle levels around the
Fermi surface. Many-body states are linear combinations of orthogonal Slater
determinants $\ket{\psi_i}$ (usually from a harmonic-oscillator basis) for
nucleons in those single-particle states,
\begin{equation}
\label{eq:sm_state}
\ket{\bar{\Phi}_i}=\sum_j c_{ij} \ket{\psi_j}
\,,
\end{equation}
with the $c_{ij}$ determined by exact diagonalization of $H_{\text{eff}}$.

The shell model describes ground-state nuclear properties such as masses,
separation energies, and charge radii quite well.  It also does a good job with
low-lying excitation spectra and with
electric moments and transitions~\cite{cau05,bro01,ots01}
if appropriate effective charges are used~\cite{duf96}.
The wide variety of successes over a broad range of
isotopes reflects the shell model's ability to capture both the excitation of a
single particle from an orbital below the Fermi surface to one above, in the
spirit of the original naive shell model~\cite{Mayer49,Haxel49}, and collective
correlations that come from the coherent motion of many nucleons in the
configuration space.  The exact diagonalization of $H_{\text{eff}}$ means that
the shell model states $\ket{\bar{\Phi_i}}$ contain all correlations (isovector
and isoscalar pairing, quadrupole collectivity, etc.) that can be induced by
$H_{\text{eff}}$.

This careful treatment of correlations, on the other hand, restricts the range
of shell model to relatively small configuration spaces, at present those for
which the Hilbert-space dimension is less than about
$(10^{11}$)~\cite{Shimizu16,Nowacki16}.  For this reason most shell model
calculations of \bbz decay have been performed in a single harmonic-oscillator
shell, consisting of four or five single-particle orbitals, not counting
degeneracies from rotational invariance, for both protons and neutrons~
\cite{caurier08,men08,men09,hor10,sen13,sen14,sen14b,Neacsu15,sen16,hor16}.
Enlarging the configuration space increases the number of active nucleons,
leading to a Hilbert-space dimension that increases combinatiorally with the
number of active-nucleon configurations in the valence space and quickly making
dimensions intractable.  (Very recently, however, the authors of Ref.\
\cite{Iwata16} performed the first calculation in two oscillator shells; we
discuss it in detail in Sec.~\ref{ss:comp}.)

Pairing correlations, which are central in nuclear structure and have large
effects on \bbz decay, may not be fully captured by $H_{\text{eff}}$ within a
single oscillator shell~\cite{vog12}.  In addition, because the one-body
spin-orbit interaction significantly lowers the energy of orbitals with spin
parallel to the orbital angular momentum in heavy nuclei, spin-orbit partners
are split by several MeV or more, and the shell-model configuration space
contains only one member of some spin-orbit pair (except in \elemA{48}Ca).  The
omission may have important consequences because the spin-isospin part of the
\bbz decay operator in Eq.\ \eqref{eq:nme_explicit}, $\bm{\sigma} \cdot
\bm{\sigma} \tau^+ \tau^+$, strongly connects spin-orbit partners. The
omission also affects single-$\beta$ decay.

Despite these limitations, the shell model reproduces experimental
single-$\beta$ decay rates well if one quenches the strength of the spin-isospin
operator $\bm{\sigma} \tau^\pm$ from its bare value
$g_A\simeq1.27$~\cite{Chou93,Wildenthal83,MartinezPinedo96}.  (Of course, that
is a big ``if,'' which we address in detail in Sec.~\ref{s:ga}.) The quenching
needed to agree with data is about 20\%-30\%, and this small
range is enough to describe
Gamow-Teller transitions in nuclei with valence nucleons in the $p$ shell (a
configuration space comprising the 0$p_{3/2}$ and 0$p_{1/2}$ single-particle
orbitals, representing the valence shell for $A\sim 10$), the $sd$ shell (the
0$d_{5/2}$, 1$s_{1/2}$, and $0d_{3/2}$ orbitals, $A\sim 30$), the $pf$ shell
(the $0f_{7/2}$, $1p_{3/2}$, $1p_{1/2}$ and $0f_{5/2}$ orbitals, $A\sim 50$),
and the space spanned by the $1p_{3/2}$, $1p_{1/2}$,$0f_{5/2}$,
and $0g_{9/2}$ orbitals, for use near $A= 70$.  With roughly the same quenching
the model also reproduces the \bbt decay rate of the only emitter in this mass
region, \elemA{48}Ca.  One shell model success, in fact, was the accurate
calculation of the \bbt decay half-life of \elemA{48}Ca~\cite{cau90,pov95}
before it was measured\footnote{The NEMO-3 collaboration has reported a longer
half-live than was measured previously, resulting in a deviation of about two
$\sigma$~ \cite{arno16} from the shell-model prediction.  If this new value is
confirmed, the shell model would overestimate the \elemA{48}Ca \bbt matrix
element.} \cite{Balysh96}.  The shell model also reproduces other \bbt decay
half-lives provided it uses a quenching factor appropriate for the configuration
space and interaction~\cite{cau12,hor13,hor16,sen14,sen14b,Neacsu15}.
It requires a similar quenching to reproduce magnetic moments
and transitions~\cite{towner87,Brown87,NeumannCosel98}, which, like $\beta$
decay, involve the spin operator.

The reliability of the shell model depends on the effective interaction
$H_{\text{eff}}$ as well as the configuration space.  The starting point for
$H_{\text{eff}}$ is usually the bare nucleon-nucleon interaction, which is fit
to two-nucleon scattering data (and discussed in a bit more detail in Sec\
\ref{ss:abinit}).  Then, many-body perturbation theory is used to obtain a
configuration-space-only Hamiltonian that takes the effects of the inert core
and neglected single-particle orbitals into account~\cite{hjorth-jensen95}.
Finally, phenomenological corrections, mainly in the monopole part of the
interaction (the part responsible for changes in single-particle-like
excitations with increasing $A$) are usually made to improve the agreement with
experimental data~\cite{cau05,ots01,bro01}. This last step in the construction
of $H_{\text{eff}}$ severely restricts our ability to quantify the theoretical
uncertainty associated with \bbz matrix-elements.  The usual fallback ---
comparing the results of calculations that use equally reasonable effective
Hamiltonians --- leads to variations off $10\%-20\%$
~\cite{men09,hor10,Neacsu15}.

As Fig.\ \ref{f:NMEcompare} shows, shell-model \bbz matrix elements are usually
smaller than all others.  We discuss possible reasons in Sec.\ \ref{ss:comp}.

\subsection{The QRPA and Some of Its Variants}
\label{ss:qrpa}

The other major thrust over the past 30 years has been the development of the
charge-changing quasiparticle random phase approximation (QRPA).  This approach
is a generalization of the ordinary random phase approximation (RPA), which has
a long history both in nuclear physics and elsewhere \cite{bohm-pines1,row68}.
The method can be derived in a number of ways.  One is through an approach we
take up in more detail later, the generator coordinate method (GCM).  Suppose
one has solved the Hartree-Fock equations, yielding a set of $\mathcal{A}$
occupied orbitals $\ket{\varphi_i}$ that when put together form the best
possible Slater determinant $\ket{\psi_0}$, and has then constructed another
set of ``nearby'' non-orthogonal Slater determinants $\ket{\psi(z)}$, each with
$\mathcal{A}$ occupied orbitals of the form $\chi_i(z)$:
\begin{equation}
\label{eq:rpa-gs}
\ket{\chi_i(z)}=\ket{\varphi_i} + \sum_{j=\mathcal{A}+1}^{\mathcal{N}_o} z^*_{ij} \ket{\varphi_j}
\,,
\end{equation}
where $\mathcal{N}_o$ is the number of orbitals included in the calculation.
The RPA ground state $\ket{\text{RPA}}$ is the (continuous) superposition of
the $\ket{\psi(z)}$ that minimizes the energy in the limit that the $z_{ij}$
are small.  In that limit, the minimization process becomes equivalent to the
solution of the Schr\"odinger equation for a multi-dimensional harmonic
oscillator, in which the $z_{ij}$ play the role of creation operators
\cite{jancovici64}.  Along with the ground state one finds a set of
``one-phonon'' normal modes, one-particle one-hole excitations that are the
only states that can be connected to the ground state by a one-body operator.
The transition strengths to those states automatically satisfy energy-weighted
sum rules.

For application to single-$\beta$ and $\beta\beta$ decay, the RPA must be modified in
two ways.  First, it must be ``charge-changing,'' that is, if $\ket{\varphi_i}$
in Eq.\ \eqref{eq:rpa-gs} is a neutron orbital then at least some of the
$\ket{\varphi_j}$ must be proton orbitals.  That condition guarantees that the
one-phonon excited states have components with one more proton and one less
neutron than the ground state.  Second, it must include the physics of pairing,
which one can add by replacing the Hartree-Fock state with the BCS or
Hartree-Fock-Bogoliubov (HFB) quasi particle vacuum, and the nearby Slater
determinants with nearby quasiparticle vacua \cite{ring80}.  The result is the
charge-changing (or proton-neutron) QRPA.

In single-$\beta$ decay QRPA excitations of the initial state produce final
states.  In \bb decay, they produce intermediate states in Eqs.\
\eqref{eq:preclosure} and \eqref{eq:two-nu-me} and one must carry out two
separate QRPA calculations, one based on the initial nucleus as the BCS vacuum,
and a second on the final nucleus. 
(The closure approximation is not required.)  In this way one obtains two
different sets of states in the intermediate nucleus, one of which must be
expressed in terms of the other. The overlaps that are needed to do that
require approximations beyond those in the QRPA.  We will return to this
subject in Sec.~\ref{sss:higher}.

The main advantage of the QRPA in comparison to the shell model is the number
of single-particle orbits that can be included in the calculation.  As we
mentioned in Sec.~\ref{ss:sm}, shell-model configuration spaces are usually
based on a few single-particle orbitals, most often in one major oscillator
shell.  In most QRPA calculations, all the orbitals within one or two
oscillator shells of the Fermi surface are treated explicitly, with those
further below assumed to be fully occupied and those further above completely
empty.  One QRPA calculation \cite{Mustonen13}, albeit a particularly demanding
one, actually included all levels between 0 and 60 MeV, with no inert core.

\begin{figure}[t]
\begin{center}
\includegraphics[width=0.5\textwidth,clip=]{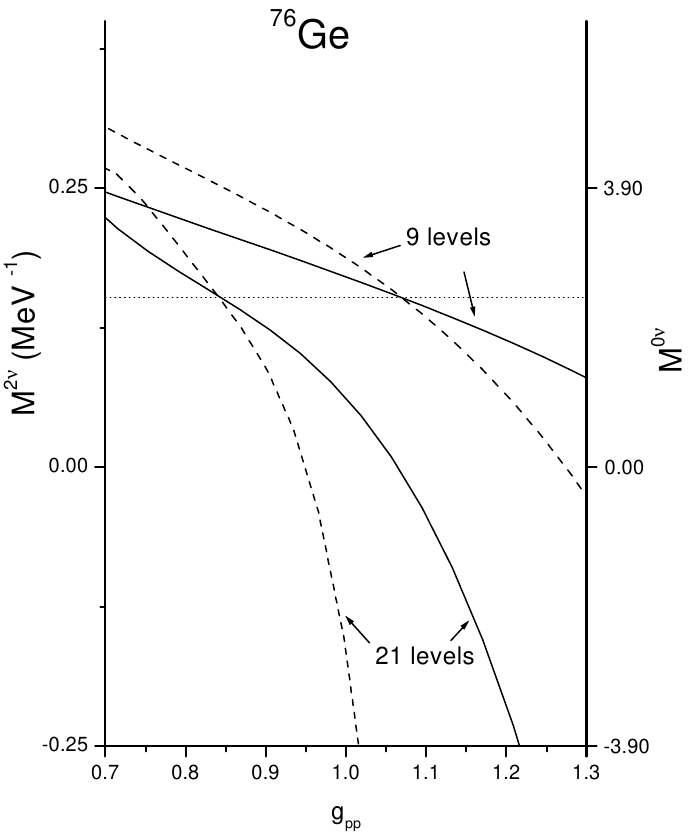}
\end{center}
\caption{Matrix elements $M^{2\nu}_{GT}$, (left scale, dashed lines) and
$M^{0\nu}$, (right scale, solid lines) for the \bbt and \bbz decay of
\elemA{76}Ge, as a function of the strength of the proton-neutron interaction
$g_{pp}$ for QRPA calculations in configuration spaces consisting of 9 and 21
single-particle orbitals. The dotted horizontal line is at the measured value
of the $M^{2\nu}_{GT}$. Figure taken from Ref.~\cite{rodin03}.}
\label{f:gpp}
\end{figure}

Being able to handle many orbitals guarantees, for example, that the $\beta^-$
and $\beta^+$ strengths obey the ``Ikeda sum rule'' [see Eq.~\eqref{eq:ikeda}
in Sec.~\ref{s:ga}].  But the price for such large single-particle spaces in
the QRPA is a restricted set of correlations.  As a result, one cannot expect a
really accurate calculation without compensating for the method's limitations
by modifying the effective nucleon-nucleon interaction used to generate the
nuclear states.  As in the shell model, the original interaction is typically a
realistic nucleon-nucleon potential, one adapted to the QRPA configuration
space through many-body perturbation theory ~\cite{hjorth-jensen95}.  The
interaction is modified more severely than in the shell model, however, usually
independently in the particle-hole and pairing channels. Practitioners often
renormalize the strengths of the interaction in the proton-neutron
particle-hole $J^\pi=1^+$ and $2^-$ channels by 10\% or so to properly
reproduce the energies of the Gamow-Teller and spin-dipole giant resonances,
altering both the \bbt and \bbz matrix elements somewhat. It is the
proton-neutron pairing interaction that is most important, however.  Its
strength, often called $g_{pp}$, is also changed by only about 10\%, but that
modest adjustment usually has a large effect on the \bb matrix elements.
Figure\ \ref{f:gpp} shows a typical plot of the \bbt and \bbz matrix
elements versus $g_{pp}$; the strong suppression near the nominally correct
value of that parameter, $g_{pp}=1$, is clear not only in the \bbt matrix
element, where it is dramatic, but also in the \bbz matrix element.  The usual
procedure in the QRPA is to fix the value of $g_{pp}$ so that the measured rate
of \bbt decay is correctly reproduced~\cite{rodin03}.  Then the same value is
used to predict the rate of \bbz decay.  As Fig.~\ref{f:gpp} shows, this
procedure almost eliminates the dependence of the \bbz matrix elements on the
number of single-particle levels in the configuration space.

The reason for the sensitivity to the proton-neutron pairing strength has been
the subject of many investigations \cite{vog86}.  Some of it is easy to
understand:  in the isoscalar channel, for example, proton-neutron pairing
mixes proton-neutron spin-one pairs, which are like in-medium deuterons, into
the usual condensate of like-particle spin-zero pairs.  The Gamow-Teller
single-$\beta$ operator, which plays a role in both the \bbt matrix element in
Eq.\ \eqref{eq:two-nu-me} and the \bbz matrix element in Eq.\
\eqref{eq:nme_explicit}, connects the two kinds of pairs, altering the matrix
elements.  (For a simple discussion of why they shrink rather than grow as a
result, see Ref.\ \cite{eng88}.) But some of the sensitivity can be an artifact
of the QRPA.  When the parameter $g_{pp}$ is made large enough the method
breaks down because of an impending phase transition from the like-particle
pairing condensate to a ``deuteron-like'' condensate.  In the region before
this (non-existent) transition, the sensitivity to $g_{pp}$ is unrealistically
high.

Several groups have tried to modify the QRPA to cure this behaviour.  The usual
approach is the ``renormalized QRPA'' (RQRPA) \cite{toi95,PhysRevC.66.051303}.
The idea is that the breakdown is due to the violation of the Pauli exclusion
principle by the small-amplitude approximation at the heart of the QRPA, and
that the violation can be remedied by modifying the BCS vacuum on which the
QRPA phonons are based.  In the ``fully renormalized'' version of Ref.\
\cite{PhysRevC.66.051303} some terms are kept beyond lowest order in the
small-amplitude expansion as well. These procedures succeed in avoiding the
infinities that mark the breakdown of the ordinary QRPA, but 
may overcompensate by eliminating all traces of the approximate phase
transitions that really take place in solvable models, where the methods have
been benchmarked \cite{eng97,hirsch97}.  In realistic calculations, these
methods alter the QRPA matrix elements by noticeable but not large
amounts~\cite{rod06}.  More comprehensive and systematic modifications to the
QRPA have been proposed; we discuss some of them in Sec.~\ref{sss:higher}.

Finally, it is possible to calculate \bbz matrix elements, in the closure
approximation, in the like-particle QRPA rather than the charge-changing
version.  With closure, any set of intermediate states that form a partition of
the identity operator will do.  To apply the like-particle QRPA, one chooses
these intermediate states to lie in the nucleus with two-more protons (or
two-less neutrons) than the initial nucleus.  Though such excitations do not
conserve particle number, or even charge in the case of two-proton addition,
they can be represented as two-like-quasiparticle excitations of the initial
state.  So far, however, only one author \cite{Terasaki15,Terasaki16} has
applied the QRPA in this way. 

Turning at last to QRPA-produced \bb matrix elements: Their most notable
feature is that they are almost uniformly larger than shell model matrix
elements (see Fig.\ \ref{f:NMEcompare}), whether or not a renormalized version
of QRPA is used.  We discuss the reasons in Sec.~\ref{ss:comp}.

\subsection{Energy-Density Functional Theory and the Generator-Coordinate
Method}
\label{ss:edf}

The term Energy-Density Functional (EDF) theory refers at its most basic level
to the process of minimizing an energy functional
$\mathcal{E}[\rho,\bm{s},\bm{j},\ldots]$ with respect to local and semi-local
densities such as the number density $\rho$, the spin density $\bm{s}$, the
current density $\bm{j}$, etc. \cite{Nogueira-dft}.  The functional
$\mathcal{E}[\rho,\bm{s},\bm{j},\ldots]$ is the minimum possible value for the
expectation value of the Hamiltonian when the densities are constrained to have
particular values.  Once the functional is obtained, minimizing it with respect
to its arguments provides the exact ground-state energy and densities, as
Hohenberg and Kohn originally showed \cite{Hohenberg64}.  Moreover, the
minimization can be formulated so that it looks like mean-field theory with
one-body potentials and orbitals, via the Kohn-Sham procedure~\cite{Kohn65}.
The independent particle or quasiparticle wave functions that result have no
meaning beyond supplying the correct energy and densities.  In nuclear physics,
approximations to the functional $\mathcal{E}$ usually derive from the
Hartree-Fock or HFB energy asociated with a ``density-dependent two-body
interaction'' of the Skyrme \cite{Vautherin72}, Gogny \cite{decharge80} or
relativistic Walecka \cite{Serot86} type, sometimes with additional
modifications.  The parameters of the interaction or functional are then fit to
ground-state properties --- masses, radii, etc.\ --- in a variety of nuclei and
used without alteration all over the nuclear chart.  The method can be extended
to EDF-based RPA or QRPA, which are adiabatic versions of time-dependent
Kohn-Sham theory.  Indeed, that approach was used profitably in the \bb decay
calculations of Ref.\ \cite{Mustonen13}.

The problem at present is that the Skyrme, Gogny, or relativistic functionals
$\mathcal{E}$, though they do a good job with collective properties such as
binding energies, radii, and $E2$ transitions, are not close enough to the
exact functional to work for all quantities in all nuclei\footnote{Nor should
they be, even in principle.  The Hohenberg-Kohn theorem does not imply that the
functional has the same form in all nuclei.}.  Something must be added to
accurately describe nuclear properties, and the easiest way to add physics is
to explicitly modify the Hartree-Fock or HFB wave functions so that they
contain explicit correlations~\cite{bender03}.  Adding correlations means
mixing the wave-function-based and EDF-based approaches, however, opening the
door to all kinds of inconsistencies~\cite{Bender09a}.  The only way to remove
these is to abandon the focus on functionals and return to a Hamiltonian.  It
is still an open question whether density-dependent Hamiltonians of the kind
used in phenomenological calculations of isotopes along the nuclear chart can
ever be made fully consistent.

The use of functionals together with explicit correlations is valuable
nonetheless, particularly within the GCM.  Although the method can be used in
conjunction with the small-amplitude approximation to derive the RPA or QRPA
(see Sec.\ \ref{ss:qrpa}) the need for it is greatest when deviations from a
single mean field have large amplitudes and the QRPA breaks down.
Large-amplitude fluctuations are often associated with collective
correlations.  Shapes with significantly different degrees of deformation often
must coexist, making the QRPA inapplicable.  The same is sometimes true of
pairing gaps.  And both deformation and pairing involve the simultaneous motion
of many nucleons in many orbitals, some of which may be outside shell-model
spaces.  The large single-particle spaces in EDF-based work and the mixing of
mean fields in the GCM make it possible for these collective correlations to be
fully captured.  

The GCM with EDFs has several steps~\cite{bender03}:
\begin{enumerate}
\item Choose one or more ``collective operators'' $\hat{O}_i$.  The most
commonly chosen is the axial quadrupole operator $r^2 Y_{20}$.  Operators that
reflect non-axial deformation and pairing gaps are other common choices. 
\item Carry out repeated mean-field calculations with the expectation values of
the $\hat{O}_i$ constrained to many different values (approximating a continuum
of values).  In other words, find the Slater determinants or HFB vacua that
minimize $\mathcal{E}$ under the constraint that the $\braket{\hat{O}_i}$ take
on particular sets of values.  
\item Project the resulting mean-field states onto states with well defined
angular momentum, particle number, and whatever other conserved quantity the
mean-field approximation does not respect.
\item Use the resulting non-orthogonal states as a basis in which to
``diagonalize the Hamiltonian.'' 
\end{enumerate}
The last phrase is in quotation marks because although the EDF is associated
with one or more density-dependent Hamiltonians (often separate ones for the
particle-hole and pairing channels), the theory says nothing about how to
evaluate matrix elements of the density or of density-dependent operators
between different Slater determinants.  The most sensible and commonly used
procedure is to replace the density by the transition density, but the approach
has been shown to produce ill-defined singularities when examined closely
\cite{Anguiano:2001in,Dobaczewski:2007ch,Bender09a} and no alternative has
gained currency. Some functionals have fewer problems than others, however, and
the Gogny and relativistic functionals, which are based strictly on
density-dependent Hamiltonians, both allow stable results, provided one does
not push numerical accuracy too far.  

The applications to \bbz decay have been relatively few so far.  Reference
\cite{rod10} contains the first, with the Gogny functional and axial quadrupole
moment $\braket{r^2 Y_{20}}$ as the generator coordinate.  In Ref.\
\cite{vaq13}, fluctuations in particle number were added as coordinates,
leading in a modest enhancement of the matrix elements from the richer pairing
correlations. References\ \cite{Song14,yao15} used the axial quadrupole moment
as a coordinate in conjunction with a relativistic functional, and Ref.\
\cite{yao16} added octupole deformation in the decay of the rare-earth nucleus
$^{150}$Nd.  The most obvious feature of all these results is that the \bbz
matrix elements are larger than those of the shell model, and usually
larger than those of the QRPA as well.  The reason seems to be missing
correlations in the GCM ansatz.  At first these were thought to be
non-collective \cite{men14}, but recent work \cite{men16} strongly suggests
that most of them come from the isoscalar pairing that we encountered when
discussing the QRPA in Sec.~\ref{ss:qrpa}. Quite recently, Ref.\ \cite{hin14}
included the isoscalar pairing amplitude as a generator coordinate together
with a Hamiltonian in one or two oscillator shells rather than an EDF.  That
coordinate allows the GCM to reproduce shell-model results quite well.  The
Hamiltonian-based GCM can easily be extended to larger spaces, and we discuss
plans to do so in Sec.\ \ref{ss:comp}. 

\subsection{The Interacting Boson Model}
\label{ss:ibm}

The strength of the shell model is the inclusion of all correlations around the
Fermi surface and that of the GCM is a careful treatment of collective motion.
The IBM aims to share both these strengths and describes excitation spectra and
electromagnetic transitions among collective states up to heavy
nuclei~\cite{Iachello87}.  The cost, perhaps, is more abstract degrees of
freedom than nucleons and a more phenomenological approach to nuclear
structure.  

The IBM leverages the algebra of boson creation and annihilation operators to
provide simple Hamiltonians that generate complex and realistic collective
spectra.  In its original form (IBM-1)~\cite{arima81}, the degrees of freedom
are $\mathcal{N}$ bosons (where $\mathcal{N}$ is usually half the number of
nucleons in the shell-model configuration space), each of which can be in six
positive parity states: an angular-momentum-zero state (in which case the boson
is labeled ``$s$'') and five angular-momentum-two states (in which case it is
labeled ``$d_\mu$,'' with $\mu$ the magnetic projection).  The Hamiltonian is
usually a combination of one- and two-boson scalar operators, of the form
$s^\dag s, (d^\dag \tilde{d})^0, ( [s^\dag \tilde{d}]^2 [d^\dag
\tilde{d}]^2)^0$, etc., where $\tilde{d}_\mu \equiv (-1)^\mu d_{-\mu}$ and the
superscripts denote the angular momentum to which the operators inside the
corresponding parentheses or square brackets are coupled.  In the
IBM-2~\cite{Iachello87}, which is used to study \bb decay, there are separate
$s$ and $d$ boson states for neutrons and protons.

The IBM is appealing because it has clear connections to both the shell model
and the collective model of Bohr and Mottelson~\cite{BM98}.  On the one hand,
the bosons represent nucleon pairs and on the other quadrupole phonons.
Because the first correspondence is hard to make precise, Hamiltonians and
effective operators are usually determined by fits to data rather than by
mappings from the shell model~\cite{Otsuka78}.  There are no data on \bbz
matrix elements, however, and the associated operators therefore must be
derived from the shell model, at least approximately.  The mapping for doing so
is described in Ref.\ \cite{bar09}.  It is approximate because it involves only
two- and four-nucleon states (which are mapped to one- and two-boson states)
and a schematic surface-delta shell-model interaction that is inconsistent with
the phenomenological boson interaction. 

The \bbz matrix elements produced by the IBM generally lie between those
of the EDF/GCM and those of the shell model, and for several isotopes are close
to the QRPA values. This result is perhaps surprising because the IBM is
supposed to be a collective approximation to the shell-model in one harmonic
oscillator shell.  The behavior may be connected to the truncation of the set of
all possible bosons to a single set of like-particle $s$ and $d$ objects, but
that is just a guess; the subject needs to be investigated more thoroughly. 

\subsection{Other Approaches}
\label{ss:other}

We will do little more than mention a couple of other methods that have been
used to calculate \bb matrix elements.  Neither gives results that are
startlingly different from those of other more modern and sophisticated
approaches.

References\ \cite{Rath08,Rath09,Rath10,Rath13} describe calculations with
projected HFB states in a valence space of between one and two shells, with a
schematic paring plus quadrupole Hamiltonian.  The method itself is a simpler
version of the GCM, which diagonalizes Hamiltonians in a space built from many
projected HFB states.

References \cite{Hirsch:1994kw,hir95} describe calculations in the pseudo-SU(3)
model of the \bbz matrix elements for $^{150}$Nd, for the very heavy nucleus
$^{238}$U, and for a few other deformed isotopes.  The model takes advantage of
a dynamical symmetry in the deformed Nilsson basis to construct states that
should be similar to those obtained from number- and angular-momentum-projected
HFB calculations of the kind discussed in the previous paragraph and in Sec.
\ref{ss:edf}.

\subsection{Tests and Comparisons}
\label{ss:comp}

Although we have mentioned the strengths and weaknesses of the many-body
approaches most commonly used to calculate \bbz matrix elements, it is worth
trying to compare them more carefully.  How important are all the correlations
included in the shell model but neglected in the QRPA, GCM, and IBM?  What about
the large configuration spaces in which the GCM and QRPA work, but which the
shell model and IBM cannot handle?  A few publications address these matters and
we discuss them here.

\begin{figure}[t]
\begin{center}
\includegraphics[width=0.5\textwidth,clip=]{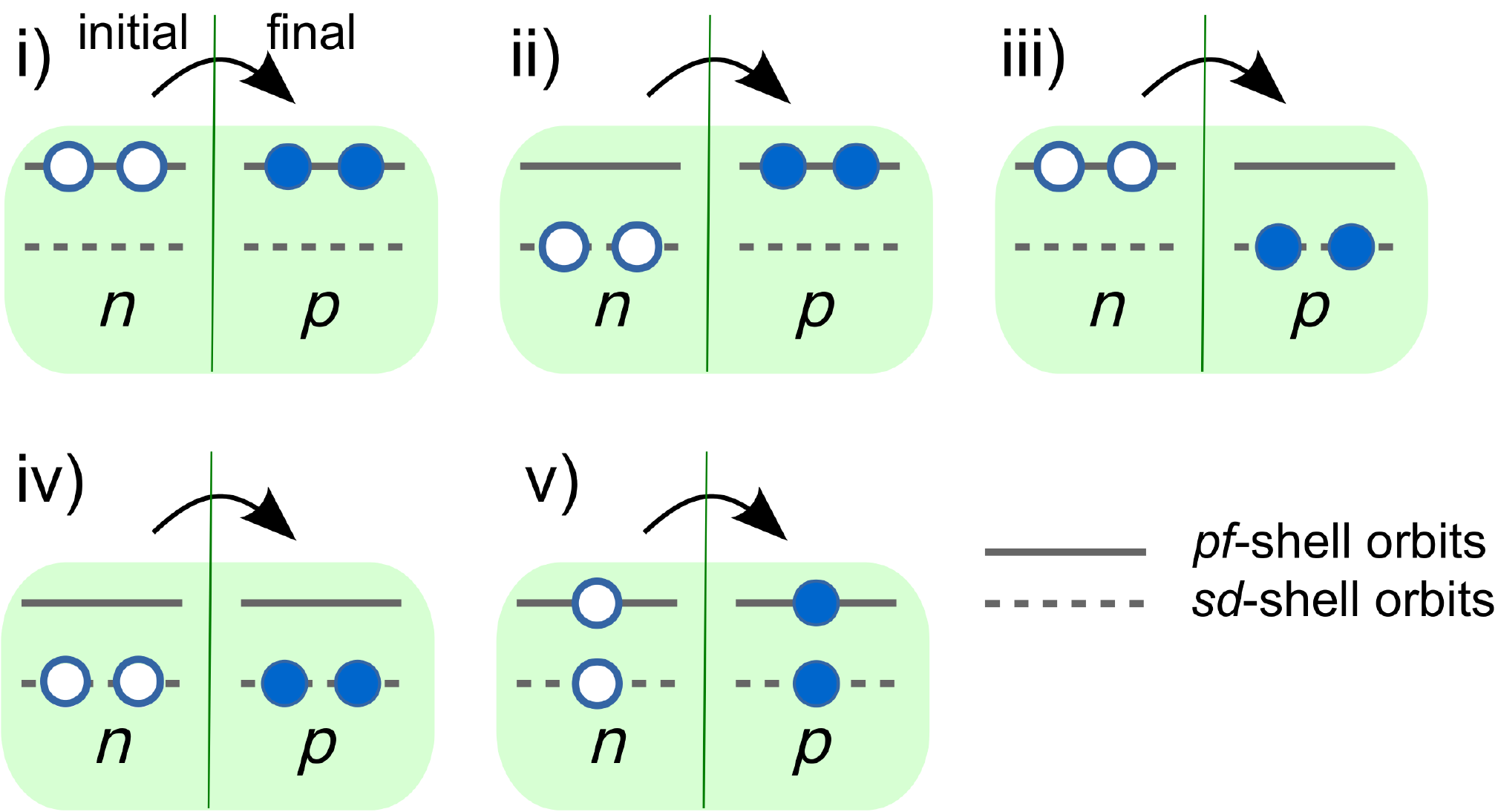}
\end{center}
\caption{Schematic contributions to shell model matrix elements from various
parts of configuration space, for neutron (n) and proton (p) orbitals.  Diagram
i) shows the contributions from within the standard shell model configuration
space.  Diagrams ii)--iv) illustrate the contributions from 2p-2h excitations
beyond this space, while diagram v) shows contributions from 1p-1h excitations.
Focus is on the $pf$ (standard configuration space) and $sd$ (additional
single-particle orbitals) shells used to study \elemA{48}Ca, but the competition
between diagrams ii)-iv) and v) is general.  Figure taken from
Ref.~\cite{Iwata16}.}
\label{f:ph_correlations}
\end{figure}

References\ \cite{caurier08a,hor13,Iwata16} have studied the effects of
enlarging the configuration space in the shell model.  The tentative conclusion
is that allowing up to one or two particles to occupy a few selected orbitals
beyond the original configuration space can have a strong effect on \bbt matrix
elements (in particular if these orbitals are the spin-orbit partners of
orbitals in the original configuration space~\cite{hor13}), but the effects on
\bbz matrix elements are more moderate.  The reason is that two distinct kinds
of correlations compete, as illustrated in Fig.~\ref{f:ph_correlations}, taken
from Ref.~\cite{Iwata16}.  On one hand, pairing-like excitations to the
additional orbitals enhance the \bbz matrix elements~\cite{caurier08a}
(diagrams ii, iii and iv in Fig.~\ref{f:ph_correlations}); on the other hand,
particle-hole excitations (1p-1h) to the additional orbitals, which in
principle require less energy than pairing-like excitations (2p-2h), tend to
reduce the matrix elements (diagram v in Fig.~\ref{f:ph_correlations}).  The
net result appears to be a moderate increase, generally by less than 50\%. A
careful calculation of the matrix element for $^{48}$Ca \cite{Iwata16} that
extended the configuration space from four to seven single-particle orbitals
found about a 30\% enhancement.  That agrees reasonably well with results of
calculations that include the effect of extra orbitals perturbatively: a 75\%
increase in the matrix element of $^{48}$Ca \cite{Kwiatkowski:2013xeq}, along
with a 20\% increase in $^{76}$Ge, and 30\% increase in $^{82}$Se
\cite{hol13c}.  (In perturbation theory configuration spaces even larger than
those in QRPA and GCM calculations can be included.) Though the enhancement is
greater in perturbation theory, the final matrix elements in
Refs.~\cite{Iwata16} and \cite{Kwiatkowski:2013xeq} differ by only 20\%.

Other publications \cite{caurier08,men09,Escuderos10} have reported studies of
the correlations included in the shell-model but not in the QRPA.  Nuclear
many-body states can be classified by a ``seniority'' quantum number ${s}$ that
labels the number of nucleons not in correlated neutron-neutron and
proton-proton $J^\pi=0^+$ pairs.  Fully paired $s=0$ states contribute the
major part of \bb matrix elements~\cite{caurier08,men09}, a fact that is
consistent with the finding that the additional pairing (2p-2h-like)
correlations from extra orbitals enhance the matrix element.  The states with
broken pairs (seniority $s>0$) contribute with opposite sign, shrinking matrix
elements~\cite{caurier08,men09}.  An overestimate of pairing correlations,
i.e.\ of the $s=0$ component in the nuclear states, thus leads to an
overestimate of the matrix elements themselves.  QRPA correlations in spherical
nuclei include states with seniority $s=4,8,12$ but not those with $s=6$ and
10, and QRPA matrix elements could be too large for that reason.  Although the
contributions of states with $s=6$ and 10 are relatively small in shell model
calculations \cite{Escuderos10}, Ref.~\cite{men11a} noted that when shell model
states are forced to have the same seniority structure as QRPA states, the
resulting shell-model matrix elements grow, implying that the shortage of
broken pairs in the QRPA makes its matrix elements too large.  On the other
hand, Ref.\ \cite{sim08} observed that when the QRPA is applied within the
small shell-model configuration space 
(which is not a natural space for the QRPA) 
the resulting matrix elements are similar to those of the shell model,
suggesting that the shell-model matrix elements are about 50\% too small.
These issues are still unresolved.

The shell model's ability to incorporate all correlations induced by the
nuclear interaction, at least in a small model space, allows it to benchmark
the GCM as well as the QRPA, and to tease out the most important kinds of
correlations for \bbz decay.  Reference\ \cite{men14} compared shell model and
EDF/GCM matrix elements for the decay of a wide range of calcium, titanium and
chromium isotopes.  These nuclei are not candidates for a \bbz decay
experiment, but their decay matrix elements can still be calculated, allowing
one to perform a systematic study in medium-mass nuclei.  The calculations of
Ref.\ \cite{men14} were in the full $pf$ shell, a useful testbed because it
includes all spin-orbit partners in the one-shell configuration space, removing
one of the shortcomings of the shell model in heavier nuclei (see
Sec.~\ref{ss:sm}).  In addition, both the shell model and EDF theory with the
GCM describe the spectra of nuclei in this region quite
well~\cite{cau05,Rodriguez07}.  The main finding of Ref.\ \cite{men14} is that
the EDF-based GCM and the shell model produce similar matrix elements when
spherical EDF configurations are kept and shell-model configurations are
restricted to those with $s=0$.  It is only when higher seniority components,
which include correlations beyond pairing, 
are permitted that the shell model matrix elements decrease.  The reduction is
much larger than that which can be induced by including axial deformation as
one of the coordinates in the EDF-based GCM.  The inclusion of high-seniority
states in the shell model but not the EDF-based GCM thus explains the results
in Fig.\ \ref{f:NMEcompare}, where shell model matrix elements are always
smaller than those produced by EDF theory.

\begin{figure}[t]
\begin{center}
\includegraphics[width=0.5\textwidth,clip=]{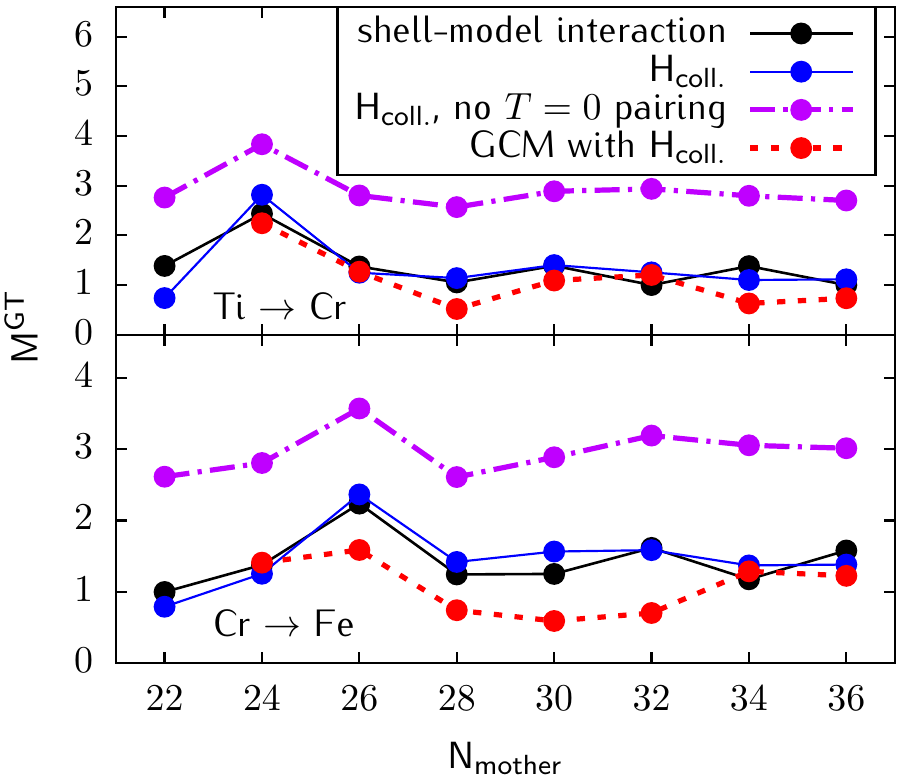}
\end{center}
\caption{Gamow-Teller part ($M^{0\nu}_{GT}$) of the nuclear matrix elements for
the decay of titanium (top) and chromium (bottom) isotopes as a function of
neutron number. Shell model results obtained with the KB3G shell model
interaction~\cite{Poves01} (black, solid lines) are compared with those 
produced by a collective Hamiltonian, $H_{\text{coll.}}$, (blue, solid lines)
and with GCM matrix elements produced
by the same collective Hamiltonian, with the isoscalar-pairing amplitude as a
generalized coordinate (red, dashed lines). Shell model results produced by the
collective Hamiltonian without isoscalar (T=0) pairing are also shown (purple
dashed-dotted lines).  Figure adapted from Ref.~\cite{men16}.}
\label{f:isos_pair}
\end{figure}

What is the nature of the higher-seniority correlations that reduce matrix
elements?  Quadrupole correlations, induced by deformation, can play a role,
particularly when the initial and final nuclei are deformed by different
amounts \cite{men11b,rod10,fang11}.  But the QRPA and shell model agree that
isoscalar pairing correlations usually play an even more crucial
role~\cite{vog86,eng88,men16}. Figure \ref{f:isos_pair}, based on work in Ref.\
\cite{men16}, compares shell-model matrix elements with and without isoscalar
pairing correlations.  Put more precisely, it compares results produced by the
full shell-model Hamiltonian with those of an approximate collective
Hamiltonian \cite{duf96} from which isoscalar pairing can be excluded.  The
collective Hamiltonian reproduces the full results nearly perfectly.  Without
isoscalar pairing, however, the collective Hamiltonian greatly overestimates
the matrix elements, always by about two units.

After noting the importance of isoscalar pairing, Ref.~\cite{men16} demonstrated
that the correlations missing from the GCM are nearly all of that kind.  The
paper tested a version of the GCM that works with Hamiltonians rather than EDFs.
This shell-model based GCM and the shell model itself used the same Hamiltonian
--- the collective approximation --- and the same configuration space; the shell
model was thus the ``exact'' solution in the test because it diagonalized the
interaction exactly.  The result, shown in Fig.~\ref{f:isos_pair}, was good
overall agreement, provided the GCM included the isoscalar
pairing amplitude as one of its generalized coordinates \cite{hin14}.
(Some small differences remain 
for decays involving closed shells because these isotopes have fewer collective
correlations.) That step allowed the method to
capture isoscalar pairing correlations quite well.  GCM calculations in larger
spaces do not yet include isoscalar pairing coordinate, but should be able to in
the near future.  The Hamiltonian-based GCM may thus be able to include both the
large model spaces of the QRPA and the important shell-model correlations,
without the drawbacks of either method.  

Comparing the many-body methods suggests that with proper attention, each of
them could obtain accurate matrix elements, which might well lie somewhere
between the current predictions of the shell model and those of the EDF/GCM and
QRPA.  We will say more about how to improve and benchmark the different methods
in the near future in Sec.~\ref{s:improving}.  First however, we turn to an
issue that plagues all calculations in heavy nuclei and vitiates the idea that
the matrix elements should be in the range spanned by current calculations: the
over-prediction of single-$\beta$ and \bbt matrix elements, sometimes referred
to as the ``$g_A$ problem".

\section{The $g_A$ Problem}
\label{s:ga}

\subsection{Systematic Over-prediction of Single-$\beta$ and \bbt Matrix
Elements}
\label{ss:over-prediction}

\begin{figure}[t]
\begin{center}
\includegraphics[width=.5\textwidth]{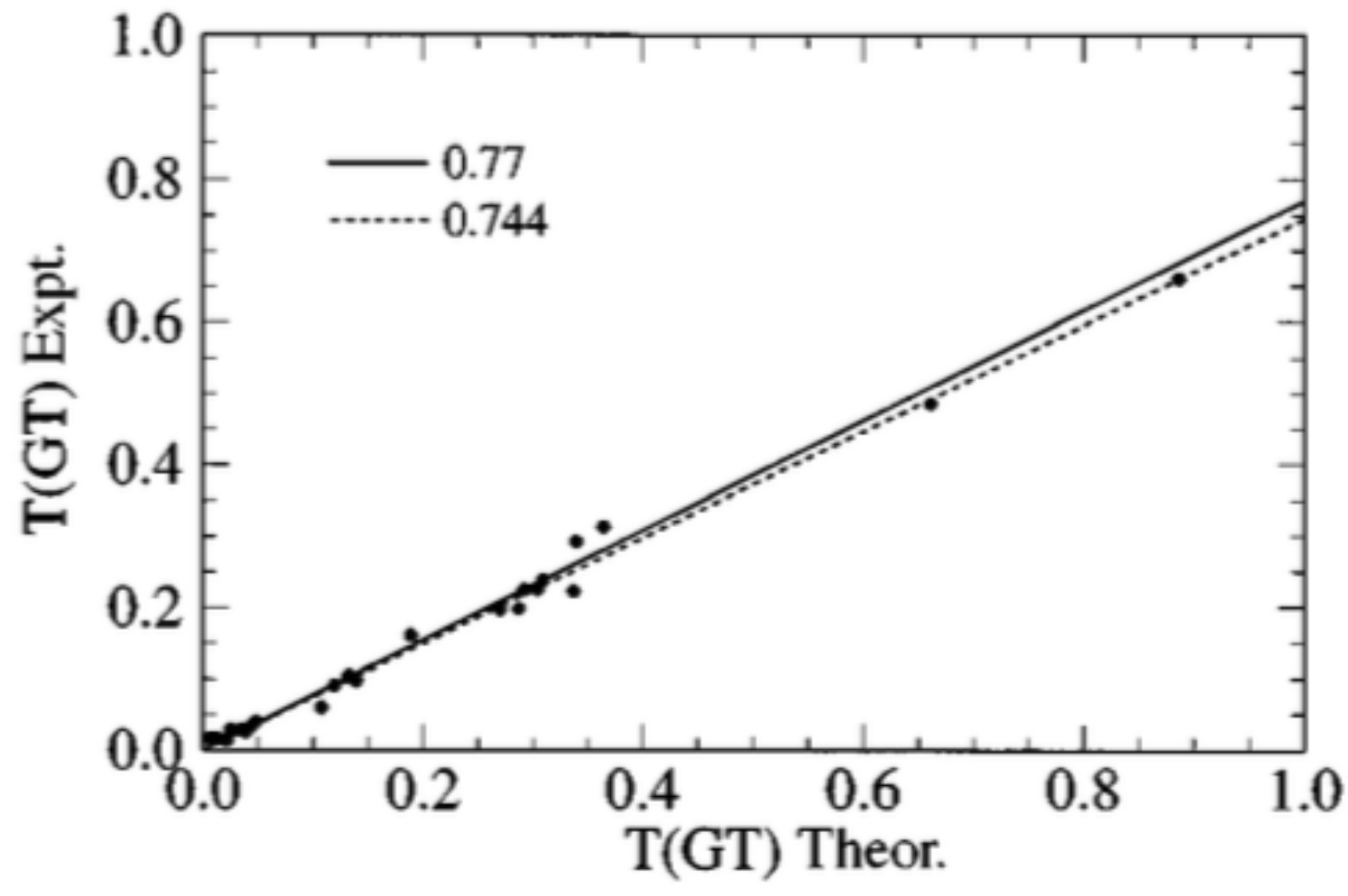}
\end{center}
\caption{Experimental versus theoretical strengths for a
compilation of Gamow-Teller transitions in nuclei with mass number between about
40 and 50.  The dashed line is a fit of the theoretical results to data,
and suggests that the effective value of $g_A^\text{eff}$ of the axial-vector
coupling is $0.74 \, g_A$.  Figure taken from Ref.\ \cite{MartinezPinedo96}.}
\label{f:gA-gabriel}
\end{figure}

For nuclei close to stability up to mass number $A\sim 60$, calculations
reproduce ground state properties, excitation spectra and
electric moments and transitions
well\footnote{The moments and transitions require effective
charges that one can obtain by treating collective states outside the
configuration space in perturbation theory~\cite{duf96}.}
\cite{cau05,ots01,bro01}.  They do not do as well with $\beta$-decay rates, at
least not without a tweak.  Figure\ \ref{f:gA-gabriel}, taken from
Ref.~\cite{MartinezPinedo96}, compares of experimental Gamow-Teller strengths
--- summed over certain low-lying states and scaled as described in that
reference --- for nuclei with mass number $A$ between about 40
and 50 versus the theoretical predictions.
The calculated strengths are generally larger than the data, but if the
Gamow-Teller operator $\bm{\sigma} \bm{\tau}$ is multiplied by 0.74 or,
equivalently, if the axial coupling $g_A \simeq 1.27$ is replaced by an
effective value $g_A^\text{eff} = 0.74 g_A$, then the
calculations reproduce most of the experimental data quite well.  A similar
phenomenological correction, $g_A^\text{eff} = 0.82 g_A$, brings predictions
into agreement with data for nuclei with $A$ less than 16~\cite{Chou93}. For $A$
between about 16 and 40, a value of $g_A^\text{eff} = 0.77
g_A$~\cite{bro88} works well, and a slightly larger
phenomenological correction $g_A^\text{eff} = 0.68 g_A$ is preferred for $A$
between 60 and 80~\cite{Kumar16}. That such a simple renormalization can
largely eliminate a theoretical problem is a remarkable fact that has resisted
attempts at explanation for several
decades~\cite{bro78,Bertsch82,arima87,cau95,towner87,Park97}.

The problem is not confined to $\beta$ decay or even electroweak operators.
The total Gamow-Teller integrated strength is governed by the Ikeda sum
rule~\cite{Ikeda63}:
\begin{equation}
\label{eq:ikeda}
\sum S_{GT}^- - S_{GT}^+ = 3(N-Z) \,,
\end{equation}
where $S_{GT}^\mp$ is the total Gamow-Teller strength from the
$\bm{\sigma}\tau^+$ (neutron to proton) or $\bm{\sigma}\tau^-$ (proton to
neutron) operators, summed over all energies, and $N$ and $Z$ are the initial
nucleus's neutron and proton numbers.  The sum rule is a simple consequence of
commutation relations and must hold to the extent that neutrons and protons (as
opposed, e.g., to $\Delta$-isobar excitations) are the only important nuclear
degree of freedom.  The summed strength can be extracted from charge-exchange
experiments in which, for example, a proton is absorbed and a neutron is
emitted, or a $^3$He ion is absorbed and a triton emitted.  The weak
interaction plays no part in these reactions but they nevertheless measure
Gamow-Teller strength because the cross-section at forward angles is determined
by the transition matrix elements of the $\bm{\sigma}\tau^\pm$
operators~\cite{Ichimura06,Fujita11,Frekers13}.  Experiments that can determine
the sum $S_{GT}^-$ of strength below about 50 MeV report considerably less than
$3(N-Z)$, typically about half that much ~\cite{Ichimura06}. ($S_{GT}^+$ is
much smaller because stable nuclei usually have more neutrons than protons and
proton-to-neutron transitions are thus Pauli blocked.) That amount would
correspond to a ``quenching" of the $\bm{\sigma}\tau^+$ operator that is
similar to what is needed for single-$\beta$ decay.

The agreement is strange.  One experiment examines weak decay and the other
tests nothing but the strong interaction.  In one experiment, nature disagrees
with complicated many-body calculations and in the other with a simple
consistency requirement (though measured strength \textit{distributions} are
also smaller than calculations at low energies, demanding a quenching that is
consistent with that used for $\beta$-decay
\cite{Yako09,Fujita13,Noji14,Iwata15}).  Some charge-exchange experiments
\cite{Yako05,Sasano09} suggest that both discrepancies are due to
$\bm{\sigma}\tau^+$ strength that spreads out above the Gamow-Teller resonance,
up to 50 MeV or more of excitation energy.  There is still no consensus about
the suggestion, however, mainly because it is hard to distinguish spin-isospin
strength from background at high energies~\cite{Ichimura06,Frekers13}.

\begin{figure}[t]
\begin{center}
\includegraphics[width=.5\textwidth]{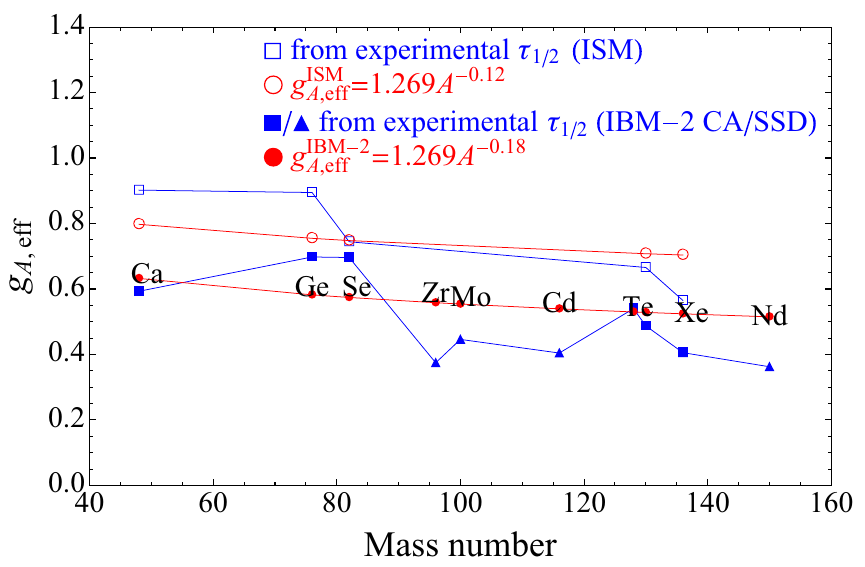}
\end{center}
\caption{Value of the effective strength of the $\bm{\sigma}\tau^+$ operator,
$g_A^\text{eff}$, required for shell model (ISM) and IBM
calculations to reproduce the experimental matrix elements for several \bbt
transitions, which are ordered by mass number. Figure taken from Ref.\
\cite{bar15}.}
\label{f:franco-ga}
\end{figure}

Given this comprehensive quenching of Gamow-Teller strength, it is not
surprising that the ``$g_A$ problem" also afflicts calculations of \bbt matrix
elements.  Figure\ \ref{f:franco-ga}, taken from Ref.\ \cite{bar15}, shows the
values of $g_A^\text{eff}$ required to reproduce measured \bbt matrix elements
for a selection of shell-model and IBM calculations (the latter of which are in
the not-always-reliable \bbt-decay closure approximation.) In both models
$g_A^\text{eff}$ decreases with mass, approaching half of $g_A$ or less in the
heavier nuclei.  Although other sets of shell-model calculations show milder
and less mass-dependent quenching \cite{cau12,Neacsu15,hor16,Neacsu15,hor16}
and the mass dependence might be due to the fixable omission of spin-orbit
partners \cite{cau12,hor13}, the main message of Fig.\ \ref{f:franco-ga} is
undeniable and its implication stark.  \bbt decay rates, which are proportional
to the fourth power of $g_A^\text{eff}$, are much smaller than shell-model and
IBM predictions\footnote{As we discussed earlier, the QRPA fits the strength of
isoscalar pairing so as to reproduce \bbt decay rates.}. If \bbz decay rates,
to which the main contribution is $M_{GT}$ in Eq.~\eqref{eq:monu}, are smaller
than predictions by a similar amount, next generation experiments will be
significantly less sensitive than we currently expect; they will not be able to
rule out the inverted hierarchy with a ton of material.  Is \bbz decay really
that quenched?

The answer depends on the source of the quenching, which is still unknown. We
discuss possibilities and their implications next.

\subsection{Possible Causes and Implications for Neutrinoless Double-Beta Decay}
\label{ss:causes}

Over the years, theorists have suggested a wide range of sources for the
quenching of Gamow-Teller strength. Almost all, in modern language, fall into
one of two classes: nuclear many-body correlations that escape
calculations~\cite{Bertsch82,arima87,towner87,cau95}, and many-nucleon weak
currents~\cite{towner87,Park97}.  The former include short-range
correlations, multi-phonon states, particle-hole excitations outside shell-model
configuration spaces, etc.  The latter stand for non-nucleonic degrees of
freedom, i.e.\ $\Delta$-isobar excitations, in-medium modification of pion
physics, partial restoration of chiral symmetry, etc.  The consequences for
\bbz decay depend on which of these two complementary sources is mostly
responsible for the renormalization of the $\bm{\sigma}\bm{\tau}$ operator.

\begin{figure}[t]
\begin{center}
\includegraphics[width=.48\textwidth]{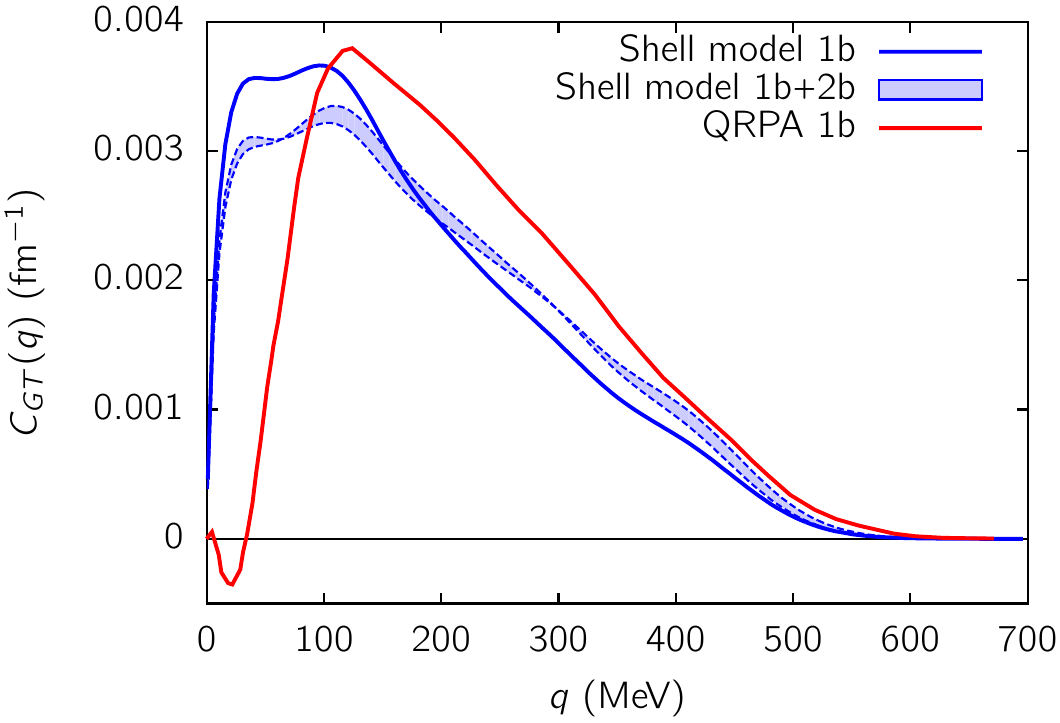}
\end{center}
\caption{Normalized momentum transfer distribution of the Gamow-Teller part of
the nuclear matrix element of \elemA{136}Xe.  The solid curves are with one-body
currents only, in the shell model (blue) and QRPA (red).  The shaded area
includes two-body contributions in the shell model.  Data are taken from
Ref.~\cite{men11} (shell model) and Ref.~\cite{eng14} (QRPA).}
\label{f:mom_trans}
\end{figure}

The reason that the relative quenching of \bbt and \bbz matrix elements can
depend on the source is that there are marked differences between the two
processes, even though both involve two virtual single-$\beta$ decays.  The
virtual neutrino that is emitted and reabsorbed in \bbz decay makes the average
momentum transfer from nucleons to leptons at each vertex much higher than in
\bbt decay.  If the neutrinos actually emerge from the decay, the momentum and
energy transfer are constrained by the $Q$-value of the transition, which is on
the order of 1 MeV.  If only electrons are emitted, however, the average
momentum is about 100 MeV, a scale set by the average distance between the two
decaying neutrons.  Fig.\ \ref{f:mom_trans} presents the
contribution of different momentum transfers $q$ to the nuclear matrix element
produced by the shell model and QRPA, for the representative mother nucleus
\elemA{136}Xe.  The momentum-transfer distribution is similar in the two
calculations.  Although the QRPA distribution is shifted to higher $q$, probably
because of the larger configuration space, it falls off slowly
in both cases so that several hundred MeV are transferred
with reasonable probability.  The higher momentum transfer means that the first
virtual $\beta^-$ transition in \bbz decay can excite virtual intermediate
nuclear states with all spins and parities, not just the $0^+$ or $1^+$
intermediate states that contribute to \bbt decay.  If the spin-isospin
renormalization depends on the momentum transfer or multipolarity of the
intermediate states, the large quenching needed to correctly predict
single-$\beta$ Gamow-Teller and \bbt decay rates may not be needed for \bbz
decay.  Although experimentalists are trying to test the momentum-transfer and
multipolarity dependence of quenching \cite{eji16}, the experiments are
difficult and the existing data inconclusive.

In the search for the cause of quenching, complex correlations that
calculations do not capture have long been a suspect.
Reference~\cite{Bertsch82} proposed in 1982 that two-particle--two hole
excitations to orbitals outside shell-model configuration spaces or beyond QRPA
correlations shift the Gamow-Teller strength to high energies. (The Ikeda sum
rule requirement means that strength does not appear or disappear, but rather
moves.)  Nuclear-structure models miss this effect and therefore need to quench
the low-energy strength.  The authors of Refs.~\cite{arima87,towner87} made a
similar argument, and Ref.~\cite{Brown87} proposed that about two thirds of the
spin-isospin quenching comes from missing particle-hole configurations.  The
authors of Ref.~\cite{cau95} argued slightly differently, suggesting that
because $\bm{\sigma}\bm{\tau}$ operates at all inter-nucleonic distances, its
matrix elements should be affected not only by the long-range (low-energy)
correlations included e.g.\ in shell-model states, but also by short-range
(high-energy) correlations, which are not included. They went on to argue that
shell-model Gamow-Teller strength should be quenched consistently with the
roughly 30\% depletion of single-particle occupancies needed to reproduce
electron scattering data \cite{Pandharipande97} because both kinds of
quenching reflect the same inability of the shell model to include short-range
correlations.

More recently, two studies have tried to use many-body perturbation theory
\cite{hjorth-jensen95} to quantify the effect of missing correlations on the
$\bm{\sigma}\bm{\tau}$ operator in the shell model.  Reference~\cite{sii01}
reported a 20\% reduction of Gamow-Teller strength for nuclei whose valence
nucleons are in the $sd$ and $pf$ shells; the result agrees well with
phenomenological fits to experimental strength.  In heavier systems the authors
found a much stronger reduction, as large as a 60\% in \elemA{100}Sn; that
resut is in reasonable agreement with the trend shown in
Fig.~\ref{f:franco-ga}.  The degree of renormalization varies by only a few
percent up to momentum transfers of about 100 MeV, suggesting similar quenching
of \bbt and \bbz matrix elements.  Reference~\cite{hol13c} studied \bb decay
within a similar perturbative framework.  While the method required the closure
approximation and so could say relatively little about \bbt decay, it produced
about a 20\% enhancement of the \bbz matrix element in $^{76}$Ge and a 30\%
enhancement in $^{82}$Se.  These results agree with the tendency of the shell
model to increase \bbz matrix elements when configuration spaces are enlarged
slightly \cite{caurier08a,Iwata16}, and argue against any suppression of \bbz
decay.  Once more, however, the argument is not conclusive: First-order
one-particle one-hole excitations strongly suppress the matrix element and it
just so happens that higher-order terms tend to counteract the suppression. But
it is not at all clear whether or how fast the perturbative expansion
converges, and neglected terms could have large effects.  

\begin{figure}[t]
\begin{center}
\includegraphics[height=3.5cm]{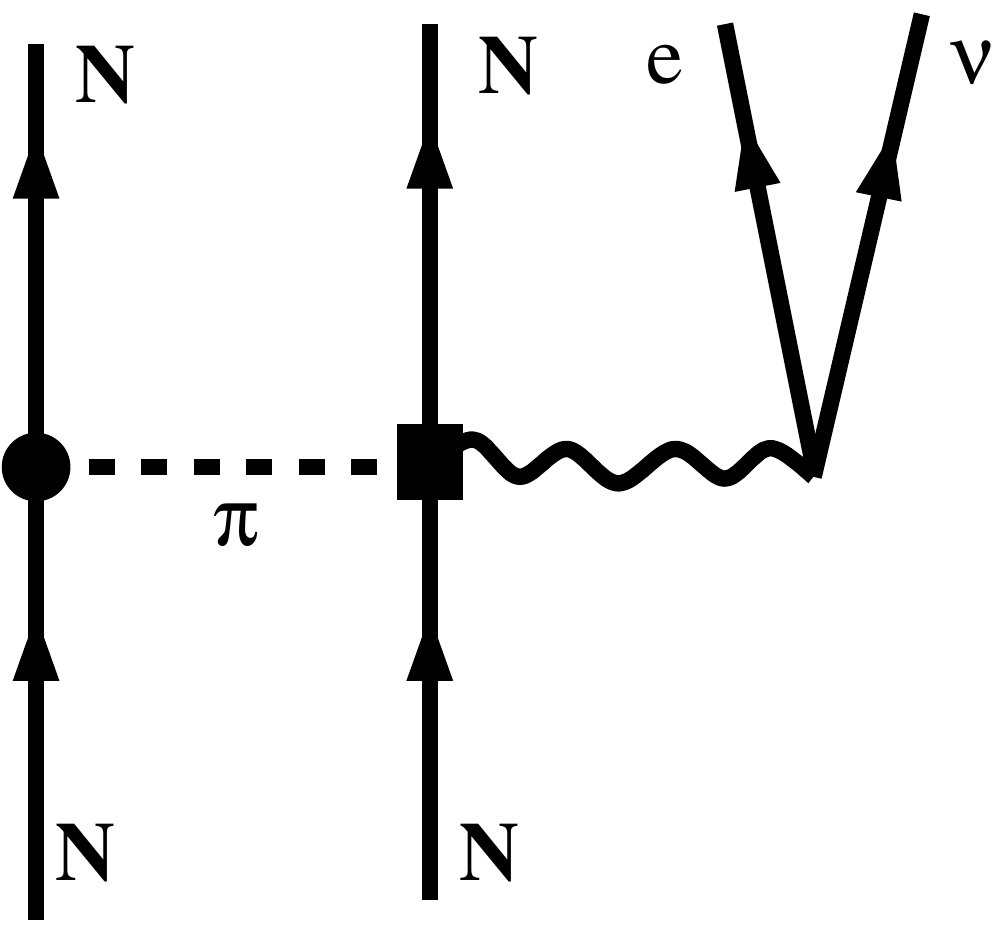}
\hspace{.5cm}
\includegraphics[height=3.5cm]{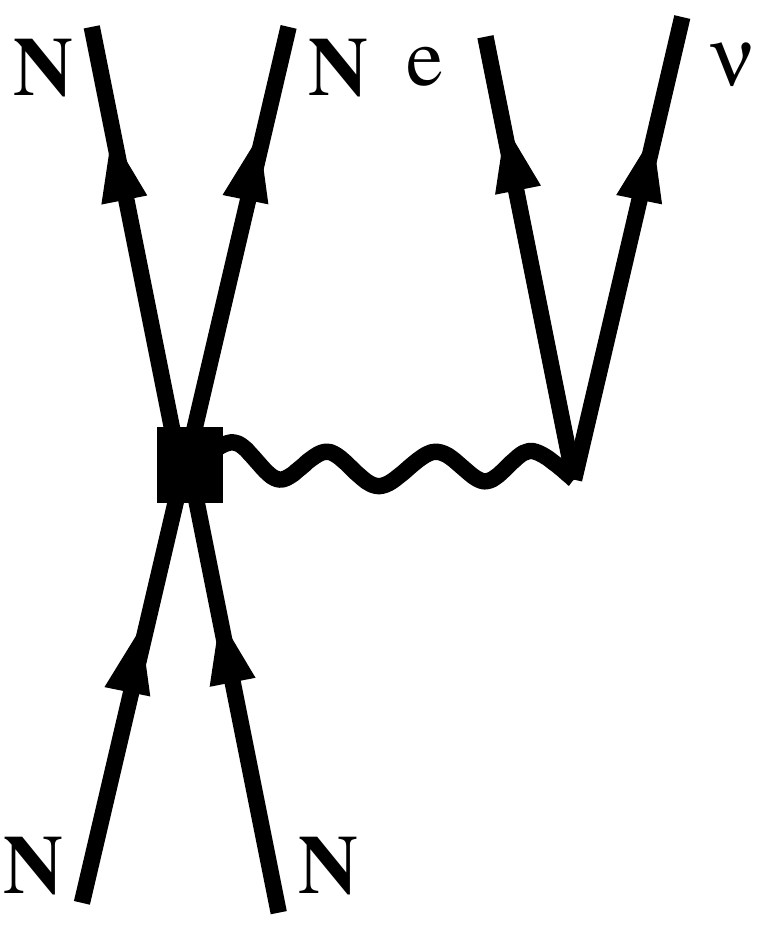}
\end{center}
\caption{Diagrams for the $\chi$EFT two-nucleon currents most important for
$\beta$-decay.}
\label{f:currents}
\end{figure}

Non-nucleonic degrees of freedom, manifested as many-nucleon currents in models
without explicit $\Delta$-isobars and pions, are also a long-term suspect in the
search to explain quenching (see e.g.\ Ref.~\cite{towner87} and references
therein). Reference~\cite{Brown87} concluded that only about one third of the
phenomenological quenching is due to meson-exchange currents, most of which
involved the $\Delta$-isobar.  The currents in use at the time were based on
models with no systematic power counting, however, and so the uncertainty in the
result is very large.  The situation has improved with the advent of $\chi$EFT,
which relates electroweak currents to nuclear interactions
\cite{Park03,Hoferichter15,Krebs16}. Several studies have exploited the relation
in nuclei with $A\lesssim 10$, and it is now clear that the two-nucleon currents
are necessary in precise calculations of
weak~\cite{Butler01,Nakamura01,Gazit09,Baroni16} and
electromagnetic~\cite{Bacca14,Marcucci16} transition rates.  The effects in
these nuclei range from a few percent to about
40\% and depend significantly on nuclear structure.  

Reference\ \cite{men11} applied many-nucleon currents, shown in
Fig.~\ref{f:currents}, to the single-$\beta$ and $\beta\beta$ decay of
medium-mass nuclei.  The authors reduced the two-body current operators to
effective one-body operators by normal ordering with respect to a
spin-isospin-symmetric Fermi-gas reference state. They found that the
single-$\beta$ and \bbt matrix elements were reduced by an amount corresponding
to $g_A^{\text{eff}}\sim(0.7-0.8) g_A$. Most of the quenching comes from the
long-range one-pion-exchange current (depicted in the left-hand diagram in
Fig.~\ref{f:currents}), which includes contributions from the $\Delta$-isobar.
Uncertainty in the $\chi$EFT couplings that appear in the two-nucleon currents,
especially in the coefficient of the contact term (depicted in the right-hand
diagram in Fig.~\ref{f:currents}), leads to a significant uncertainty in the
quenching, and even allows a slight enhancement.

Interestingly, the corresponding reduction in \bbz matrix elements turned out to
be about 30\%~\cite{men11} as well, much less than one would get by squaring the
renormalization factor from single-$\beta$ decay, as one does to get the
quenching of \bbt matrix elements.  Later corrections to the currents
\cite{Klos14,men16a} did not change this fact.  The two-body currents have a
smaller effect at high momentum transfer, softening the quenching of \bbz matrix
elements.  QRPA calculations that used the same currents obtained even less
quenching (about 20\% \cite{eng14}) partly because the average momentum transfer
is slightly higher in QRPA calculations and partly because the coefficient of
the isoscalar pairing interaction was readjusted after adding the two-body
currents to ensure that \bbt matrix elements were still correctly reproduced.
In any case, these studies suggest that whatever renormalization is due to
many-nucleon currents will be milder in \bbz decay than in single-$\beta$ and
\bbt decay.

The smaller quenching at momentum transfers near the pion mass is consistent
with the studies of muon capture, where the evidence for Gamow-Teller quenching
is weaker than in $\beta$ decay~\cite{Zinner06}.  Exclusive
charged-current electron-neutrino scattering from \elemA{12}C to the ground
state of \elemA{12}N also shows little evidence of quenching, at momentum
transfer even smaller than the pion mass \cite{Hayes00,Volpe00,Suzuki06}.  The
inclusive cross-section, for which additional multipoles are relevant, is harder
to calculate and shell model calculations disagree with one another by a factor
of about two \cite{Hayes00,Volpe00,Suzuki06}.  Improved calculations, using {\it
ab initio} methods such as those initiated in Ref.~\cite{Hayes03,Lovato16}, are
clearly needed.  Further measurements, preferably in targets that complement
\elemA{12}C, would also provide valuable information about quenching at non-zero
momentum transfer.

Recently, Ref.\ \cite{Ekstrom:2014iya} included two-nucleon currents in
coupled-cluster calculations of single-$\beta$ decay.  The authors
normal-ordered the operator with respect to a Hartree-Fock reference state
rather than the simpler Fermi gas of Ref.\ \cite{men11}.  Focusing on carbon and
oxygen isotopes, they found that the two-nucleon currents reduce the strength of
the ${\bm \sigma}{\bm \tau}$ operator by about 10\%.  Though this result
suggests a very small quenching of \bbz matrix elements, the
potentially coherent contribution of the nuclear core could make many-body
currents more effective in the much heavier nuclei that will be used in
$\beta\beta$ decay experiments.(The normal ordering
approximation that makes the coherence apparent should be carefully explored,
however.)

The discussion in this section should make it clear just how important it is to
characterize and untangle the sources of the $g_A$ problem.  If they lie mainly
in complicated many-body effects, the \bbz matrix elements may be significantly
too large (though that conclusion is not uniformly supported by theoretical
work in perturbation theory).  On the other hand, if the problem is mostly due
to many-nucleon currents, then our matrix elements are probably too large by
less than a factor of two.

To determine which of the contributions is more important and plumb the
consequences for \bbz decay, we need calculations that treat many-body
correlations in a comprehensive way and include many-nucleon currents
consistently. It should be possible to do all this in the next five or so
years, and we discuss how in the next section.  We raise just one more
consideration now.  As suggested by the discussion of electron-scattering
strengths just above, we may be able to learn about the source of
$\bm{\sigma}{\bm \tau}$ quenching by studying other electroweak processes, for
instance magnetic moments and transitions.  The operators for these observables
also need phenomenological renormalization to agree with experimental
data~\cite{towner87,Brown87,NeumannCosel98}.  The missing many-body effects are
expected to be similar to those in $\beta$ decay because they emerge from the
same nuclear states and similar non-relativistic operators (${\bm \sigma}$ and
$\bm{\sigma} \tau^z$ vs.\ $\bm{\sigma}\tau^\pm$).  The many-nucleon corrections
to the current operators can be quite different, however, because the photon
has no axial-vector coupling.

\section{Improving Matrix Element Calculations in the  Next Few Years}
\label{s:improving}

\subsection{General Ideas}
\label{ss:ideas}

We have seen that the important \bbz matrix elements still carry significant
theoretical uncertainty. We believe, however, that recent developments in the
shell model, QRPA, GCM, and especially in \textit{ab initio} nuclear structure
methods and $\chi$EFT will finally allow the community of nuclear theorists to
produce more accurate matrix elements with real estimates of uncertainty.  The
approach to that task will have at least three prongs:
\begin{itemize}
\item the improvement of current methods, in particular to accommodate all the
collective correlations we know to be important, 
\item the development and application of \textit{ab initio} methods to the
initial and final \bb-decay nuclei,
\item a systematic assessment of theoretical uncertainty.
\end{itemize}
In what follows we address each of these in turn.  

\subsection{Improving Present Models}
\label{ss:imp_present}

\subsubsection{Extending Shell Model Configuration Spaces}
\label{sss:sm_extend}

The main limitation of the shell model is that it is restricted to small
configuration spaces.  We have already mentioned that the effects of the
neglected single-particle orbitals can be estimated
perturbatively~\cite{Kwiatkowski:2013xeq,hol13c}, and that calculations in
two oscillator shells are now becoming feasible~\cite{Iwata16}.  To add more
than a few single-particle levels in a nonperturbative way, however, requires
new ideas. 

One way to mitigate the shell model's shortcomings is to discard many-body
configurations that have little effect on the observables one is
interested in.  The most established approach for doing this is the Monte Carlo
shell model (MCSM)~\cite{Shimizu12}, which uses statistical sampling to select
deformed Slater determinants and then projection to restore angular-momentum
symmetry.  More recent work uses importance truncation~\cite{Stumpf16} or the
density-matrix renormalization group \cite{Legeza15}, though not yet in a
systematic way.  These schemes allow configuration spaces with dimensions
several orders of magnitude larger than those that can be handled with exact
diagonalization.  MCSM calculations now can to work with configuration spaces
of dimension $10^{23}$ \cite{Togashi16}, while exact diagonalization currently
requires a space of dimension $10^{11}$ or less.  The MCSM number is large
enough to allow the extension of shell model configuration spaces so that they
include all spin-orbit partners in $\beta\beta$-decaying nuclei and all
single-particle orbitals found relevant in QRPA or EDF
calculations~\cite{sim09a,Rodriguez16}.  The Monte Carlo approach could also
facilitate \textit{ab initio} no-core shell model calculations~\cite{Abe12},
which are currently limited to nuclei with less than about 20 nucleons
\cite{Navratil16,Barrett13}, in isotopes closer to those used in \bb-decay
experiments.  Calculations in light nuclei are useful for other reasons as
well, and we discuss them further in Sec.~\ref{ss:abinit}.

Extended shell-model spaces will require suitable nuclear interactions.  The
phenomenological modification of $H_{\text{eff}}$ that accurate calculations
usually require makes it difficult to obtain reliable effective interactions in
larger configuration spaces; the number of two-body matrix elements to be
constrained phenomenologically increases with the size of the space.  The
situation would be better if we could dispense with the adjustment, the need
for which may be due to the absence of three-nucleon forces in the original
interaction~\cite{Zuker03}.  Recent promising work
\cite{ots10,Jansen14,bog14,Dikmen15,Tsunoda16} avoids any phenomenological
tuning by using $\chi$EFT two- and three-body interactions as a starting point
to generate $H_{\text{eff}}$. These \textit{ab initio} interactions allow a
better assessment of uncertainty and appear to be usable in larger
configuration spaces.  One can obtain transition operators the same way.  We
discuss the use of $\chi$EFT and nonperturbative many-body methods to create
shell model interactions and operators in Sec.~\ref{sss:effop}.

\subsubsection{Adding Correlations to the EDF and the IBM}
\label{sss:corr}

The EDF matrix elements, as we noted in Sec. \ref{ss:edf}, are almost
universally larger than matrix elements calculated in any other approach.  The
reason, as discussed in Refs.\ \cite{men14,men16}, is missing correlations,
particularly isoscalar pairing correlations.  It is actually not hard to add
proton-neutron pairing to the GCM; Ref.\ \cite{hin14} did precisely that within
a two-shell calculation that used a multi-separable interaction.  There is no
reason that the same physics cannot be added to EDF-based GCM.  The main
ingredient is a set of HFB calculations that allow proton-neutron mixing, i.e.\
quasiparticles that are a combination of a neutron and a proton as well as a
particle and a hole.  Computer codes to do these calculations are nearly
complete \cite{sheikh14}.  Problems with the EDF-based GGM, as noted earlier,
arise only when one tries to superpose the projected HFB quasiparticle vacua.
Here, with a few limited exceptions, one needs a Hamiltonian rather than a
functional.  That does not mean, however, that EDF theory can play no role.
One might, for example, use the EDF-based HFB, with proton-neutron pairing,
simply to generate the projected HFB vacua.  These states can be constructed so
that they include all possible collective correlations, and in particular all
correlations relevant to \bb decay.  One then can use them to form a basis in
which to diagonalize a real Hamiltonian such as that produced by $\chi$EFT
rather than the original density functional.  Functionals were designed to work
well in mean-field calculations and thus should produce a good correlated basis
with contributions from many more single-particle states than are in the shell
model.  The Hamiltonian will then determine how these states are mixed.  The
result should combine many of the virtues of shell-model calculations with
those of current EDF and QRPA calculations.

Even so, some correlations, e.g.\ those from high relative momentum, may still
be absent from EDF-based approaches. Within the shell model, as we discuss in
Sec.~\ref{sss:effop}, one can construct an effective Hamiltonian and a
consistent \bbz decay operator that captures such effects. It has not been as
clear how to do that with EDF or QRPA-based calculations, in which the full
many-body Hilbert space does not cleanly separate into included and neglected
pieces.  Recently, however, techniques to add missing correlations to
approximate state vectors such as those produced by the GCM have nonetheless
been developed; we discuss them in Sec.\ \ref{sss:multiref}.  

\begin{figure}[t]
\begin{center}
\includegraphics[width=.5\textwidth]{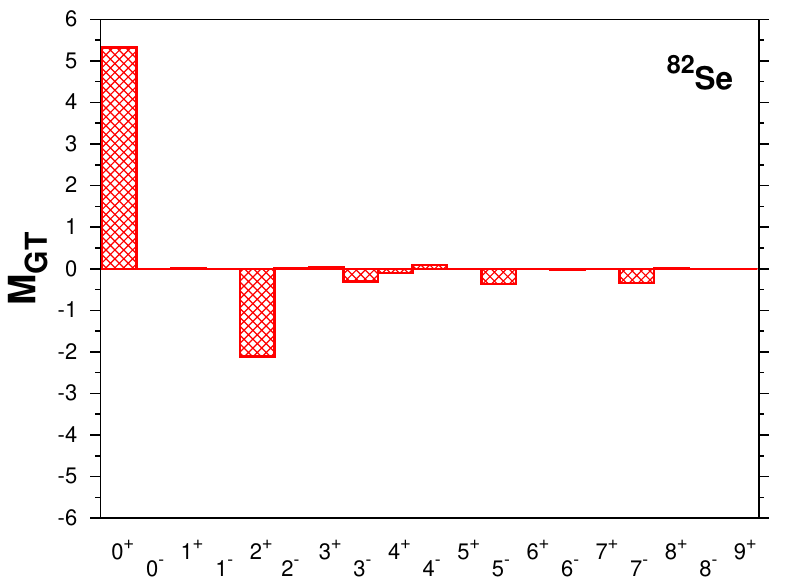}
\end{center}
\caption{Contributions of the decay of nucleon-nucleon (or proton-proton)
pairs with given angular momentum and parity to the Gamow-Teller \bbz decay
matrix element $M_{GT}$ for the \elemA{82}Se transition.
Figure taken from Ref. \cite{caurier08}}.
\label{f:pairs}
\end{figure}

Finally, the IBM-2 \bbz decay calculations done so far, like those in EDF
theory, have no explicit proton-neutron pair degrees of freedom.  Bosons that
represent isoscalar pairs can certainly be added --- a boson model with both
isovector and isoscalar proton-neutron pairs is known as IBM-4 --- and work is
in progress to see the effects on \bb decay \cite{isa16}.  While it is not
obvious that these effects will be large, the new pairs may bring the IBM
matrix elements closer to those of the shell model.  The IBM-2, however,
already includes bosons that represent like-particle monopole and quadrupole
pairs.  The coupling of two spin-1 isoscalar pairs to total angular momentum
zero can be represented as a combination of two like-particle $J=0$ or $J=2$
pairs, and these are exactly the degrees of freedom the usual IBM-2 bosons
represent.  It may be, then, that present IBM results capture most of the
effects of isoscalar pairing without explicit isoscalar bosons. On the other
hand, some facts indicate that isoscalar pairing effects are still missing.
Figure\ \ref{f:pairs}, which is typical of shell model results, shows that
almost the entire matrix element arises from the decay of like-particle $J=0$
pairs or $J=2$ pairs, with a sizable cancellation between these two
contributions. In the IBM-2, by contrast, the $d$-boson contribution is much
smaller than that of the $s$ boson, so that there is very little cancellation
\cite{bar09}.  The dominance of $J=0$ pairs is much like what one obtains in
the shell model when isoscalar pairing is omitted ~\cite{men16}.  

\subsubsection{Higher QRPA and the Overlap Problem}
\label{sss:higher}

The QRPA can also be improved.  It is currently limited by the simplicity of
its correlations, which can be represented as one-boson states; the bosons are
simplified versions of correlated quasiparticle pairs.  Without many-boson
states, the Pauli principle is violated and the transition strength to
intermediate states overly concentrated.  Accuracy requires that the many-boson
states be included.  

Many studies advance one scheme or another for doing that.  The second QRPA and
the quasiparticle time-blocking approximation (QTBA) are two representative
examples.  The second QRPA systematically adds all four-quasiparticle
excitations to the two-quasiparticle excitations that enter the usual QRPA.
Outside of preliminary investigations in Refs.\ \cite{rad91,sto01}, however, it
has not been systematically developed.  Second RPA, without quasiparticles, is
further along, both in an \textit{ab inito} form \cite{Papakonstantinou09} and
in conjunction with EDF theory~\cite{gam10,gam15}. Though these schemes are
computationally intensive, their quasiparticle versions should not be
completely intractable and may be both accurate and flexible enough to allow
\textit{ab initio} calculations.

The QTBA is a Green's-function approach to linear response that supplements the
iterated ring and ladder diagrams that are equivalent to the QRPA with a subset
of diagrams that contain the emission and re-absorption of QRPA phonons by
quasiparticles and the exchange of phonons between
quasiparticles~\cite{Litvinova08}.  When used in conjunction with a
relativistic EDF, the QTBA appears to significantly improve predictions of
single-$\beta$ decay strength functions~\cite{Robin16}.  This method has the
potential to significantly reduce the important physics omitted by the QRPA.

Any RPA-like method designed to calculate linear response faces a problem when
applied to \bb decay, however. As mentioned in \ref{ss:qrpa}, \bb matrix
elements involve separate transition matrix elements to each
intermediate-nucleus state from the initial and the final ground states.  These
two contributions must be multiplied and then summed over all intermediate
states.  But the two QRPA ground states are unrelated, so there is no way to
match intermediate states produced by the excitation of one nucleus with those
from the excitation of the other.  Intermediate-state energies are different in
the two cases, and the QRPA provides only transition densities, not full wave
functions.  Thus even attempts to compute the overlap of two different sets of
intermediate-state wave functions must rely on prescriptions.

The only way to solve this problem, at least partially, is to extend the QRPA
so that it does provide wave functions.  One solution is to use wave functions
from the simpler quasi-Tamm Dancoff approximation.  Another track, taken by
Refs.\ \cite{Terasaki12} and \cite{Terasaki13}, is to represent the ground
states of the initial and final nuclei by quasiparticle coupled-cluster wave
functions of the schematic form
\begin{equation}
\label{eq:QRPA-wf}
\ket{\Psi} \propto \exp \sum_{\nu,abcd} \left( [Y^\nu_{ab}/ X^\nu_{cd}] \alpha^\dag_a
\alpha^\dag_b \alpha^\dag_c \alpha^\dag_d \right) \,,
\end{equation}
where the $X^\nu$'s and $Y^\nu$'s are the usual QRPA amplitudes for the
$\nu^\text{th}$ excited state.  This form is a natural extension of the
quasi-boson wave function that the QRPA does provide when pairs of fermionic
operators are replaced by bosonic operators (with respect to a particular
vacuum).  The quality of the approximate wave function in Eq.\
\eqref{eq:QRPA-wf} is not yet known, however.  In any event, every extension of
the QRPA must face this issue.  The method is a small-oscillations
approximation, and the initial and final nuclei in \bb decay are often quite
different in structure, so they cannot both be only a little different from
states in the intermediate nucleus.

\subsection{\textit{Ab Initio} Approaches}

\label{ss:abinit}
\subsubsection{Kinds of \textit{Ab Initio} Calculations}
\label{sss:kinds}
    
The term \textit{ab initio}, literally ``from the beginning,'' is a little vague
in the nuclear-physics context.  A truly first-principles calculation of nuclei
involves solving the underlying theory, quantum chromodynamics (QCD), with
quarks and gluon degrees of freedom.  The ambitious framework for doing so is
lattice QCD~\cite{Beane11,Aoki12,Briceno15}. In spite of rapid progress, only
the lightest nuclei with two, three and four nucleons have been studied thus
far, and even these are treated with significant approximations, e.g.\ at
larger-than-realistic values for the pion mass.  We will not see QCD
calculations of the structure of medium-mass and heavy nuclei for some time.
Nevertheless, lattice QCD is valuable for the \bbz decay program.  It can
connect the parameters in the $\chi$EFT diagrams discussed in Sec.~\ref{sss:bsm}
with the underlying fundamental beyond-standard-model physics, for example
\cite{Nicholson16}.  More generally, it provides a way of determining all
$\chi$EFT parameters, which are often poorly constrained by data.
Though doing so is computationally demanding and computed
parameters are not accurate enough at
present~\cite{Beane11,Aoki12,Abdel-Rehim15}, researchers are working to improve
their calculations.

In nuclear structure the term \textit{ab initio} usually refers to calculations
that 1) take nucleons --- all of them --- as the degrees of freedom, and 2) use
nuclear interactions and currents obtained from fits to nucleon-nucleon
scattering data and properties of the lightest nuclei: the
deuteron and isotopes of hydrogen and helium with 3 or 4 nucleons, and as few
(slightly) heavier nuclei as possible.  These fits can proceed through a
meson-exchange phenomenology that describes the elastic channel of
nucleon-nucleon scattering up to and perhaps beyond the pion-production
threshold, leading, e.g., to the ``Argonne'' potentials and other similar
interactions \cite{Wir95,cdbonn,Stoks:1994wp}.  Phenomenological potentials of
this sort work quite well in light nuclei but seem to need improvement for use
in medium-mass nuclei~\cite{gan14}.

The fits that determine nuclear interactions can also be based on $\chi$EFT
\cite{Meissner16,chiral,Hammer13,Machleidt11}, discussed already in connection
with heavy-particle exchange (Sec.~\ref{sss:bsm}) and many-nucleon currents
(Sec.~\ref{ss:causes}).  Here we give a very brief description of its use in
\textit{ab initio} nuclear structure.  $\chi$EFT is an effective theory, based
on the symmetries of QCD, that provides a perturbative framework for
interactions and currents.  It yields a systematic expansion of two- and
many-nucleon forces and consistent one-, two-, and many-nucleon currents in
powers of the parameters $p/\Lambda_b$ and $m_{\pi}/\Lambda_b$, where $p$ is a
typical nucleon momentum and $\Lambda_b$ is the chiral-symmetry
breaking scale defined in Sec.~\ref{sss:bsm}.  Once the
interactions are fixed, an accurate many-body method is used to calculate
nuclear binding energies, radii, excitation spectra, electromagnetic
transitions, decay rates, and other observables, with error estimates inferred
from the power counting and tests of the many-body method.  

$\chi$EFT is not without its problems.  Though they sometimes produce excellent
results~\cite{Hebeler15}, the most widely-used $\chi$EFT nuclear interactions
frequently predict binding energies that are too large and radii that are too
small, especially in heavy systems~\cite{Binder14,Hergert16,Soma14}.  The
overbinding can be partly cured by fitting the $\chi$EFT couplings to properties
of nuclei as heavy as oxygen~\cite{Ekstrom:2015rta}, even while omitting
interactions beyond the three-nucleon level (and thus limiting the ability to
estimate uncertainties).  But the theory may still work without that step; some
interactions that are obtained without it predict saturation
properties~\cite{Hebeler11} and agree better with experimental radii than
others~\cite{GarciaRuiz16}.  And fits that include consistent two- and
three-nucleon forces to fourth-order in the chiral expansion are now
underway~\cite{Hebeler15a}.  Reference\ \cite{Drischler16} recently presented
the initial calculations in neutron matter with these interactions.

\textit{Ab initio} calculations are not yet successful everywhere in the
isotopic chart, but progress has been impressive.  Green function Monte Carlo
(GFMC) \cite{carlson15} and the no-core shell model
(NCSM)~\cite{Navratil16,Barrett13} have extended the scope of {\it ab initio}
studies light nuclei to detailed spectroscopy, reactions, and nuclear response.
\textit{Ab initio} work on heavy nuclei has a fairly long history; milestones
include coupled-clusters (CC) calculations in $^{16}$O
and $^{40}$Ca \cite{Kummel78,PhysRevC.61.054309,PhysRevC.76.044305},
the introduction of the now essential
normal-ordering procedure for treating three-body interactions
\cite{PhysRevC.76.034302} and the use of $\chi$EFT two- and three-nucleon
interactions together with many-body perturbation theory to derive a shell-model
effective interaction for oxygen isotopes~\cite{ots10}.  The same ideas were
later used in heavier nuclei~\cite{Hebeler15}.  Non-perturbative {\it ab initio}
methods, including CC theory~\cite{hagen14}, the in-medium similarity
renormalization group (IMSRG)~\cite{Hergert16}, the self-consistent Green's
function method~\cite{Soma14}, and nuclear lattice simulations~\cite{Lee09,Meissner16}
have developed quickly as computers have become more powerful.  We describe some
of them here.

The application of GFMC~\cite{carlson15} to nuclear structure was a true turning
point for the field \cite{pudliner97}. The method starts with an approximate
wave function that includes two and three-body correlations in all spin and
isospin channels and is usually obtained through a variational Monte Carlo
calculation.  This wave function is then evolved in imaginary time to filter out
spurious components.  The procedure, which exploits the locality of Argonne
two-body and ``Illinois'' three-body interactions, reproduces binding energies
with an r.m.s.\ error of 0.36~MeV and the energies of the lowest excited states
with an r.m.s.\ error of 0.5~MeV in nuclei up to $^{12}$C~\cite{carlson15}.  The
approach also works with local versions of $\chi$EFT interactions \cite{Lynn15}
and is valuable for the computation of electromagnetic and weak
response~\cite{Bacca14,carlson15,Lovato16}.  Unfortunately, computation time
scales exponentially with the number of particles, and the method cannot soon be
extended to the heavy nuclei in \bb decay experiments.  Although other forms of
quantum Monte Carlo, e.g.\ auxiliary diffusion Monte Carlo, should reach heavier
systems~\cite{gan14}, a treatment of open shell nuclei appears to lie at least a
few years in the future. The promise of quantum Monte Carlo, however, is
exceptional. And the GFMC can already be used in light nuclei to help diagnose
the source of the $g_A$ quenching we discussed in Sec.~\ref{s:ga}.

The NCSM~\cite{Navratil16,Barrett13}, already discussed briefly in Sec.\
\ref{sss:sm_extend}, is a shell model in which none of the $A$ nucleons are
forced to occupy particular single-particle orbitals, and in which one can
explicitly check the convergence of results as the configuration space grows.
The space is usually truncated so as to keep all configurations with total
oscillator excitation energy $\mathcal{N} \hbar \omega$ or below (for some
$\mathcal{N}$) compared to the lowest-energy configuration.  The parameter
$\mathcal{N}$ ranges from hundreds for the deuteron to just a few for nuclei
with a valence $sd$ shell.  Whether any particular $\mathcal{N}$ is sufficient
depends on the nuclear interaction employed. For that reason, it is helpful to
``evolve" or ``soften" the interaction via the SRG, so that the calculation
converges in a smaller configuration space (with the drawback that the evolution
generates many-body interactions).  The initial unevolved interaction can come
from a phenomenological model or from $\chi$EFT.  

In general, the NCSM produces energies that are comparable in accuracy to those
of the GFMC, but can be extended to somewhat heavier nuclei by using importance
truncation to exclude irrelevant configurations~\cite{Roth09}. Unfortunately, as
with the GFMC, computation time scales exponentially with the number of
particles and an extension to the medium mass \bb-decaying nuclei will be
difficult. NCSM calculations will still be useful in \bb decay research,
however.  In addition to investigating $g_A$ in single-$\beta$
decay~\cite{Maris11}, they allow tests of schemes to generate effective
operators for the usual shell model (with a core)~\cite{engel11}. 

The {\it ab initio} methods that can soon be applied in heavier nuclei
generally have the benefit of being explicitly ``size extensive''.  That term
refers to the correct (roughly linear) scaling of binding energy with the mass
number $A$ at any level of truncation.  The computation time for these methods,
which include CC, the IMSRG, and the self-consistent Green's function approach,
is a polynomial in $A$ rather than an exponential.  A basic CC calculation, for
example, takes a time that is proportional to $A^4$.

The starting point for CC theory~\cite{hagen14} is an exponential ansatz for the
ground state $\ket{\Psi_0}$ of a closed-shell even-even nucleus:
\begin{equation}
\ket{\Psi_0} = e^T \ket{\varphi_0}\,,
\end{equation}
where $\ket{\varphi}$ is a Slater determinant and 
\begin{equation}
T = \frac{1}{4}\sum_{i,a} t^a_i a^\dag_a a_i + \frac{1}{36}
\sum_{ij,ab} t^{ab}_{ij}
a^\dag_a a^\dag_b a_i a_j + \ldots \nonumber\\
\end{equation}
with the indices $a,b$ denoting particle orbitals (above the Fermi surface) and
$i,j$ hole orbitals (below the Fermi surface).  The $t$'s are amplitudes
determined by projecting the Schr\"{o}dinger equation onto one-particle one-hole
excitations (generating an equation for the $t^a_i$), two-particle two-hole
excitations (generating one for the $T^{ab}_{ij}$), etc. If the operator $T$ is
truncated at that point, the method is called CC-SD, where SD stands for
``singles and doubles,'' and if it is continued to the next order it is called
CC-SDT, where T stands for ``triples.'' Still higher-order terms, involving
four-particle clusters, are rarely considered. CC theory has been applied to
closed-shell nuclei as heavy as \elemA{132}Sn~\cite{Binder14} and to compute not
only energies but also charge and matter radii, single-$\beta$ decay rates, etc.
\cite{Hagen15,GarciaRuiz16,Ekstrom:2014iya}. 

To calculate \bb decay elements in a closed shell nucleus such as $^{48}$Ca, one
needs, in addition to the ground-state $\ket{\Psi_0}$ of that nucleus, the
ground state $\ket{\Psi_0'}$ of the final nucleus $^{48}$Ti, which has both
valence protons and neutrons.  There one uses the ``equations-of-motion''
method~\cite{hagen14}.  The ground state of the final nucleus is represented as
$\ket{\Psi_0'} = \mathcal{R} \ket{\Psi_0}$, where
\begin{equation}
\mathcal{R} = \sum_{ijab} r^{ab}_{ij} p^\dagger_a p^\dagger_b n_i n_j  +
\frac{1}{6} \sum_{ijkabc} r^{abc}_{ijk}  p^\dagger_a p^\dagger_b N^\dagger_c
N_k n_i n_j +\ldots\,
\end{equation}
The operator $\mathcal{R}$ creates two-particle two-hole excitations,
three-particle three-hole, excitations, etc.\ of the CC $^{48}$Ca ground state.
Equations for the corresponding $r$ amplitudes, like those for the $t$
amplitudes, follow from projection of the Schr\"odinger equation.  The
application of CC theory in mid-shell nuclei requires a more elaborate
construct, which we discuss in Sec.~\ref{sss:effop}.

The IMSRG~\cite{Hergert16} is a set of flow equations for a unitary
transformation that in the asymptotic limit decouples some pre-selected subspace
of states from the rest. In its simplest version, the flow equations for the
Hamiltonian take the form
\begin{equation}
\label{eq:flow}
\frac{d}{ds} H(s) = [\eta(s),H(s)]\,, \quad H(\infty)=H_\text{eff}
\end{equation}
with $s$ the flow parameter and the ``generator'' $\eta(s)$ given by
\begin{equation}
\label{eq:generator}
\eta(s) = [H_d(s),H_{od}(s)] \,.
\end{equation}
Here, $H_d$, (where $d$ stands for diagonal, though its meaning can be more
general) is the part of the Hamiltonian that does not couple the space one is
interested in to the rest, and $H_{od}$ is the part that does.  It is not hard to
show that as the flow progresses, $H_{od}$ is driven to zero.  In a closed-shell
nucleus, if one chooses to decouple a reasonable zeroth-order approximation to
the ground state --- e.g. a Hartree-Fock Slater determinant --- one ends up with
a Hamiltonian $H(s=\infty)$ for which that approximation is the exact ground
state.  Because the evolution equation is unitary, the eigenvalue is the
ground-state energy. A consistent flow equation governs the evolution of
transition operators other than the Hamiltonian.

Application of the method in practice is more complicated than this simple
description indicates, in a number of ways.  First, the flow equations generate
three-body interactions, four-body interactions, etc., (as in any SRG flow) even
if the initial Hamiltonian has only one- and two-body terms.  In practice, the
equations are usually truncated to include only normal-ordered one- and two-body
pieces.  But the normal-ordering, carried out with respect to the initial Slater
determinant in the example above, incorporates the most important effects of
higher-body terms, just as it includes the most important pieces of the two-body
interaction in a Hartree-Fock single-particle Hamiltonian.  Second, the flow
equations above are stiff and solutions hard to obtain.  Other forms for the
generator $\eta$ are better behaved and still drive $H_{od}$ to zero. Also, the
evolution of transition operators is easiest with a single set of equations for
the unitary transformation that does the decoupling rather than a separate set
for every operator~\cite{Morris15}.  Finally, in open-shell systems a Slater
determinant is not a good zeroth-order approximation to the exact ground state
and the method fails.  

The approach described here has been used to calculate energies, charge radii,
and matter radii in nuclei that range from the lightest to nickel isotopes
~\cite{Hergert16,Lapoux16}.  In Secs.~\ref{sss:effop} and \ref{sss:multiref} we
describe ways the method can be extended.

\subsubsection{Effective Operators for the Shell Model from \textit{Ab Initio}
Calculations}
\label{sss:effop}

The computation times for CC theory and the IMSRG both scale in a polynomial
way with $A$, but both methods need to be modified in open-shell nuclei like
those that are used in \bb decay experiments.  One approach to these nuclei is
to use the methods indirectly, to construct effective interactions and
operators for use in shell-model configuration spaces. Then one can simply
diagonalize the effective interaction $H_{\text{eff}}$ and compute the
transition matrix element of the effective \bb decay operator in exactly the
same way as with the usual (phenomenological) shell model.

One of the oldest procedures for constructing effective operators, and the one
that is used in conjunction with CC theory, employs what is known as an
Okubo-Lee-Suzuki mapping~\cite{oku54,lee80,suzuki80}. The idea is to do
\textit{ab initio} calculations in the closed-shell nucleus --- the shell-model
``core'' --- in the nucleus with one nucleon outside the core, and in the
nucleus with two nucleons outside the core \cite{lis08}. (The procedure can be
continued, though it gets increasingly complicated).  CC theory uses the
equations-of-motion method in the nuclei with one or two nucleons outside the
core.  With the lowest few states in these nuclei computed, one obtains the
energy of the shell-model core from the ground-state energy in the CC
closed-shell calculation, and the shell-model single-particle energies from the
lowest CC energies in the nuclei with one additional nucleon.  To get
the two-body matrix elements that define an effective interaction
$H_{\text{eff}}$ in the dimension-$d$ shell-model configuration space, one
projects $d$ CC eigenstates $\ket{k}$ in the nuclei with two particles outside
the core onto the shell-model space, and then orthogonalizes the projected
states to get the shell-model states $\ket{\tilde{k}}$.  The orthogonalization
procedure, sometimes called ``symmetric orthogonalization'' minimizes the
quantity
\begin{equation}
\label{eq:lowdin}
\sum_{k=1}^d \left( \bra{k}-\bra{\tilde{k}} \right) \left(
\ket{k}-\ket{\tilde{k}} \right)\,,
\end{equation}
so that the shell-model states are, on average, as close as possible to the
original CC sates.  With the mapping specified, one simply sets the matrix
elements of all two-body effective operators between the shell model
$\ket{\tilde{k}}$'s to equal those of the ``bare'' operators between the CC
$\ket{k}$'s.  The program has been carried out in nuclei with a valence $sd$
shell~\cite{Jansen14,Jansen16}.  

In the IMSRG, the procedure for obtaining effective shell-model operators is a
little different~\cite{Hergert16}. The idea is to construct the generator $\eta$
in Eq.\ \eqref{eq:generator} so that the Hamiltonian does not connect states in
which the core is filled and the ``active" particles are all in the shell-model
configuration space to the rest of the Hilbert space.  The resulting Hamiltonian
$H(s=\infty)$ is the effective shell-model interaction $H_{\text{eff}}$.  To
capture residual three-nucleon interactions among active particles (generated by
the SRG flow) an ensemble ``reference state,'' yielding partially occupied
orbitals, is used for the normal-ordering~\cite{Stroberg16a}.  Since the
ensemble depends on the nucleus being studied, a different $H_{\text{eff}}$
results for each isotope.  As with the simpler version of the IMSRG discussed
above, one obtains the effective \bbz decay operator by applying the same
unitary transformation that generates $H_{\text{eff}}$ to the bare \bbz decay
operator. IMSRG calculations of this sort have shown the ability to compute
binding energies and spectra in nuclei with valence $sd$- and $pf$-shell
nucleons~\cite{bog14,Stroberg16,Stroberg16a}.

\begin{figure}[t]
\begin{center}
\includegraphics[width=.48\textwidth]{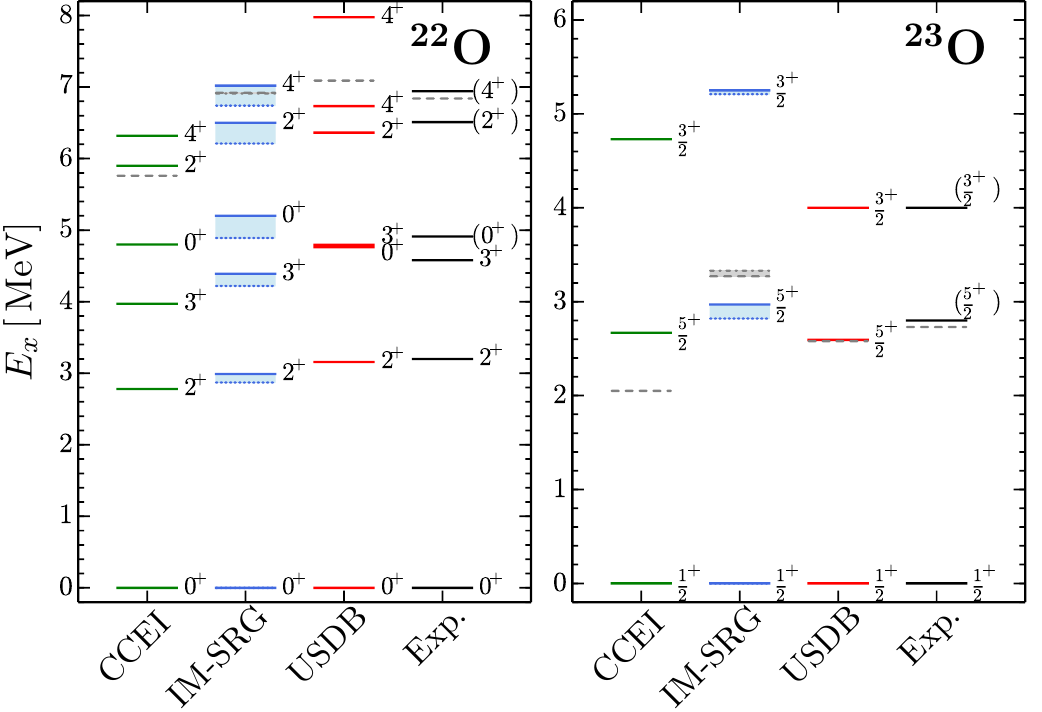}
\end{center}
\caption{Excitation spectra of two neutron-rich oxygen isotopes, with effective
interactions obtained in coupled-clusters theory (CCEI) and the IMSRG (IM-SRG),
compared with spectra produced by a good phenomenological shell-model
interaction (USDB~\cite{brown06}), and with experiment.
The bands in the IM-SRG results indicate the uncertainty
associated with the choice of frequency for the harmonic oscillator potential
that generates basis states.
Figure taken from Ref.~\cite{Hergert16a}.}
\label{f:effh}
\end{figure}

The \textit{ab initio} shell model program is still in its early stages, both in
CC theory and the IMSRG, but has already generated promising results.  Figure
\ref{f:effh} shows spectra computed by the shell-model effective interactions
that result from both procedures. The spectra are not exact for a number of
reasons: 1) High-order pieces of the $\chi$EFT Hamiltonian are neglected, 2) the
CC expansion and the IMSRG flow equation are truncated, 3) the Okubo-Lee-Suzuki
mapping is carried out only up to two valence particles (in CC theory).
Nevertheless, the results are comparable to those obtained with a standard
phenomenological shell-model interaction, which was fit to spectroscopic data in
the same mass region.  The \textit{ab initio} approaches have clear promise.

The computation of matrix elements for other operators is just beginning, and it
remains to be seen how successful it will be.  There is no reason in principle,
however, that matrix elements of single-$\beta$ and \bb decay operators should
be much less accurate than those of the effective interaction.  The framework
offers an opportunity to make the shell model a much more rigorous tool for \bb
decay calculations.  Researchers will pursue it vigorously.

\subsubsection{Multi-reference IMSRG}
\label{sss:multiref}

To the extent that the closure approximation is good (see Sec.~\ref{sss:LNE}),
only ground states in the initial and final nuclei are required to calculate the
\bbz decay transition rate. The effective-operator methods for use in the
shell-model, just described in Sec.~\ref{sss:effop}, are quite general and
produce information not only about the ground state but also about a whole set
of low-lying states.  An approach that restricts itself to ground states might be more efficient.  Both the CC and IMSRG procedures, as described in
Sec.~\ref{sss:kinds}, begin with a Slater determinant as an approximate ground
state, a reference state to which they add correlations.  With the exception of
$^{48}$Ca, however, none of the \bb emitters of potential use can be described
even approximately by a Slater determinant.

Fortunately, both the CC and IMSRG approaches can be extended to more general
reference states.  The extension is perhaps easier in the IMSRG, which relies
on the reference only for the normal ordering of many-body operators that
includes their most important effects within zero-, one-, and two-body
operators.  And normal ordering has, in fact, been generalized to arbitrary
reference states, most prominently in Refs.\ \cite{Kutzelnigg97} and
\cite{Mukherjee97}.  The generalized normal ordering has most of the benefits
of the usual normal ordering, including an associated (generalized) Wick's
theorem that expresses the product of an $\mathcal{N}$-body normal-ordered
operator with an $\mathcal{M}$-body normal-ordered operator as a linear
combination of normal-ordered operators of rank less than or equal to
$\mathcal{N}+\mathcal{M}$.  The only drawback is that the contractions that
enable this procedure generate up to $(\mathcal{N}+\mathcal{M})$-body density
matrices, rather than just the one-body density matrices produced by the usual
Wick's theorem.  But that is a small price for the benefit:  The generalized
normal ordering allows one to include most of the effects of three- and
higher-body operators generated by the flow equations by discarding just the
normal-ordered pieces of these operators, making the equations as tractable as
in the ordinary IMSRG~\cite{Hergert16a}.  The use of a generalized reference
state in this way is usually called the multi-reference IMSRG.

Thus far, the method has been applied only to spherical nuclei (e.g.\ calcium
and nickel isotopes) with reference states obtained from spherical HFB
calculations~\cite{Hergert16}, but there is no reason it cannot be applied more
generally, with a shell-model reference state or one from an EDF.  That would
allow important collective correlations to be included directly in the
reference, with the flow equations adding the most important parts of other
less identifiable correlations.  Work in this direction is in
progress~\cite{Hergert16a}.

\subsection{Assessing Error}
\label{ss:error}

At present, nothing resembling a solid quantitative measure of uncertainty in
the \bbz matrix element exists. In Fig.~\ref{f:NMEcompare} the error bars on
some of the QRPA values represent only the spread in results coming from the
use of two different nucleon-nucleon interactions, a crude sort of statistical
uncertainty. The bars on some of the shell-model values come from the use of
two prescriptions for including short-range correlations and represent a crude
estimate of a part of what we will call the systematic uncertainty associated
with the model.

With the lack of real uncertainty estimates, experimentalists usually rely on
the range of numbers obtained by seemingly reliable models (e.g. the one spanned
in Fig.~\ref{f:NMEcompare}).  The problems with that approach to uncertainty are
obvious: on the one hand the range is large and may include the results of poor
calculations and, on the other, all the models may be omitting the same
important physics (such as that associated with the $g_A$ problem discussed in
Sec.~\ref{s:ga}).  The point we have made in this review is that models must
incorporate all known correlations relevant for \bbz decay and reproduce related
observables.  Capturing the important correlations is easiest in an \textit{ab
initio} calculation that systematically controls error at each step, even if
that control is imperfect. Phenomenological models can be extended to include
more correlations in the ways discussed in Sec.~\ref{ss:imp_present}, and can be
benchmarked against \textit{ab initio} efforts. In this way all the many-body
methods we have discussed can be useful in improving the accuracy of matrix
elements and estimating their uncertainty. 

The absence of error quantification in \bbz decay is not an exception.  Because
it has been traditionally been dominated by phenomenological approaches,
nuclear-structure theory has traditionally neglected even to try to estimate
uncertainty.  This situation is slowly changing, however, and {\it ab initio}
methods and $\chi$EFT forces provide natural tools to quantify theoretical
error.  References~\cite{Epelbaum15,Furnstahl:2015rha,Carlsson16} have taken the
first steps in quantifying error, and Ref.\ \cite{Simonis16} provides initial
estimates of theoretical uncertainty in $sd$-shell nuclei.  We can expect this sort of work to progress rapidly. 

Dividing error estimates into two kinds --- statistical and systematic --- is
useful.  Statistical error is probably easier to assess.  We discuss both
below, in turn.  

\subsubsection{Statistical Error}
\label{sss:params}

By statistical error we mean, roughly, the uncertainty due to the choice of
parameters within a given model or method.  The matrix elements of the
effective interaction $H_{\text{eff}}$ for the shell model, the value of the
parameter $g_{pp}$ in the QRPA, the effective value of the spin-isospin
coupling $g_A$ in all phenomenological models, and the low-energy constants
that specify the $\chi$EFT forces and currents in \textit{ab initio}
calculations are all examples of model parameters.  No matter what statistical
protocols are used, the statistical uncertainty in \bbz matrix elements will
emerge from an analysis of the correlation between them and other observables
for which we have data.  Rates of single-$\beta$ decay and \bbt decay will be
strongly correlated.  Gamow-Teller transition strengths, excitation spectra,
occupation numbers, and two-neutron transfer
probabilities~\cite{Freeman12,Brown14} can also be correlated. These
observables will constrain parameters and, in turn, the predictions for \bbz
matrix elements.

Some analysis of this kind has been carried out in the QRPA~\cite{rod06,sim13};
the \bbz matrix elements tend to vary by about one unit when parameters are
changed.  Reference~\cite{fae09} examined the correlations among matrix elements
in different isotopes.  References\ \cite{hor10,men11a} examined the sensitivity
of matrix elements to $H_{\text{eff}}$ in the shell model.  Again, as can be
seen in Fig.~\ref{f:NMEcompare}, results vary by about one unit.  This amount is
similar to the average difference between matrix elements calculated with
non-relativistic and relativistic EDFs. (Only in \elemA{150}Nd is the difference
much larger.)  All these results suggest that systematic uncertainty is larger
than statistical uncertainty.

The statistical uncertainty related to $g_A$ has been explored via the
correlation of this parameter with rates of electron capture, single-$\beta$
decay, and \bbt decay data, first for selected isotopes~\cite{fae08} and more
recently in a more complete study for a wide range of nuclei~\cite{Deppisch16}.
References\ \cite{men11,eng14,men16a} studied the sensitivity of $g_A$ to
variations of the couplings in the $\chi$EFT two-body currents. More
comprehensive work in the same vein is important.  We need to examine
sensitivity in other computational approaches and correlations with other
observables in the same way.

\subsubsection{Systematic Error and $g_A$}
\label{sss:systematics}

Systematic error, the uncertainty related to the insufficiency of models, is
harder to assess.  Once a model incorporates as many of the
known-to-be-important correlations as possible --- isoscalar pairing, or
like-particle pairing correlations involving single-particle orbitals away from
the Fermi level, for example --- its deficiencies can only be assessed by a
confrontation with data and with other models.  Here there is a lot to do.  The
statistical error analysis just discussed can shed light on systematic error as
well; if there are no sensible sets of parameters that lead to an accurate
description of the available data, we know that our model is systematically
deficient~\cite{dobaczewski14}.

Benchmarking is another important tool.  The \textit{ab initio} methods reviewed
here unavoidably entail some kind of truncation, but a different kind in each.
For instance, CC theory and the IMSRG both impose a cutoff on the maximum
numbers of particles within a cluster (roughly speaking) or the maximum number
of single-particle oscillator orbitals.  Meanwhile, the NCSM instead allows only
configurations with excitation energy below a certain cutoff and the GFMC
discretizes time and constrains paths to cope with the fermion
sign problem.  Comparing the predictions of these approaches in nuclei for
which they are all implementable --- e.g.\ nuclei with a valence $sd$ shell ---
can reveal a lot.  Agreement between ground state energies of oxygen isotopes in
this kind of inter-method test is at the few percent level \cite{Hebeler15}.
There is no reason not to test calculations of \bbz matrix elements in the same
way, even though the corresponding transitions are not observable.  Benchmarking
complementary methods can reveal the importance of each kind of truncation, and
agreement among methods could reduce uncertainty considerably. 

The renormalization of the spin-isospin operator ($g_A$ quenching) is
particularly ripe for analysis in light nuclei.  The \textit{ab initio} methods
appear powerful enough to include all important correlations as well as the
leading two-body currents in those isotopes, for which there is no shortage of
measured single-$\beta$ decay rates.  The analysis can also include magnetic
moments and transitions, which, like $\beta$-decay, involve the spin operator
and sometimes require that its strength be quenched to agree with data.  If {\it
ab initio} calculations all fail to reproduce experimental $\beta$-decay rates,
then there will have to be some other source of systematic uncertainty that we
cannot currently imagine.  If, on the other hand, one or more succeed, the
result will be a reduction of systematic uncertainty in the \bbz decay matrix
elements and a clear path to \textit{ab initio} \bb decay calculations in the
heavier nuclei used or contemplated by experimentalists. Understanding the cause
of quenching will also allow the development of realistic procedures for
improving the more phenomenological methods. 

These tests and similar ones in heavier nuclei, where data for \bbt decay can
be directly confronted, will not in themselves yield a precise error bar for
\bbz matrix elements, especially for those in \elemA{76}Ge or 
\elemA{136}Xe. One can never be sure that all sources of systematic
error are understood.  But a good protocol (e.g.\ based on popular Bayesian
methods) should lead to error bars in which we can be reasonably confident,
much more confident than we are in the ``uncertainty'' associated with the
spread in the predictions of only partially tested models.  A good protocol
will require collaboration among all kinds of theorists.  The necessary
groundwork for such collaboration now exists.

\section{Summary and Prospects}
\label{s:conclusion}

Next-generation experiments could well be in a position to observe \bbz decay if
neutrinos are Majorana particles and neutrino masses are arranged according to
the inverted hierarchy.  They may also be able to discover new fundamental
physics even if the mass hierarchy is normal.  The rate of any kind of \bb
decay, however, depends on nuclear matrix elements, and informed decisions about
which and how much material to use in experiments rely on our ability to
calculate them accurately.  If \bbz decay is actually observed, the matrix
elements will play a key role in identifying the mechanism responsible and/or
extracting information about neutrino masses.

At present, a handful of many-body methods produce matrix elements that differ
from one another by factors of up to about three.  The range suggests that each
approach is missing important physics.  We have identified the most obvious
weaknesses of each approach --- usually either a restricted configuration space
or simplified correlations --- and have discussed programs to improve each of
them.  Phenomenological models are not in themselves sufficient to produce truly
reliable matrix elements, however, as the longstanding need to renormalize $g_A$
makes clear.  Without a more systematic approach to nuclear structure we are
unlikely either to understand the implications of $g_A$ quenching for \bbz decay
or to assign meaningful theoretical uncertainty to matrix elements. 

We are hopeful that \textit{ab initio} nuclear-structure theory will help with
these tasks.  Once limited to very light nuclei, \textit{ab initio} methods have
progressed rapidly and will soon allow the matrix element of the lightest \bb
emitter, \elemA{48}Ca, to be computed. They promise to do the same in the
heavier isotopes that are currently used in experiments: $^{76}$Ge, $^{130}$Te,
and $^{136}$Xe.  Comparing the by now wide variety of \textit{ab initio} methods
to one another in lighter nuclei will facilitate the assignment of uncertainty.
The \textit{ab initio} methods can also revitalize traditional models, e.g.\ by
supplying input for the shell model or using states produced by EDF calculations
as references.  And they provide tools that should let us, finally, solve the
$g_A$ problem.  That step alone will significantly reduce the systematic
uncertainty in \bbz matrix elements.

We are in an exciting time.  After years of development, \bb decay experiments
are poised to scale up, opening new windows into fundamental physics.  We are
optimistic that nuclear theory will soon produce the accurate matrix elements
and controlled theoretical error that these experiments demand.

\begin{acknowledgments}
We would like to thank B. A. Brown, E. Caurier, S. R. Elliott, D. Gazit, R.
Henning, N. Hinohara, J. D. Holt, M. Horoi, F. Iachello, C. F.\ Jiao, Y. Iwata,
G. Mart{\'inez}-Pinedo, M. T.\ Mustonen, F. Nowacki, T. Otsuka, A. Poves, T. R.
Rodr{\'i}guez, F. {\v S}imkovic, A. Schwenk, N. Shimizu, P. Vogel, and J. M. Yao
for many useful discussions on the topics we covered in this review, and H.
Hergert, T. Papenbrock, S. Pastore, T. R. Rodr{\'i}guez and S. R. Stroberg for
useful comments on the manuscript.  J. E.\ acknowledges support from DOE grants
DE-FG02-97ER41019, DE-SC0008641, and DE-SC0004142. J. M.\ was supported by an
International Research Fellowship from JSPS and JSPS Grant-in-Aid for Scientific
Research No.\ 26$\cdot$04323. 
\end{acknowledgments}

\bibliography{bb-ropp}

\begin{thebibliography}{311}%
\makeatletter
\providecommand \@ifxundefined [1]{%
 \@ifx{#1\undefined}
}%
\providecommand \@ifnum [1]{%
 \ifnum #1\expandafter \@firstoftwo
 \else \expandafter \@secondoftwo
 \fi
}%
\providecommand \@ifx [1]{%
 \ifx #1\expandafter \@firstoftwo
 \else \expandafter \@secondoftwo
 \fi
}%
\providecommand \natexlab [1]{#1}%
\providecommand \enquote  [1]{``#1''}%
\providecommand \bibnamefont  [1]{#1}%
\providecommand \bibfnamefont [1]{#1}%
\providecommand \citenamefont [1]{#1}%
\providecommand \href@noop [0]{\@secondoftwo}%
\providecommand \href [0]{\begingroup \@sanitize@url \@href}%
\providecommand \@href[1]{\@@startlink{#1}\@@href}%
\providecommand \@@href[1]{\endgroup#1\@@endlink}%
\providecommand \@sanitize@url [0]{\catcode `\\12\catcode `\$12\catcode
  `\&12\catcode `\#12\catcode `\^12\catcode `\_12\catcode `\%12\relax}%
\providecommand \@@startlink[1]{}%
\providecommand \@@endlink[0]{}%
\providecommand \url  [0]{\begingroup\@sanitize@url \@url }%
\providecommand \@url [1]{\endgroup\@href {#1}{\urlprefix }}%
\providecommand \urlprefix  [0]{URL }%
\providecommand \Eprint [0]{\href }%
\providecommand \doibase [0]{http://dx.doi.org/}%
\providecommand \selectlanguage [0]{\@gobble}%
\providecommand \bibinfo  [0]{\@secondoftwo}%
\providecommand \bibfield  [0]{\@secondoftwo}%
\providecommand \translation [1]{[#1]}%
\providecommand \BibitemOpen [0]{}%
\providecommand \bibitemStop [0]{}%
\providecommand \bibitemNoStop [0]{.\EOS\space}%
\providecommand \EOS [0]{\spacefactor3000\relax}%
\providecommand \BibitemShut  [1]{\csname bibitem#1\endcsname}%
\let\auto@bib@innerbib\@empty
\bibitem [{\citenamefont {Henning}(2016)}]{Henning16}%
  \BibitemOpen
  \bibfield  {author} {\bibinfo {author} {\bibfnamefont {Reyco}\ \bibnamefont
  {Henning}},\ }\bibfield  {title} {\enquote {\bibinfo {title} {{Current status
  of neutrinoless double-beta decay searches}},}\ }\href {\doibase
  10.1016/j.revip.2016.03.001} {\bibfield  {journal} {\bibinfo  {journal} {Rev.
  Phys.}\ }\textbf {\bibinfo {volume} {1}},\ \bibinfo {pages} {29--35}
  (\bibinfo {year} {2016})}\BibitemShut {NoStop}%
\bibitem [{\citenamefont {Dell'Oro}\ \emph {et~al.}(2016)\citenamefont
  {Dell'Oro}, \citenamefont {Marcocci}, \citenamefont {Viel},\ and\
  \citenamefont {Vissani}}]{DellOro16}%
  \BibitemOpen
  \bibfield  {author} {\bibinfo {author} {\bibfnamefont {Stefano}\ \bibnamefont
  {Dell'Oro}}, \bibinfo {author} {\bibfnamefont {Simone}\ \bibnamefont
  {Marcocci}}, \bibinfo {author} {\bibfnamefont {Matteo}\ \bibnamefont {Viel}},
  \ and\ \bibinfo {author} {\bibfnamefont {Francesco}\ \bibnamefont
  {Vissani}},\ }\bibfield  {title} {\enquote {\bibinfo {title} {{Neutrinoless
  double beta decay: 2015 review}},}\ }\href {\doibase 10.1155/2016/2162659}
  {\bibfield  {journal} {\bibinfo  {journal} {Adv. High Energy Phys.}\ }\textbf
  {\bibinfo {volume} {2016}},\ \bibinfo {pages} {2162659} (\bibinfo {year}
  {2016})}\BibitemShut {NoStop}%
\bibitem [{\citenamefont {Cremonesi}\ and\ \citenamefont
  {Pavan}(2014)}]{Cremonesi14}%
  \BibitemOpen
  \bibfield  {author} {\bibinfo {author} {\bibfnamefont {O.}~\bibnamefont
  {Cremonesi}}\ and\ \bibinfo {author} {\bibfnamefont {M.}~\bibnamefont
  {Pavan}},\ }\bibfield  {title} {\enquote {\bibinfo {title} {{Challenges in
  Double Beta Decay}},}\ }\href {\doibase 10.1155/2014/951432} {\bibfield
  {journal} {\bibinfo  {journal} {Adv. High Energy Phys.}\ }\textbf {\bibinfo
  {volume} {2014}},\ \bibinfo {pages} {951432} (\bibinfo {year}
  {2014})}\BibitemShut {NoStop}%
\bibitem [{\citenamefont {G{\'o}mez-Cadenas}\ \emph {et~al.}(2012)\citenamefont
  {G{\'o}mez-Cadenas}, \citenamefont {Mart{\'i}n-Albo}, \citenamefont
  {Mezzetto}, \citenamefont {Monrabal},\ and\ \citenamefont
  {Sorel}}]{GomezCadenas12}%
  \BibitemOpen
  \bibfield  {author} {\bibinfo {author} {\bibfnamefont {J.~J.}\ \bibnamefont
  {G{\'o}mez-Cadenas}}, \bibinfo {author} {\bibfnamefont {J.}~\bibnamefont
  {Mart{\'i}n-Albo}}, \bibinfo {author} {\bibfnamefont {M.}~\bibnamefont
  {Mezzetto}}, \bibinfo {author} {\bibfnamefont {F.}~\bibnamefont {Monrabal}},
  \ and\ \bibinfo {author} {\bibfnamefont {M.}~\bibnamefont {Sorel}},\
  }\bibfield  {title} {\enquote {\bibinfo {title} {{The search for neutrinoless
  double beta decay}},}\ }\href {\doibase 10.1393/ncr/i2012-10074-9} {\bibfield
   {journal} {\bibinfo  {journal} {Riv. Nuovo Cim.}\ }\textbf {\bibinfo
  {volume} {35}},\ \bibinfo {pages} {29--98} (\bibinfo {year}
  {2012})}\BibitemShut {NoStop}%
\bibitem [{\citenamefont {Gando}\ \emph {et~al.}(2016)\citenamefont {Gando}
  \emph {et~al.}}]{KamLAND-Zen16}%
  \BibitemOpen
  \bibfield  {author} {\bibinfo {author} {\bibfnamefont {A.}~\bibnamefont
  {Gando}} \emph {et~al.} (\bibinfo {collaboration} {KamLAND-Zen
  Collaboration}),\ }\bibfield  {title} {\enquote {\bibinfo {title} {{Search
  for Majorana Neutrinos near the Inverted Mass Hierarchy Region with
  KamLAND-Zen}},}\ }\href {\doibase 10.1103/PhysRevLett.117.109903} {\bibfield
  {journal} {\bibinfo  {journal} {Phys. Rev. Lett.}\ }\textbf {\bibinfo
  {volume} {117}},\ \bibinfo {pages} {082503} (\bibinfo {year}
  {2016})}\BibitemShut {NoStop}%
\bibitem [{\citenamefont {Albert}\ \emph {et~al.}(2014)\citenamefont {Albert}
  \emph {et~al.}}]{EXO14}%
  \BibitemOpen
  \bibfield  {author} {\bibinfo {author} {\bibfnamefont {J.~B.}\ \bibnamefont
  {Albert}} \emph {et~al.} (\bibinfo {collaboration} {EXO-200 Collaboration}),\
  }\bibfield  {title} {\enquote {\bibinfo {title} {{Search for Majorana
  neutrinos with the first two years of EXO-200 data}},}\ }\href {\doibase
  10.1038/nature13432} {\bibfield  {journal} {\bibinfo  {journal} {Nature}\
  }\textbf {\bibinfo {volume} {510}},\ \bibinfo {pages} {229--234} (\bibinfo
  {year} {2014})}\BibitemShut {NoStop}%
\bibitem [{\citenamefont {Mart{\'i}n-Albo}\ \emph {et~al.}(2016)\citenamefont
  {Mart{\'i}n-Albo} \emph {et~al.}}]{Next16}%
  \BibitemOpen
  \bibfield  {author} {\bibinfo {author} {\bibfnamefont {J.}~\bibnamefont
  {Mart{\'i}n-Albo}} \emph {et~al.} (\bibinfo {collaboration} {NEXT
  Collaboration}),\ }\bibfield  {title} {\enquote {\bibinfo {title}
  {{Sensitivity of NEXT-100 to neutrinoless double beta decay}},}\ }\href
  {\doibase 10.1007/JHEP05(2016)159} {\bibfield  {journal} {\bibinfo  {journal}
  {JHEP}\ }\textbf {\bibinfo {volume} {05}},\ \bibinfo {pages} {159} (\bibinfo
  {year} {2016})}\BibitemShut {NoStop}%
\bibitem [{\citenamefont {Agostini}\ \emph {et~al.}(2013)\citenamefont
  {Agostini} \emph {et~al.}}]{GERDA13}%
  \BibitemOpen
  \bibfield  {author} {\bibinfo {author} {\bibfnamefont {M.}~\bibnamefont
  {Agostini}} \emph {et~al.} (\bibinfo {collaboration} {GERDA Collaboration}),\
  }\bibfield  {title} {\enquote {\bibinfo {title} {{Results on Neutrinoless
  Double-$\beta$ Decay of $^{76}$Ge from Phase I of the GERDA Experiment}},}\
  }\href {\doibase 10.1103/PhysRevLett.111.122503} {\bibfield  {journal}
  {\bibinfo  {journal} {Phys. Rev. Lett.}\ }\textbf {\bibinfo {volume} {111}},\
  \bibinfo {pages} {122503} (\bibinfo {year} {2013})}\BibitemShut {NoStop}%
\bibitem [{\citenamefont {Abgrall}\ \emph {et~al.}(2014)\citenamefont {Abgrall}
  \emph {et~al.}}]{MAJORANA14}%
  \BibitemOpen
  \bibfield  {author} {\bibinfo {author} {\bibfnamefont {N.}~\bibnamefont
  {Abgrall}} \emph {et~al.} (\bibinfo {collaboration} {Majorana
  Collaboration}),\ }\bibfield  {title} {\enquote {\bibinfo {title} {{The
  Majorana Demonstrator Neutrinoless Double-Beta Decay Experiment}},}\ }\href
  {\doibase 10.1155/2014/365432} {\bibfield  {journal} {\bibinfo  {journal}
  {Adv. High Energy Phys.}\ }\textbf {\bibinfo {volume} {2014}},\ \bibinfo
  {pages} {365432} (\bibinfo {year} {2014})}\BibitemShut {NoStop}%
\bibitem [{\citenamefont {Alfonso}\ \emph {et~al.}(2015)\citenamefont {Alfonso}
  \emph {et~al.}}]{CUORE15}%
  \BibitemOpen
  \bibfield  {author} {\bibinfo {author} {\bibfnamefont {K.}~\bibnamefont
  {Alfonso}} \emph {et~al.} (\bibinfo {collaboration} {CUORE Collaboration}),\
  }\bibfield  {title} {\enquote {\bibinfo {title} {{Search for Neutrinoless
  Double-Beta Decay of $^{130}$Te with CUORE-0}},}\ }\href {\doibase
  10.1103/PhysRevLett.115.102502} {\bibfield  {journal} {\bibinfo  {journal}
  {Phys. Rev. Lett.}\ }\textbf {\bibinfo {volume} {115}},\ \bibinfo {pages}
  {102502} (\bibinfo {year} {2015})}\BibitemShut {NoStop}%
\bibitem [{\citenamefont {Andringa}\ \emph {et~al.}(2016)\citenamefont
  {Andringa} \emph {et~al.}}]{SNO+16}%
  \BibitemOpen
  \bibfield  {author} {\bibinfo {author} {\bibfnamefont {S.}~\bibnamefont
  {Andringa}} \emph {et~al.} (\bibinfo {collaboration} {SNO+ Collaboration}),\
  }\bibfield  {title} {\enquote {\bibinfo {title} {{Current Status and Future
  Prospects of the SNO+ Experiment}},}\ }\href {\doibase 10.1155/2016/6194250}
  {\bibfield  {journal} {\bibinfo  {journal} {Adv. High Energy Phys.}\ }\textbf
  {\bibinfo {volume} {2016}},\ \bibinfo {pages} {6194250} (\bibinfo {year}
  {2016})}\BibitemShut {NoStop}%
\bibitem [{\citenamefont {Blot}\ \emph {et~al.}(2016)\citenamefont {Blot} \emph
  {et~al.}}]{SUperNEMO16}%
  \BibitemOpen
  \bibfield  {author} {\bibinfo {author} {\bibfnamefont {S.}~\bibnamefont
  {Blot}} \emph {et~al.} (\bibinfo {collaboration} {NEMO-3 and SuperNEMO
  Collaborations}),\ }\bibfield  {title} {\enquote {\bibinfo {title}
  {{Investigating ββ decay with the NEMO-3 and SuperNEMO experiments}},}\
  }\href {\doibase 10.1088/1742-6596/718/6/062006} {\bibfield  {journal}
  {\bibinfo  {journal} {J. Phys. Conf. Ser.}\ }\textbf {\bibinfo {volume}
  {718}},\ \bibinfo {pages} {062006} (\bibinfo {year} {2016})}\BibitemShut
  {NoStop}%
\bibitem [{\citenamefont {Beeman}\ \emph {et~al.}(2015)\citenamefont {Beeman}
  \emph {et~al.}}]{LUCIFER15}%
  \BibitemOpen
  \bibfield  {author} {\bibinfo {author} {\bibfnamefont {J.~W.}\ \bibnamefont
  {Beeman}} \emph {et~al.} (\bibinfo {collaboration} {LUCIFER Collaboration}),\
  }\bibfield  {title} {\enquote {\bibinfo {title} {{Double-beta decay
  investigation with highly pure enriched $^{82}$Se for the LUCIFER
  experiment}},}\ }\href {\doibase 10.1140/epjc/s10052-015-3822-x} {\bibfield
  {journal} {\bibinfo  {journal} {Eur. Phys. J. C}\ }\textbf {\bibinfo {volume}
  {75}},\ \bibinfo {pages} {591} (\bibinfo {year} {2015})}\BibitemShut
  {NoStop}%
\bibitem [{\citenamefont {Iida}\ \emph {et~al.}(2016)\citenamefont {Iida} \emph
  {et~al.}}]{CANDLES16}%
  \BibitemOpen
  \bibfield  {author} {\bibinfo {author} {\bibfnamefont {T.}~\bibnamefont
  {Iida}} \emph {et~al.} (\bibinfo {collaboration} {CANDLES Collaboration}),\
  }\bibfield  {title} {\enquote {\bibinfo {title} {{Status and future prospect
  of $^{48}$Ca double beta decay search in CANDLES}},}\ }\href {\doibase
  10.1088/1742-6596/718/6/062026} {\bibfield  {journal} {\bibinfo  {journal}
  {J. Phys. Conf. Ser.}\ }\textbf {\bibinfo {volume} {718}},\ \bibinfo {pages}
  {062026} (\bibinfo {year} {2016})}\BibitemShut {NoStop}%
\bibitem [{\citenamefont {Park}(2016)}]{AMoRE16}%
  \BibitemOpen
  \bibfield  {author} {\bibinfo {author} {\bibfnamefont {HyangKyu}\
  \bibnamefont {Park}} (\bibinfo {collaboration} {AMoRE Collaboration}),\
  }\bibfield  {title} {\enquote {\bibinfo {title} {{The AMoRE: Search for
  Neutrinoless Double Beta Decay in 100 Mo}},}\ }\href {\doibase
  10.1016/j.nuclphysbps.2015.10.012} {\bibfield  {journal} {\bibinfo  {journal}
  {Nucl. Part. Phys. Proc.}\ }\textbf {\bibinfo {volume} {273-275}},\ \bibinfo
  {pages} {2630--2632} (\bibinfo {year} {2016})}\BibitemShut {NoStop}%
\bibitem [{\citenamefont {Ebert}\ \emph {et~al.}(2016)\citenamefont {Ebert}
  \emph {et~al.}}]{COBRA16}%
  \BibitemOpen
  \bibfield  {author} {\bibinfo {author} {\bibfnamefont {J.}~\bibnamefont
  {Ebert}} \emph {et~al.} (\bibinfo {collaboration} {COBRA Collaboration}),\
  }\bibfield  {title} {\enquote {\bibinfo {title} {{The COBRA demonstrator at
  the LNGS underground laboratory}},}\ }\href {\doibase
  10.1016/j.nima.2015.10.079} {\bibfield  {journal} {\bibinfo  {journal} {Nucl.
  Instrum. Meth. A}\ }\textbf {\bibinfo {volume} {807}},\ \bibinfo {pages}
  {114--120} (\bibinfo {year} {2016})}\BibitemShut {NoStop}%
\bibitem [{\citenamefont {Dokania}\ \emph {et~al.}()\citenamefont {Dokania},
  \citenamefont {Singh}, \citenamefont {Ghosh}, \citenamefont {Mathimalar},
  \citenamefont {Garai}, \citenamefont {Pal}, \citenamefont {Nanal},
  \citenamefont {Pillay}, \citenamefont {Shrivastava},\ and\ \citenamefont
  {Bhushan}}]{TINTIN15}%
  \BibitemOpen
  \bibfield  {author} {\bibinfo {author} {\bibfnamefont {Neha}\ \bibnamefont
  {Dokania}}, \bibinfo {author} {\bibfnamefont {V.}~\bibnamefont {Singh}},
  \bibinfo {author} {\bibfnamefont {C.}~\bibnamefont {Ghosh}}, \bibinfo
  {author} {\bibfnamefont {S.}~\bibnamefont {Mathimalar}}, \bibinfo {author}
  {\bibfnamefont {A.}~\bibnamefont {Garai}}, \bibinfo {author} {\bibfnamefont
  {S.}~\bibnamefont {Pal}}, \bibinfo {author} {\bibfnamefont {V.}~\bibnamefont
  {Nanal}}, \bibinfo {author} {\bibfnamefont {R.~G.}\ \bibnamefont {Pillay}},
  \bibinfo {author} {\bibfnamefont {A.}~\bibnamefont {Shrivastava}}, \ and\
  \bibinfo {author} {\bibfnamefont {K.~G.}\ \bibnamefont {Bhushan}},\
  }\bibfield  {title} {\enquote {\bibinfo {title} {{Radiation Background
  Studies for 0$\nu\beta\beta$ decay in $^{124}$Sn}},}\ }\href@noop {} {\
  }\Eprint {http://arxiv.org/abs/1504.05433} {arXiv:1504.05433} \BibitemShut
  {NoStop}%
\bibitem [{\citenamefont {Fukuda}(2016)}]{ZICOS16}%
  \BibitemOpen
  \bibfield  {author} {\bibinfo {author} {\bibfnamefont {Yoshiyuki}\
  \bibnamefont {Fukuda}},\ }\bibfield  {title} {\enquote {\bibinfo {title}
  {{ZICOS - New project for neutrinoless double beta decay experiment using
  zirconium complex in liquid scintillator}},}\ }\href {\doibase
  10.1088/1742-6596/718/6/062019} {\bibfield  {journal} {\bibinfo  {journal}
  {J. Phys. Conf. Ser.}\ }\textbf {\bibinfo {volume} {718}},\ \bibinfo {pages}
  {062019} (\bibinfo {year} {2016})}\BibitemShut {NoStop}%
\bibitem [{\citenamefont {Epelbaum}\ \emph {et~al.}(2009)\citenamefont
  {Epelbaum}, \citenamefont {Hammer},\ and\ \citenamefont
  {Mei{\ss}ner}}]{chiral}%
  \BibitemOpen
  \bibfield  {author} {\bibinfo {author} {\bibfnamefont {E.}~\bibnamefont
  {Epelbaum}}, \bibinfo {author} {\bibfnamefont {H.-W.}\ \bibnamefont
  {Hammer}}, \ and\ \bibinfo {author} {\bibfnamefont {U.-G.}\ \bibnamefont
  {Mei{\ss}ner}},\ }\bibfield  {title} {\enquote {\bibinfo {title} {Modern
  theory of nuclear forces},}\ }\href {\doibase 10.1103/RevModPhys.81.1773}
  {\bibfield  {journal} {\bibinfo  {journal} {Rev. Mod. Phys.}\ }\textbf
  {\bibinfo {volume} {81}},\ \bibinfo {pages} {1773} (\bibinfo {year}
  {2009})}\BibitemShut {NoStop}%
\bibitem [{\citenamefont {Machleidt}\ and\ \citenamefont
  {Entem}(2011)}]{Machleidt11}%
  \BibitemOpen
  \bibfield  {author} {\bibinfo {author} {\bibfnamefont {R.}~\bibnamefont
  {Machleidt}}\ and\ \bibinfo {author} {\bibfnamefont {D.~R.}\ \bibnamefont
  {Entem}},\ }\bibfield  {title} {\enquote {\bibinfo {title} {{Chiral effective
  field theory and nuclear forces}},}\ }\href {\doibase
  10.1016/j.physrep.2011.02.001} {\bibfield  {journal} {\bibinfo  {journal}
  {Phys. Rept.}\ }\textbf {\bibinfo {volume} {503}},\ \bibinfo {pages} {1--75}
  (\bibinfo {year} {2011})}\BibitemShut {NoStop}%
\bibitem [{\citenamefont {Hammer}\ \emph {et~al.}(2013)\citenamefont {Hammer},
  \citenamefont {Nogga},\ and\ \citenamefont {Schwenk}}]{Hammer13}%
  \BibitemOpen
  \bibfield  {author} {\bibinfo {author} {\bibfnamefont {Hans-Werner}\
  \bibnamefont {Hammer}}, \bibinfo {author} {\bibfnamefont {Andreas}\
  \bibnamefont {Nogga}}, \ and\ \bibinfo {author} {\bibfnamefont {Achim}\
  \bibnamefont {Schwenk}},\ }\bibfield  {title} {\enquote {\bibinfo {title}
  {{Three-body forces: From cold atoms to nuclei}},}\ }\href {\doibase
  10.1103/RevModPhys.85.197} {\bibfield  {journal} {\bibinfo  {journal} {Rev.
  Mod. Phys.}\ }\textbf {\bibinfo {volume} {85}},\ \bibinfo {pages} {197}
  (\bibinfo {year} {2013})}\BibitemShut {NoStop}%
\bibitem [{\citenamefont {Machleidt}\ and\ \citenamefont
  {Sammarruca}(2016)}]{Machleidt:2016rvv}%
  \BibitemOpen
  \bibfield  {author} {\bibinfo {author} {\bibfnamefont {R.}~\bibnamefont
  {Machleidt}}\ and\ \bibinfo {author} {\bibfnamefont {F.}~\bibnamefont
  {Sammarruca}},\ }\bibfield  {title} {\enquote {\bibinfo {title} {{Chiral EFT
  based nuclear forces: Achievements and challenges}},}\ }\href {\doibase
  10.1088/0031-8949/91/8/083007} {\bibfield  {journal} {\bibinfo  {journal}
  {Phys. Scr.}\ }\textbf {\bibinfo {volume} {91}},\ \bibinfo {pages} {083007}
  (\bibinfo {year} {2016})}\BibitemShut {NoStop}%
\bibitem [{\citenamefont {Carlson}\ \emph {et~al.}(2015)\citenamefont
  {Carlson}, \citenamefont {Gandolfi}, \citenamefont {Pederiva}, \citenamefont
  {Pieper}, \citenamefont {Schiavilla}, \citenamefont {Schmidt},\ and\
  \citenamefont {Wiringa}}]{carlson15}%
  \BibitemOpen
  \bibfield  {author} {\bibinfo {author} {\bibfnamefont {J.}~\bibnamefont
  {Carlson}}, \bibinfo {author} {\bibfnamefont {S.}~\bibnamefont {Gandolfi}},
  \bibinfo {author} {\bibfnamefont {F.}~\bibnamefont {Pederiva}}, \bibinfo
  {author} {\bibfnamefont {Steven~C.}\ \bibnamefont {Pieper}}, \bibinfo
  {author} {\bibfnamefont {R.}~\bibnamefont {Schiavilla}}, \bibinfo {author}
  {\bibfnamefont {K.~E.}\ \bibnamefont {Schmidt}}, \ and\ \bibinfo {author}
  {\bibfnamefont {R.~B.}\ \bibnamefont {Wiringa}},\ }\bibfield  {title}
  {\enquote {\bibinfo {title} {Quantum monte carlo methods for nuclear
  physics},}\ }\href {\doibase 10.1103/RevModPhys.87.1067} {\bibfield
  {journal} {\bibinfo  {journal} {Rev. Mod. Phys.}\ }\textbf {\bibinfo {volume}
  {87}},\ \bibinfo {pages} {1067--1118} (\bibinfo {year} {2015})}\BibitemShut
  {NoStop}%
\bibitem [{\citenamefont {Hebeler}\ \emph
  {et~al.}(2015{\natexlab{a}})\citenamefont {Hebeler}, \citenamefont {Holt},
  \citenamefont {Men{\'e}ndez},\ and\ \citenamefont {Schwenk}}]{Hebeler15}%
  \BibitemOpen
  \bibfield  {author} {\bibinfo {author} {\bibfnamefont {K.}~\bibnamefont
  {Hebeler}}, \bibinfo {author} {\bibfnamefont {J.~D.}\ \bibnamefont {Holt}},
  \bibinfo {author} {\bibfnamefont {J.}~\bibnamefont {Men{\'e}ndez}}, \ and\
  \bibinfo {author} {\bibfnamefont {A.}~\bibnamefont {Schwenk}},\ }\bibfield
  {title} {\enquote {\bibinfo {title} {{Nuclear forces and their impact on
  neutron-rich nuclei and neutron-rich matter}},}\ }\href {\doibase
  10.1146/annurev-nucl-102313-025446} {\bibfield  {journal} {\bibinfo
  {journal} {Ann. Rev. Nucl. Part. Sci.}\ }\textbf {\bibinfo {volume} {65}},\
  \bibinfo {pages} {457--484} (\bibinfo {year}
  {2015}{\natexlab{a}})}\BibitemShut {NoStop}%
\bibitem [{\citenamefont {Hagen}\ \emph {et~al.}(2014)\citenamefont {Hagen},
  \citenamefont {Papenbrock}, \citenamefont {Hjorth-Jensen},\ and\
  \citenamefont {Dean}}]{hagen14}%
  \BibitemOpen
  \bibfield  {author} {\bibinfo {author} {\bibfnamefont {G.}~\bibnamefont
  {Hagen}}, \bibinfo {author} {\bibfnamefont {T.}~\bibnamefont {Papenbrock}},
  \bibinfo {author} {\bibfnamefont {M.}~\bibnamefont {Hjorth-Jensen}}, \ and\
  \bibinfo {author} {\bibfnamefont {D.~J.}\ \bibnamefont {Dean}},\ }\bibfield
  {title} {\enquote {\bibinfo {title} {{Coupled-cluster computations of atomic
  nuclei}},}\ }\href {\doibase 10.1088/0034-4885/77/9/096302} {\bibfield
  {journal} {\bibinfo  {journal} {Rept. Prog. Phys.}\ }\textbf {\bibinfo
  {volume} {77}},\ \bibinfo {pages} {096302} (\bibinfo {year}
  {2014})}\BibitemShut {NoStop}%
\bibitem [{\citenamefont {Hergert}\ \emph {et~al.}(2016)\citenamefont
  {Hergert}, \citenamefont {Bogner}, \citenamefont {Morris}, \citenamefont
  {Schwenk},\ and\ \citenamefont {Tsukiyama}}]{Hergert16}%
  \BibitemOpen
  \bibfield  {author} {\bibinfo {author} {\bibfnamefont {H.}~\bibnamefont
  {Hergert}}, \bibinfo {author} {\bibfnamefont {S.~K.}\ \bibnamefont {Bogner}},
  \bibinfo {author} {\bibfnamefont {T.~D.}\ \bibnamefont {Morris}}, \bibinfo
  {author} {\bibfnamefont {A.}~\bibnamefont {Schwenk}}, \ and\ \bibinfo
  {author} {\bibfnamefont {K.}~\bibnamefont {Tsukiyama}},\ }\bibfield  {title}
  {\enquote {\bibinfo {title} {{The In-Medium Similarity Renormalization Group:
  A Novel Ab Initio Method for Nuclei}},}\ }\href {\doibase
  10.1016/j.physrep.2015.12.007} {\bibfield  {journal} {\bibinfo  {journal}
  {Phys. Rept.}\ }\textbf {\bibinfo {volume} {621}},\ \bibinfo {pages}
  {165--222} (\bibinfo {year} {2016})}\BibitemShut {NoStop}%
\bibitem [{\citenamefont {Som{\`a}}\ \emph {et~al.}(2014)\citenamefont
  {Som{\`a}}, \citenamefont {Cipollone}, \citenamefont {Barbieri},
  \citenamefont {Navr{\'a}til},\ and\ \citenamefont {Duguet}}]{Soma14}%
  \BibitemOpen
  \bibfield  {author} {\bibinfo {author} {\bibfnamefont {V.}~\bibnamefont
  {Som{\`a}}}, \bibinfo {author} {\bibfnamefont {A.}~\bibnamefont {Cipollone}},
  \bibinfo {author} {\bibfnamefont {C.}~\bibnamefont {Barbieri}}, \bibinfo
  {author} {\bibfnamefont {P.}~\bibnamefont {Navr{\'a}til}}, \ and\ \bibinfo
  {author} {\bibfnamefont {T.}~\bibnamefont {Duguet}},\ }\bibfield  {title}
  {\enquote {\bibinfo {title} {{Chiral two- and three-nucleon forces along
  medium-mass isotope chains}},}\ }\href {\doibase 10.1103/PhysRevC.89.061301}
  {\bibfield  {journal} {\bibinfo  {journal} {Phys. Rev. C}\ }\textbf {\bibinfo
  {volume} {89}},\ \bibinfo {pages} {061301} (\bibinfo {year}
  {2014})}\BibitemShut {NoStop}%
\bibitem [{\citenamefont {Mei{\ss}ner}(2016)}]{Meissner16}%
  \BibitemOpen
  \bibfield  {author} {\bibinfo {author} {\bibfnamefont {Ulf-G}\ \bibnamefont
  {Mei{\ss}ner}},\ }\bibfield  {title} {\enquote {\bibinfo {title} {{The long
  and winding road from chiral effective Lagrangians to nuclear structure}},}\
  }\href {\doibase 10.1088/0031-8949/91/3/033005} {\bibfield  {journal}
  {\bibinfo  {journal} {Phys. Scr.}\ }\textbf {\bibinfo {volume} {91}},\
  \bibinfo {pages} {033005} (\bibinfo {year} {2016})}\BibitemShut {NoStop}%
\bibitem [{\citenamefont {{Avignone{ }III}}\ \emph {et~al.}(2008)\citenamefont
  {{Avignone{ }III}}, \citenamefont {Elliott},\ and\ \citenamefont
  {Engel}}]{avi08}%
  \BibitemOpen
  \bibfield  {author} {\bibinfo {author} {\bibfnamefont {F.~T.}\ \bibnamefont
  {{Avignone{ }III}}}, \bibinfo {author} {\bibfnamefont {Steven~R.}\
  \bibnamefont {Elliott}}, \ and\ \bibinfo {author} {\bibfnamefont {Jonathan}\
  \bibnamefont {Engel}},\ }\bibfield  {title} {\enquote {\bibinfo {title}
  {{Double beta decay, Majorana neutrinos, and neutrino mass}},}\ }\href
  {\doibase 10.1103/RevModPhys.80.481} {\bibfield  {journal} {\bibinfo
  {journal} {Rev.\ Mod.\ Phys.}\ }\textbf {\bibinfo {volume} {80}},\ \bibinfo
  {pages} {481} (\bibinfo {year} {2008})}\BibitemShut {NoStop}%
\bibitem [{\citenamefont {Vergados}\ \emph {et~al.}(2012)\citenamefont
  {Vergados}, \citenamefont {Ejiri},\ and\ \citenamefont {{\v
  S}imkovic}}]{ver12}%
  \BibitemOpen
  \bibfield  {author} {\bibinfo {author} {\bibfnamefont {J.D.}\ \bibnamefont
  {Vergados}}, \bibinfo {author} {\bibfnamefont {H.}~\bibnamefont {Ejiri}}, \
  and\ \bibinfo {author} {\bibfnamefont {F.}~\bibnamefont {{\v S}imkovic}},\
  }\bibfield  {title} {\enquote {\bibinfo {title} {Theory of neutrinoless
  double beta decay},}\ }\href {\doibase 10.1088/0034-4885/75/10/106301}
  {\bibfield  {journal} {\bibinfo  {journal} {Rept.\ Prog.\ Phys.}\ }\textbf
  {\bibinfo {volume} {75}},\ \bibinfo {pages} {106301} (\bibinfo {year}
  {2012})}\BibitemShut {NoStop}%
\bibitem [{\citenamefont {Fukuda}\ \emph {et~al.}(1998)\citenamefont {Fukuda}
  \emph {et~al.}}]{SuperKamiokande98}%
  \BibitemOpen
  \bibfield  {author} {\bibinfo {author} {\bibfnamefont {Y.}~\bibnamefont
  {Fukuda}} \emph {et~al.} (\bibinfo {collaboration} {Super-Kamiokande
  Collaboration}),\ }\bibfield  {title} {\enquote {\bibinfo {title} {{Evidence
  for oscillation of atmospheric neutrinos}},}\ }\href {\doibase
  10.1103/PhysRevLett.81.1562} {\bibfield  {journal} {\bibinfo  {journal}
  {Phys. Rev. Lett.}\ }\textbf {\bibinfo {volume} {81}},\ \bibinfo {pages}
  {1562--1567} (\bibinfo {year} {1998})}\BibitemShut {NoStop}%
\bibitem [{\citenamefont {Ahmad}\ \emph {et~al.}(2002)\citenamefont {Ahmad}
  \emph {et~al.}}]{SNO02}%
  \BibitemOpen
  \bibfield  {author} {\bibinfo {author} {\bibfnamefont {Q.~R.}\ \bibnamefont
  {Ahmad}} \emph {et~al.} (\bibinfo {collaboration} {SNO Collaboration}),\
  }\bibfield  {title} {\enquote {\bibinfo {title} {{Direct evidence for
  neutrino flavor transformation from neutral current interactions in the
  Sudbury Neutrino Observatory}},}\ }\href {\doibase
  10.1103/PhysRevLett.89.011301} {\bibfield  {journal} {\bibinfo  {journal}
  {Phys. Rev. Lett.}\ }\textbf {\bibinfo {volume} {89}},\ \bibinfo {pages}
  {011301} (\bibinfo {year} {2002})}\BibitemShut {NoStop}%
\bibitem [{\citenamefont {Eguchi}\ \emph {et~al.}(2003)\citenamefont {Eguchi}
  \emph {et~al.}}]{KamLAND03}%
  \BibitemOpen
  \bibfield  {author} {\bibinfo {author} {\bibfnamefont {K.}~\bibnamefont
  {Eguchi}} \emph {et~al.} (\bibinfo {collaboration} {KamLAND Collaboration}),\
  }\bibfield  {title} {\enquote {\bibinfo {title} {{First results from KamLAND:
  Evidence for reactor anti-neutrino disappearance}},}\ }\href {\doibase
  10.1103/PhysRevLett.90.021802} {\bibfield  {journal} {\bibinfo  {journal}
  {Phys. Rev. Lett.}\ }\textbf {\bibinfo {volume} {90}},\ \bibinfo {pages}
  {021802} (\bibinfo {year} {2003})}\BibitemShut {NoStop}%
\bibitem [{\citenamefont {Olive}\ \emph {et~al.}(2014)\citenamefont {Olive}
  \emph {et~al.}}]{pdg14}%
  \BibitemOpen
  \bibfield  {author} {\bibinfo {author} {\bibfnamefont {K.~A.}\ \bibnamefont
  {Olive}} \emph {et~al.} (\bibinfo {collaboration} {Particle Data Group}),\
  }\bibfield  {title} {\enquote {\bibinfo {title} {{Review of Particle
  Physics}},}\ }\href {\doibase 10.1088/1674-1137/38/9/090001} {\bibfield
  {journal} {\bibinfo  {journal} {Chin. Phys. C}\ }\textbf {\bibinfo {volume}
  {38}},\ \bibinfo {pages} {090001} (\bibinfo {year} {2014})}\BibitemShut
  {NoStop}%
\bibitem [{\citenamefont {Gonzalez-Garcia}\ \emph {et~al.}(2014)\citenamefont
  {Gonzalez-Garcia}, \citenamefont {Maltoni},\ and\ \citenamefont
  {Schwetz}}]{GonzalezGarcia14}%
  \BibitemOpen
  \bibfield  {author} {\bibinfo {author} {\bibfnamefont {M.~C.}\ \bibnamefont
  {Gonzalez-Garcia}}, \bibinfo {author} {\bibfnamefont {Michele}\ \bibnamefont
  {Maltoni}}, \ and\ \bibinfo {author} {\bibfnamefont {Thomas}\ \bibnamefont
  {Schwetz}},\ }\bibfield  {title} {\enquote {\bibinfo {title} {{Updated fit to
  three neutrino mixing: status of leptonic CP violation}},}\ }\href {\doibase
  10.1007/JHEP11(2014)052} {\bibfield  {journal} {\bibinfo  {journal} {JHEP}\
  }\textbf {\bibinfo {volume} {11}},\ \bibinfo {pages} {052} (\bibinfo {year}
  {2014})}\BibitemShut {NoStop}%
\bibitem [{\citenamefont {Abe}\ \emph {et~al.}(2015)\citenamefont {Abe} \emph
  {et~al.}}]{T2K15}%
  \BibitemOpen
  \bibfield  {author} {\bibinfo {author} {\bibfnamefont {K.}~\bibnamefont
  {Abe}} \emph {et~al.} (\bibinfo {collaboration} {T2K Collaboration}),\
  }\bibfield  {title} {\enquote {\bibinfo {title} {{Measurements of neutrino
  oscillation in appearance and disappearance channels by the T2K experiment
  with $6.6\times 10^{20}$ protons on target}},}\ }\href {\doibase
  10.1103/PhysRevD.91.072010} {\bibfield  {journal} {\bibinfo  {journal} {Phys.
  Rev. D}\ }\textbf {\bibinfo {volume} {91}},\ \bibinfo {pages} {072010}
  (\bibinfo {year} {2015})}\BibitemShut {NoStop}%
\bibitem [{\citenamefont {Adamson}\ \emph {et~al.}(2016)\citenamefont {Adamson}
  \emph {et~al.}}]{NOvA16}%
  \BibitemOpen
  \bibfield  {author} {\bibinfo {author} {\bibfnamefont {P.}~\bibnamefont
  {Adamson}} \emph {et~al.} (\bibinfo {collaboration} {NOvA Collaboration}),\
  }\bibfield  {title} {\enquote {\bibinfo {title} {{First measurement of
  electron neutrino appearance in NOvA}},}\ }\href {\doibase
  10.1103/PhysRevLett.116.151806} {\bibfield  {journal} {\bibinfo  {journal}
  {Phys. Rev. Lett.}\ }\textbf {\bibinfo {volume} {116}},\ \bibinfo {pages}
  {151806} (\bibinfo {year} {2016})}\BibitemShut {NoStop}%
\bibitem [{\citenamefont {Bilenky}\ \emph {et~al.}(2003)\citenamefont
  {Bilenky}, \citenamefont {Giunti}, \citenamefont {Grifols},\ and\
  \citenamefont {Masso}}]{bil03}%
  \BibitemOpen
  \bibfield  {author} {\bibinfo {author} {\bibfnamefont {S.~M.}\ \bibnamefont
  {Bilenky}}, \bibinfo {author} {\bibfnamefont {C.}~\bibnamefont {Giunti}},
  \bibinfo {author} {\bibfnamefont {J.~A.}\ \bibnamefont {Grifols}}, \ and\
  \bibinfo {author} {\bibfnamefont {E.}~\bibnamefont {Masso}},\ }\bibfield
  {title} {\enquote {\bibinfo {title} {{Absolute values of neutrino masses:
  Status and prospects}},}\ }\href {\doibase 10.1016/S0370-1573(03)00102-9}
  {\bibfield  {journal} {\bibinfo  {journal} {Phys. Rep.}\ }\textbf {\bibinfo
  {volume} {379}},\ \bibinfo {pages} {69} (\bibinfo {year} {2003})}\BibitemShut
  {NoStop}%
\bibitem [{\citenamefont {Doi}\ \emph {et~al.}(1985)\citenamefont {Doi},
  \citenamefont {Kotani},\ and\ \citenamefont {Takasugi}}]{doi85}%
  \BibitemOpen
  \bibfield  {author} {\bibinfo {author} {\bibfnamefont {M.}~\bibnamefont
  {Doi}}, \bibinfo {author} {\bibfnamefont {T.}~\bibnamefont {Kotani}}, \ and\
  \bibinfo {author} {\bibfnamefont {E.}~\bibnamefont {Takasugi}},\ }\bibfield
  {title} {\enquote {\bibinfo {title} {Double beta decay and majorana
  neutrino},}\ }\href {\doibase 10.1143/PTPS.83.1} {\bibfield  {journal}
  {\bibinfo  {journal} {Prog.\ Theor.\ Phys.}\ }\textbf {\bibinfo {volume}
  {83}},\ \bibinfo {pages} {1} (\bibinfo {year} {1985})}\BibitemShut {NoStop}%
\bibitem [{\citenamefont {Haxton}\ and\ \citenamefont {{G.~J.~Stephenson{
  Jr.}}}(1984)}]{hax84}%
  \BibitemOpen
  \bibfield  {author} {\bibinfo {author} {\bibfnamefont {W.~C.}\ \bibnamefont
  {Haxton}}\ and\ \bibinfo {author} {\bibnamefont {{G.~J.~Stephenson{ Jr.}}}},\
  }\bibfield  {title} {\enquote {\bibinfo {title} {Double beta decay},}\ }\href
  {\doibase 10.1016/0146-6410(84)90006-1} {\bibfield  {journal} {\bibinfo
  {journal} {Prog.\ Part.\ and Nucl.\ Phys.}\ }\textbf {\bibinfo {volume}
  {12}},\ \bibinfo {pages} {409} (\bibinfo {year} {1984})}\BibitemShut
  {NoStop}%
\bibitem [{\citenamefont {Tomoda}(1991)}]{tom91}%
  \BibitemOpen
  \bibfield  {author} {\bibinfo {author} {\bibfnamefont {T.}~\bibnamefont
  {Tomoda}},\ }\bibfield  {title} {\enquote {\bibinfo {title} {Double beta
  decay},}\ }\href {\doibase 10.1088/0034-4885/54/1/002} {\bibfield  {journal}
  {\bibinfo  {journal} {Rep.\ Prog.\ Phys.}\ }\textbf {\bibinfo {volume}
  {54}},\ \bibinfo {pages} {53} (\bibinfo {year} {1991})}\BibitemShut {NoStop}%
\bibitem [{\citenamefont {Mendenhall}\ \emph {et~al.}(2013)\citenamefont
  {Mendenhall} \emph {et~al.}}]{UCNA13}%
  \BibitemOpen
  \bibfield  {author} {\bibinfo {author} {\bibfnamefont {M.~P.}\ \bibnamefont
  {Mendenhall}} \emph {et~al.} (\bibinfo {collaboration} {UCNA
  Collaboration}),\ }\bibfield  {title} {\enquote {\bibinfo {title} {Precision
  measurement of the neutron $\ensuremath{\beta}$-decay asymmetry},}\ }\href
  {\doibase 10.1103/PhysRevC.87.032501} {\bibfield  {journal} {\bibinfo
  {journal} {Phys. Rev. C}\ }\textbf {\bibinfo {volume} {87}},\ \bibinfo
  {pages} {032501} (\bibinfo {year} {2013})}\BibitemShut {NoStop}%
\bibitem [{\citenamefont {Dumbrajs}\ \emph {et~al.}(1983)\citenamefont
  {Dumbrajs}, \citenamefont {Koch}, \citenamefont {Pilkuhn}, \citenamefont
  {Oades}, \citenamefont {Behrens}, \citenamefont {De~Swart},\ and\
  \citenamefont {Kroll}}]{Dumbrajs83}%
  \BibitemOpen
  \bibfield  {author} {\bibinfo {author} {\bibfnamefont {O.}~\bibnamefont
  {Dumbrajs}}, \bibinfo {author} {\bibfnamefont {R.}~\bibnamefont {Koch}},
  \bibinfo {author} {\bibfnamefont {H.}~\bibnamefont {Pilkuhn}}, \bibinfo
  {author} {\bibfnamefont {G.~C.}\ \bibnamefont {Oades}}, \bibinfo {author}
  {\bibfnamefont {H.}~\bibnamefont {Behrens}}, \bibinfo {author} {\bibfnamefont
  {J.~J.}\ \bibnamefont {De~Swart}}, \ and\ \bibinfo {author} {\bibfnamefont
  {P.}~\bibnamefont {Kroll}},\ }\bibfield  {title} {\enquote {\bibinfo {title}
  {{Compilation of Coupling Constants and Low-Energy Parameters}},}\ }\href
  {\doibase 10.1016/0550-3213(83)90288-2} {\bibfield  {journal} {\bibinfo
  {journal} {Nucl. Phys. B}\ }\textbf {\bibinfo {volume} {216}},\ \bibinfo
  {pages} {277--335} (\bibinfo {year} {1983})}\BibitemShut {NoStop}%
\bibitem [{\citenamefont {Bernard}\ \emph {et~al.}(2002)\citenamefont
  {Bernard}, \citenamefont {Elouadrhiri},\ and\ \citenamefont
  {Meissner}}]{Bernard01}%
  \BibitemOpen
  \bibfield  {author} {\bibinfo {author} {\bibfnamefont {Veronique}\
  \bibnamefont {Bernard}}, \bibinfo {author} {\bibfnamefont {Latifa}\
  \bibnamefont {Elouadrhiri}}, \ and\ \bibinfo {author} {\bibfnamefont
  {Ulf-G.}\ \bibnamefont {Meissner}},\ }\bibfield  {title} {\enquote {\bibinfo
  {title} {{Axial structure of the nucleon}},}\ }\href {\doibase
  10.1088/0954-3899/28/1/201} {\bibfield  {journal} {\bibinfo  {journal} {J.
  Phys. G: Nucl. Part. Phys.}\ }\textbf {\bibinfo {volume} {28}},\ \bibinfo
  {pages} {R1--R35} (\bibinfo {year} {2002})}\BibitemShut {NoStop}%
\bibitem [{\citenamefont {Muto}(1994)}]{mut94}%
  \BibitemOpen
  \bibfield  {author} {\bibinfo {author} {\bibfnamefont {K.}~\bibnamefont
  {Muto}},\ }\bibfield  {title} {\enquote {\bibinfo {title} {{Neutrinoless
  double beta decay beyond closure approximation}},}\ }\href {\doibase
  10.1016/0375-9474(94)90890-7} {\bibfield  {journal} {\bibinfo  {journal}
  {Nucl. Phys. A}\ }\textbf {\bibinfo {volume} {577}},\ \bibinfo {pages}
  {415C--420C} (\bibinfo {year} {1994})}\BibitemShut {NoStop}%
\bibitem [{\citenamefont {{\v S}imkovic}\ \emph {et~al.}(2011)\citenamefont
  {{\v S}imkovic}, \citenamefont {Hodak}, \citenamefont {Faessler},\ and\
  \citenamefont {Vogel}}]{sim11}%
  \BibitemOpen
  \bibfield  {author} {\bibinfo {author} {\bibfnamefont {Fedor}\ \bibnamefont
  {{\v S}imkovic}}, \bibinfo {author} {\bibfnamefont {Rastislav}\ \bibnamefont
  {Hodak}}, \bibinfo {author} {\bibfnamefont {Amand}\ \bibnamefont {Faessler}},
  \ and\ \bibinfo {author} {\bibfnamefont {Petr}\ \bibnamefont {Vogel}},\
  }\bibfield  {title} {\enquote {\bibinfo {title} {{Relation between the
  $0\nu\beta\beta$ and $2\nu\beta\beta$ nuclear matrix elements revisited}},}\
  }\href {\doibase 10.1103/PhysRevC.83.015502} {\bibfield  {journal} {\bibinfo
  {journal} {Phys. Rev. C}\ }\textbf {\bibinfo {volume} {83}},\ \bibinfo
  {pages} {015502} (\bibinfo {year} {2011})}\BibitemShut {NoStop}%
\bibitem [{\citenamefont {Sen'kov}\ and\ \citenamefont {Horoi}(2013)}]{sen13}%
  \BibitemOpen
  \bibfield  {author} {\bibinfo {author} {\bibfnamefont {R.~A.}\ \bibnamefont
  {Sen'kov}}\ and\ \bibinfo {author} {\bibfnamefont {M.}~\bibnamefont
  {Horoi}},\ }\bibfield  {title} {\enquote {\bibinfo {title} {{Neutrinoless
  double-$\beta$ decay of $^{48}$Ca in the shell model: Closure versus
  nonclosure approximation}},}\ }\href {\doibase 10.1103/PhysRevC.88.064312}
  {\bibfield  {journal} {\bibinfo  {journal} {Phys. Rev. C}\ }\textbf {\bibinfo
  {volume} {88}},\ \bibinfo {pages} {064312} (\bibinfo {year}
  {2013})}\BibitemShut {NoStop}%
\bibitem [{\citenamefont {Sen'kov}\ \emph {et~al.}(2014)\citenamefont
  {Sen'kov}, \citenamefont {Horoi},\ and\ \citenamefont {Brown}}]{sen14}%
  \BibitemOpen
  \bibfield  {author} {\bibinfo {author} {\bibfnamefont {R.~A.}\ \bibnamefont
  {Sen'kov}}, \bibinfo {author} {\bibfnamefont {M.}~\bibnamefont {Horoi}}, \
  and\ \bibinfo {author} {\bibfnamefont {B.~A.}\ \bibnamefont {Brown}},\
  }\bibfield  {title} {\enquote {\bibinfo {title} {{Neutrinoless double-$\beta$
  decay of ${}^{82}$Se in the shell model: beyond closure approximation}},}\
  }\href {\doibase 10.1103/PhysRevC.89.054304} {\bibfield  {journal} {\bibinfo
  {journal} {Phys. Rev. C}\ }\textbf {\bibinfo {volume} {89}},\ \bibinfo
  {pages} {054304} (\bibinfo {year} {2014})}\BibitemShut {NoStop}%
\bibitem [{\citenamefont {Sen'kov}\ and\ \citenamefont {Horoi}(2014)}]{sen14b}%
  \BibitemOpen
  \bibfield  {author} {\bibinfo {author} {\bibfnamefont {R.~A.}\ \bibnamefont
  {Sen'kov}}\ and\ \bibinfo {author} {\bibfnamefont {M.}~\bibnamefont
  {Horoi}},\ }\bibfield  {title} {\enquote {\bibinfo {title} {{Accurate
  shell-model nuclear matrix elements for neutrinoless double-$\beta$
  decay}},}\ }\href {\doibase 10.1103/PhysRevC.90.051301} {\bibfield  {journal}
  {\bibinfo  {journal} {Phys. Rev. C}\ }\textbf {\bibinfo {volume} {90}},\
  \bibinfo {pages} {051301} (\bibinfo {year} {2014})}\BibitemShut {NoStop}%
\bibitem [{\citenamefont {Kotila}\ and\ \citenamefont
  {Iachello}(2012)}]{kot12}%
  \BibitemOpen
  \bibfield  {author} {\bibinfo {author} {\bibfnamefont {J.}~\bibnamefont
  {Kotila}}\ and\ \bibinfo {author} {\bibfnamefont {F.}~\bibnamefont
  {Iachello}},\ }\bibfield  {title} {\enquote {\bibinfo {title} {Phase space
  factors for double-$\beta$ decay},}\ }\href {\doibase
  10.1103/PhysRevC.85.034316} {\bibfield  {journal} {\bibinfo  {journal}
  {Phys.\ Rev.\ C}\ }\textbf {\bibinfo {volume} {85}},\ \bibinfo {pages}
  {034316} (\bibinfo {year} {2012})}\BibitemShut {NoStop}%
\bibitem [{\citenamefont {Stoica}\ and\ \citenamefont
  {Mirea}(2013)}]{Stoica13}%
  \BibitemOpen
  \bibfield  {author} {\bibinfo {author} {\bibfnamefont {Sabin}\ \bibnamefont
  {Stoica}}\ and\ \bibinfo {author} {\bibfnamefont {Mihail}\ \bibnamefont
  {Mirea}},\ }\bibfield  {title} {\enquote {\bibinfo {title} {{New calculations
  for phase space factors involved in double-$\beta$ decay}},}\ }\href
  {\doibase 10.1103/PhysRevC.88.037303} {\bibfield  {journal} {\bibinfo
  {journal} {Phys. Rev. C}\ }\textbf {\bibinfo {volume} {88}},\ \bibinfo
  {pages} {037303} (\bibinfo {year} {2013})}\BibitemShut {NoStop}%
\bibitem [{\citenamefont {{{\protect \v{S}}imkovic}}\ \emph
  {et~al.}(1999)\citenamefont {{{\protect \v{S}}imkovic}}, \citenamefont
  {Pantis}, \citenamefont {Vergados},\ and\ \citenamefont {Faessler}}]{sim99}%
  \BibitemOpen
  \bibfield  {author} {\bibinfo {author} {\bibfnamefont {F.}~\bibnamefont
  {{{\protect \v{S}}imkovic}}}, \bibinfo {author} {\bibfnamefont
  {G.}~\bibnamefont {Pantis}}, \bibinfo {author} {\bibfnamefont {J.~D.}\
  \bibnamefont {Vergados}}, \ and\ \bibinfo {author} {\bibfnamefont
  {A.}~\bibnamefont {Faessler}},\ }\bibfield  {title} {\enquote {\bibinfo
  {title} {Additional nucleon current contributions to neutrinoless double
  $\beta$ decay},}\ }\href {\doibase 10.1103/PhysRevC.60.055502} {\bibfield
  {journal} {\bibinfo  {journal} {Phys.\ Rev. C}\ }\textbf {\bibinfo {volume}
  {60}},\ \bibinfo {pages} {055502} (\bibinfo {year} {1999})}\BibitemShut
  {NoStop}%
\bibitem [{\citenamefont {Rodin}\ \emph {et~al.}(2006)\citenamefont {Rodin},
  \citenamefont {Faessler}, \citenamefont {\v{S}imkovic},\ and\ \citenamefont
  {Vogel}}]{rod06}%
  \BibitemOpen
  \bibfield  {author} {\bibinfo {author} {\bibfnamefont {V.~A.}\ \bibnamefont
  {Rodin}}, \bibinfo {author} {\bibfnamefont {A.}~\bibnamefont {Faessler}},
  \bibinfo {author} {\bibfnamefont {F.}~\bibnamefont {\v{S}imkovic}}, \ and\
  \bibinfo {author} {\bibfnamefont {P.}~\bibnamefont {Vogel}},\ }\bibfield
  {title} {\enquote {\bibinfo {title} {{Assessment of uncertainties in QRPA
  $0\nu\beta\beta$-decay nuclear matrix elements}},}\ }\href {\doibase
  10.1016/j.nuclphysa.2005.12.004} {\bibfield  {journal} {\bibinfo  {journal}
  {Nucl.\ Phys. A}\ }\textbf {\bibinfo {volume} {766}},\ \bibinfo {pages} {107}
  (\bibinfo {year} {2006})},\ \bibinfo {note} {[Erratum: Nucl. Phys. A 793, 213
  (2007)]}\BibitemShut {NoStop}%
\bibitem [{\citenamefont {Horoi}\ and\ \citenamefont {Stoica}(2010)}]{hor10}%
  \BibitemOpen
  \bibfield  {author} {\bibinfo {author} {\bibfnamefont {Mihai}\ \bibnamefont
  {Horoi}}\ and\ \bibinfo {author} {\bibfnamefont {Sabin}\ \bibnamefont
  {Stoica}},\ }\bibfield  {title} {\enquote {\bibinfo {title} {Shell model
  analysis of the neutrinoless double-$\beta$ decay of $^{48}$\protect{C}a},}\
  }\href {\doibase 10.1103/PhysRevC.81.024321} {\bibfield  {journal} {\bibinfo
  {journal} {Phys. Rev. C}\ }\textbf {\bibinfo {volume} {81}},\ \bibinfo
  {pages} {024321} (\bibinfo {year} {2010})}\BibitemShut {NoStop}%
\bibitem [{\citenamefont {Miller}\ and\ \citenamefont
  {Spencer}(1976)}]{miller76}%
  \BibitemOpen
  \bibfield  {author} {\bibinfo {author} {\bibfnamefont {G.A.}\ \bibnamefont
  {Miller}}\ and\ \bibinfo {author} {\bibfnamefont {J.E.}\ \bibnamefont
  {Spencer}},\ }\bibfield  {title} {\enquote {\bibinfo {title} {A survey of
  pion charge-exchange reactions with nuclei},}\ }\href {\doibase
  10.1016/0003-4916(76)90073-7} {\bibfield  {journal} {\bibinfo  {journal}
  {Ann.\ Phys.}\ }\textbf {\bibinfo {volume} {100}},\ \bibinfo {pages} {562}
  (\bibinfo {year} {1976})}\BibitemShut {NoStop}%
\bibitem [{\citenamefont {Roth}\ \emph {et~al.}(2005)\citenamefont {Roth},
  \citenamefont {Hergert}, \citenamefont {Papakonstantinou}, \citenamefont
  {Neff},\ and\ \citenamefont {Feldmeier}}]{rot05}%
  \BibitemOpen
  \bibfield  {author} {\bibinfo {author} {\bibfnamefont {R.}~\bibnamefont
  {Roth}}, \bibinfo {author} {\bibfnamefont {H.}~\bibnamefont {Hergert}},
  \bibinfo {author} {\bibfnamefont {P.}~\bibnamefont {Papakonstantinou}},
  \bibinfo {author} {\bibfnamefont {T.}~\bibnamefont {Neff}}, \ and\ \bibinfo
  {author} {\bibfnamefont {H.}~\bibnamefont {Feldmeier}},\ }\bibfield  {title}
  {\enquote {\bibinfo {title} {Matrix elements and few-body calculations within
  the unitary correlation operator method},}\ }\href {\doibase
  10.1103/PhysRevC.72.034002} {\bibfield  {journal} {\bibinfo  {journal} {Phys.
  Rev. C}\ }\textbf {\bibinfo {volume} {72}},\ \bibinfo {pages} {034002}
  (\bibinfo {year} {2005})}\BibitemShut {NoStop}%
\bibitem [{\citenamefont {Brueckner}(1955)}]{Brueckner55}%
  \BibitemOpen
  \bibfield  {author} {\bibinfo {author} {\bibfnamefont {K.~A.}\ \bibnamefont
  {Brueckner}},\ }\bibfield  {title} {\enquote {\bibinfo {title} {Many-body
  problem for strongly interacting particles. \protect{II}. \protect{L}inked
  cluster expansion},}\ }\href {\doibase 10.1103/PhysRev.100.36} {\bibfield
  {journal} {\bibinfo  {journal} {Phys. Rev.}\ }\textbf {\bibinfo {volume}
  {100}},\ \bibinfo {pages} {36--45} (\bibinfo {year} {1955})}\BibitemShut
  {NoStop}%
\bibitem [{\citenamefont {M{\"u}ther}\ and\ \citenamefont
  {Polls}(2000)}]{Muther00}%
  \BibitemOpen
  \bibfield  {author} {\bibinfo {author} {\bibfnamefont {H.}~\bibnamefont
  {M{\"u}ther}}\ and\ \bibinfo {author} {\bibfnamefont {A.}~\bibnamefont
  {Polls}},\ }\bibfield  {title} {\enquote {\bibinfo {title} {{Two-body
  correlations in nuclear systems}},}\ }\href {\doibase
  10.1016/S0146-6410(00)00105-8} {\bibfield  {journal} {\bibinfo  {journal}
  {Prog. Part. Nucl. Phys.}\ }\textbf {\bibinfo {volume} {45}},\ \bibinfo
  {pages} {243--334} (\bibinfo {year} {2000})}\BibitemShut {NoStop}%
\bibitem [{\citenamefont {Benhar}\ \emph {et~al.}(2014)\citenamefont {Benhar},
  \citenamefont {Biondi},\ and\ \citenamefont {Speranza}}]{Benhar14}%
  \BibitemOpen
  \bibfield  {author} {\bibinfo {author} {\bibfnamefont {Omar}\ \bibnamefont
  {Benhar}}, \bibinfo {author} {\bibfnamefont {Riccardo}\ \bibnamefont
  {Biondi}}, \ and\ \bibinfo {author} {\bibfnamefont {Enrico}\ \bibnamefont
  {Speranza}},\ }\bibfield  {title} {\enquote {\bibinfo {title} {{Short-range
  correlation effects on the nuclear matrix element of neutrinoless
  double-$\beta$ decay}},}\ }\href {\doibase 10.1103/PhysRevC.90.065504}
  {\bibfield  {journal} {\bibinfo  {journal} {Phys. Rev. C}\ }\textbf {\bibinfo
  {volume} {90}},\ \bibinfo {pages} {065504} (\bibinfo {year}
  {2014})}\BibitemShut {NoStop}%
\bibitem [{\citenamefont {Engel}\ \emph {et~al.}(2011)\citenamefont {Engel},
  \citenamefont {Carlson},\ and\ \citenamefont {Wiringa}}]{engel11}%
  \BibitemOpen
  \bibfield  {author} {\bibinfo {author} {\bibfnamefont {J.}~\bibnamefont
  {Engel}}, \bibinfo {author} {\bibfnamefont {J.}~\bibnamefont {Carlson}}, \
  and\ \bibinfo {author} {\bibfnamefont {R.B.}\ \bibnamefont {Wiringa}},\
  }\bibfield  {title} {\enquote {\bibinfo {title} {Jastrow functions in
  double-$\beta$ decay},}\ }\href {\doibase 10.1103/PhysRevC.83.034317}
  {\bibfield  {journal} {\bibinfo  {journal} {Phys. Rev. C}\ }\textbf {\bibinfo
  {volume} {83}},\ \bibinfo {pages} {034617} (\bibinfo {year}
  {2011})}\BibitemShut {NoStop}%
\bibitem [{\citenamefont {Kortelainen}\ \emph {et~al.}(2007)\citenamefont
  {Kortelainen}, \citenamefont {Civitarese}, \citenamefont {Suhonen},\ and\
  \citenamefont {Toivanen}}]{kor07}%
  \BibitemOpen
  \bibfield  {author} {\bibinfo {author} {\bibfnamefont {M.}~\bibnamefont
  {Kortelainen}}, \bibinfo {author} {\bibfnamefont {O.}~\bibnamefont
  {Civitarese}}, \bibinfo {author} {\bibfnamefont {J.}~\bibnamefont {Suhonen}},
  \ and\ \bibinfo {author} {\bibfnamefont {J.}~\bibnamefont {Toivanen}},\
  }\bibfield  {title} {\enquote {\bibinfo {title} {Short-range correlations and
  neutrinoless double beta decay},}\ }\href {\doibase
  10.1016/j.physletb.2007.01.054} {\bibfield  {journal} {\bibinfo  {journal}
  {Phys. Lett. B}\ }\textbf {\bibinfo {volume} {647}},\ \bibinfo {pages} {128}
  (\bibinfo {year} {2007})}\BibitemShut {NoStop}%
\bibitem [{\citenamefont {{{\protect \v{S}}imkovic}}\ \emph
  {et~al.}(2009)\citenamefont {{{\protect \v{S}}imkovic}}, \citenamefont
  {Faessler}, \citenamefont {M{\"u}ther}, \citenamefont {Rodin},\ and\
  \citenamefont {Stauf}}]{sim09}%
  \BibitemOpen
  \bibfield  {author} {\bibinfo {author} {\bibfnamefont {Fedor}\ \bibnamefont
  {{{\protect \v{S}}imkovic}}}, \bibinfo {author} {\bibfnamefont {Amand}\
  \bibnamefont {Faessler}}, \bibinfo {author} {\bibfnamefont {Herbert}\
  \bibnamefont {M{\"u}ther}}, \bibinfo {author} {\bibfnamefont {Vadim}\
  \bibnamefont {Rodin}}, \ and\ \bibinfo {author} {\bibfnamefont {Markus}\
  \bibnamefont {Stauf}},\ }\bibfield  {title} {\enquote {\bibinfo {title} {The
  $0\nu\beta\beta$-decay nuclear matrix elements with self-consistent
  short-range correlations},}\ }\href {\doibase 10.1103/PhysRevC.79.055501}
  {\bibfield  {journal} {\bibinfo  {journal} {Phys. Rev. C}\ }\textbf {\bibinfo
  {volume} {79}},\ \bibinfo {pages} {055501} (\bibinfo {year}
  {2009})}\BibitemShut {NoStop}%
\bibitem [{\citenamefont {Schechter}\ and\ \citenamefont
  {Valle}(1982)}]{sch82}%
  \BibitemOpen
  \bibfield  {author} {\bibinfo {author} {\bibfnamefont {J.}~\bibnamefont
  {Schechter}}\ and\ \bibinfo {author} {\bibfnamefont {J.~W.~F.}\ \bibnamefont
  {Valle}},\ }\bibfield  {title} {\enquote {\bibinfo {title} {Neutrino decay
  and spontaneous violation of lepton number},}\ }\href {\doibase
  10.1103/PhysRevD.25.774} {\bibfield  {journal} {\bibinfo  {journal} {Phys.\
  Rev. D}\ }\textbf {\bibinfo {volume} {25}},\ \bibinfo {pages} {25091}
  (\bibinfo {year} {1982})}\BibitemShut {NoStop}%
\bibitem [{\citenamefont {Rodejohann}(2011)}]{Rodejohann11}%
  \BibitemOpen
  \bibfield  {author} {\bibinfo {author} {\bibfnamefont {Werner}\ \bibnamefont
  {Rodejohann}},\ }\bibfield  {title} {\enquote {\bibinfo {title}
  {{Neutrino-less Double Beta Decay and Particle Physics}},}\ }\href {\doibase
  10.1142/S0218301311020186} {\bibfield  {journal} {\bibinfo  {journal} {Int.
  J. Mod. Phys. E}\ }\textbf {\bibinfo {volume} {20}},\ \bibinfo {pages}
  {1833--1930} (\bibinfo {year} {2011})}\BibitemShut {NoStop}%
\bibitem [{\citenamefont {Deppisch}\ \emph {et~al.}(2012)\citenamefont
  {Deppisch}, \citenamefont {Hirsch},\ and\ \citenamefont {Pas}}]{Deppisch12}%
  \BibitemOpen
  \bibfield  {author} {\bibinfo {author} {\bibfnamefont {Frank~F.}\
  \bibnamefont {Deppisch}}, \bibinfo {author} {\bibfnamefont {Martin}\
  \bibnamefont {Hirsch}}, \ and\ \bibinfo {author} {\bibfnamefont {Heinrich}\
  \bibnamefont {Pas}},\ }\bibfield  {title} {\enquote {\bibinfo {title}
  {{Neutrinoless Double Beta Decay and Physics Beyond the Standard Model}},}\
  }\href {\doibase 10.1088/0954-3899/39/12/124007} {\bibfield  {journal}
  {\bibinfo  {journal} {J. Phys. G: Nucl. Part. Phys.}\ }\textbf {\bibinfo
  {volume} {39}},\ \bibinfo {pages} {124007} (\bibinfo {year}
  {2012})}\BibitemShut {NoStop}%
\bibitem [{\citenamefont {de~Gouvea}\ and\ \citenamefont
  {Vogel}(2013)}]{Gouvea13}%
  \BibitemOpen
  \bibfield  {author} {\bibinfo {author} {\bibfnamefont {Andre}\ \bibnamefont
  {de~Gouvea}}\ and\ \bibinfo {author} {\bibfnamefont {Petr}\ \bibnamefont
  {Vogel}},\ }\bibfield  {title} {\enquote {\bibinfo {title} {{Lepton Flavor
  and Number Conservation, and Physics Beyond the Standard Model}},}\ }\href
  {\doibase 10.1016/j.ppnp.2013.03.006} {\bibfield  {journal} {\bibinfo
  {journal} {Prog. Part. Nucl. Phys.}\ }\textbf {\bibinfo {volume} {71}},\
  \bibinfo {pages} {75--92} (\bibinfo {year} {2013})}\BibitemShut {NoStop}%
\bibitem [{\citenamefont {Helo}\ \emph {et~al.}(2013)\citenamefont {Helo},
  \citenamefont {Hirsch}, \citenamefont {Kovalenko},\ and\ \citenamefont
  {Pas}}]{Helo13}%
  \BibitemOpen
  \bibfield  {author} {\bibinfo {author} {\bibfnamefont {J.~C.}\ \bibnamefont
  {Helo}}, \bibinfo {author} {\bibfnamefont {M.}~\bibnamefont {Hirsch}},
  \bibinfo {author} {\bibfnamefont {S.~G.}\ \bibnamefont {Kovalenko}}, \ and\
  \bibinfo {author} {\bibfnamefont {H.}~\bibnamefont {Pas}},\ }\bibfield
  {title} {\enquote {\bibinfo {title} {{Neutrinoless double beta decay and
  lepton number violation at the LHC}},}\ }\href {\doibase
  10.1103/PhysRevD.88.011901} {\bibfield  {journal} {\bibinfo  {journal} {Phys.
  Rev. D}\ }\textbf {\bibinfo {volume} {88}},\ \bibinfo {pages} {011901}
  (\bibinfo {year} {2013})}\BibitemShut {NoStop}%
\bibitem [{\citenamefont {P{\"a}s}\ and\ \citenamefont
  {Rodejohann}(2015)}]{Pas15}%
  \BibitemOpen
  \bibfield  {author} {\bibinfo {author} {\bibfnamefont {Heinrich}\
  \bibnamefont {P{\"a}s}}\ and\ \bibinfo {author} {\bibfnamefont {Werner}\
  \bibnamefont {Rodejohann}},\ }\bibfield  {title} {\enquote {\bibinfo {title}
  {{Neutrinoless Double Beta Decay}},}\ }\href {\doibase
  10.1088/1367-2630/17/11/115010} {\bibfield  {journal} {\bibinfo  {journal}
  {New J. Phys.}\ }\textbf {\bibinfo {volume} {17}},\ \bibinfo {pages} {115010}
  (\bibinfo {year} {2015})}\BibitemShut {NoStop}%
\bibitem [{\citenamefont {Peng}\ \emph {et~al.}(2016)\citenamefont {Peng},
  \citenamefont {Ramsey-Musolf},\ and\ \citenamefont {Winslow}}]{Peng16}%
  \BibitemOpen
  \bibfield  {author} {\bibinfo {author} {\bibfnamefont {Tao}\ \bibnamefont
  {Peng}}, \bibinfo {author} {\bibfnamefont {Michael~J.}\ \bibnamefont
  {Ramsey-Musolf}}, \ and\ \bibinfo {author} {\bibfnamefont {Peter}\
  \bibnamefont {Winslow}},\ }\bibfield  {title} {\enquote {\bibinfo {title}
  {{TeV lepton number violation: From neutrinoless double-$\beta$ decay to the
  LHC}},}\ }\href {\doibase 10.1103/PhysRevD.93.093002} {\bibfield  {journal}
  {\bibinfo  {journal} {Phys. Rev. D}\ }\textbf {\bibinfo {volume} {93}},\
  \bibinfo {pages} {093002} (\bibinfo {year} {2016})}\BibitemShut {NoStop}%
\bibitem [{\citenamefont {Blennow}\ \emph {et~al.}(2010)\citenamefont
  {Blennow}, \citenamefont {Fernandez-Martinez}, \citenamefont {Lopez-Pavon},\
  and\ \citenamefont {Men{\'e}ndez}}]{Blennow10}%
  \BibitemOpen
  \bibfield  {author} {\bibinfo {author} {\bibfnamefont {Mattias}\ \bibnamefont
  {Blennow}}, \bibinfo {author} {\bibfnamefont {Enrique}\ \bibnamefont
  {Fernandez-Martinez}}, \bibinfo {author} {\bibfnamefont {Jacobo}\
  \bibnamefont {Lopez-Pavon}}, \ and\ \bibinfo {author} {\bibfnamefont
  {Javier}\ \bibnamefont {Men{\'e}ndez}},\ }\bibfield  {title} {\enquote
  {\bibinfo {title} {{Neutrinoless double beta decay in seesaw models}},}\
  }\href {\doibase 10.1007/JHEP07(2010)096} {\bibfield  {journal} {\bibinfo
  {journal} {JHEP}\ }\textbf {\bibinfo {volume} {07}},\ \bibinfo {pages} {096}
  (\bibinfo {year} {2010})}\BibitemShut {NoStop}%
\bibitem [{\citenamefont {Mohapatra}\ and\ \citenamefont
  {Senjanovic}(1980)}]{moh80}%
  \BibitemOpen
  \bibfield  {author} {\bibinfo {author} {\bibfnamefont {R.}~\bibnamefont
  {Mohapatra}}\ and\ \bibinfo {author} {\bibfnamefont {G.}~\bibnamefont
  {Senjanovic}},\ }\bibfield  {title} {\enquote {\bibinfo {title} {Neutrino
  mass and spontaneous parity nonconservation},}\ }\href {\doibase
  10.1103/PhysRevLett.44.912} {\bibfield  {journal} {\bibinfo  {journal}
  {Phys.\ Rev.\ Lett.}\ }\textbf {\bibinfo {volume} {44}},\ \bibinfo {pages}
  {912} (\bibinfo {year} {1980})}\BibitemShut {NoStop}%
\bibitem [{\citenamefont {Mohapatra}\ and\ \citenamefont
  {Vergados}(1981)}]{moh81}%
  \BibitemOpen
  \bibfield  {author} {\bibinfo {author} {\bibfnamefont {R.}~\bibnamefont
  {Mohapatra}}\ and\ \bibinfo {author} {\bibfnamefont {J.~D.}\ \bibnamefont
  {Vergados}},\ }\bibfield  {title} {\enquote {\bibinfo {title} {New
  contribution to neutrinoless double beta decay in gauge models},}\ }\href
  {\doibase 10.1103/PhysRevLett.47.1713} {\bibfield  {journal} {\bibinfo
  {journal} {Phys.\ Rev.\ Lett.}\ }\textbf {\bibinfo {volume} {47}},\ \bibinfo
  {pages} {1713} (\bibinfo {year} {1981})}\BibitemShut {NoStop}%
\bibitem [{\citenamefont {Mohapatra}(1986)}]{moh86}%
  \BibitemOpen
  \bibfield  {author} {\bibinfo {author} {\bibfnamefont {R.}~\bibnamefont
  {Mohapatra}},\ }\bibfield  {title} {\enquote {\bibinfo {title} {New
  contributions to neutrinoless double-beta decay in supersymmetric
  theories},}\ }\href {\doibase 10.1103/PhysRevD.34.3457} {\bibfield  {journal}
  {\bibinfo  {journal} {Phys.\ Rev.\ D}\ }\textbf {\bibinfo {volume} {34}},\
  \bibinfo {pages} {3457} (\bibinfo {year} {1986})}\BibitemShut {NoStop}%
\bibitem [{\citenamefont {Vergados}(1987)}]{ver87}%
  \BibitemOpen
  \bibfield  {author} {\bibinfo {author} {\bibfnamefont {J.~D.}\ \bibnamefont
  {Vergados}},\ }\bibfield  {title} {\enquote {\bibinfo {title} {{Neutrinoless
  Double Beta Decay Without Majorana Neutrinos in Supersymmetric Theories}},}\
  }\href {\doibase 10.1016/0370-2693(87)90487-4} {\bibfield  {journal}
  {\bibinfo  {journal} {Phys. Lett. B}\ }\textbf {\bibinfo {volume} {184}},\
  \bibinfo {pages} {55--62} (\bibinfo {year} {1987})}\BibitemShut {NoStop}%
\bibitem [{\citenamefont {Chikashigi}\ \emph {et~al.}(1981)\citenamefont
  {Chikashigi}, \citenamefont {Mohapatra},\ and\ \citenamefont
  {Peccei}}]{chi81}%
  \BibitemOpen
  \bibfield  {author} {\bibinfo {author} {\bibfnamefont {Y.}~\bibnamefont
  {Chikashigi}}, \bibinfo {author} {\bibfnamefont {R.~N.}\ \bibnamefont
  {Mohapatra}}, \ and\ \bibinfo {author} {\bibfnamefont {R.~D.}\ \bibnamefont
  {Peccei}},\ }\bibfield  {title} {\enquote {\bibinfo {title} {Are there real
  goldstone bosons associated with broken lepton number?}}\ }\href {\doibase
  10.1016/0370-2693(81)90011-3} {\bibfield  {journal} {\bibinfo  {journal}
  {Phys. Lett. B}\ }\textbf {\bibinfo {volume} {98}},\ \bibinfo {pages} {265}
  (\bibinfo {year} {1981})}\BibitemShut {NoStop}%
\bibitem [{\citenamefont {Gelmini}\ and\ \citenamefont
  {Roncadelli}(1981)}]{gel81}%
  \BibitemOpen
  \bibfield  {author} {\bibinfo {author} {\bibfnamefont {G.~B.}\ \bibnamefont
  {Gelmini}}\ and\ \bibinfo {author} {\bibfnamefont {M.}~\bibnamefont
  {Roncadelli}},\ }\bibfield  {title} {\enquote {\bibinfo {title} {Left-handed
  neutrino mass scale and spontaneously broken lepton number},}\ }\href
  {\doibase 10.1016/0370-2693(81)90559-1} {\bibfield  {journal} {\bibinfo
  {journal} {Phys. Lett. B}\ }\textbf {\bibinfo {volume} {99}},\ \bibinfo
  {pages} {411} (\bibinfo {year} {1981})}\BibitemShut {NoStop}%
\bibitem [{\citenamefont {Georgi}\ \emph {et~al.}(1981)\citenamefont {Georgi},
  \citenamefont {Glashow},\ and\ \citenamefont {Nussinov}}]{geo81}%
  \BibitemOpen
  \bibfield  {author} {\bibinfo {author} {\bibfnamefont {H.~M.}\ \bibnamefont
  {Georgi}}, \bibinfo {author} {\bibfnamefont {S.~L.}\ \bibnamefont {Glashow}},
  \ and\ \bibinfo {author} {\bibfnamefont {S.}~\bibnamefont {Nussinov}},\
  }\bibfield  {title} {\enquote {\bibinfo {title} {Unconventional model of
  neutrino masses},}\ }\href {\doibase 10.1016/0550-3213(81)90336-9} {\bibfield
   {journal} {\bibinfo  {journal} {Nucl. Phys. B}\ }\textbf {\bibinfo {volume}
  {193}},\ \bibinfo {pages} {297} (\bibinfo {year} {1981})}\BibitemShut
  {NoStop}%
\bibitem [{\citenamefont {{\v S}tef{\'a}nik}\ \emph {et~al.}(2015)\citenamefont
  {{\v S}tef{\'a}nik}, \citenamefont {Dvornicky}, \citenamefont {{\v
  S}imkovic},\ and\ \citenamefont {Vogel}}]{Stefanik15}%
  \BibitemOpen
  \bibfield  {author} {\bibinfo {author} {\bibfnamefont {Dusan}\ \bibnamefont
  {{\v S}tef{\'a}nik}}, \bibinfo {author} {\bibfnamefont {Rastislav}\
  \bibnamefont {Dvornicky}}, \bibinfo {author} {\bibfnamefont {Fedor}\
  \bibnamefont {{\v S}imkovic}}, \ and\ \bibinfo {author} {\bibfnamefont
  {Petr}\ \bibnamefont {Vogel}},\ }\bibfield  {title} {\enquote {\bibinfo
  {title} {{Reexamining the light neutrino exchange mechanism of the
  $0\nu\beta\beta$ decay with left- and right-handed leptonic and hadronic
  currents}},}\ }\href {\doibase 10.1103/PhysRevC.92.055502} {\bibfield
  {journal} {\bibinfo  {journal} {Phys. Rev. C}\ }\textbf {\bibinfo {volume}
  {92}},\ \bibinfo {pages} {055502} (\bibinfo {year} {2015})}\BibitemShut
  {NoStop}%
\bibitem [{\citenamefont {Faessler}\ \emph {et~al.}(2014)\citenamefont
  {Faessler}, \citenamefont {Gonz{\'a}lez}, \citenamefont {Kovalenko},\ and\
  \citenamefont {{\v S}imkovic}}]{fae14}%
  \BibitemOpen
  \bibfield  {author} {\bibinfo {author} {\bibfnamefont {Amand}\ \bibnamefont
  {Faessler}}, \bibinfo {author} {\bibfnamefont {Marcela}\ \bibnamefont
  {Gonz{\'a}lez}}, \bibinfo {author} {\bibfnamefont {Sergey}\ \bibnamefont
  {Kovalenko}}, \ and\ \bibinfo {author} {\bibfnamefont {Fedor}\ \bibnamefont
  {{\v S}imkovic}},\ }\bibfield  {title} {\enquote {\bibinfo {title}
  {{Arbitrary mass Majorana neutrinos in neutrinoless double beta decay}},}\
  }\href {\doibase 10.1103/PhysRevD.90.096010} {\bibfield  {journal} {\bibinfo
  {journal} {Phys. Rev. D}\ }\textbf {\bibinfo {volume} {90}},\ \bibinfo
  {pages} {096010} (\bibinfo {year} {2014})}\BibitemShut {NoStop}%
\bibitem [{\citenamefont {Barea}\ \emph
  {et~al.}(2015{\natexlab{a}})\citenamefont {Barea}, \citenamefont {Kotila},\
  and\ \citenamefont {Iachello}}]{bar15a}%
  \BibitemOpen
  \bibfield  {author} {\bibinfo {author} {\bibfnamefont {J.}~\bibnamefont
  {Barea}}, \bibinfo {author} {\bibfnamefont {J.}~\bibnamefont {Kotila}}, \
  and\ \bibinfo {author} {\bibfnamefont {F.}~\bibnamefont {Iachello}},\
  }\bibfield  {title} {\enquote {\bibinfo {title} {{Limits on sterile neutrino
  contributions to neutrinoless double beta decay}},}\ }\href {\doibase
  10.1103/PhysRevD.92.093001} {\bibfield  {journal} {\bibinfo  {journal} {Phys.
  Rev. D}\ }\textbf {\bibinfo {volume} {92}},\ \bibinfo {pages} {093001}
  (\bibinfo {year} {2015}{\natexlab{a}})}\BibitemShut {NoStop}%
\bibitem [{\citenamefont {Hyv{\"a}rinen}\ and\ \citenamefont
  {Suhonen}(2015)}]{Hyvarinen15}%
  \BibitemOpen
  \bibfield  {author} {\bibinfo {author} {\bibfnamefont {Juhani}\ \bibnamefont
  {Hyv{\"a}rinen}}\ and\ \bibinfo {author} {\bibfnamefont {Jouni}\ \bibnamefont
  {Suhonen}},\ }\bibfield  {title} {\enquote {\bibinfo {title} {{Nuclear matrix
  elements for $0\nu\beta\beta$ decays with light or heavy Majorana-neutrino
  exchange}},}\ }\href {\doibase 10.1103/PhysRevC.91.024613} {\bibfield
  {journal} {\bibinfo  {journal} {Phys. Rev. C}\ }\textbf {\bibinfo {volume}
  {91}},\ \bibinfo {pages} {024613} (\bibinfo {year} {2015})}\BibitemShut
  {NoStop}%
\bibitem [{\citenamefont {Horoi}\ and\ \citenamefont
  {Neacsu}(2016{\natexlab{a}})}]{hor16}%
  \BibitemOpen
  \bibfield  {author} {\bibinfo {author} {\bibfnamefont {Mihai}\ \bibnamefont
  {Horoi}}\ and\ \bibinfo {author} {\bibfnamefont {Andrei}\ \bibnamefont
  {Neacsu}},\ }\bibfield  {title} {\enquote {\bibinfo {title} {Shell model
  predictions for $^{124}$\protect{S}n double-$\beta$ decay},}\ }\href
  {\doibase 10.1103/PhysRevC.93.024308} {\bibfield  {journal} {\bibinfo
  {journal} {Phys. Rev. C}\ }\textbf {\bibinfo {volume} {93}},\ \bibinfo
  {pages} {024308} (\bibinfo {year} {2016}{\natexlab{a}})}\BibitemShut
  {NoStop}%
\bibitem [{\citenamefont {Retamosa}\ \emph {et~al.}(1995)\citenamefont
  {Retamosa}, \citenamefont {Caurier},\ and\ \citenamefont
  {Nowacki}}]{Retamosa95}%
  \BibitemOpen
  \bibfield  {author} {\bibinfo {author} {\bibfnamefont {J.}~\bibnamefont
  {Retamosa}}, \bibinfo {author} {\bibfnamefont {E.}~\bibnamefont {Caurier}}, \
  and\ \bibinfo {author} {\bibfnamefont {F.}~\bibnamefont {Nowacki}},\
  }\bibfield  {title} {\enquote {\bibinfo {title} {{Neutrinoless double beta
  decay of $^{48}$Ca}},}\ }\href {\doibase 10.1103/PhysRevC.51.371} {\bibfield
  {journal} {\bibinfo  {journal} {Phys. Rev. C}\ }\textbf {\bibinfo {volume}
  {51}},\ \bibinfo {pages} {371--378} (\bibinfo {year} {1995})}\BibitemShut
  {NoStop}%
\bibitem [{\citenamefont {Caurier}\ \emph {et~al.}(1996)\citenamefont
  {Caurier}, \citenamefont {Nowacki}, \citenamefont {Poves},\ and\
  \citenamefont {Retamosa}}]{cau96}%
  \BibitemOpen
  \bibfield  {author} {\bibinfo {author} {\bibfnamefont {E.}~\bibnamefont
  {Caurier}}, \bibinfo {author} {\bibfnamefont {F.}~\bibnamefont {Nowacki}},
  \bibinfo {author} {\bibfnamefont {A.}~\bibnamefont {Poves}}, \ and\ \bibinfo
  {author} {\bibfnamefont {J.}~\bibnamefont {Retamosa}},\ }\bibfield  {title}
  {\enquote {\bibinfo {title} {{Shell Model Studies of the Double Beta Decays
  of $^{76}$Ge, $^{82}$Se, and $^{136}$Xe}},}\ }\href {\doibase
  10.1103/PhysRevLett.77.1954} {\bibfield  {journal} {\bibinfo  {journal}
  {Phys. Rev. Lett.}\ }\textbf {\bibinfo {volume} {77}},\ \bibinfo {pages}
  {1954--1957} (\bibinfo {year} {1996})}\BibitemShut {NoStop}%
\bibitem [{\citenamefont {Horoi}\ and\ \citenamefont
  {Neacsu}(2016{\natexlab{b}})}]{hor16a}%
  \BibitemOpen
  \bibfield  {author} {\bibinfo {author} {\bibfnamefont {Mihai}\ \bibnamefont
  {Horoi}}\ and\ \bibinfo {author} {\bibfnamefont {Andrei}\ \bibnamefont
  {Neacsu}},\ }\bibfield  {title} {\enquote {\bibinfo {title} {{Analysis of
  mechanisms that could contribute to neutrinoless double-beta decay}},}\
  }\href {\doibase 10.1103/PhysRevD.93.113014} {\bibfield  {journal} {\bibinfo
  {journal} {Phys. Rev. D}\ }\textbf {\bibinfo {volume} {93}},\ \bibinfo
  {pages} {113014} (\bibinfo {year} {2016}{\natexlab{b}})}\BibitemShut
  {NoStop}%
\bibitem [{\citenamefont {Hirsch}\ \emph
  {et~al.}(1995{\natexlab{a}})\citenamefont {Hirsch}, \citenamefont
  {Klapdor-Kleingrothaus},\ and\ \citenamefont {Kovalenko}}]{Hirsch1995}%
  \BibitemOpen
  \bibfield  {author} {\bibinfo {author} {\bibfnamefont {M.}~\bibnamefont
  {Hirsch}}, \bibinfo {author} {\bibfnamefont {H.~V.}\ \bibnamefont
  {Klapdor-Kleingrothaus}}, \ and\ \bibinfo {author} {\bibfnamefont {S.~G.}\
  \bibnamefont {Kovalenko}},\ }\bibfield  {title} {\enquote {\bibinfo {title}
  {{New constraints on $R$-parity broken supersymmetry from neutrinoless double
  beta decay}},}\ }\href {\doibase 10.1103/PhysRevLett.75.17} {\bibfield
  {journal} {\bibinfo  {journal} {Phys. Rev. Lett.}\ }\textbf {\bibinfo
  {volume} {75}},\ \bibinfo {pages} {17--20} (\bibinfo {year}
  {1995}{\natexlab{a}})}\BibitemShut {NoStop}%
\bibitem [{\citenamefont {Hirsch}\ \emph {et~al.}(1996)\citenamefont {Hirsch},
  \citenamefont {Klapdor-Kleingrothaus},\ and\ \citenamefont
  {Panella}}]{hir96}%
  \BibitemOpen
  \bibfield  {author} {\bibinfo {author} {\bibfnamefont {M.}~\bibnamefont
  {Hirsch}}, \bibinfo {author} {\bibfnamefont {H.V.}\ \bibnamefont
  {Klapdor-Kleingrothaus}}, \ and\ \bibinfo {author} {\bibfnamefont
  {O.}~\bibnamefont {Panella}},\ }\bibfield  {title} {\enquote {\bibinfo
  {title} {Double beta decay in left-right symmetric models},}\ }\href
  {\doibase 10.1016/0370-2693(96)00185-2} {\bibfield  {journal} {\bibinfo
  {journal} {Phys. Lett. B}\ }\textbf {\bibinfo {volume} {374}},\ \bibinfo
  {pages} {7} (\bibinfo {year} {1996})}\BibitemShut {NoStop}%
\bibitem [{\citenamefont {Horoi}(2013)}]{hor13b}%
  \BibitemOpen
  \bibfield  {author} {\bibinfo {author} {\bibfnamefont {Mihai}\ \bibnamefont
  {Horoi}},\ }\bibfield  {title} {\enquote {\bibinfo {title} {{Shell model
  analysis of competing contributions to the double-$\beta$ decay of
  $^{48}$Ca}},}\ }\href {\doibase 10.1103/PhysRevC.87.014320} {\bibfield
  {journal} {\bibinfo  {journal} {Phys. Rev. C}\ }\textbf {\bibinfo {volume}
  {87}},\ \bibinfo {pages} {014320} (\bibinfo {year} {2013})}\BibitemShut
  {NoStop}%
\bibitem [{\citenamefont {Meroni}\ \emph {et~al.}(2013)\citenamefont {Meroni},
  \citenamefont {Petcov},\ and\ \citenamefont {{\v S}imkovic}}]{Meroni13}%
  \BibitemOpen
  \bibfield  {author} {\bibinfo {author} {\bibfnamefont {A.}~\bibnamefont
  {Meroni}}, \bibinfo {author} {\bibfnamefont {S.~T.}\ \bibnamefont {Petcov}},
  \ and\ \bibinfo {author} {\bibfnamefont {F.}~\bibnamefont {{\v S}imkovic}},\
  }\bibfield  {title} {\enquote {\bibinfo {title} {{Multiple CP non-conserving
  mechanisms of $(\beta\beta)_{0\nu}$-decay and nuclei with largely different
  nuclear matrix elements}},}\ }\href {\doibase 10.1007/JHEP02(2013)025}
  {\bibfield  {journal} {\bibinfo  {journal} {JHEP}\ }\textbf {\bibinfo
  {volume} {02}},\ \bibinfo {pages} {025} (\bibinfo {year} {2013})}\BibitemShut
  {NoStop}%
\bibitem [{\citenamefont {Pr{\'e}zeau}\ \emph {et~al.}(2003)\citenamefont
  {Pr{\'e}zeau}, \citenamefont {Ramsey-Musolf},\ and\ \citenamefont
  {Vogel}}]{pre03}%
  \BibitemOpen
  \bibfield  {author} {\bibinfo {author} {\bibfnamefont {G.}~\bibnamefont
  {Pr{\'e}zeau}}, \bibinfo {author} {\bibfnamefont {M.}~\bibnamefont
  {Ramsey-Musolf}}, \ and\ \bibinfo {author} {\bibfnamefont {P.}~\bibnamefont
  {Vogel}},\ }\bibfield  {title} {\enquote {\bibinfo {title} {Neutrinoless
  double $\beta$ decay and effective field theory},}\ }\href {\doibase
  10.1103/PhysRevD.68.034016} {\bibfield  {journal} {\bibinfo  {journal}
  {Phys.\ Rev. D}\ }\textbf {\bibinfo {volume} {68}},\ \bibinfo {pages}
  {034016} (\bibinfo {year} {2003})}\BibitemShut {NoStop}%
\bibitem [{\citenamefont {Faessler}\ \emph {et~al.}(1997)\citenamefont
  {Faessler}, \citenamefont {Kovalenko}, \citenamefont {\v{S}imkovic},\ and\
  \citenamefont {Schwieger}}]{fae97}%
  \BibitemOpen
  \bibfield  {author} {\bibinfo {author} {\bibfnamefont {A.}~\bibnamefont
  {Faessler}}, \bibinfo {author} {\bibfnamefont {S.}~\bibnamefont {Kovalenko}},
  \bibinfo {author} {\bibfnamefont {F.}~\bibnamefont {\v{S}imkovic}}, \ and\
  \bibinfo {author} {\bibfnamefont {J.}~\bibnamefont {Schwieger}},\ }\bibfield
  {title} {\enquote {\bibinfo {title} {Dominance of pion exchange in
  \protect{$R$}-parity-violating supersymmetric contributions to neutrinoless
  double beta decay},}\ }\href {\doibase 10.1103/PhysRevLett.78.183} {\bibfield
   {journal} {\bibinfo  {journal} {Phys.\ Rev.\ Lett.}\ }\textbf {\bibinfo
  {volume} {78}},\ \bibinfo {pages} {183} (\bibinfo {year} {1997})}\BibitemShut
  {NoStop}%
\bibitem [{\citenamefont {Faessler}\ \emph {et~al.}(1998)\citenamefont
  {Faessler}, \citenamefont {Kovalenko},\ and\ \citenamefont {{\v
  S}imkovic}}]{fae98a}%
  \BibitemOpen
  \bibfield  {author} {\bibinfo {author} {\bibfnamefont {Amand}\ \bibnamefont
  {Faessler}}, \bibinfo {author} {\bibfnamefont {Sergey}\ \bibnamefont
  {Kovalenko}}, \ and\ \bibinfo {author} {\bibfnamefont {Fedor}\ \bibnamefont
  {{\v S}imkovic}},\ }\bibfield  {title} {\enquote {\bibinfo {title} {{Pions in
  nuclei and manifestations of supersymmetry in neutrinoless double beta
  decay}},}\ }\href {\doibase 10.1103/PhysRevD.58.115004} {\bibfield  {journal}
  {\bibinfo  {journal} {Phys. Rev. D}\ }\textbf {\bibinfo {volume} {58}},\
  \bibinfo {pages} {115004} (\bibinfo {year} {1998})}\BibitemShut {NoStop}%
\bibitem [{\citenamefont {Faessler}\ \emph
  {et~al.}(2008{\natexlab{a}})\citenamefont {Faessler}, \citenamefont
  {Gutsche}, \citenamefont {Kovalenko},\ and\ \citenamefont {{\v
  S}imkovic}}]{fae08a}%
  \BibitemOpen
  \bibfield  {author} {\bibinfo {author} {\bibfnamefont {Amand}\ \bibnamefont
  {Faessler}}, \bibinfo {author} {\bibfnamefont {Thomas}\ \bibnamefont
  {Gutsche}}, \bibinfo {author} {\bibfnamefont {Sergey}\ \bibnamefont
  {Kovalenko}}, \ and\ \bibinfo {author} {\bibfnamefont {Fedor}\ \bibnamefont
  {{\v S}imkovic}},\ }\bibfield  {title} {\enquote {\bibinfo {title} {{Pion
  dominance in RPV SUSY induced neutrinoless double beta decay}},}\ }\href
  {\doibase 10.1103/PhysRevD.77.113012} {\bibfield  {journal} {\bibinfo
  {journal} {Phys. Rev. D}\ }\textbf {\bibinfo {volume} {77}},\ \bibinfo
  {pages} {113012} (\bibinfo {year} {2008}{\natexlab{a}})}\BibitemShut
  {NoStop}%
\bibitem [{\citenamefont {Savage}(1999)}]{Savage99}%
  \BibitemOpen
  \bibfield  {author} {\bibinfo {author} {\bibfnamefont {Martin~J.}\
  \bibnamefont {Savage}},\ }\bibfield  {title} {\enquote {\bibinfo {title}
  {{Pionic matrix elements in neutrinoless double beta decay}},}\ }\href
  {\doibase 10.1103/PhysRevC.59.2293} {\bibfield  {journal} {\bibinfo
  {journal} {Phys. Rev. C}\ }\textbf {\bibinfo {volume} {59}},\ \bibinfo
  {pages} {2293--2296} (\bibinfo {year} {1999})}\BibitemShut {NoStop}%
\bibitem [{\citenamefont {Nicholson}\ \emph {et~al.}()\citenamefont
  {Nicholson}, \citenamefont {Berkowitz}, \citenamefont {Chang}, \citenamefont
  {Clark}, \citenamefont {Joo}, \citenamefont {Kurth}, \citenamefont {Rinaldi},
  \citenamefont {Tiburzi}, \citenamefont {Vranas},\ and\ \citenamefont
  {Walker-Loud}}]{Nicholson16}%
  \BibitemOpen
  \bibfield  {author} {\bibinfo {author} {\bibfnamefont {Amy}\ \bibnamefont
  {Nicholson}}, \bibinfo {author} {\bibfnamefont {Evan}\ \bibnamefont
  {Berkowitz}}, \bibinfo {author} {\bibfnamefont {Chia~Cheng}\ \bibnamefont
  {Chang}}, \bibinfo {author} {\bibfnamefont {M.~A.}\ \bibnamefont {Clark}},
  \bibinfo {author} {\bibfnamefont {Balint}\ \bibnamefont {Joo}}, \bibinfo
  {author} {\bibfnamefont {Thorsten}\ \bibnamefont {Kurth}}, \bibinfo {author}
  {\bibfnamefont {Enrico}\ \bibnamefont {Rinaldi}}, \bibinfo {author}
  {\bibfnamefont {Brian}\ \bibnamefont {Tiburzi}}, \bibinfo {author}
  {\bibfnamefont {Pavlos}\ \bibnamefont {Vranas}}, \ and\ \bibinfo {author}
  {\bibfnamefont {Andre}\ \bibnamefont {Walker-Loud}},\ }\bibfield  {title}
  {\enquote {\bibinfo {title} {{Neutrinoless double beta decay from lattice
  QCD}},}\ }\Eprint {http://arxiv.org/abs/1608.04793} {arXiv:1608.04793}
  \BibitemShut {NoStop}%
\bibitem [{\citenamefont {Bogner}\ \emph {et~al.}(2010)\citenamefont {Bogner},
  \citenamefont {Furnstahl},\ and\ \citenamefont {Schwenk}}]{bog10}%
  \BibitemOpen
  \bibfield  {author} {\bibinfo {author} {\bibfnamefont {S.~K.}\ \bibnamefont
  {Bogner}}, \bibinfo {author} {\bibfnamefont {R.~J.}\ \bibnamefont
  {Furnstahl}}, \ and\ \bibinfo {author} {\bibfnamefont {A.}~\bibnamefont
  {Schwenk}},\ }\bibfield  {title} {\enquote {\bibinfo {title} {{From
  low-momentum interactions to nuclear structure}},}\ }\href {\doibase
  10.1016/j.ppnp.2010.03.001} {\bibfield  {journal} {\bibinfo  {journal} {Prog.
  Part. Nucl. Phys.}\ }\textbf {\bibinfo {volume} {65}},\ \bibinfo {pages}
  {94--147} (\bibinfo {year} {2010})}\BibitemShut {NoStop}%
\bibitem [{\citenamefont {Furnstahl}\ and\ \citenamefont
  {Hebeler}(2013)}]{Furnstahl13}%
  \BibitemOpen
  \bibfield  {author} {\bibinfo {author} {\bibfnamefont {R.~J.}\ \bibnamefont
  {Furnstahl}}\ and\ \bibinfo {author} {\bibfnamefont {K.}~\bibnamefont
  {Hebeler}},\ }\bibfield  {title} {\enquote {\bibinfo {title} {{New
  applications of renormalization group methods in nuclear physics}},}\ }\href
  {\doibase 10.1088/0034-4885/76/12/126301} {\bibfield  {journal} {\bibinfo
  {journal} {Rept. Prog. Phys.}\ }\textbf {\bibinfo {volume} {76}},\ \bibinfo
  {pages} {126301} (\bibinfo {year} {2013})}\BibitemShut {NoStop}%
\bibitem [{\citenamefont {Pr{\'e}zeau}(2006)}]{pre06}%
  \BibitemOpen
  \bibfield  {author} {\bibinfo {author} {\bibfnamefont {Gary}\ \bibnamefont
  {Pr{\'e}zeau}},\ }\bibfield  {title} {\enquote {\bibinfo {title} {Light
  neutrino and heavy particle exchange in $0\nu\beta\beta$-decay},}\ }\href
  {\doibase 10.1016/j.physletb.2005.11.048} {\bibfield  {journal} {\bibinfo
  {journal} {Phys. Lett. B}\ }\textbf {\bibinfo {volume} {633}},\ \bibinfo
  {pages} {93} (\bibinfo {year} {2006})}\BibitemShut {NoStop}%
\bibitem [{\citenamefont {Cirigliano}\ \emph {et~al.}(2004)\citenamefont
  {Cirigliano}, \citenamefont {Kurylov}, \citenamefont {Ramsey-Musolf},\ and\
  \citenamefont {Vogel}}]{cir04}%
  \BibitemOpen
  \bibfield  {author} {\bibinfo {author} {\bibfnamefont {V.}~\bibnamefont
  {Cirigliano}}, \bibinfo {author} {\bibfnamefont {A.}~\bibnamefont {Kurylov}},
  \bibinfo {author} {\bibfnamefont {M.~J.}\ \bibnamefont {Ramsey-Musolf}}, \
  and\ \bibinfo {author} {\bibfnamefont {P.}~\bibnamefont {Vogel}},\ }\bibfield
   {title} {\enquote {\bibinfo {title} {Neutrinoless double beta decay and
  lepton flavor violation},}\ }\href {\doibase 10.1103/PhysRevLett.93.231802}
  {\bibfield  {journal} {\bibinfo  {journal} {Phys. Rev. Lett.}\ }\textbf
  {\bibinfo {volume} {93}},\ \bibinfo {pages} {231802} (\bibinfo {year}
  {2004})}\BibitemShut {NoStop}%
\bibitem [{\citenamefont {Arnold}\ \emph {et~al.}(2010)\citenamefont {Arnold}
  \emph {et~al.}}]{arno10}%
  \BibitemOpen
  \bibfield  {author} {\bibinfo {author} {\bibfnamefont {R.}~\bibnamefont
  {Arnold}} \emph {et~al.} (\bibinfo {collaboration} {SuperNEMO
  Collaboration}),\ }\bibfield  {title} {\enquote {\bibinfo {title} {{Probing
  New Physics Models of Neutrinoless Double Beta Decay with SuperNEMO}},}\
  }\href {\doibase 10.1140/epjc/s10052-010-1481-5} {\bibfield  {journal}
  {\bibinfo  {journal} {Eur. Phys. J. C}\ }\textbf {\bibinfo {volume} {70}},\
  \bibinfo {pages} {927--943} (\bibinfo {year} {2010})}\BibitemShut {NoStop}%
\bibitem [{\citenamefont {Ali}\ \emph {et~al.}(2007)\citenamefont {Ali},
  \citenamefont {Borisov},\ and\ \citenamefont {Zhuridov}}]{ali06}%
  \BibitemOpen
  \bibfield  {author} {\bibinfo {author} {\bibfnamefont {A.}~\bibnamefont
  {Ali}}, \bibinfo {author} {\bibfnamefont {A.~V.}\ \bibnamefont {Borisov}}, \
  and\ \bibinfo {author} {\bibfnamefont {D.~V.}\ \bibnamefont {Zhuridov}},\
  }\bibfield  {title} {\enquote {\bibinfo {title} {{Probing new physics in the
  neutrinoless double beta decay using electron angular correlation}},}\ }\href
  {\doibase 10.1103/PhysRevD.76.093009} {\bibfield  {journal} {\bibinfo
  {journal} {Phys. Rev. D}\ }\textbf {\bibinfo {volume} {76}},\ \bibinfo
  {pages} {093009} (\bibinfo {year} {2007})}\BibitemShut {NoStop}%
\bibitem [{\citenamefont {Deppisch}\ and\ \citenamefont
  {P{\protect{\"a}}s}(2007)}]{dep06}%
  \BibitemOpen
  \bibfield  {author} {\bibinfo {author} {\bibfnamefont {Frank}\ \bibnamefont
  {Deppisch}}\ and\ \bibinfo {author} {\bibfnamefont {Heinrich}\ \bibnamefont
  {P{\protect{\"a}}s}},\ }\bibfield  {title} {\enquote {\bibinfo {title}
  {Pinning down the mechanism of neutrinoless double $\beta$ decay with
  measurements in different nuclei},}\ }\href {\doibase
  10.1103/PhysRevLett.98.232501} {\bibfield  {journal} {\bibinfo  {journal}
  {Phys. Rev. Lett.}\ }\textbf {\bibinfo {volume} {98}},\ \bibinfo {pages}
  {232501} (\bibinfo {year} {2007})}\BibitemShut {NoStop}%
\bibitem [{\citenamefont {Gehman}\ and\ \citenamefont {Elliott}(2007)}]{geh07}%
  \BibitemOpen
  \bibfield  {author} {\bibinfo {author} {\bibfnamefont {V.~M.}\ \bibnamefont
  {Gehman}}\ and\ \bibinfo {author} {\bibfnamefont {S.~R.}\ \bibnamefont
  {Elliott}},\ }\bibfield  {title} {\enquote {\bibinfo {title}
  {Multiple-isotope comparison for determining $0\nu\beta\beta$ mechanisms},}\
  }\href {\doibase 10.1088/0954-3899/34/4/006} {\bibfield  {journal} {\bibinfo
  {journal} {J. Phys. G: Nucl. Part. Phys.}\ }\textbf {\bibinfo {volume}
  {34}},\ \bibinfo {pages} {667} (\bibinfo {year} {2007})},\ \bibinfo {note}
  {[Erratum: J. Phys. G 35, 029701 (2008)]}\BibitemShut {NoStop}%
\bibitem [{\citenamefont {{\v S}imkovic}\ \emph {et~al.}(2010)\citenamefont
  {{\v S}imkovic}, \citenamefont {Vergados},\ and\ \citenamefont
  {Faessler}}]{sim10}%
  \BibitemOpen
  \bibfield  {author} {\bibinfo {author} {\bibfnamefont {Fedor}\ \bibnamefont
  {{\v S}imkovic}}, \bibinfo {author} {\bibfnamefont {John}\ \bibnamefont
  {Vergados}}, \ and\ \bibinfo {author} {\bibfnamefont {Amand}\ \bibnamefont
  {Faessler}},\ }\bibfield  {title} {\enquote {\bibinfo {title} {{Few active
  mechanisms of the neutrinoless double beta-decay and effective mass of
  Majorana neutrinos}},}\ }\href {\doibase 10.1103/PhysRevD.82.113015}
  {\bibfield  {journal} {\bibinfo  {journal} {Phys. Rev. D}\ }\textbf {\bibinfo
  {volume} {82}},\ \bibinfo {pages} {113015} (\bibinfo {year}
  {2010})}\BibitemShut {NoStop}%
\bibitem [{\citenamefont {Faessler}\ \emph {et~al.}(2011)\citenamefont
  {Faessler}, \citenamefont {Fogli}, \citenamefont {Lisi}, \citenamefont
  {Rotunno},\ and\ \citenamefont {\v{S}imkovic}}]{fae11}%
  \BibitemOpen
  \bibfield  {author} {\bibinfo {author} {\bibfnamefont {Amand}\ \bibnamefont
  {Faessler}}, \bibinfo {author} {\bibfnamefont {G.~L.}\ \bibnamefont {Fogli}},
  \bibinfo {author} {\bibfnamefont {E.}~\bibnamefont {Lisi}}, \bibinfo {author}
  {\bibfnamefont {A.M.}\ \bibnamefont {Rotunno}}, \ and\ \bibinfo {author}
  {\bibfnamefont {F.}~\bibnamefont {\v{S}imkovic}},\ }\bibfield  {title}
  {\enquote {\bibinfo {title} {Multi-isotope degeneracy of neutrinoless double
  beta decay mechanisms in the quasi-particle random phase approximation},}\
  }\href {\doibase 10.1103/PhysRevD.83.113015} {\bibfield  {journal} {\bibinfo
  {journal} {Phys.\ Rev.\ D}\ }\textbf {\bibinfo {volume} {83}},\ \bibinfo
  {pages} {113015} (\bibinfo {year} {2011})}\BibitemShut {NoStop}%
\bibitem [{\citenamefont {Lisi}\ \emph {et~al.}(2015)\citenamefont {Lisi},
  \citenamefont {Rotunno},\ and\ \citenamefont {{\v S}imkovic}}]{Lisi15}%
  \BibitemOpen
  \bibfield  {author} {\bibinfo {author} {\bibfnamefont {E.}~\bibnamefont
  {Lisi}}, \bibinfo {author} {\bibfnamefont {A.}~\bibnamefont {Rotunno}}, \
  and\ \bibinfo {author} {\bibfnamefont {F.}~\bibnamefont {{\v S}imkovic}},\
  }\bibfield  {title} {\enquote {\bibinfo {title} {{Degeneracies of particle
  and nuclear physics uncertainties in neutrinoless $\beta \beta$ decay}},}\
  }\href {\doibase 10.1103/PhysRevD.92.093004} {\bibfield  {journal} {\bibinfo
  {journal} {Phys. Rev. D}\ }\textbf {\bibinfo {volume} {92}},\ \bibinfo
  {pages} {093004} (\bibinfo {year} {2015})}\BibitemShut {NoStop}%
\bibitem [{\citenamefont {{\v S}imkovic}\ \emph {et~al.}(2001)\citenamefont
  {{\v S}imkovic}, \citenamefont {Nowak}, \citenamefont {Kaminski},
  \citenamefont {Raduta},\ and\ \citenamefont {Faessler}}]{sim01a}%
  \BibitemOpen
  \bibfield  {author} {\bibinfo {author} {\bibfnamefont {F.}~\bibnamefont {{\v
  S}imkovic}}, \bibinfo {author} {\bibfnamefont {M.}~\bibnamefont {Nowak}},
  \bibinfo {author} {\bibfnamefont {W.~A.}\ \bibnamefont {Kaminski}}, \bibinfo
  {author} {\bibfnamefont {A.~A.}\ \bibnamefont {Raduta}}, \ and\ \bibinfo
  {author} {\bibfnamefont {Amand}\ \bibnamefont {Faessler}},\ }\bibfield
  {title} {\enquote {\bibinfo {title} {{Neutrinoless double beta decay of
  $^{76}$Ge, $^{82}$Se, $^{100}$Mo and $^{136}$Xe to excited $0^+$ states}},}\
  }\href {\doibase 10.1103/PhysRevC.64.035501} {\bibfield  {journal} {\bibinfo
  {journal} {Phys. Rev. C}\ }\textbf {\bibinfo {volume} {64}},\ \bibinfo
  {pages} {035501} (\bibinfo {year} {2001})}\BibitemShut {NoStop}%
\bibitem [{\citenamefont {Men\'{e}ndez}\ \emph
  {et~al.}(2009{\natexlab{a}})\citenamefont {Men\'{e}ndez}, \citenamefont
  {Poves}, \citenamefont {Caurier},\ and\ \citenamefont {Nowacki}}]{men09}%
  \BibitemOpen
  \bibfield  {author} {\bibinfo {author} {\bibfnamefont {J.}~\bibnamefont
  {Men\'{e}ndez}}, \bibinfo {author} {\bibfnamefont {A.}~\bibnamefont {Poves}},
  \bibinfo {author} {\bibfnamefont {E.}~\bibnamefont {Caurier}}, \ and\
  \bibinfo {author} {\bibfnamefont {F.}~\bibnamefont {Nowacki}},\ }\bibfield
  {title} {\enquote {\bibinfo {title} {Occupancies of individual orbits, and
  the nuclear matrix element of the $^{76}$\protect{G}e neutrinoless
  $\beta\beta$ decay},}\ }\href {\doibase 10.1103/PhysRevC.80.048501}
  {\bibfield  {journal} {\bibinfo  {journal} {Phys. Rev. C}\ }\textbf {\bibinfo
  {volume} {80}},\ \bibinfo {pages} {048501} (\bibinfo {year}
  {2009}{\natexlab{a}})}\BibitemShut {NoStop}%
\bibitem [{\citenamefont {Barea}\ \emph
  {et~al.}(2015{\natexlab{b}})\citenamefont {Barea}, \citenamefont {Kotila},\
  and\ \citenamefont {Iachello}}]{bar15}%
  \BibitemOpen
  \bibfield  {author} {\bibinfo {author} {\bibfnamefont {J.}~\bibnamefont
  {Barea}}, \bibinfo {author} {\bibfnamefont {J.}~\bibnamefont {Kotila}}, \
  and\ \bibinfo {author} {\bibfnamefont {F.}~\bibnamefont {Iachello}},\
  }\bibfield  {title} {\enquote {\bibinfo {title} {{$0\nu\beta\beta$ and
  $2\nu\beta\beta$ nuclear matrix elements in the interacting boson model with
  isospin restoration}},}\ }\href {\doibase 10.1103/PhysRevC.91.034304}
  {\bibfield  {journal} {\bibinfo  {journal} {Phys. Rev. C}\ }\textbf {\bibinfo
  {volume} {91}},\ \bibinfo {pages} {034304} (\bibinfo {year}
  {2015}{\natexlab{b}})}\BibitemShut {NoStop}%
\bibitem [{\citenamefont {Hyv{\"a}rinen}\ and\ \citenamefont
  {Suhonen}(2016)}]{Hyvarinen16}%
  \BibitemOpen
  \bibfield  {author} {\bibinfo {author} {\bibfnamefont {Juhani}\ \bibnamefont
  {Hyv{\"a}rinen}}\ and\ \bibinfo {author} {\bibfnamefont {Jouni}\ \bibnamefont
  {Suhonen}},\ }\bibfield  {title} {\enquote {\bibinfo {title} {{Neutrinoless
  $\beta\beta$ decays to excited 0$^+$ states and the Majorana-neutrino
  mass}},}\ }\href {\doibase 10.1103/PhysRevC.93.064306} {\bibfield  {journal}
  {\bibinfo  {journal} {Phys. Rev. C}\ }\textbf {\bibinfo {volume} {93}},\
  \bibinfo {pages} {064306} (\bibinfo {year} {2016})}\BibitemShut {NoStop}%
\bibitem [{\citenamefont {Gomez-Cadenas}\ \emph {et~al.}(2011)\citenamefont
  {Gomez-Cadenas}, \citenamefont {Martin-Albo}, \citenamefont {Sorel},
  \citenamefont {Ferrario}, \citenamefont {Monrabal}, \citenamefont
  {Munoz-Vidal}, \citenamefont {Novella},\ and\ \citenamefont
  {Poves}}]{GomezCadenas11}%
  \BibitemOpen
  \bibfield  {author} {\bibinfo {author} {\bibfnamefont {J.~J.}\ \bibnamefont
  {Gomez-Cadenas}}, \bibinfo {author} {\bibfnamefont {J.}~\bibnamefont
  {Martin-Albo}}, \bibinfo {author} {\bibfnamefont {M.}~\bibnamefont {Sorel}},
  \bibinfo {author} {\bibfnamefont {P.}~\bibnamefont {Ferrario}}, \bibinfo
  {author} {\bibfnamefont {F.}~\bibnamefont {Monrabal}}, \bibinfo {author}
  {\bibfnamefont {J.}~\bibnamefont {Munoz-Vidal}}, \bibinfo {author}
  {\bibfnamefont {P.}~\bibnamefont {Novella}}, \ and\ \bibinfo {author}
  {\bibfnamefont {A.}~\bibnamefont {Poves}},\ }\bibfield  {title} {\enquote
  {\bibinfo {title} {{Sense and sensitivity of double beta decay
  experiments}},}\ }\href {\doibase 10.1088/1475-7516/2011/06/007} {\bibfield
  {journal} {\bibinfo  {journal} {JCAP}\ }\textbf {\bibinfo {volume} {1106}},\
  \bibinfo {pages} {007} (\bibinfo {year} {2011})}\BibitemShut {NoStop}%
\bibitem [{\citenamefont {Robertson}(2013)}]{hamish13}%
  \BibitemOpen
  \bibfield  {author} {\bibinfo {author} {\bibfnamefont {R.~G.~H.}\
  \bibnamefont {Robertson}},\ }\bibfield  {title} {\enquote {\bibinfo {title}
  {Empirical survey of neutrinoless double beta decay matrix elements},}\
  }\href {\doibase 10.1142/S0217732313500211} {\bibfield  {journal} {\bibinfo
  {journal} {Mod. Phys. Lett. A}\ }\textbf {\bibinfo {volume} {28}},\ \bibinfo
  {pages} {1350021} (\bibinfo {year} {2013})}\BibitemShut {NoStop}%
\bibitem [{\citenamefont {Men\'{e}ndez}\ \emph
  {et~al.}(2009{\natexlab{b}})\citenamefont {Men\'{e}ndez}, \citenamefont
  {Poves}, \citenamefont {Caurier},\ and\ \citenamefont {Nowacki}}]{men08}%
  \BibitemOpen
  \bibfield  {author} {\bibinfo {author} {\bibfnamefont {J.}~\bibnamefont
  {Men\'{e}ndez}}, \bibinfo {author} {\bibfnamefont {A.}~\bibnamefont {Poves}},
  \bibinfo {author} {\bibfnamefont {E.}~\bibnamefont {Caurier}}, \ and\
  \bibinfo {author} {\bibfnamefont {F.}~\bibnamefont {Nowacki}},\ }\bibfield
  {title} {\enquote {\bibinfo {title} {Disassembling the nuclear matrix
  elements of the neutrinoless double beta decay},}\ }\href {\doibase
  10.1016/j.nuclphysa.2008.12.005} {\bibfield  {journal} {\bibinfo  {journal}
  {Nucl. Phys. A}\ }\textbf {\bibinfo {volume} {818}},\ \bibinfo {pages}
  {139--151} (\bibinfo {year} {2009}{\natexlab{b}})}\BibitemShut {NoStop}%
\bibitem [{\citenamefont {Iwata}\ \emph {et~al.}(2016)\citenamefont {Iwata},
  \citenamefont {Shimizu}, \citenamefont {Otsuka}, \citenamefont {Utsuno},
  \citenamefont {Men\'{e}ndez}, \citenamefont {Honma},\ and\ \citenamefont
  {Abe}}]{Iwata16}%
  \BibitemOpen
  \bibfield  {author} {\bibinfo {author} {\bibfnamefont {Y.}~\bibnamefont
  {Iwata}}, \bibinfo {author} {\bibfnamefont {N.}~\bibnamefont {Shimizu}},
  \bibinfo {author} {\bibfnamefont {T.}~\bibnamefont {Otsuka}}, \bibinfo
  {author} {\bibfnamefont {Y.}~\bibnamefont {Utsuno}}, \bibinfo {author}
  {\bibfnamefont {J.}~\bibnamefont {Men\'{e}ndez}}, \bibinfo {author}
  {\bibfnamefont {M.}~\bibnamefont {Honma}}, \ and\ \bibinfo {author}
  {\bibfnamefont {T.}~\bibnamefont {Abe}},\ }\bibfield  {title} {\enquote
  {\bibinfo {title} {{Large-scale shell-model analysis of the neutrinoless
  $\beta\beta$ decay of $^{48}$Ca}},}\ }\href {\doibase
  10.1103/PhysRevLett.116.112502} {\bibfield  {journal} {\bibinfo  {journal}
  {Phys. Rev. Lett.}\ }\textbf {\bibinfo {volume} {116}},\ \bibinfo {pages}
  {112502} (\bibinfo {year} {2016})}\BibitemShut {NoStop}%
\bibitem [{\citenamefont {{{\protect \v{S}}imkovic}}\ \emph
  {et~al.}(2013)\citenamefont {{{\protect \v{S}}imkovic}}, \citenamefont
  {Rodin}, \citenamefont {Faessler},\ and\ \citenamefont {Vogel}}]{sim13}%
  \BibitemOpen
  \bibfield  {author} {\bibinfo {author} {\bibfnamefont {Fedor}\ \bibnamefont
  {{{\protect \v{S}}imkovic}}}, \bibinfo {author} {\bibfnamefont {Vadim}\
  \bibnamefont {Rodin}}, \bibinfo {author} {\bibfnamefont {Amand}\ \bibnamefont
  {Faessler}}, \ and\ \bibinfo {author} {\bibfnamefont {Petr}\ \bibnamefont
  {Vogel}},\ }\bibfield  {title} {\enquote {\bibinfo {title} {$0\nu\beta\beta$
  and $2\nu\beta\beta$ nuclear matrix elements, quasiparticle ramdom-phase
  approximation, and isospin symmetry restoration},}\ }\href {\doibase
  10.1103/PhysRevC.87.045501} {\bibfield  {journal} {\bibinfo  {journal} {Phys.
  Rev. C}\ }\textbf {\bibinfo {volume} {87}},\ \bibinfo {pages} {045501}
  (\bibinfo {year} {2013})}\BibitemShut {NoStop}%
\bibitem [{\citenamefont {Fang}\ \emph {et~al.}(2015)\citenamefont {Fang},
  \citenamefont {Faessler},\ and\ \citenamefont {{\v S}imkovic}}]{fang15}%
  \BibitemOpen
  \bibfield  {author} {\bibinfo {author} {\bibfnamefont {Dong-Liang}\
  \bibnamefont {Fang}}, \bibinfo {author} {\bibfnamefont {Amand}\ \bibnamefont
  {Faessler}}, \ and\ \bibinfo {author} {\bibfnamefont {Fedor}\ \bibnamefont
  {{\v S}imkovic}},\ }\bibfield  {title} {\enquote {\bibinfo {title} {{Partial
  restoration of isospin symmetry for neutrinoless double $\beta$ decay in the
  deformed nuclear system of $^{150}$Nd}},}\ }\href {\doibase
  10.1103/PhysRevC.92.044301} {\bibfield  {journal} {\bibinfo  {journal} {Phys.
  Rev. C}\ }\textbf {\bibinfo {volume} {92}},\ \bibinfo {pages} {044301}
  (\bibinfo {year} {2015})}\BibitemShut {NoStop}%
\bibitem [{\citenamefont {Mustonen}\ and\ \citenamefont
  {Engel}(2013)}]{Mustonen13}%
  \BibitemOpen
  \bibfield  {author} {\bibinfo {author} {\bibfnamefont {M.~T.}\ \bibnamefont
  {Mustonen}}\ and\ \bibinfo {author} {\bibfnamefont {J.}~\bibnamefont
  {Engel}},\ }\bibfield  {title} {\enquote {\bibinfo {title} {Large-scale
  calculations of the double-$\beta$ decay of $^{76}$\protect{G}e,
  $^{128}$\protect{T}e, $^{136}$\protect{X}e, and $^{150}$\protect{N}d in the
  deformed self-consistent \protect{S}kyrme quasiparticle random-phase
  approximation},}\ }\href {\doibase 10.1103/PhysRevC.87.064302} {\bibfield
  {journal} {\bibinfo  {journal} {Phys. Rev. C}\ }\textbf {\bibinfo {volume}
  {87}},\ \bibinfo {pages} {064302} (\bibinfo {year} {2013})}\BibitemShut
  {NoStop}%
\bibitem [{\citenamefont {Yao}\ \emph {et~al.}(2015)\citenamefont {Yao},
  \citenamefont {Song}, \citenamefont {Hagino}, \citenamefont {Ring},\ and\
  \citenamefont {Meng}}]{yao15}%
  \BibitemOpen
  \bibfield  {author} {\bibinfo {author} {\bibfnamefont {J.~M.}\ \bibnamefont
  {Yao}}, \bibinfo {author} {\bibfnamefont {L.~S.}\ \bibnamefont {Song}},
  \bibinfo {author} {\bibfnamefont {K.}~\bibnamefont {Hagino}}, \bibinfo
  {author} {\bibfnamefont {P.}~\bibnamefont {Ring}}, \ and\ \bibinfo {author}
  {\bibfnamefont {J.}~\bibnamefont {Meng}},\ }\bibfield  {title} {\enquote
  {\bibinfo {title} {{Systematic study of nuclear matrix elements in
  neutrinoless double-$\beta$ decay with a beyond-mean-field covariant density
  functional theory}},}\ }\href {\doibase 10.1103/PhysRevC.91.024316}
  {\bibfield  {journal} {\bibinfo  {journal} {Phys. Rev. C}\ }\textbf {\bibinfo
  {volume} {91}},\ \bibinfo {pages} {024316} (\bibinfo {year}
  {2015})}\BibitemShut {NoStop}%
\bibitem [{\citenamefont {Yao}\ and\ \citenamefont {Engel}(2016)}]{yao16}%
  \BibitemOpen
  \bibfield  {author} {\bibinfo {author} {\bibfnamefont {J.~M.}\ \bibnamefont
  {Yao}}\ and\ \bibinfo {author} {\bibfnamefont {J.}~\bibnamefont {Engel}},\
  }\bibfield  {title} {\enquote {\bibinfo {title} {{Octupole correlations in
  low-lying states of $^{150}$Nd and $^{150}$Sm and their impact on
  neutrinoless double-$\beta$ decay}},}\ }\href {\doibase
  10.1103/PhysRevC.94.014306} {\bibfield  {journal} {\bibinfo  {journal} {Phys.
  Rev. C}\ }\textbf {\bibinfo {volume} {94}},\ \bibinfo {pages} {014306}
  (\bibinfo {year} {2016})}\BibitemShut {NoStop}%
\bibitem [{\citenamefont {Vaquero}\ \emph {et~al.}(2013)\citenamefont
  {Vaquero}, \citenamefont {Rodr\'iguez},\ and\ \citenamefont {Egido}}]{vaq13}%
  \BibitemOpen
  \bibfield  {author} {\bibinfo {author} {\bibfnamefont {Nuria~L{\'o}pez}\
  \bibnamefont {Vaquero}}, \bibinfo {author} {\bibfnamefont {Tom{\'a}s~R.}\
  \bibnamefont {Rodr\'iguez}}, \ and\ \bibinfo {author} {\bibfnamefont
  {J.~Luis}\ \bibnamefont {Egido}},\ }\bibfield  {title} {\enquote {\bibinfo
  {title} {Shape and pairing fluctuations effects on neutrinoless double beta
  decay},}\ }\href {\doibase 10.1103/PhysRevLett.111.142501} {\bibfield
  {journal} {\bibinfo  {journal} {Phys. Rev. Lett.}\ }\textbf {\bibinfo
  {volume} {111}},\ \bibinfo {pages} {142501} (\bibinfo {year}
  {2013})}\BibitemShut {NoStop}%
\bibitem [{\citenamefont {Caurier}\ \emph {et~al.}(2005)\citenamefont
  {Caurier}, \citenamefont {Mart{\'i}nez-Pinedo}, \citenamefont {Nowacki},
  \citenamefont {Poves},\ and\ \citenamefont {Zuker}}]{cau05}%
  \BibitemOpen
  \bibfield  {author} {\bibinfo {author} {\bibfnamefont {E.}~\bibnamefont
  {Caurier}}, \bibinfo {author} {\bibfnamefont {G.}~\bibnamefont
  {Mart{\'i}nez-Pinedo}}, \bibinfo {author} {\bibfnamefont {F.}~\bibnamefont
  {Nowacki}}, \bibinfo {author} {\bibfnamefont {A.}~\bibnamefont {Poves}}, \
  and\ \bibinfo {author} {\bibfnamefont {A.~P.}\ \bibnamefont {Zuker}},\
  }\bibfield  {title} {\enquote {\bibinfo {title} {{The shell model as a
  unified view of nuclear structure}},}\ }\href {\doibase
  10.1103/RevModPhys.77.427} {\bibfield  {journal} {\bibinfo  {journal} {Rev.
  Mod. Phys.}\ }\textbf {\bibinfo {volume} {77}},\ \bibinfo {pages} {427}
  (\bibinfo {year} {2005})}\BibitemShut {NoStop}%
\bibitem [{\citenamefont {Barea}\ \emph {et~al.}(2013)\citenamefont {Barea},
  \citenamefont {Kotila},\ and\ \citenamefont {Iachello}}]{bar13}%
  \BibitemOpen
  \bibfield  {author} {\bibinfo {author} {\bibfnamefont {J.}~\bibnamefont
  {Barea}}, \bibinfo {author} {\bibfnamefont {J.}~\bibnamefont {Kotila}}, \
  and\ \bibinfo {author} {\bibfnamefont {F.}~\bibnamefont {Iachello}},\
  }\bibfield  {title} {\enquote {\bibinfo {title} {Nuclear matrix elements for
  double-$\beta$ decay},}\ }\href {\doibase 10.1103/PhysRevC.87.014315}
  {\bibfield  {journal} {\bibinfo  {journal} {Phys.\ Rev.\ C}\ }\textbf
  {\bibinfo {volume} {87}},\ \bibinfo {pages} {014315} (\bibinfo {year}
  {2013})}\BibitemShut {NoStop}%
\bibitem [{\citenamefont {Pirinen}\ and\ \citenamefont
  {Suhonen}(2015)}]{Pirinen15}%
  \BibitemOpen
  \bibfield  {author} {\bibinfo {author} {\bibfnamefont {Pekka}\ \bibnamefont
  {Pirinen}}\ and\ \bibinfo {author} {\bibfnamefont {Jouni}\ \bibnamefont
  {Suhonen}},\ }\bibfield  {title} {\enquote {\bibinfo {title} {{Systematic
  approach to $\beta$ and 2$\nu\beta\beta$ decays of mass $A=100-136$
  nuclei}},}\ }\href {\doibase 10.1103/PhysRevC.91.054309} {\bibfield
  {journal} {\bibinfo  {journal} {Phys. Rev. C}\ }\textbf {\bibinfo {volume}
  {91}},\ \bibinfo {pages} {054309} (\bibinfo {year} {2015})}\BibitemShut
  {NoStop}%
\bibitem [{\citenamefont {Brown}(2001)}]{bro01}%
  \BibitemOpen
  \bibfield  {author} {\bibinfo {author} {\bibfnamefont {B.~A.}\ \bibnamefont
  {Brown}},\ }\bibfield  {title} {\enquote {\bibinfo {title} {The nuclear shell
  model towards the drip lines},}\ }\href {\doibase
  10.1016/S0146-6410(01)00159-4} {\bibfield  {journal} {\bibinfo  {journal}
  {Prog. Part. Nucl. Phys.}\ }\textbf {\bibinfo {volume} {47}},\ \bibinfo
  {pages} {517 -- 599} (\bibinfo {year} {2001})}\BibitemShut {NoStop}%
\bibitem [{\citenamefont {Otsuka}\ \emph {et~al.}(2001)\citenamefont {Otsuka},
  \citenamefont {Honma}, \citenamefont {Mizusaki}, \citenamefont {Shimizu},\
  and\ \citenamefont {Utsuno}}]{ots01}%
  \BibitemOpen
  \bibfield  {author} {\bibinfo {author} {\bibfnamefont {T.}~\bibnamefont
  {Otsuka}}, \bibinfo {author} {\bibfnamefont {M.}~\bibnamefont {Honma}},
  \bibinfo {author} {\bibfnamefont {T.}~\bibnamefont {Mizusaki}}, \bibinfo
  {author} {\bibfnamefont {N.}~\bibnamefont {Shimizu}}, \ and\ \bibinfo
  {author} {\bibfnamefont {Y.}~\bibnamefont {Utsuno}},\ }\bibfield  {title}
  {\enquote {\bibinfo {title} {Monte carlo shell model for atomic nuclei},}\
  }\href {\doibase 10.1016/S0146-6410(01)00157-0} {\bibfield  {journal}
  {\bibinfo  {journal} {Prog. Part. Nucl. Phys.}\ }\textbf {\bibinfo {volume}
  {47}},\ \bibinfo {pages} {319 -- 400} (\bibinfo {year} {2001})}\BibitemShut
  {NoStop}%
\bibitem [{\citenamefont {{\v C}{\'a}rsky}(1998)}]{carsky98}%
  \BibitemOpen
  \bibfield  {author} {\bibinfo {author} {\bibfnamefont {P.}~\bibnamefont {{\v
  C}{\'a}rsky}},\ }\bibfield  {title} {\enquote {\bibinfo {title}
  {Configuration interaction},}\ }in\ \href {\doibase
  10.1002/0470845015.cca036} {\emph {\bibinfo {booktitle} {Encyclopedia of
  Computation Chemistry}}},\ \bibinfo {editor} {edited by\ \bibinfo {editor}
  {\bibfnamefont {P.}~\bibnamefont {Schleyer}}, \bibinfo {editor}
  {\bibfnamefont {N.~L.}\ \bibnamefont {Allinger}}, \bibinfo {editor}
  {\bibfnamefont {T.}~\bibnamefont {Clark}}, \bibinfo {editor} {\bibfnamefont
  {J.}~\bibnamefont {Gasteiger}}, \bibinfo {editor} {\bibfnamefont {P.~A.}\
  \bibnamefont {Kollman}}, \bibinfo {editor} {\bibfnamefont {H.~F.}\
  \bibnamefont {Schaefer}}, \ and\ \bibinfo {editor} {\bibfnamefont {P.~R.}\
  \bibnamefont {Schreiner}}}\ (\bibinfo  {publisher} {John Wiley and Sons},\
  \bibinfo {address} {Chichester},\ \bibinfo {year} {1998})\ pp.\ \bibinfo
  {pages} {485--497}\BibitemShut {NoStop}%
\bibitem [{\citenamefont {Sherrill}\ and\ \citenamefont
  {{Schaefer}}(1999)}]{Sherrill99}%
  \BibitemOpen
  \bibfield  {author} {\bibinfo {author} {\bibfnamefont {C.~David}\
  \bibnamefont {Sherrill}}\ and\ \bibinfo {author} {\bibfnamefont {Henry~F.}\
  \bibnamefont {{Schaefer}}},\ }\bibfield  {title} {\enquote {\bibinfo {title}
  {The configuration interaction method: Advances in highly correlated
  approaches},}\ }\href {\doibase 10.1016/S0065-3276(08)60532-8} {\bibfield
  {journal} {\bibinfo  {journal} {Adv. Quantum Chem.}\ }\textbf {\bibinfo
  {volume} {34}},\ \bibinfo {pages} {143 -- 269} (\bibinfo {year}
  {1999})}\BibitemShut {NoStop}%
\bibitem [{\citenamefont {Dufour}\ and\ \citenamefont {Zuker}(1996)}]{duf96}%
  \BibitemOpen
  \bibfield  {author} {\bibinfo {author} {\bibfnamefont {Marianne}\
  \bibnamefont {Dufour}}\ and\ \bibinfo {author} {\bibfnamefont
  {Andr\'{e}s~P.}\ \bibnamefont {Zuker}},\ }\bibfield  {title} {\enquote
  {\bibinfo {title} {The realistic collective nuclear hamiltonian},}\ }\href
  {\doibase 10.1103/PhysRevC.54.1641} {\bibfield  {journal} {\bibinfo
  {journal} {Phys. Rev. C}\ }\textbf {\bibinfo {volume} {54}},\ \bibinfo
  {pages} {1641--1660} (\bibinfo {year} {1996})}\BibitemShut {NoStop}%
\bibitem [{\citenamefont {Mayer}(1949)}]{Mayer49}%
  \BibitemOpen
  \bibfield  {author} {\bibinfo {author} {\bibfnamefont {M.~G.}\ \bibnamefont
  {Mayer}},\ }\bibfield  {title} {\enquote {\bibinfo {title} {On closed shells
  in nuclei. \protect{II}},}\ }\href {\doibase 10.1103/PhysRev.75.1969}
  {\bibfield  {journal} {\bibinfo  {journal} {Phys. Rev.}\ }\textbf {\bibinfo
  {volume} {75}},\ \bibinfo {pages} {1969} (\bibinfo {year}
  {1949})}\BibitemShut {NoStop}%
\bibitem [{\citenamefont {Haxel}\ \emph {et~al.}(1949)\citenamefont {Haxel},
  \citenamefont {Jensen},\ and\ \citenamefont {Suess}}]{Haxel49}%
  \BibitemOpen
  \bibfield  {author} {\bibinfo {author} {\bibfnamefont {O.}~\bibnamefont
  {Haxel}}, \bibinfo {author} {\bibfnamefont {J.~H.~D.}\ \bibnamefont
  {Jensen}}, \ and\ \bibinfo {author} {\bibfnamefont {H.~E.}\ \bibnamefont
  {Suess}},\ }\bibfield  {title} {\enquote {\bibinfo {title} {On the "magic
  numbers" in nuclear structure},}\ }\href {\doibase 10.1103/PhysRev.75.1766.2}
  {\bibfield  {journal} {\bibinfo  {journal} {Phys. Rev.}\ }\textbf {\bibinfo
  {volume} {75}},\ \bibinfo {pages} {1766} (\bibinfo {year}
  {1949})}\BibitemShut {NoStop}%
\bibitem [{\citenamefont {Shimizu}\ \emph {et~al.}(2016)\citenamefont
  {Shimizu}, \citenamefont {Utsuno}, \citenamefont {Futamura}, \citenamefont
  {Sakurai}, \citenamefont {Mizusaki},\ and\ \citenamefont
  {Otsuka}}]{Shimizu16}%
  \BibitemOpen
  \bibfield  {author} {\bibinfo {author} {\bibfnamefont {Noritaka}\
  \bibnamefont {Shimizu}}, \bibinfo {author} {\bibfnamefont {Yutaka}\
  \bibnamefont {Utsuno}}, \bibinfo {author} {\bibfnamefont {Yasunori}\
  \bibnamefont {Futamura}}, \bibinfo {author} {\bibfnamefont {Tetsuya}\
  \bibnamefont {Sakurai}}, \bibinfo {author} {\bibfnamefont {Takahiro}\
  \bibnamefont {Mizusaki}}, \ and\ \bibinfo {author} {\bibfnamefont {Takaharu}\
  \bibnamefont {Otsuka}},\ }\bibfield  {title} {\enquote {\bibinfo {title}
  {{Stochastic estimation of nuclear level density in the nuclear shell model:
  An application to parity-dependent level density in $^{58}$Ni}},}\ }\href
  {\doibase 10.1016/j.physletb.2015.12.005} {\bibfield  {journal} {\bibinfo
  {journal} {Phys. Lett. B}\ }\textbf {\bibinfo {volume} {753}},\ \bibinfo
  {pages} {13} (\bibinfo {year} {2016})}\BibitemShut {NoStop}%
\bibitem [{\citenamefont {Nowacki}\ \emph {et~al.}(2016)\citenamefont
  {Nowacki}, \citenamefont {Poves}, \citenamefont {Caurier},\ and\
  \citenamefont {Bounthong}}]{Nowacki16}%
  \BibitemOpen
  \bibfield  {author} {\bibinfo {author} {\bibfnamefont {F.}~\bibnamefont
  {Nowacki}}, \bibinfo {author} {\bibfnamefont {A.}~\bibnamefont {Poves}},
  \bibinfo {author} {\bibfnamefont {E.}~\bibnamefont {Caurier}}, \ and\
  \bibinfo {author} {\bibfnamefont {B.}~\bibnamefont {Bounthong}},\ }\bibfield
  {title} {\enquote {\bibinfo {title} {{Shape Coexistence in $^{78}$Ni as the
  Portal to the Fifth Island of Inversion}},}\ }\href {\doibase
  10.1103/PhysRevLett.117.272501} {\bibfield  {journal} {\bibinfo  {journal}
  {Phys. Rev. Lett.}\ }\textbf {\bibinfo {volume} {117}},\ \bibinfo {pages}
  {272501} (\bibinfo {year} {2016})}\BibitemShut {NoStop}%
\bibitem [{\citenamefont {Caurier}\ \emph
  {et~al.}(2008{\natexlab{a}})\citenamefont {Caurier}, \citenamefont
  {Men\'{e}ndez}, \citenamefont {Nowacki},\ and\ \citenamefont
  {Poves}}]{caurier08}%
  \BibitemOpen
  \bibfield  {author} {\bibinfo {author} {\bibfnamefont {E.}~\bibnamefont
  {Caurier}}, \bibinfo {author} {\bibfnamefont {J.}~\bibnamefont
  {Men\'{e}ndez}}, \bibinfo {author} {\bibfnamefont {F.}~\bibnamefont
  {Nowacki}}, \ and\ \bibinfo {author} {\bibfnamefont {A.}~\bibnamefont
  {Poves}},\ }\bibfield  {title} {\enquote {\bibinfo {title} {The influence of
  pairing on the nuclear matrix elements of the neutrinoless double beta
  decays},}\ }\href {\doibase 10.1103/PhysRevLett.100.052503} {\bibfield
  {journal} {\bibinfo  {journal} {Phys.\ Rev.\ Lett.}\ }\textbf {\bibinfo
  {volume} {100}},\ \bibinfo {pages} {052503} (\bibinfo {year}
  {2008}{\natexlab{a}})}\BibitemShut {NoStop}%
\bibitem [{\citenamefont {Neacsu}\ and\ \citenamefont
  {Horoi}(2015)}]{Neacsu15}%
  \BibitemOpen
  \bibfield  {author} {\bibinfo {author} {\bibfnamefont {Andrei}\ \bibnamefont
  {Neacsu}}\ and\ \bibinfo {author} {\bibfnamefont {Mihai}\ \bibnamefont
  {Horoi}},\ }\bibfield  {title} {\enquote {\bibinfo {title} {{Shell model
  studies of the $^{130}$\protect{T}e neutrinoless double-beta decay}},}\
  }\href {\doibase 10.1103/PhysRevC.91.024309} {\bibfield  {journal} {\bibinfo
  {journal} {Phys. Rev. C}\ }\textbf {\bibinfo {volume} {91}},\ \bibinfo
  {pages} {024309} (\bibinfo {year} {2015})}\BibitemShut {NoStop}%
\bibitem [{\citenamefont {Sen'kov}\ and\ \citenamefont {Horoi}(2016)}]{sen16}%
  \BibitemOpen
  \bibfield  {author} {\bibinfo {author} {\bibfnamefont {R.~A.}\ \bibnamefont
  {Sen'kov}}\ and\ \bibinfo {author} {\bibfnamefont {M.}~\bibnamefont
  {Horoi}},\ }\bibfield  {title} {\enquote {\bibinfo {title} {{Shell-model
  calculation of neutrinoless double-$\beta$ decay of $^{76}$Ge}},}\ }\href
  {\doibase 10.1103/PhysRevC.93.044334} {\bibfield  {journal} {\bibinfo
  {journal} {Phys. Rev. C}\ }\textbf {\bibinfo {volume} {93}},\ \bibinfo
  {pages} {044334} (\bibinfo {year} {2016})}\BibitemShut {NoStop}%
\bibitem [{\citenamefont {Vogel}(2012)}]{vog12}%
  \BibitemOpen
  \bibfield  {author} {\bibinfo {author} {\bibfnamefont {Petr}\ \bibnamefont
  {Vogel}},\ }\bibfield  {title} {\enquote {\bibinfo {title} {{Nuclear
  structure and double beta decay}},}\ }\href {\doibase
  10.1088/0954-3899/39/12/124002} {\bibfield  {journal} {\bibinfo  {journal}
  {J. Phys. G: Nucl. Part. Phys.}\ }\textbf {\bibinfo {volume} {39}},\ \bibinfo
  {pages} {124002} (\bibinfo {year} {2012})}\BibitemShut {NoStop}%
\bibitem [{\citenamefont {Chou}\ \emph {et~al.}(1993)\citenamefont {Chou},
  \citenamefont {Warburton},\ and\ \citenamefont {Brown}}]{Chou93}%
  \BibitemOpen
  \bibfield  {author} {\bibinfo {author} {\bibfnamefont {W.~T.}\ \bibnamefont
  {Chou}}, \bibinfo {author} {\bibfnamefont {E.~K.}\ \bibnamefont {Warburton}},
  \ and\ \bibinfo {author} {\bibfnamefont {B.~Alex}\ \bibnamefont {Brown}},\
  }\bibfield  {title} {\enquote {\bibinfo {title} {{Gamow-Teller beta-decay
  rates for A $\leq$ 18 nuclei}},}\ }\href {\doibase 10.1103/PhysRevC.47.163}
  {\bibfield  {journal} {\bibinfo  {journal} {Phys. Rev. C}\ }\textbf {\bibinfo
  {volume} {47}},\ \bibinfo {pages} {163--177} (\bibinfo {year}
  {1993})}\BibitemShut {NoStop}%
\bibitem [{\citenamefont {Wildenthal}\ \emph {et~al.}(1983)\citenamefont
  {Wildenthal}, \citenamefont {Curtin},\ and\ \citenamefont
  {Brown}}]{Wildenthal83}%
  \BibitemOpen
  \bibfield  {author} {\bibinfo {author} {\bibfnamefont {B.~H.}\ \bibnamefont
  {Wildenthal}}, \bibinfo {author} {\bibfnamefont {M.~S.}\ \bibnamefont
  {Curtin}}, \ and\ \bibinfo {author} {\bibfnamefont {B.~Alex}\ \bibnamefont
  {Brown}},\ }\bibfield  {title} {\enquote {\bibinfo {title} {{Predicted
  features of the beta decay of neutron-rich sd-shell nuclei}},}\ }\href
  {\doibase 10.1103/PhysRevC.28.1343} {\bibfield  {journal} {\bibinfo
  {journal} {Phys. Rev. C}\ }\textbf {\bibinfo {volume} {28}},\ \bibinfo
  {pages} {1343--1366} (\bibinfo {year} {1983})}\BibitemShut {NoStop}%
\bibitem [{\citenamefont {Mart{\'i}nez-Pinedo}\ \emph
  {et~al.}(1996)\citenamefont {Mart{\'i}nez-Pinedo}, \citenamefont {Poves},
  \citenamefont {Caurier},\ and\ \citenamefont {Zuker}}]{MartinezPinedo96}%
  \BibitemOpen
  \bibfield  {author} {\bibinfo {author} {\bibfnamefont {G.}~\bibnamefont
  {Mart{\'i}nez-Pinedo}}, \bibinfo {author} {\bibfnamefont {A.}~\bibnamefont
  {Poves}}, \bibinfo {author} {\bibfnamefont {E.}~\bibnamefont {Caurier}}, \
  and\ \bibinfo {author} {\bibfnamefont {A.~P.}\ \bibnamefont {Zuker}},\
  }\bibfield  {title} {\enquote {\bibinfo {title} {{Effective $g_A$ in the pf
  shell}},}\ }\href {\doibase 10.1103/PhysRevC.53.R2602} {\bibfield  {journal}
  {\bibinfo  {journal} {Phys. Rev. C}\ }\textbf {\bibinfo {volume} {53}},\
  \bibinfo {pages} {R2602} (\bibinfo {year} {1996})}\BibitemShut {NoStop}%
\bibitem [{\citenamefont {Caurier}\ \emph {et~al.}(1990)\citenamefont
  {Caurier}, \citenamefont {Zuker},\ and\ \citenamefont {Poves}}]{cau90}%
  \BibitemOpen
  \bibfield  {author} {\bibinfo {author} {\bibfnamefont {E.}~\bibnamefont
  {Caurier}}, \bibinfo {author} {\bibfnamefont {A.~P.}\ \bibnamefont {Zuker}},
  \ and\ \bibinfo {author} {\bibfnamefont {A.}~\bibnamefont {Poves}},\
  }\bibfield  {title} {\enquote {\bibinfo {title} {{A Full $0\hbar\omega$
  description of the $2\nu\beta\beta$ decay of $^{48}$\protect{C}a}},}\ }\href
  {\doibase 10.1016/0370-2693(90)91071-I} {\bibfield  {journal} {\bibinfo
  {journal} {Phys. Lett. B}\ }\textbf {\bibinfo {volume} {252}},\ \bibinfo
  {pages} {13--17} (\bibinfo {year} {1990})}\BibitemShut {NoStop}%
\bibitem [{\citenamefont {Poves}\ \emph {et~al.}(1995)\citenamefont {Poves},
  \citenamefont {Bahukutumbi}, \citenamefont {Langanke},\ and\ \citenamefont
  {Vogel}}]{pov95}%
  \BibitemOpen
  \bibfield  {author} {\bibinfo {author} {\bibfnamefont {A.}~\bibnamefont
  {Poves}}, \bibinfo {author} {\bibfnamefont {R.~P.}\ \bibnamefont
  {Bahukutumbi}}, \bibinfo {author} {\bibfnamefont {K.}~\bibnamefont
  {Langanke}}, \ and\ \bibinfo {author} {\bibfnamefont {P.}~\bibnamefont
  {Vogel}},\ }\bibfield  {title} {\enquote {\bibinfo {title} {{Double beta
  decay of $^{48}$Ca revisited}},}\ }\href {\doibase
  10.1016/0370-2693(95)01134-C} {\bibfield  {journal} {\bibinfo  {journal}
  {Phys. Lett. B}\ }\textbf {\bibinfo {volume} {361}},\ \bibinfo {pages} {1--4}
  (\bibinfo {year} {1995})}\BibitemShut {NoStop}%
\bibitem [{\citenamefont {Arnold}\ \emph {et~al.}(2016)\citenamefont {Arnold}
  \emph {et~al.}}]{arno16}%
  \BibitemOpen
  \bibfield  {author} {\bibinfo {author} {\bibfnamefont {R.}~\bibnamefont
  {Arnold}} \emph {et~al.} (\bibinfo {collaboration} {NEMO-3 Collaboration}),\
  }\bibfield  {title} {\enquote {\bibinfo {title} {{Measurement of the
  double-beta decay half-life and search for the neutrinoless double-beta decay
  of $^{48}$Ca with the NEMO-3 detector}},}\ }\href {\doibase
  10.1103/PhysRevD.93.112008} {\bibfield  {journal} {\bibinfo  {journal} {Phys.
  Rev. D}\ }\textbf {\bibinfo {volume} {93}},\ \bibinfo {pages} {112008}
  (\bibinfo {year} {2016})}\BibitemShut {NoStop}%
\bibitem [{\citenamefont {Balysh}\ \emph {et~al.}(1996)\citenamefont {Balysh},
  \citenamefont {De~Silva}, \citenamefont {Lebedev}, \citenamefont {Lou},
  \citenamefont {Moe}, \citenamefont {Nelson}, \citenamefont {Piepke},
  \citenamefont {Pronsky}, \citenamefont {Vient},\ and\ \citenamefont
  {Vogel}}]{Balysh96}%
  \BibitemOpen
  \bibfield  {author} {\bibinfo {author} {\bibfnamefont {A.}~\bibnamefont
  {Balysh}}, \bibinfo {author} {\bibfnamefont {Asoka~S.}\ \bibnamefont
  {De~Silva}}, \bibinfo {author} {\bibfnamefont {V.~I.}\ \bibnamefont
  {Lebedev}}, \bibinfo {author} {\bibfnamefont {K.}~\bibnamefont {Lou}},
  \bibinfo {author} {\bibfnamefont {M.~K.}\ \bibnamefont {Moe}}, \bibinfo
  {author} {\bibfnamefont {M.~A.}\ \bibnamefont {Nelson}}, \bibinfo {author}
  {\bibfnamefont {A.}~\bibnamefont {Piepke}}, \bibinfo {author} {\bibfnamefont
  {A.}~\bibnamefont {Pronsky}}, \bibinfo {author} {\bibfnamefont {M.~A.}\
  \bibnamefont {Vient}}, \ and\ \bibinfo {author} {\bibfnamefont
  {P.}~\bibnamefont {Vogel}},\ }\bibfield  {title} {\enquote {\bibinfo {title}
  {{Double beta decay of $^{48}$Ca}},}\ }\href {\doibase
  10.1103/PhysRevLett.77.5186} {\bibfield  {journal} {\bibinfo  {journal}
  {Phys. Rev. Lett.}\ }\textbf {\bibinfo {volume} {77}},\ \bibinfo {pages}
  {5186--5189} (\bibinfo {year} {1996})}\BibitemShut {NoStop}%
\bibitem [{\citenamefont {Caurier}\ \emph {et~al.}(2012)\citenamefont
  {Caurier}, \citenamefont {Nowacki},\ and\ \citenamefont {Poves}}]{cau12}%
  \BibitemOpen
  \bibfield  {author} {\bibinfo {author} {\bibfnamefont {Etienne}\ \bibnamefont
  {Caurier}}, \bibinfo {author} {\bibfnamefont {Frederic}\ \bibnamefont
  {Nowacki}}, \ and\ \bibinfo {author} {\bibfnamefont {Alfredo}\ \bibnamefont
  {Poves}},\ }\bibfield  {title} {\enquote {\bibinfo {title} {{Shell Model
  description of the $\beta\beta$ decay of $^{136}$Xe}},}\ }\href {\doibase
  10.1016/j.physletb.2012.03.076} {\bibfield  {journal} {\bibinfo  {journal}
  {Phys. Lett. B}\ }\textbf {\bibinfo {volume} {711}},\ \bibinfo {pages}
  {62--64} (\bibinfo {year} {2012})}\BibitemShut {NoStop}%
\bibitem [{\citenamefont {Horoi}\ and\ \citenamefont {Brown}(2013)}]{hor13}%
  \BibitemOpen
  \bibfield  {author} {\bibinfo {author} {\bibfnamefont {M.}~\bibnamefont
  {Horoi}}\ and\ \bibinfo {author} {\bibfnamefont {B.A.}\ \bibnamefont
  {Brown}},\ }\bibfield  {title} {\enquote {\bibinfo {title} {Shell-model
  analysis of the $^{136}$\protect{X}e double beta decay nuclear matrix
  elements},}\ }\href {\doibase 10.1103/PhysRevLett.110.222502} {\bibfield
  {journal} {\bibinfo  {journal} {Phys. Rev. Lett.}\ }\textbf {\bibinfo
  {volume} {110}},\ \bibinfo {pages} {222502} (\bibinfo {year}
  {2013})}\BibitemShut {NoStop}%
\bibitem [{\citenamefont {Towner}(1997)}]{towner87}%
  \BibitemOpen
  \bibfield  {author} {\bibinfo {author} {\bibfnamefont {I.S.}\ \bibnamefont
  {Towner}},\ }\bibfield  {title} {\enquote {\bibinfo {title} {Quenching of
  spin matrix elements in nuclei},}\ }\href {\doibase
  10.1016/0370-1573(87)90138-4} {\bibfield  {journal} {\bibinfo  {journal}
  {Phys.\ Rep.}\ }\textbf {\bibinfo {volume} {155}},\ \bibinfo {pages} {263}
  (\bibinfo {year} {1997})}\BibitemShut {NoStop}%
\bibitem [{\citenamefont {Brown}\ and\ \citenamefont
  {Wildenthal}(1987)}]{Brown87}%
  \BibitemOpen
  \bibfield  {author} {\bibinfo {author} {\bibfnamefont {B.~A.}\ \bibnamefont
  {Brown}}\ and\ \bibinfo {author} {\bibfnamefont {B.~H.}\ \bibnamefont
  {Wildenthal}},\ }\bibfield  {title} {\enquote {\bibinfo {title} {{Empirically
  optimum M1 operator for sd-shell nuclei}},}\ }\href {\doibase
  10.1016/0375-9474(87)90619-1} {\bibfield  {journal} {\bibinfo  {journal}
  {Nucl. Phys. A}\ }\textbf {\bibinfo {volume} {474}},\ \bibinfo {pages}
  {290--306} (\bibinfo {year} {1987})}\BibitemShut {NoStop}%
\bibitem [{\citenamefont {von Neumann-Cosel}\ \emph {et~al.}(1998)\citenamefont
  {von Neumann-Cosel}, \citenamefont {Poves}, \citenamefont {Retamosa},\ and\
  \citenamefont {Richter}}]{NeumannCosel98}%
  \BibitemOpen
  \bibfield  {author} {\bibinfo {author} {\bibfnamefont {P.}~\bibnamefont {von
  Neumann-Cosel}}, \bibinfo {author} {\bibfnamefont {A.}~\bibnamefont {Poves}},
  \bibinfo {author} {\bibfnamefont {J.}~\bibnamefont {Retamosa}}, \ and\
  \bibinfo {author} {\bibfnamefont {A.}~\bibnamefont {Richter}},\ }\bibfield
  {title} {\enquote {\bibinfo {title} {{Magnetic dipole response in nuclei at
  the $N=28$ shell}},}\ }\href {\doibase 10.1016/S0370-2693(98)01298-2}
  {\bibfield  {journal} {\bibinfo  {journal} {Phys. Lett. B}\ }\textbf
  {\bibinfo {volume} {443}},\ \bibinfo {pages} {1--6} (\bibinfo {year}
  {1998})}\BibitemShut {NoStop}%
\bibitem [{\citenamefont {Hjorth-Jensen}\ \emph {et~al.}(1995)\citenamefont
  {Hjorth-Jensen}, \citenamefont {Kuo},\ and\ \citenamefont
  {Osnes}}]{hjorth-jensen95}%
  \BibitemOpen
  \bibfield  {author} {\bibinfo {author} {\bibfnamefont {M.}~\bibnamefont
  {Hjorth-Jensen}}, \bibinfo {author} {\bibfnamefont {T.T.S.}\ \bibnamefont
  {Kuo}}, \ and\ \bibinfo {author} {\bibfnamefont {E.}~\bibnamefont {Osnes}},\
  }\bibfield  {title} {\enquote {\bibinfo {title} {Realistic effective
  interactions for nuclear systems},}\ }\href {\doibase
  10.1016/0370-1573(95)00012-6} {\bibfield  {journal} {\bibinfo  {journal}
  {Phys.\ Rep.}\ }\textbf {\bibinfo {volume} {261}},\ \bibinfo {pages} {125}
  (\bibinfo {year} {1995})}\BibitemShut {NoStop}%
\bibitem [{\citenamefont {Bohm}\ and\ \citenamefont
  {Pines}(1951)}]{bohm-pines1}%
  \BibitemOpen
  \bibfield  {author} {\bibinfo {author} {\bibfnamefont {David}\ \bibnamefont
  {Bohm}}\ and\ \bibinfo {author} {\bibfnamefont {David}\ \bibnamefont
  {Pines}},\ }\bibfield  {title} {\enquote {\bibinfo {title} {{A Collective
  Description of Electron Interactions. \protect{I}. Magnetic Interactions}},}\
  }\href {\doibase 10.1103/PhysRev.82.625} {\bibfield  {journal} {\bibinfo
  {journal} {Phys. Rev.}\ }\textbf {\bibinfo {volume} {82}},\ \bibinfo {pages}
  {625--634} (\bibinfo {year} {1951})}\BibitemShut {NoStop}%
\bibitem [{\citenamefont {Rowe}(1968)}]{row68}%
  \BibitemOpen
  \bibfield  {author} {\bibinfo {author} {\bibfnamefont {D.~J.}\ \bibnamefont
  {Rowe}},\ }\bibfield  {title} {\enquote {\bibinfo {title}
  {Equations-of-motion method and the extended shell model},}\ }\href {\doibase
  10.1103/RevModPhys.40.153} {\bibfield  {journal} {\bibinfo  {journal} {Rev.\
  Mod.\ Phys.}\ }\textbf {\bibinfo {volume} {40}},\ \bibinfo {pages} {153}
  (\bibinfo {year} {1968})}\BibitemShut {NoStop}%
\bibitem [{\citenamefont {Jancovici}\ and\ \citenamefont
  {Schiff}(1964)}]{jancovici64}%
  \BibitemOpen
  \bibfield  {author} {\bibinfo {author} {\bibfnamefont {B.}~\bibnamefont
  {Jancovici}}\ and\ \bibinfo {author} {\bibfnamefont {D.~H.}\ \bibnamefont
  {Schiff}},\ }\bibfield  {title} {\enquote {\bibinfo {title} {The collective
  vibrations of a many-fermion system},}\ }\href {\doibase
  10.1016/0029-5582(64)90578-4} {\bibfield  {journal} {\bibinfo  {journal}
  {Nucl. Phys.}\ }\textbf {\bibinfo {volume} {58}},\ \bibinfo {pages}
  {678--686} (\bibinfo {year} {1964})}\BibitemShut {NoStop}%
\bibitem [{\citenamefont {Ring}\ and\ \citenamefont {Schuck}(1980)}]{ring80}%
  \BibitemOpen
  \bibfield  {author} {\bibinfo {author} {\bibfnamefont {Peter}\ \bibnamefont
  {Ring}}\ and\ \bibinfo {author} {\bibfnamefont {Peter}\ \bibnamefont
  {Schuck}},\ }\href@noop {} {\emph {\bibinfo {title} {The Nuclear Many-Body
  Problem}}}\ (\bibinfo  {publisher} {Springer-Verlag},\ \bibinfo {address}
  {New York, Heidelberg, Berlin},\ \bibinfo {year} {1980})\BibitemShut
  {NoStop}%
\bibitem [{\citenamefont {Rodin}\ \emph {et~al.}(2003)\citenamefont {Rodin},
  \citenamefont {Faessler}, \citenamefont {\v{S}imkovic},\ and\ \citenamefont
  {Vogel}}]{rodin03}%
  \BibitemOpen
  \bibfield  {author} {\bibinfo {author} {\bibfnamefont {V.A.}\ \bibnamefont
  {Rodin}}, \bibinfo {author} {\bibfnamefont {Amand}\ \bibnamefont {Faessler}},
  \bibinfo {author} {\bibfnamefont {F.}~\bibnamefont {\v{S}imkovic}}, \ and\
  \bibinfo {author} {\bibfnamefont {Petr}\ \bibnamefont {Vogel}},\ }\bibfield
  {title} {\enquote {\bibinfo {title} {Uncertainty in the $0\nu\beta\beta$
  decay nuclear matrix elements},}\ }\href {\doibase
  10.1103/PhysRevC.68.044302} {\bibfield  {journal} {\bibinfo  {journal}
  {Phys.\ Rev.\ C}\ }\textbf {\bibinfo {volume} {68}},\ \bibinfo {pages}
  {044302} (\bibinfo {year} {2003})}\BibitemShut {NoStop}%
\bibitem [{\citenamefont {Vogel}\ and\ \citenamefont
  {Zirnbauer}(1986)}]{vog86}%
  \BibitemOpen
  \bibfield  {author} {\bibinfo {author} {\bibfnamefont {P.}~\bibnamefont
  {Vogel}}\ and\ \bibinfo {author} {\bibfnamefont {M.~R.}\ \bibnamefont
  {Zirnbauer}},\ }\bibfield  {title} {\enquote {\bibinfo {title} {Suppression
  of the two-neutrino double-beta decay by nuclear-structure effects},}\ }\href
  {\doibase 10.1103/PhysRevLett.57.3148} {\bibfield  {journal} {\bibinfo
  {journal} {Phys. Rev. Lett.}\ }\textbf {\bibinfo {volume} {57}},\ \bibinfo
  {pages} {3148--3151} (\bibinfo {year} {1986})}\BibitemShut {NoStop}%
\bibitem [{\citenamefont {Engel}\ \emph {et~al.}(1988)\citenamefont {Engel},
  \citenamefont {Vogel},\ and\ \citenamefont {Zirnbauer}}]{eng88}%
  \BibitemOpen
  \bibfield  {author} {\bibinfo {author} {\bibfnamefont {J.}~\bibnamefont
  {Engel}}, \bibinfo {author} {\bibfnamefont {P.}~\bibnamefont {Vogel}}, \ and\
  \bibinfo {author} {\bibfnamefont {M.~R.}\ \bibnamefont {Zirnbauer}},\
  }\bibfield  {title} {\enquote {\bibinfo {title} {Nuclear structure effects in
  double-beta decay},}\ }\href {\doibase 10.1103/PhysRevC.37.731} {\bibfield
  {journal} {\bibinfo  {journal} {Phys.\ Rev. C}\ }\textbf {\bibinfo {volume}
  {37}},\ \bibinfo {pages} {731} (\bibinfo {year} {1988})}\BibitemShut
  {NoStop}%
\bibitem [{\citenamefont {Toivanen}\ and\ \citenamefont
  {Suhonen}(1995)}]{toi95}%
  \BibitemOpen
  \bibfield  {author} {\bibinfo {author} {\bibfnamefont {J.}~\bibnamefont
  {Toivanen}}\ and\ \bibinfo {author} {\bibfnamefont {J.}~\bibnamefont
  {Suhonen}},\ }\bibfield  {title} {\enquote {\bibinfo {title} {Renormalized
  proton-neutron quasiparticle random-phase approximation and its application
  to double beta decay},}\ }\href {\doibase 10.1103/PhysRevLett.75.410}
  {\bibfield  {journal} {\bibinfo  {journal} {Phys. Rev. Lett.}\ }\textbf
  {\bibinfo {volume} {75}},\ \bibinfo {pages} {410} (\bibinfo {year}
  {1995})}\BibitemShut {NoStop}%
\bibitem [{\citenamefont {Rodin}\ and\ \citenamefont
  {Faessler}(2002)}]{PhysRevC.66.051303}%
  \BibitemOpen
  \bibfield  {author} {\bibinfo {author} {\bibfnamefont {Vadim}\ \bibnamefont
  {Rodin}}\ and\ \bibinfo {author} {\bibfnamefont {Amand}\ \bibnamefont
  {Faessler}},\ }\bibfield  {title} {\enquote {\bibinfo {title} {Fully
  renormalized quasiparticle random-phase approximation fulfills
  \protect{I}keda sum rule exactly},}\ }\href {\doibase
  10.1103/PhysRevC.66.051303} {\bibfield  {journal} {\bibinfo  {journal} {Phys.
  Rev. C}\ }\textbf {\bibinfo {volume} {66}},\ \bibinfo {pages} {051303}
  (\bibinfo {year} {2002})}\BibitemShut {NoStop}%
\bibitem [{\citenamefont {Engel}\ \emph {et~al.}(1997)\citenamefont {Engel},
  \citenamefont {Pittel}, \citenamefont {Stoitsov}, \citenamefont {Vogel},\
  and\ \citenamefont {Dukelsky}}]{eng97}%
  \BibitemOpen
  \bibfield  {author} {\bibinfo {author} {\bibfnamefont {J.}~\bibnamefont
  {Engel}}, \bibinfo {author} {\bibfnamefont {S.}~\bibnamefont {Pittel}},
  \bibinfo {author} {\bibfnamefont {M.}~\bibnamefont {Stoitsov}}, \bibinfo
  {author} {\bibfnamefont {P.}~\bibnamefont {Vogel}}, \ and\ \bibinfo {author}
  {\bibfnamefont {J.}~\bibnamefont {Dukelsky}},\ }\bibfield  {title} {\enquote
  {\bibinfo {title} {Neutron-proton correlations in an exactly solvable
  model},}\ }\href {\doibase 10.1103/PhysRevC.55.1781} {\bibfield  {journal}
  {\bibinfo  {journal} {Phys.\ Rev.\ C}\ }\textbf {\bibinfo {volume} {55}},\
  \bibinfo {pages} {1781} (\bibinfo {year} {1997})}\BibitemShut {NoStop}%
\bibitem [{\citenamefont {Hirsch}\ \emph {et~al.}(1997)\citenamefont {Hirsch},
  \citenamefont {Hess},\ and\ \citenamefont {Civitarese}}]{hirsch97}%
  \BibitemOpen
  \bibfield  {author} {\bibinfo {author} {\bibfnamefont {Jorge~G.}\
  \bibnamefont {Hirsch}}, \bibinfo {author} {\bibfnamefont {Peter~O.}\
  \bibnamefont {Hess}}, \ and\ \bibinfo {author} {\bibfnamefont {Osvaldo}\
  \bibnamefont {Civitarese}},\ }\bibfield  {title} {\enquote {\bibinfo {title}
  {{Single- and Double-Beta Decay Fermi Transitions in an Exactly Solvable
  Model}},}\ }\href {\doibase 10.1103/PhysRevC.56.199} {\bibfield  {journal}
  {\bibinfo  {journal} {Phys.\ Rev.\ C}\ }\textbf {\bibinfo {volume} {56}},\
  \bibinfo {pages} {199} (\bibinfo {year} {1997})}\BibitemShut {NoStop}%
\bibitem [{\citenamefont {Terasaki}(2015)}]{Terasaki15}%
  \BibitemOpen
  \bibfield  {author} {\bibinfo {author} {\bibfnamefont {J.}~\bibnamefont
  {Terasaki}},\ }\bibfield  {title} {\enquote {\bibinfo {title} {{Many-body
  correlations of quasiparticle random-phase approximation in nuclear matrix
  elements of neutrinoless double-$\beta$ decay}},}\ }\href {\doibase
  10.1103/PhysRevC.91.034318} {\bibfield  {journal} {\bibinfo  {journal} {Phys.
  Rev. C}\ }\textbf {\bibinfo {volume} {91}},\ \bibinfo {pages} {034318}
  (\bibinfo {year} {2015})}\BibitemShut {NoStop}%
\bibitem [{\citenamefont {Terasaki}(2016)}]{Terasaki16}%
  \BibitemOpen
  \bibfield  {author} {\bibinfo {author} {\bibfnamefont {Jun}\ \bibnamefont
  {Terasaki}},\ }\bibfield  {title} {\enquote {\bibinfo {title} {{Two decay
  paths for calculating the nuclear matrix element of neutrinoless
  double-$\beta$ decay using quasiparticle random-phase approximation}},}\
  }\href {\doibase 10.1103/PhysRevC.93.024317} {\bibfield  {journal} {\bibinfo
  {journal} {Phys. Rev. C}\ }\textbf {\bibinfo {volume} {93}},\ \bibinfo
  {pages} {024317} (\bibinfo {year} {2016})}\BibitemShut {NoStop}%
\bibitem [{\citenamefont {Nogueira}\ \emph {et~al.}(2003)\citenamefont
  {Nogueira}, \citenamefont {Marques},\ and\ \citenamefont
  {Fiolhais}}]{Nogueira-dft}%
  \BibitemOpen
  \bibfield  {author} {\bibinfo {author} {\bibfnamefont {Fernando}\
  \bibnamefont {Nogueira}}, \bibinfo {author} {\bibfnamefont {Miguel A~L}\
  \bibnamefont {Marques}}, \ and\ \bibinfo {author} {\bibfnamefont {Carlos}\
  \bibnamefont {Fiolhais}},\ }\href {https://cds.cern.ch/record/1391332} {\emph
  {\bibinfo {title} {{A primer in density functional theory}}}},\ Lecture Notes
  in Physics\ (\bibinfo  {publisher} {Springer},\ \bibinfo {address} {Berlin},\
  \bibinfo {year} {2003})\BibitemShut {NoStop}%
\bibitem [{\citenamefont {Hohenberg}\ and\ \citenamefont
  {Kohn}(1964)}]{Hohenberg64}%
  \BibitemOpen
  \bibfield  {author} {\bibinfo {author} {\bibfnamefont {P.}~\bibnamefont
  {Hohenberg}}\ and\ \bibinfo {author} {\bibfnamefont {W.}~\bibnamefont
  {Kohn}},\ }\bibfield  {title} {\enquote {\bibinfo {title} {{Inhomogeneous
  Electron Gas}},}\ }\href {\doibase 10.1103/PhysRev.136.B864} {\bibfield
  {journal} {\bibinfo  {journal} {Phys. Rev.}\ }\textbf {\bibinfo {volume}
  {136}},\ \bibinfo {pages} {B864--B871} (\bibinfo {year} {1964})}\BibitemShut
  {NoStop}%
\bibitem [{\citenamefont {Kohn}\ and\ \citenamefont {Sham}(1965)}]{Kohn65}%
  \BibitemOpen
  \bibfield  {author} {\bibinfo {author} {\bibfnamefont {W.}~\bibnamefont
  {Kohn}}\ and\ \bibinfo {author} {\bibfnamefont {L.~J.}\ \bibnamefont
  {Sham}},\ }\bibfield  {title} {\enquote {\bibinfo {title} {{Self-Consistent
  Equations Including Exchange and Correlation Effects}},}\ }\href {\doibase
  10.1103/PhysRev.140.A1133} {\bibfield  {journal} {\bibinfo  {journal} {Phys.
  Rev.}\ }\textbf {\bibinfo {volume} {140}},\ \bibinfo {pages} {A1133--A1138}
  (\bibinfo {year} {1965})}\BibitemShut {NoStop}%
\bibitem [{\citenamefont {Vautherin}\ and\ \citenamefont
  {Brink}(1972)}]{Vautherin72}%
  \BibitemOpen
  \bibfield  {author} {\bibinfo {author} {\bibfnamefont {D.}~\bibnamefont
  {Vautherin}}\ and\ \bibinfo {author} {\bibfnamefont {D.~M.}\ \bibnamefont
  {Brink}},\ }\bibfield  {title} {\enquote {\bibinfo {title} {{Hartree-Fock
  calculations with Skyrme's interaction. 1. Spherical nuclei}},}\ }\href
  {\doibase 10.1103/PhysRevC.5.626} {\bibfield  {journal} {\bibinfo  {journal}
  {Phys. Rev. C}\ }\textbf {\bibinfo {volume} {5}},\ \bibinfo {pages}
  {626--647} (\bibinfo {year} {1972})}\BibitemShut {NoStop}%
\bibitem [{\citenamefont {Decharg\'{e}}\ and\ \citenamefont
  {Gogny}(1980)}]{decharge80}%
  \BibitemOpen
  \bibfield  {author} {\bibinfo {author} {\bibfnamefont {J.}~\bibnamefont
  {Decharg\'{e}}}\ and\ \bibinfo {author} {\bibfnamefont {D.}~\bibnamefont
  {Gogny}},\ }\bibfield  {title} {\enquote {\bibinfo {title}
  {Hartre-fock-bogolyubov calculations with the d1 effective interaction on
  spherical nuclei},}\ }\href {\doibase 10.1103/PhysRevC.21.1568} {\bibfield
  {journal} {\bibinfo  {journal} {Phys. Rev. C}\ }\textbf {\bibinfo {volume}
  {21}},\ \bibinfo {pages} {1568--1593} (\bibinfo {year} {1980})}\BibitemShut
  {NoStop}%
\bibitem [{\citenamefont {Serot}\ and\ \citenamefont
  {Walecka}(1986)}]{Serot86}%
  \BibitemOpen
  \bibfield  {author} {\bibinfo {author} {\bibfnamefont {Brian~D.}\
  \bibnamefont {Serot}}\ and\ \bibinfo {author} {\bibfnamefont {John~Dirk}\
  \bibnamefont {Walecka}},\ }\bibfield  {title} {\enquote {\bibinfo {title}
  {{The Relativistic Nuclear Many Body Problem}},}\ }\href@noop {} {\bibfield
  {journal} {\bibinfo  {journal} {Adv. Nucl. Phys.}\ }\textbf {\bibinfo
  {volume} {16}},\ \bibinfo {pages} {1--327} (\bibinfo {year}
  {1986})}\BibitemShut {NoStop}%
\bibitem [{\citenamefont {Bender}\ \emph {et~al.}(2003)\citenamefont {Bender},
  \citenamefont {Heenen},\ and\ \citenamefont {Reinhard}}]{bender03}%
  \BibitemOpen
  \bibfield  {author} {\bibinfo {author} {\bibfnamefont {Michael}\ \bibnamefont
  {Bender}}, \bibinfo {author} {\bibfnamefont {Paul-Henri}\ \bibnamefont
  {Heenen}}, \ and\ \bibinfo {author} {\bibfnamefont {Paul-Gerhard}\
  \bibnamefont {Reinhard}},\ }\bibfield  {title} {\enquote {\bibinfo {title}
  {Self-consistent mean-field models for nuclear structure},}\ }\href {\doibase
  10.1103/RevModPhys.75.121} {\bibfield  {journal} {\bibinfo  {journal} {Rev.
  Mod. Phys.}\ }\textbf {\bibinfo {volume} {75}},\ \bibinfo {pages} {121--180}
  (\bibinfo {year} {2003})}\BibitemShut {NoStop}%
\bibitem [{\citenamefont {Bender}\ \emph {et~al.}(2009)\citenamefont {Bender},
  \citenamefont {Duguet},\ and\ \citenamefont {Lacroix}}]{Bender09a}%
  \BibitemOpen
  \bibfield  {author} {\bibinfo {author} {\bibfnamefont {M.}~\bibnamefont
  {Bender}}, \bibinfo {author} {\bibfnamefont {T.}~\bibnamefont {Duguet}}, \
  and\ \bibinfo {author} {\bibfnamefont {D.}~\bibnamefont {Lacroix}},\
  }\bibfield  {title} {\enquote {\bibinfo {title} {{Particle-Number Restoration
  within the Energy Density Functional Formalism}},}\ }\href {\doibase
  10.1103/PhysRevC.79.044319} {\bibfield  {journal} {\bibinfo  {journal} {Phys.
  Rev. C}\ }\textbf {\bibinfo {volume} {79}},\ \bibinfo {pages} {044319}
  (\bibinfo {year} {2009})}\BibitemShut {NoStop}%
\bibitem [{\citenamefont {Anguiano}\ \emph {et~al.}(2001)\citenamefont
  {Anguiano}, \citenamefont {Egido},\ and\ \citenamefont
  {Robledo}}]{Anguiano:2001in}%
  \BibitemOpen
  \bibfield  {author} {\bibinfo {author} {\bibfnamefont {M.}~\bibnamefont
  {Anguiano}}, \bibinfo {author} {\bibfnamefont {J.~L.}\ \bibnamefont {Egido}},
  \ and\ \bibinfo {author} {\bibfnamefont {L.~M.}\ \bibnamefont {Robledo}},\
  }\bibfield  {title} {\enquote {\bibinfo {title} {{Particle number projection
  with effective forces}},}\ }\href {\doibase 10.1016/S0375-9474(01)01219-2}
  {\bibfield  {journal} {\bibinfo  {journal} {Nucl. Phys. A}\ }\textbf
  {\bibinfo {volume} {696}},\ \bibinfo {pages} {467--493} (\bibinfo {year}
  {2001})}\BibitemShut {NoStop}%
\bibitem [{\citenamefont {Dobaczewski}\ \emph {et~al.}(2007)\citenamefont
  {Dobaczewski}, \citenamefont {Stoitsov}, \citenamefont {Nazarewicz},\ and\
  \citenamefont {Reinhard}}]{Dobaczewski:2007ch}%
  \BibitemOpen
  \bibfield  {author} {\bibinfo {author} {\bibfnamefont {J.}~\bibnamefont
  {Dobaczewski}}, \bibinfo {author} {\bibfnamefont {M.~V.}\ \bibnamefont
  {Stoitsov}}, \bibinfo {author} {\bibfnamefont {W.}~\bibnamefont
  {Nazarewicz}}, \ and\ \bibinfo {author} {\bibfnamefont {P.~G.}\ \bibnamefont
  {Reinhard}},\ }\bibfield  {title} {\enquote {\bibinfo {title}
  {{Particle-Number Projection and the Density Functional Theory}},}\ }\href
  {\doibase 10.1103/PhysRevC.76.054315} {\bibfield  {journal} {\bibinfo
  {journal} {Phys. Rev. C}\ }\textbf {\bibinfo {volume} {76}},\ \bibinfo
  {pages} {054315} (\bibinfo {year} {2007})}\BibitemShut {NoStop}%
\bibitem [{\citenamefont {Rodr\'iguez}\ and\ \citenamefont
  {Mart{\'i}nez-Pinedo}(2010)}]{rod10}%
  \BibitemOpen
  \bibfield  {author} {\bibinfo {author} {\bibfnamefont {Tom{\'a}s~R.}\
  \bibnamefont {Rodr\'iguez}}\ and\ \bibinfo {author} {\bibfnamefont
  {G.}~\bibnamefont {Mart{\'i}nez-Pinedo}},\ }\bibfield  {title} {\enquote
  {\bibinfo {title} {Energy density functional study of nuclear matrix elements
  for neutrinoless $\beta\beta$ decay},}\ }\href {\doibase
  10.1103/PhysRevLett.105.252503} {\bibfield  {journal} {\bibinfo  {journal}
  {Phys. Rev. Lett.}\ }\textbf {\bibinfo {volume} {105}},\ \bibinfo {pages}
  {252503} (\bibinfo {year} {2010})}\BibitemShut {NoStop}%
\bibitem [{\citenamefont {Song}\ \emph {et~al.}(2014)\citenamefont {Song},
  \citenamefont {Yao}, \citenamefont {Ring},\ and\ \citenamefont
  {Meng}}]{Song14}%
  \BibitemOpen
  \bibfield  {author} {\bibinfo {author} {\bibfnamefont {L.~S.}\ \bibnamefont
  {Song}}, \bibinfo {author} {\bibfnamefont {J.~M.}\ \bibnamefont {Yao}},
  \bibinfo {author} {\bibfnamefont {P.}~\bibnamefont {Ring}}, \ and\ \bibinfo
  {author} {\bibfnamefont {J.}~\bibnamefont {Meng}},\ }\bibfield  {title}
  {\enquote {\bibinfo {title} {{Relativistic description of nuclear matrix
  elements in neutrinoless double-$\beta$ decay}},}\ }\href {\doibase
  10.1103/PhysRevC.90.054309} {\bibfield  {journal} {\bibinfo  {journal} {Phys.
  Rev. C}\ }\textbf {\bibinfo {volume} {90}},\ \bibinfo {pages} {054309}
  (\bibinfo {year} {2014})}\BibitemShut {NoStop}%
\bibitem [{\citenamefont {Men{\'e}ndez}\ \emph {et~al.}(2014)\citenamefont
  {Men{\'e}ndez}, \citenamefont {Rodr{\'i}guez}, \citenamefont
  {Mart{\'i}nez-Pinedo},\ and\ \citenamefont {Poves}}]{men14}%
  \BibitemOpen
  \bibfield  {author} {\bibinfo {author} {\bibfnamefont {Javier}\ \bibnamefont
  {Men{\'e}ndez}}, \bibinfo {author} {\bibfnamefont {Tom{\'a}s~R.}\
  \bibnamefont {Rodr{\'i}guez}}, \bibinfo {author} {\bibfnamefont {Gabriel}\
  \bibnamefont {Mart{\'i}nez-Pinedo}}, \ and\ \bibinfo {author} {\bibfnamefont
  {Alfredo}\ \bibnamefont {Poves}},\ }\bibfield  {title} {\enquote {\bibinfo
  {title} {{Correlations and neutrinoless $\beta\beta$ decay nuclear matrix
  elements of $pf$-shell nuclei}},}\ }\href {\doibase
  10.1103/PhysRevC.90.024311} {\bibfield  {journal} {\bibinfo  {journal} {Phys.
  Rev. C}\ }\textbf {\bibinfo {volume} {90}},\ \bibinfo {pages} {024311}
  (\bibinfo {year} {2014})}\BibitemShut {NoStop}%
\bibitem [{\citenamefont {Men{\'e}ndez}\ \emph {et~al.}(2016)\citenamefont
  {Men{\'e}ndez}, \citenamefont {Hinohara}, \citenamefont {Engel},
  \citenamefont {Mart{\'i}nez-Pinedo},\ and\ \citenamefont
  {Rodr{\'i}guez}}]{men16}%
  \BibitemOpen
  \bibfield  {author} {\bibinfo {author} {\bibfnamefont {Javier}\ \bibnamefont
  {Men{\'e}ndez}}, \bibinfo {author} {\bibfnamefont {Nobuo}\ \bibnamefont
  {Hinohara}}, \bibinfo {author} {\bibfnamefont {Jonathan}\ \bibnamefont
  {Engel}}, \bibinfo {author} {\bibfnamefont {Gabriel}\ \bibnamefont
  {Mart{\'i}nez-Pinedo}}, \ and\ \bibinfo {author} {\bibfnamefont
  {Tom{\'a}s~R.}\ \bibnamefont {Rodr{\'i}guez}},\ }\bibfield  {title} {\enquote
  {\bibinfo {title} {{Testing the importance of collective correlations in
  neutrinoless ββ decay}},}\ }\href {\doibase 10.1103/PhysRevC.93.014305}
  {\bibfield  {journal} {\bibinfo  {journal} {Phys. Rev. C}\ }\textbf {\bibinfo
  {volume} {93}},\ \bibinfo {pages} {014305} (\bibinfo {year}
  {2016})}\BibitemShut {NoStop}%
\bibitem [{\citenamefont {Hinohara}\ and\ \citenamefont {Engel}(2014)}]{hin14}%
  \BibitemOpen
  \bibfield  {author} {\bibinfo {author} {\bibfnamefont {Nobuo}\ \bibnamefont
  {Hinohara}}\ and\ \bibinfo {author} {\bibfnamefont {Jonathan}\ \bibnamefont
  {Engel}},\ }\bibfield  {title} {\enquote {\bibinfo {title} {{Proton-Neutron
  Pairing Amplitude as a Generator Coordinate for Double-Beta Decay}},}\ }\href
  {\doibase 10.1103/PhysRevC.90.031301} {\bibfield  {journal} {\bibinfo
  {journal} {Phys. Rev. C}\ }\textbf {\bibinfo {volume} {90}},\ \bibinfo
  {pages} {031301} (\bibinfo {year} {2014})}\BibitemShut {NoStop}%
\bibitem [{\citenamefont {Iachello}\ and\ \citenamefont
  {Arima}(1987)}]{Iachello87}%
  \BibitemOpen
  \bibfield  {author} {\bibinfo {author} {\bibfnamefont {Franceso}\
  \bibnamefont {Iachello}}\ and\ \bibinfo {author} {\bibfnamefont {Akito}\
  \bibnamefont {Arima}},\ }\href@noop {} {\emph {\bibinfo {title} {The
  interacting boson model}}}\ (\bibinfo  {publisher} {Cambridge University
  Press},\ \bibinfo {address} {Cambridge},\ \bibinfo {year} {1987})\BibitemShut
  {NoStop}%
\bibitem [{\citenamefont {Arima}\ and\ \citenamefont
  {Iachello}(1981)}]{arima81}%
  \BibitemOpen
  \bibfield  {author} {\bibinfo {author} {\bibfnamefont {A.}~\bibnamefont
  {Arima}}\ and\ \bibinfo {author} {\bibfnamefont {F.}~\bibnamefont
  {Iachello}},\ }\bibfield  {title} {\enquote {\bibinfo {title} {The
  interacting boson model},}\ }\href {\doibase
  10.1146/annurev.ns.31.120181.000451} {\bibfield  {journal} {\bibinfo
  {journal} {Ann.\ Rev.\ Nucl.\ Part.\ Sci.}\ }\textbf {\bibinfo {volume}
  {31}},\ \bibinfo {pages} {75--105} (\bibinfo {year} {1981})}\BibitemShut
  {NoStop}%
\bibitem [{\citenamefont {Bohr}\ and\ \citenamefont {Mottelson}(1998)}]{BM98}%
  \BibitemOpen
  \bibfield  {author} {\bibinfo {author} {\bibfnamefont {Aage}\ \bibnamefont
  {Bohr}}\ and\ \bibinfo {author} {\bibfnamefont {Ben~R.}\ \bibnamefont
  {Mottelson}},\ }\href@noop {} {\emph {\bibinfo {title} {Nuclear
  Structure}}},\ Vol.~\bibinfo {volume} {2}\ (\bibinfo  {publisher} {World
  Scientific Pub Co Inc},\ \bibinfo {year} {1998})\BibitemShut {NoStop}%
\bibitem [{\citenamefont {Otsuka}\ \emph {et~al.}(1978)\citenamefont {Otsuka},
  \citenamefont {Arima},\ and\ \citenamefont {Iachello}}]{Otsuka78}%
  \BibitemOpen
  \bibfield  {author} {\bibinfo {author} {\bibfnamefont {T.}~\bibnamefont
  {Otsuka}}, \bibinfo {author} {\bibfnamefont {A.}~\bibnamefont {Arima}}, \
  and\ \bibinfo {author} {\bibfnamefont {F.}~\bibnamefont {Iachello}},\
  }\bibfield  {title} {\enquote {\bibinfo {title} {{Nuclear shell model and
  interacting bosons}},}\ }\href {\doibase 10.1016/0375-9474(78)90532-8}
  {\bibfield  {journal} {\bibinfo  {journal} {Nucl. Phys. A}\ }\textbf
  {\bibinfo {volume} {309}},\ \bibinfo {pages} {1--33} (\bibinfo {year}
  {1978})}\BibitemShut {NoStop}%
\bibitem [{\citenamefont {Barea}\ and\ \citenamefont {Iachello}(2009)}]{bar09}%
  \BibitemOpen
  \bibfield  {author} {\bibinfo {author} {\bibfnamefont {J.}~\bibnamefont
  {Barea}}\ and\ \bibinfo {author} {\bibfnamefont {F.}~\bibnamefont
  {Iachello}},\ }\bibfield  {title} {\enquote {\bibinfo {title} {Neutrinoless
  double-$\beta$ decay in the microscopic interacting boson model},}\ }\href
  {\doibase 10.1103/PhysRevC.79.044301} {\bibfield  {journal} {\bibinfo
  {journal} {Phys.\ Rev.\ C}\ }\textbf {\bibinfo {volume} {79}},\ \bibinfo
  {pages} {044301} (\bibinfo {year} {2009})}\BibitemShut {NoStop}%
\bibitem [{\citenamefont {Chaturvedi}\ \emph {et~al.}(2008)\citenamefont
  {Chaturvedi}, \citenamefont {Chandra}, \citenamefont {Rath}, \citenamefont
  {Raina},\ and\ \citenamefont {Hirsch}}]{Rath08}%
  \BibitemOpen
  \bibfield  {author} {\bibinfo {author} {\bibfnamefont {K.}~\bibnamefont
  {Chaturvedi}}, \bibinfo {author} {\bibfnamefont {R.}~\bibnamefont {Chandra}},
  \bibinfo {author} {\bibfnamefont {P.~K.}\ \bibnamefont {Rath}}, \bibinfo
  {author} {\bibfnamefont {P.~K.}\ \bibnamefont {Raina}}, \ and\ \bibinfo
  {author} {\bibfnamefont {J.~G.}\ \bibnamefont {Hirsch}},\ }\bibfield  {title}
  {\enquote {\bibinfo {title} {{Nuclear deformation and neutrinoless
  double-$\beta$ decay of $^{94}$Zr, $^{96}$Zr, $^{98}$Mo, $^{100}$Mo,
  $^{104}$Ru, $^{110}$Pd, $^{128}$Te, $^{130}$Te, and $^{150}$Nd nuclei within
  a mechanism involving neutrino mass}},}\ }\href {\doibase
  10.1103/PhysRevC.78.054302} {\bibfield  {journal} {\bibinfo  {journal} {Phys.
  Rev. C}\ }\textbf {\bibinfo {volume} {78}},\ \bibinfo {pages} {054302}
  (\bibinfo {year} {2008})}\BibitemShut {NoStop}%
\bibitem [{\citenamefont {Chandra}\ \emph {et~al.}(2009)\citenamefont
  {Chandra}, \citenamefont {Chaturvedi}, \citenamefont {Rath}, \citenamefont
  {Raina},\ and\ \citenamefont {Hirsch}}]{Rath09}%
  \BibitemOpen
  \bibfield  {author} {\bibinfo {author} {\bibfnamefont {R.}~\bibnamefont
  {Chandra}}, \bibinfo {author} {\bibfnamefont {K.}~\bibnamefont {Chaturvedi}},
  \bibinfo {author} {\bibfnamefont {P.~K.}\ \bibnamefont {Rath}}, \bibinfo
  {author} {\bibfnamefont {P.~K.}\ \bibnamefont {Raina}}, \ and\ \bibinfo
  {author} {\bibfnamefont {J.~G.}\ \bibnamefont {Hirsch}},\ }\bibfield  {title}
  {\enquote {\bibinfo {title} {{Multipolar correlations and deformation effect
  on nuclear transition matrix elements of double-beta decay}},}\ }\href
  {\doibase 10.1209/0295-5075/86/32001} {\bibfield  {journal} {\bibinfo
  {journal} {Europhys. Lett.}\ }\textbf {\bibinfo {volume} {86}},\ \bibinfo
  {pages} {32001} (\bibinfo {year} {2009})}\BibitemShut {NoStop}%
\bibitem [{\citenamefont {Rath}\ \emph {et~al.}(2010)\citenamefont {Rath},
  \citenamefont {Chandra}, \citenamefont {Chaturvedi}, \citenamefont {Raina},\
  and\ \citenamefont {Hirsch}}]{Rath10}%
  \BibitemOpen
  \bibfield  {author} {\bibinfo {author} {\bibfnamefont {P.~K.}\ \bibnamefont
  {Rath}}, \bibinfo {author} {\bibfnamefont {R.}~\bibnamefont {Chandra}},
  \bibinfo {author} {\bibfnamefont {K.}~\bibnamefont {Chaturvedi}}, \bibinfo
  {author} {\bibfnamefont {P.~K.}\ \bibnamefont {Raina}}, \ and\ \bibinfo
  {author} {\bibfnamefont {J.~G.}\ \bibnamefont {Hirsch}},\ }\bibfield  {title}
  {\enquote {\bibinfo {title} {{Uncertainties in nuclear transition matrix
  elements for neutrinoless $\beta \beta $ decay within the PHFB model}},}\
  }\href {\doibase 10.1103/PhysRevC.82.064310} {\bibfield  {journal} {\bibinfo
  {journal} {Phys. Rev. C}\ }\textbf {\bibinfo {volume} {82}},\ \bibinfo
  {pages} {064310} (\bibinfo {year} {2010})}\BibitemShut {NoStop}%
\bibitem [{\citenamefont {Rath}\ \emph {et~al.}(2013)\citenamefont {Rath},
  \citenamefont {Chandra}, \citenamefont {Chaturvedi}, \citenamefont {Lohani},
  \citenamefont {Raina},\ and\ \citenamefont {Hirsch}}]{Rath13}%
  \BibitemOpen
  \bibfield  {author} {\bibinfo {author} {\bibfnamefont {P.~K.}\ \bibnamefont
  {Rath}}, \bibinfo {author} {\bibfnamefont {R.}~\bibnamefont {Chandra}},
  \bibinfo {author} {\bibfnamefont {K.}~\bibnamefont {Chaturvedi}}, \bibinfo
  {author} {\bibfnamefont {P.}~\bibnamefont {Lohani}}, \bibinfo {author}
  {\bibfnamefont {P.~K.}\ \bibnamefont {Raina}}, \ and\ \bibinfo {author}
  {\bibfnamefont {J.~G.}\ \bibnamefont {Hirsch}},\ }\bibfield  {title}
  {\enquote {\bibinfo {title} {{Neutrinoless ββ decay transition matrix
  elements within mechanisms involving light Majorana neutrinos, classical
  Majorons, and sterile neutrinos}},}\ }\href {\doibase
  10.1103/PhysRevC.88.064322} {\bibfield  {journal} {\bibinfo  {journal} {Phys.
  Rev. C}\ }\textbf {\bibinfo {volume} {88}},\ \bibinfo {pages} {064322}
  (\bibinfo {year} {2013})}\BibitemShut {NoStop}%
\bibitem [{\citenamefont {Hirsch}\ \emph {et~al.}(1994)\citenamefont {Hirsch},
  \citenamefont {Casta{\~n}os}, \citenamefont {Hess},\ and\ \citenamefont
  {Civitarese}}]{Hirsch:1994kw}%
  \BibitemOpen
  \bibfield  {author} {\bibinfo {author} {\bibfnamefont {J.~G.}\ \bibnamefont
  {Hirsch}}, \bibinfo {author} {\bibfnamefont {O.}~\bibnamefont
  {Casta{\~n}os}}, \bibinfo {author} {\bibfnamefont {P.~O.}\ \bibnamefont
  {Hess}}, \ and\ \bibinfo {author} {\bibfnamefont {O.}~\bibnamefont
  {Civitarese}},\ }\bibfield  {title} {\enquote {\bibinfo {title} {{Pseudo
  SU(3) approach to the $\beta\beta$ decay}},}\ }\href {\doibase
  10.1016/0146-6410(94)90031-0} {\bibfield  {journal} {\bibinfo  {journal}
  {Prog. Part. Nucl. Phys.}\ }\textbf {\bibinfo {volume} {32}},\ \bibinfo
  {pages} {333--334} (\bibinfo {year} {1994})}\BibitemShut {NoStop}%
\bibitem [{\citenamefont {Hirsch}\ \emph
  {et~al.}(1995{\natexlab{b}})\citenamefont {Hirsch}, \citenamefont
  {Casta{\~n}os},\ and\ \citenamefont {Hess}}]{hir95}%
  \BibitemOpen
  \bibfield  {author} {\bibinfo {author} {\bibfnamefont {J.~G.}\ \bibnamefont
  {Hirsch}}, \bibinfo {author} {\bibfnamefont {O.}~\bibnamefont
  {Casta{\~n}os}}, \ and\ \bibinfo {author} {\bibfnamefont {P.~O.}\
  \bibnamefont {Hess}},\ }\bibfield  {title} {\enquote {\bibinfo {title}
  {Neutrinoless double beta decay in heavy deformed nuclei},}\ }\href {\doibase
  10.1016/0375-9474(94)00464-X} {\bibfield  {journal} {\bibinfo  {journal}
  {Nucl.\ Phys. A}\ }\textbf {\bibinfo {volume} {582}},\ \bibinfo {pages} {124}
  (\bibinfo {year} {1995}{\natexlab{b}})}\BibitemShut {NoStop}%
\bibitem [{\citenamefont {Caurier}\ \emph
  {et~al.}(2008{\natexlab{b}})\citenamefont {Caurier}, \citenamefont
  {Nowacki},\ and\ \citenamefont {Poves}}]{caurier08a}%
  \BibitemOpen
  \bibfield  {author} {\bibinfo {author} {\bibfnamefont {E.}~\bibnamefont
  {Caurier}}, \bibinfo {author} {\bibfnamefont {F.}~\bibnamefont {Nowacki}}, \
  and\ \bibinfo {author} {\bibfnamefont {A.}~\bibnamefont {Poves}},\ }\bibfield
   {title} {\enquote {\bibinfo {title} {{Nuclear Structure Aspects of the
  Neutrinoless Double Beta Decay}},}\ }\href {\doibase
  10.1140/epja/i2007-10527-x} {\bibfield  {journal} {\bibinfo  {journal} {Eur.
  Phys. J. A}\ }\textbf {\bibinfo {volume} {36}},\ \bibinfo {pages} {195--200}
  (\bibinfo {year} {2008}{\natexlab{b}})}\BibitemShut {NoStop}%
\bibitem [{\citenamefont {Kwiatkowski}\ \emph {et~al.}(2014)\citenamefont
  {Kwiatkowski} \emph {et~al.}}]{Kwiatkowski:2013xeq}%
  \BibitemOpen
  \bibfield  {author} {\bibinfo {author} {\bibfnamefont {A.~A.}\ \bibnamefont
  {Kwiatkowski}} \emph {et~al.},\ }\bibfield  {title} {\enquote {\bibinfo
  {title} {{New Determination of Double-Beta-Decay Properties in
  $^{48}$\protect{C}a: High-Precision Q-Value Measurement and Improved Nuclear
  Matrix Element Calculations}},}\ }\href {\doibase 10.1103/PhysRevC.89.045502}
  {\bibfield  {journal} {\bibinfo  {journal} {Phys.\ Rev.\ C}\ }\textbf
  {\bibinfo {volume} {89}},\ \bibinfo {pages} {045502} (\bibinfo {year}
  {2014})}\BibitemShut {NoStop}%
\bibitem [{\citenamefont {Holt}\ and\ \citenamefont {Engel}(2013)}]{hol13c}%
  \BibitemOpen
  \bibfield  {author} {\bibinfo {author} {\bibfnamefont {Jason~D.}\
  \bibnamefont {Holt}}\ and\ \bibinfo {author} {\bibfnamefont {Jonathan}\
  \bibnamefont {Engel}},\ }\bibfield  {title} {\enquote {\bibinfo {title}
  {Effective double-$\beta$-decay operator for $^{76}$\protect{G}e and
  $^{82}$\protect{S}e},}\ }\href {\doibase 10.1103/PhysRevC.87.064315}
  {\bibfield  {journal} {\bibinfo  {journal} {Phys.\ Rev.\ C}\ }\textbf
  {\bibinfo {volume} {87}},\ \bibinfo {pages} {064315} (\bibinfo {year}
  {2013})}\BibitemShut {NoStop}%
\bibitem [{\citenamefont {Escuderos}\ \emph {et~al.}(2010)\citenamefont
  {Escuderos}, \citenamefont {Faessler}, \citenamefont {Rodin},\ and\
  \citenamefont {{\v S}imkovic}}]{Escuderos10}%
  \BibitemOpen
  \bibfield  {author} {\bibinfo {author} {\bibfnamefont {Alberto}\ \bibnamefont
  {Escuderos}}, \bibinfo {author} {\bibfnamefont {Amand}\ \bibnamefont
  {Faessler}}, \bibinfo {author} {\bibfnamefont {Vadim}\ \bibnamefont {Rodin}},
  \ and\ \bibinfo {author} {\bibfnamefont {Fedor}\ \bibnamefont {{\v
  S}imkovic}},\ }\bibfield  {title} {\enquote {\bibinfo {title} {{Contributions
  of different neutron pairs in different approaches for neutrinoless double
  beta decay}},}\ }\href {\doibase 10.1088/0954-3899/37/12/125108} {\bibfield
  {journal} {\bibinfo  {journal} {J. Phys. G: Nucl. Part. Phys.}\ }\textbf
  {\bibinfo {volume} {37}},\ \bibinfo {pages} {125108} (\bibinfo {year}
  {2010})}\BibitemShut {NoStop}%
\bibitem [{\citenamefont {Men{\'e}ndez}\ \emph
  {et~al.}(2011{\natexlab{a}})\citenamefont {Men{\'e}ndez}, \citenamefont
  {Poves}, \citenamefont {Caurier},\ and\ \citenamefont {Nowacki}}]{men11a}%
  \BibitemOpen
  \bibfield  {author} {\bibinfo {author} {\bibfnamefont {J.}~\bibnamefont
  {Men{\'e}ndez}}, \bibinfo {author} {\bibfnamefont {A.}~\bibnamefont {Poves}},
  \bibinfo {author} {\bibfnamefont {E.}~\bibnamefont {Caurier}}, \ and\
  \bibinfo {author} {\bibfnamefont {F.}~\bibnamefont {Nowacki}},\ }\bibfield
  {title} {\enquote {\bibinfo {title} {{Neutrinoless double beta decay: The
  nuclear matrix elements revisited}},}\ }\href {\doibase
  10.1088/1742-6596/312/7/072005} {\bibfield  {journal} {\bibinfo  {journal}
  {J. Phys. Conf. Ser.}\ }\textbf {\bibinfo {volume} {312}},\ \bibinfo {pages}
  {072005} (\bibinfo {year} {2011}{\natexlab{a}})}\BibitemShut {NoStop}%
\bibitem [{\citenamefont {\v{S}imkovic}\ \emph {et~al.}(2008)\citenamefont
  {\v{S}imkovic}, \citenamefont {Faessler}, \citenamefont {Rodin},
  \citenamefont {Vogel},\ and\ \citenamefont {Engel}}]{sim08}%
  \BibitemOpen
  \bibfield  {author} {\bibinfo {author} {\bibfnamefont {Fedor}\ \bibnamefont
  {\v{S}imkovic}}, \bibinfo {author} {\bibfnamefont {Amand}\ \bibnamefont
  {Faessler}}, \bibinfo {author} {\bibfnamefont {Vadim}\ \bibnamefont {Rodin}},
  \bibinfo {author} {\bibfnamefont {Petr}\ \bibnamefont {Vogel}}, \ and\
  \bibinfo {author} {\bibfnamefont {Jonathan}\ \bibnamefont {Engel}},\
  }\bibfield  {title} {\enquote {\bibinfo {title} {Anatomy of the
  $0\nu\beta\beta$ nuclear matrix elements},}\ }\href {\doibase
  10.1103/PhysRevC.77.045503} {\bibfield  {journal} {\bibinfo  {journal}
  {Phys.\ Rev.\ C}\ }\textbf {\bibinfo {volume} {77}},\ \bibinfo {pages}
  {045503} (\bibinfo {year} {2008})}\BibitemShut {NoStop}%
\bibitem [{\citenamefont {Rodr{\'i}guez}\ and\ \citenamefont
  {Egido}(2007)}]{Rodriguez07}%
  \BibitemOpen
  \bibfield  {author} {\bibinfo {author} {\bibfnamefont {Tom{\'a}s~R.}\
  \bibnamefont {Rodr{\'i}guez}}\ and\ \bibinfo {author} {\bibfnamefont
  {J.~Luis}\ \bibnamefont {Egido}},\ }\bibfield  {title} {\enquote {\bibinfo
  {title} {{New Beyond-Mean-Field Theories: Examination of the Potential Shell
  Closures at N=32 or 34}},}\ }\href {\doibase 10.1103/PhysRevLett.99.062501}
  {\bibfield  {journal} {\bibinfo  {journal} {Phys. Rev. Lett.}\ }\textbf
  {\bibinfo {volume} {99}},\ \bibinfo {pages} {062501} (\bibinfo {year}
  {2007})}\BibitemShut {NoStop}%
\bibitem [{\citenamefont {Poves}\ \emph {et~al.}(2001)\citenamefont {Poves},
  \citenamefont {S{\'a}nchez-Solano}, \citenamefont {Caurier},\ and\
  \citenamefont {Nowacki}}]{Poves01}%
  \BibitemOpen
  \bibfield  {author} {\bibinfo {author} {\bibfnamefont {A.}~\bibnamefont
  {Poves}}, \bibinfo {author} {\bibfnamefont {J.}~\bibnamefont
  {S{\'a}nchez-Solano}}, \bibinfo {author} {\bibfnamefont {E.}~\bibnamefont
  {Caurier}}, \ and\ \bibinfo {author} {\bibfnamefont {F.}~\bibnamefont
  {Nowacki}},\ }\bibfield  {title} {\enquote {\bibinfo {title} {{Shell model
  study of the isobaric chains A = 50, A = 51 and A = 52}},}\ }\href {\doibase
  10.1016/S0375-9474(01)00967-8} {\bibfield  {journal} {\bibinfo  {journal}
  {Nucl. Phys. A}\ }\textbf {\bibinfo {volume} {694}},\ \bibinfo {pages}
  {157--198} (\bibinfo {year} {2001})}\BibitemShut {NoStop}%
\bibitem [{\citenamefont {Men{\'e}ndez}\ \emph
  {et~al.}(2011{\natexlab{b}})\citenamefont {Men{\'e}ndez}, \citenamefont
  {Poves}, \citenamefont {Caurier},\ and\ \citenamefont {Nowacki}}]{men11b}%
  \BibitemOpen
  \bibfield  {author} {\bibinfo {author} {\bibfnamefont {J.}~\bibnamefont
  {Men{\'e}ndez}}, \bibinfo {author} {\bibfnamefont {A.}~\bibnamefont {Poves}},
  \bibinfo {author} {\bibfnamefont {E.}~\bibnamefont {Caurier}}, \ and\
  \bibinfo {author} {\bibfnamefont {F.}~\bibnamefont {Nowacki}},\ }\bibfield
  {title} {\enquote {\bibinfo {title} {Novel nuclear structure aspects of the
  $0 \nu \beta \beta$-decay},}\ }\href {\doibase
  10.1088/1742-6596/267/1/012058} {\bibfield  {journal} {\bibinfo  {journal}
  {J. Phys. Conf. Ser.}\ }\textbf {\bibinfo {volume} {267}},\ \bibinfo {pages}
  {012058} (\bibinfo {year} {2011}{\natexlab{b}})}\BibitemShut {NoStop}%
\bibitem [{\citenamefont {Fang}\ \emph {et~al.}(2011)\citenamefont {Fang},
  \citenamefont {Faessler}, \citenamefont {Rodin},\ and\ \citenamefont {{\v
  S}imkovic}}]{fang11}%
  \BibitemOpen
  \bibfield  {author} {\bibinfo {author} {\bibfnamefont {Dong-Liang}\
  \bibnamefont {Fang}}, \bibinfo {author} {\bibfnamefont {Amand}\ \bibnamefont
  {Faessler}}, \bibinfo {author} {\bibfnamefont {Vadim}\ \bibnamefont {Rodin}},
  \ and\ \bibinfo {author} {\bibfnamefont {Fedor}\ \bibnamefont {{\v
  S}imkovic}},\ }\bibfield  {title} {\enquote {\bibinfo {title} {{Neutrinoless
  Double Beta Decay of Deformed Nuclei within QRPA with Realistic
  Interaction}},}\ }\href {\doibase 10.1103/PhysRevC.83.034320} {\bibfield
  {journal} {\bibinfo  {journal} {Phys. Rev. C}\ }\textbf {\bibinfo {volume}
  {83}},\ \bibinfo {pages} {034329} (\bibinfo {year} {2011})}\BibitemShut
  {NoStop}%
\bibitem [{\citenamefont {Brown}\ and\ \citenamefont
  {Wildenthal}(1988)}]{bro88}%
  \BibitemOpen
  \bibfield  {author} {\bibinfo {author} {\bibfnamefont {B.~Alex}\ \bibnamefont
  {Brown}}\ and\ \bibinfo {author} {\bibfnamefont {B.~H.}\ \bibnamefont
  {Wildenthal}},\ }\bibfield  {title} {\enquote {\bibinfo {title} {{Status of
  the nuclear shell model}},}\ }\href {\doibase
  10.1146/annurev.ns.38.120188.000333} {\bibfield  {journal} {\bibinfo
  {journal} {Ann. Rev. Nucl. Part. Sci.}\ }\textbf {\bibinfo {volume} {38}},\
  \bibinfo {pages} {29--66} (\bibinfo {year} {1988})}\BibitemShut {NoStop}%
\bibitem [{\citenamefont {Kumar}\ \emph {et~al.}(2016)\citenamefont {Kumar},
  \citenamefont {Srivastava},\ and\ \citenamefont {Li}}]{Kumar16}%
  \BibitemOpen
  \bibfield  {author} {\bibinfo {author} {\bibfnamefont {Vikas}\ \bibnamefont
  {Kumar}}, \bibinfo {author} {\bibfnamefont {P.~C.}\ \bibnamefont
  {Srivastava}}, \ and\ \bibinfo {author} {\bibfnamefont {Hantao}\ \bibnamefont
  {Li}},\ }\bibfield  {title} {\enquote {\bibinfo {title} {{Nuclear
  $\beta^-$-decay half-lives for $fp$ and $fpg$ shell nuclei}},}\ }\href
  {\doibase 10.1088/0954-3899/43/10/105104} {\bibfield  {journal} {\bibinfo
  {journal} {J. Phys. G: Nucl. Part. Phys.}\ }\textbf {\bibinfo {volume}
  {43}},\ \bibinfo {pages} {105104} (\bibinfo {year} {2016})}\BibitemShut
  {NoStop}%
\bibitem [{\citenamefont {Brown}\ \emph {et~al.}(1978)\citenamefont {Brown},
  \citenamefont {Chung},\ and\ \citenamefont {Wildenthal}}]{bro78}%
  \BibitemOpen
  \bibfield  {author} {\bibinfo {author} {\bibfnamefont {B.~Alex}\ \bibnamefont
  {Brown}}, \bibinfo {author} {\bibfnamefont {W.}~\bibnamefont {Chung}}, \ and\
  \bibinfo {author} {\bibfnamefont {B.~H.}\ \bibnamefont {Wildenthal}},\
  }\bibfield  {title} {\enquote {\bibinfo {title} {{Empirical Renormalization
  of the One-Body Gamow-Teller beta-Decay Matrix Elements in the 1s-0d
  Shell}},}\ }\href {\doibase 10.1103/PhysRevLett.40.1631} {\bibfield
  {journal} {\bibinfo  {journal} {Phys. Rev. Lett.}\ }\textbf {\bibinfo
  {volume} {40}},\ \bibinfo {pages} {1631--1635} (\bibinfo {year}
  {1978})}\BibitemShut {NoStop}%
\bibitem [{\citenamefont {Bertsch}\ and\ \citenamefont
  {Hamamoto}(1982)}]{Bertsch82}%
  \BibitemOpen
  \bibfield  {author} {\bibinfo {author} {\bibfnamefont {G.~F.}\ \bibnamefont
  {Bertsch}}\ and\ \bibinfo {author} {\bibfnamefont {I.}~\bibnamefont
  {Hamamoto}},\ }\bibfield  {title} {\enquote {\bibinfo {title} {{Gamow-Teller
  strength at high excitations}},}\ }\href {\doibase 10.1103/PhysRevC.26.1323}
  {\bibfield  {journal} {\bibinfo  {journal} {Phys. Rev. C}\ }\textbf {\bibinfo
  {volume} {26}},\ \bibinfo {pages} {1323--1326} (\bibinfo {year}
  {1982})}\BibitemShut {NoStop}%
\bibitem [{\citenamefont {Arima}\ \emph {et~al.}(1987)\citenamefont {Arima},
  \citenamefont {Shimizu}, \citenamefont {Bentz},\ and\ \citenamefont
  {Hyuga}}]{arima87}%
  \BibitemOpen
  \bibfield  {author} {\bibinfo {author} {\bibfnamefont {A.}~\bibnamefont
  {Arima}}, \bibinfo {author} {\bibfnamefont {K.}~\bibnamefont {Shimizu}},
  \bibinfo {author} {\bibfnamefont {W.}~\bibnamefont {Bentz}}, \ and\ \bibinfo
  {author} {\bibfnamefont {H.}~\bibnamefont {Hyuga}},\ }\bibfield  {title}
  {\enquote {\bibinfo {title} {{Nuclear Magnetic Properties and Gamow-teller
  Transitions}},}\ }\href@noop {} {\bibfield  {journal} {\bibinfo  {journal}
  {Adv. Nucl. Phys.}\ }\textbf {\bibinfo {volume} {18}},\ \bibinfo {pages}
  {1--106} (\bibinfo {year} {1987})}\BibitemShut {NoStop}%
\bibitem [{\citenamefont {Caurier}\ \emph {et~al.}(1995)\citenamefont
  {Caurier}, \citenamefont {Poves},\ and\ \citenamefont {Zuker}}]{cau95}%
  \BibitemOpen
  \bibfield  {author} {\bibinfo {author} {\bibfnamefont {E.}~\bibnamefont
  {Caurier}}, \bibinfo {author} {\bibfnamefont {A.}~\bibnamefont {Poves}}, \
  and\ \bibinfo {author} {\bibfnamefont {A.~P.}\ \bibnamefont {Zuker}},\
  }\bibfield  {title} {\enquote {\bibinfo {title} {{Missing and quenched
  Gamow-Teller strength}},}\ }\href {\doibase 10.1103/PhysRevLett.74.1517}
  {\bibfield  {journal} {\bibinfo  {journal} {Phys. Rev. Lett.}\ }\textbf
  {\bibinfo {volume} {74}},\ \bibinfo {pages} {1517--1520} (\bibinfo {year}
  {1995})}\BibitemShut {NoStop}%
\bibitem [{\citenamefont {Park}\ \emph {et~al.}(1997)\citenamefont {Park},
  \citenamefont {Jung},\ and\ \citenamefont {Min}}]{Park97}%
  \BibitemOpen
  \bibfield  {author} {\bibinfo {author} {\bibfnamefont {T.-S.}\ \bibnamefont
  {Park}}, \bibinfo {author} {\bibfnamefont {H.}~\bibnamefont {Jung}}, \ and\
  \bibinfo {author} {\bibfnamefont {D.-P.}\ \bibnamefont {Min}},\ }\bibfield
  {title} {\enquote {\bibinfo {title} {{In-medium effective axial-vector
  coupling constant}},}\ }\href {\doibase 10.1016/S0370-2693(97)00880-0}
  {\bibfield  {journal} {\bibinfo  {journal} {Phys. Lett. B}\ }\textbf
  {\bibinfo {volume} {409}},\ \bibinfo {pages} {26--32} (\bibinfo {year}
  {1997})}\BibitemShut {NoStop}%
\bibitem [{\citenamefont {Ikeda}\ \emph {et~al.}(1963)\citenamefont {Ikeda},
  \citenamefont {Fujii},\ and\ \citenamefont {Fujita}}]{Ikeda63}%
  \BibitemOpen
  \bibfield  {author} {\bibinfo {author} {\bibfnamefont {K.}~\bibnamefont
  {Ikeda}}, \bibinfo {author} {\bibfnamefont {S.}~\bibnamefont {Fujii}}, \ and\
  \bibinfo {author} {\bibfnamefont {J.~I.}\ \bibnamefont {Fujita}},\ }\bibfield
   {title} {\enquote {\bibinfo {title} {The (p,n) reactions and beta decays},}\
  }\href {\doibase 10.1016/0031-9163(63)90255-5} {\bibfield  {journal}
  {\bibinfo  {journal} {Phys. Lett.}\ }\textbf {\bibinfo {volume} {3}},\
  \bibinfo {pages} {271 -- 272} (\bibinfo {year} {1963})}\BibitemShut {NoStop}%
\bibitem [{\citenamefont {Ichimura}\ \emph {et~al.}(2006)\citenamefont
  {Ichimura}, \citenamefont {Sakai},\ and\ \citenamefont
  {Wakasa}}]{Ichimura06}%
  \BibitemOpen
  \bibfield  {author} {\bibinfo {author} {\bibfnamefont {M.}~\bibnamefont
  {Ichimura}}, \bibinfo {author} {\bibfnamefont {H.}~\bibnamefont {Sakai}}, \
  and\ \bibinfo {author} {\bibfnamefont {T.}~\bibnamefont {Wakasa}},\
  }\bibfield  {title} {\enquote {\bibinfo {title} {{Spin-isospin responses via
  (p,n) and (n,p) reactions}},}\ }\href {\doibase 10.1016/j.ppnp.2005.09.001}
  {\bibfield  {journal} {\bibinfo  {journal} {Prog. Part. Nucl. Phys.}\
  }\textbf {\bibinfo {volume} {56}},\ \bibinfo {pages} {446--531} (\bibinfo
  {year} {2006})}\BibitemShut {NoStop}%
\bibitem [{\citenamefont {Fujita}\ \emph {et~al.}(2011)\citenamefont {Fujita},
  \citenamefont {Rubio},\ and\ \citenamefont {Gelletly}}]{Fujita11}%
  \BibitemOpen
  \bibfield  {author} {\bibinfo {author} {\bibfnamefont {Y.}~\bibnamefont
  {Fujita}}, \bibinfo {author} {\bibfnamefont {B.}~\bibnamefont {Rubio}}, \
  and\ \bibinfo {author} {\bibfnamefont {W.}~\bibnamefont {Gelletly}},\
  }\bibfield  {title} {\enquote {\bibinfo {title} {{Spin-isospin excitations
  probed by strong, weak and electro-magnetic interactions}},}\ }\href
  {\doibase 10.1016/j.ppnp.2011.01.056} {\bibfield  {journal} {\bibinfo
  {journal} {Prog. Part. Nucl. Phys.}\ }\textbf {\bibinfo {volume} {66}},\
  \bibinfo {pages} {549--606} (\bibinfo {year} {2011})}\BibitemShut {NoStop}%
\bibitem [{\citenamefont {Frekers}\ \emph {et~al.}(2013)\citenamefont
  {Frekers}, \citenamefont {Puppe}, \citenamefont {Thies},\ and\ \citenamefont
  {Ejiri}}]{Frekers13}%
  \BibitemOpen
  \bibfield  {author} {\bibinfo {author} {\bibfnamefont {D.}~\bibnamefont
  {Frekers}}, \bibinfo {author} {\bibfnamefont {P.}~\bibnamefont {Puppe}},
  \bibinfo {author} {\bibfnamefont {J.~H.}\ \bibnamefont {Thies}}, \ and\
  \bibinfo {author} {\bibfnamefont {H.}~\bibnamefont {Ejiri}},\ }\bibfield
  {title} {\enquote {\bibinfo {title} {{Gamow-Teller strength extraction from
  $(^{3}$He, $t)$ reactions}},}\ }\href {\doibase
  10.1016/j.nuclphysa.2013.08.006} {\bibfield  {journal} {\bibinfo  {journal}
  {Nucl. Phys. A}\ }\textbf {\bibinfo {volume} {916}},\ \bibinfo {pages}
  {219--240} (\bibinfo {year} {2013})}\BibitemShut {NoStop}%
\bibitem [{\citenamefont {Yako}\ \emph {et~al.}(2009)\citenamefont {Yako} \emph
  {et~al.}}]{Yako09}%
  \BibitemOpen
  \bibfield  {author} {\bibinfo {author} {\bibfnamefont {K.}~\bibnamefont
  {Yako}} \emph {et~al.},\ }\bibfield  {title} {\enquote {\bibinfo {title}
  {{Gamow-Teller Strength Distributions in $^{48}$Sc by the $^{48}$Ca(p,n) and
  $^{48}$Ti(n,p) Reactions and Two-Neutrino Double-$\beta$ Decay Nuclear Matrix
  Elements}},}\ }\href {\doibase 10.1103/PhysRevLett.103.012503} {\bibfield
  {journal} {\bibinfo  {journal} {Phys. Rev. Lett.}\ }\textbf {\bibinfo
  {volume} {103}},\ \bibinfo {pages} {012503} (\bibinfo {year}
  {2009})}\BibitemShut {NoStop}%
\bibitem [{\citenamefont {Fujita}\ \emph {et~al.}(2013)\citenamefont {Fujita}
  \emph {et~al.}}]{Fujita13}%
  \BibitemOpen
  \bibfield  {author} {\bibinfo {author} {\bibfnamefont {Y.}~\bibnamefont
  {Fujita}} \emph {et~al.},\ }\bibfield  {title} {\enquote {\bibinfo {title}
  {{High-resolution study of $T_z= +2 \rightarrow +1$ Gamow-Teller transitions
  in the $^{44}$Ca($^3$He,t)$^{44}$Sc reaction}},}\ }\href {\doibase
  10.1103/PhysRevC.88.014308} {\bibfield  {journal} {\bibinfo  {journal} {Phys.
  Rev. C}\ }\textbf {\bibinfo {volume} {88}},\ \bibinfo {pages} {014308}
  (\bibinfo {year} {2013})}\BibitemShut {NoStop}%
\bibitem [{\citenamefont {Noji}\ \emph {et~al.}(2014)\citenamefont {Noji} \emph
  {et~al.}}]{Noji14}%
  \BibitemOpen
  \bibfield  {author} {\bibinfo {author} {\bibfnamefont {S.}~\bibnamefont
  {Noji}} \emph {et~al.},\ }\bibfield  {title} {\enquote {\bibinfo {title}
  {{$\beta^+$ Gamow-Teller Transition Strengths from $^{46}$\protect{T}i and
  Stellar Electron-Capture Rates}},}\ }\href {\doibase
  10.1103/PhysRevLett.112.252501} {\bibfield  {journal} {\bibinfo  {journal}
  {Phys. Rev. Lett.}\ }\textbf {\bibinfo {volume} {112}},\ \bibinfo {pages}
  {252501} (\bibinfo {year} {2014})}\BibitemShut {NoStop}%
\bibitem [{\citenamefont {Iwata}\ \emph {et~al.}(2015)\citenamefont {Iwata},
  \citenamefont {Shimizu}, \citenamefont {Utsuno}, \citenamefont {Honma},
  \citenamefont {Abe},\ and\ \citenamefont {Otsuka}}]{Iwata15}%
  \BibitemOpen
  \bibfield  {author} {\bibinfo {author} {\bibfnamefont {Y.}~\bibnamefont
  {Iwata}}, \bibinfo {author} {\bibfnamefont {N.}~\bibnamefont {Shimizu}},
  \bibinfo {author} {\bibfnamefont {Y.}~\bibnamefont {Utsuno}}, \bibinfo
  {author} {\bibfnamefont {M.}~\bibnamefont {Honma}}, \bibinfo {author}
  {\bibfnamefont {T.}~\bibnamefont {Abe}}, \ and\ \bibinfo {author}
  {\bibfnamefont {T.}~\bibnamefont {Otsuka}},\ }\bibfield  {title} {\enquote
  {\bibinfo {title} {{Ingredients of nuclear matrix element for two-neutrino
  double-beta decay of $^{48}$\protect{C}a}},}\ }\href {\doibase
  10.7566/JPSCP.6.030057} {\bibfield  {journal} {\bibinfo  {journal} {JPS Conf.
  Proc.}\ }\textbf {\bibinfo {volume} {6}},\ \bibinfo {pages} {030057}
  (\bibinfo {year} {2015})}\BibitemShut {NoStop}%
\bibitem [{\citenamefont {Yako}\ \emph {et~al.}(2005)\citenamefont {Yako} \emph
  {et~al.}}]{Yako05}%
  \BibitemOpen
  \bibfield  {author} {\bibinfo {author} {\bibfnamefont {K.}~\bibnamefont
  {Yako}} \emph {et~al.},\ }\bibfield  {title} {\enquote {\bibinfo {title}
  {{Determination of the Gamow-Teller quenching factor from charge exchange
  reactions on $^{90}$Zr}},}\ }\href {\doibase 10.1016/j.physletb.2005.04.032}
  {\bibfield  {journal} {\bibinfo  {journal} {Phys. Lett. B}\ }\textbf
  {\bibinfo {volume} {615}},\ \bibinfo {pages} {193--199} (\bibinfo {year}
  {2005})}\BibitemShut {NoStop}%
\bibitem [{\citenamefont {Sasano}\ \emph {et~al.}(2009)\citenamefont {Sasano}
  \emph {et~al.}}]{Sasano09}%
  \BibitemOpen
  \bibfield  {author} {\bibinfo {author} {\bibfnamefont {M.}~\bibnamefont
  {Sasano}} \emph {et~al.},\ }\bibfield  {title} {\enquote {\bibinfo {title}
  {{Gamow-Teller unit cross sections of the (p, n) reaction at 198 and 297 MeV
  on medium-heavy nuclei}},}\ }\href {\doibase 10.1103/PhysRevC.79.024602}
  {\bibfield  {journal} {\bibinfo  {journal} {Phys. Rev. C}\ }\textbf {\bibinfo
  {volume} {79}},\ \bibinfo {pages} {024602} (\bibinfo {year}
  {2009})}\BibitemShut {NoStop}%
\bibitem [{\citenamefont {Ejiri}\ and\ \citenamefont {Frekers}(2016)}]{eji16}%
  \BibitemOpen
  \bibfield  {author} {\bibinfo {author} {\bibfnamefont {H.}~\bibnamefont
  {Ejiri}}\ and\ \bibinfo {author} {\bibfnamefont {D.}~\bibnamefont
  {Frekers}},\ }\bibfield  {title} {\enquote {\bibinfo {title} {Spin dipole
  nuclear matrix elements for double beta decay nuclei by charge-exchange
  reactions},}\ }\href {http://stacks.iop.org/0954-3899/43/i=11/a=11LT01}
  {\bibfield  {journal} {\bibinfo  {journal} {J. Phys. G: Nucl. Part. Phys.}\
  }\textbf {\bibinfo {volume} {43}},\ \bibinfo {pages} {11LT01} (\bibinfo
  {year} {2016})}\BibitemShut {NoStop}%
\bibitem [{\citenamefont {Men{\'e}ndez}\ \emph
  {et~al.}(2011{\natexlab{c}})\citenamefont {Men{\'e}ndez}, \citenamefont
  {Gazit},\ and\ \citenamefont {Schwenk}}]{men11}%
  \BibitemOpen
  \bibfield  {author} {\bibinfo {author} {\bibfnamefont {J.}~\bibnamefont
  {Men{\'e}ndez}}, \bibinfo {author} {\bibfnamefont {D.}~\bibnamefont {Gazit}},
  \ and\ \bibinfo {author} {\bibfnamefont {A.}~\bibnamefont {Schwenk}},\
  }\bibfield  {title} {\enquote {\bibinfo {title} {Chiral two-body currents in
  nuclei: Gamow-teller transitions and neutrinoless double-beta decay},}\
  }\href {\doibase 10.1103/PhysRevLett.107.062501} {\bibfield  {journal}
  {\bibinfo  {journal} {Phys. Rev. Lett.}\ }\textbf {\bibinfo {volume} {107}},\
  \bibinfo {pages} {062501} (\bibinfo {year} {2011}{\natexlab{c}})}\BibitemShut
  {NoStop}%
\bibitem [{\citenamefont {Engel}\ \emph {et~al.}(2014)\citenamefont {Engel},
  \citenamefont {{\v S}imkovic},\ and\ \citenamefont {Vogel}}]{eng14}%
  \BibitemOpen
  \bibfield  {author} {\bibinfo {author} {\bibfnamefont {J.}~\bibnamefont
  {Engel}}, \bibinfo {author} {\bibfnamefont {F.}~\bibnamefont {{\v
  S}imkovic}}, \ and\ \bibinfo {author} {\bibfnamefont {P.}~\bibnamefont
  {Vogel}},\ }\bibfield  {title} {\enquote {\bibinfo {title} {{Chiral Two-Body
  Currents and Neutrinoless Double-Beta Decay in the QRPA}},}\ }\href {\doibase
  10.1103/PhysRevC.89.064308} {\bibfield  {journal} {\bibinfo  {journal}
  {Phys.\ Rev.\ C}\ }\textbf {\bibinfo {volume} {89}},\ \bibinfo {pages}
  {064308} (\bibinfo {year} {2014})}\BibitemShut {NoStop}%
\bibitem [{\citenamefont {Pandharipande}\ \emph {et~al.}(1997)\citenamefont
  {Pandharipande}, \citenamefont {Sick},\ and\ \citenamefont
  {Huberts}}]{Pandharipande97}%
  \BibitemOpen
  \bibfield  {author} {\bibinfo {author} {\bibfnamefont {Vijay~R.}\
  \bibnamefont {Pandharipande}}, \bibinfo {author} {\bibfnamefont {Ingo}\
  \bibnamefont {Sick}}, \ and\ \bibinfo {author} {\bibfnamefont {Peter K.
  A.~deWitt}\ \bibnamefont {Huberts}},\ }\bibfield  {title} {\enquote {\bibinfo
  {title} {{Independent particle motion and correlations in fermion
  systems}},}\ }\href {\doibase 10.1103/RevModPhys.69.981} {\bibfield
  {journal} {\bibinfo  {journal} {Rev. Mod. Phys.}\ }\textbf {\bibinfo {volume}
  {69}},\ \bibinfo {pages} {981--991} (\bibinfo {year} {1997})}\BibitemShut
  {NoStop}%
\bibitem [{\citenamefont {Siiskonen}\ \emph {et~al.}(2001)\citenamefont
  {Siiskonen}, \citenamefont {Hjorth-Jensen},\ and\ \citenamefont
  {Suhonen}}]{sii01}%
  \BibitemOpen
  \bibfield  {author} {\bibinfo {author} {\bibfnamefont {T.}~\bibnamefont
  {Siiskonen}}, \bibinfo {author} {\bibfnamefont {M.}~\bibnamefont
  {Hjorth-Jensen}}, \ and\ \bibinfo {author} {\bibfnamefont {J.}~\bibnamefont
  {Suhonen}},\ }\bibfield  {title} {\enquote {\bibinfo {title} {Renormalization
  of the weak hadronic current in the nuclear medium},}\ }\href {\doibase
  10.1103/PhysRevC.63.055501} {\bibfield  {journal} {\bibinfo  {journal}
  {Phys.\ Rev. C}\ }\textbf {\bibinfo {volume} {63}},\ \bibinfo {pages}
  {055501} (\bibinfo {year} {2001})}\BibitemShut {NoStop}%
\bibitem [{\citenamefont {Park}\ \emph {et~al.}(2003)\citenamefont {Park},
  \citenamefont {Marcucci}, \citenamefont {Schiavilla}, \citenamefont
  {Viviani}, \citenamefont {Kievsky}, \citenamefont {Rosati}, \citenamefont
  {Kubodera}, \citenamefont {Min},\ and\ \citenamefont {Rho}}]{Park03}%
  \BibitemOpen
  \bibfield  {author} {\bibinfo {author} {\bibfnamefont {T.~S.}\ \bibnamefont
  {Park}}, \bibinfo {author} {\bibfnamefont {L.~E.}\ \bibnamefont {Marcucci}},
  \bibinfo {author} {\bibfnamefont {R.}~\bibnamefont {Schiavilla}}, \bibinfo
  {author} {\bibfnamefont {M.}~\bibnamefont {Viviani}}, \bibinfo {author}
  {\bibfnamefont {A.}~\bibnamefont {Kievsky}}, \bibinfo {author} {\bibfnamefont
  {S.}~\bibnamefont {Rosati}}, \bibinfo {author} {\bibfnamefont
  {K.}~\bibnamefont {Kubodera}}, \bibinfo {author} {\bibfnamefont {D.~P.}\
  \bibnamefont {Min}}, \ and\ \bibinfo {author} {\bibfnamefont
  {M.}~\bibnamefont {Rho}},\ }\bibfield  {title} {\enquote {\bibinfo {title}
  {{Parameter free effective field theory calculation for the solar proton
  fusion and hep processes}},}\ }\href {\doibase 10.1103/PhysRevC.67.055206}
  {\bibfield  {journal} {\bibinfo  {journal} {Phys. Rev. C}\ }\textbf {\bibinfo
  {volume} {67}},\ \bibinfo {pages} {055206} (\bibinfo {year}
  {2003})}\BibitemShut {NoStop}%
\bibitem [{\citenamefont {Hoferichter}\ \emph {et~al.}(2015)\citenamefont
  {Hoferichter}, \citenamefont {Klos},\ and\ \citenamefont
  {Schwenk}}]{Hoferichter15}%
  \BibitemOpen
  \bibfield  {author} {\bibinfo {author} {\bibfnamefont {Martin}\ \bibnamefont
  {Hoferichter}}, \bibinfo {author} {\bibfnamefont {Philipp}\ \bibnamefont
  {Klos}}, \ and\ \bibinfo {author} {\bibfnamefont {Achim}\ \bibnamefont
  {Schwenk}},\ }\bibfield  {title} {\enquote {\bibinfo {title} {{Chiral power
  counting of one- and two-body currents in direct detection of dark
  matter}},}\ }\href {\doibase 10.1016/j.physletb.2015.05.041} {\bibfield
  {journal} {\bibinfo  {journal} {Phys. Lett. B}\ }\textbf {\bibinfo {volume}
  {746}},\ \bibinfo {pages} {410--416} (\bibinfo {year} {2015})}\BibitemShut
  {NoStop}%
\bibitem [{\citenamefont {Krebs}\ \emph {et~al.}(2017)\citenamefont {Krebs},
  \citenamefont {Epelbaum},\ and\ \citenamefont {Meißner}}]{Krebs16}%
  \BibitemOpen
  \bibfield  {author} {\bibinfo {author} {\bibfnamefont {H.}~\bibnamefont
  {Krebs}}, \bibinfo {author} {\bibfnamefont {E.}~\bibnamefont {Epelbaum}}, \
  and\ \bibinfo {author} {\bibfnamefont {U.~G.}\ \bibnamefont {Meißner}},\
  }\bibfield  {title} {\enquote {\bibinfo {title} {{Nuclear axial current
  operators to fourth order in chiral effective field theory}},}\ }\href
  {\doibase 10.1016/j.aop.2017.01.021} {\bibfield  {journal} {\bibinfo
  {journal} {Annals Phys.}\ }\textbf {\bibinfo {volume} {378}},\ \bibinfo
  {pages} {317--395} (\bibinfo {year} {2017})}\BibitemShut {NoStop}%
\bibitem [{\citenamefont {Butler}\ \emph {et~al.}(2001)\citenamefont {Butler},
  \citenamefont {Chen},\ and\ \citenamefont {Kong}}]{Butler01}%
  \BibitemOpen
  \bibfield  {author} {\bibinfo {author} {\bibfnamefont {Malcolm}\ \bibnamefont
  {Butler}}, \bibinfo {author} {\bibfnamefont {Jiunn-Wei}\ \bibnamefont
  {Chen}}, \ and\ \bibinfo {author} {\bibfnamefont {Xinwei}\ \bibnamefont
  {Kong}},\ }\bibfield  {title} {\enquote {\bibinfo {title} {{Neutrino deuteron
  scattering in effective field theory at next-to-next-to-leading order}},}\
  }\href {\doibase 10.1103/PhysRevC.63.035501} {\bibfield  {journal} {\bibinfo
  {journal} {Phys. Rev. C}\ }\textbf {\bibinfo {volume} {63}},\ \bibinfo
  {pages} {035501} (\bibinfo {year} {2001})}\BibitemShut {NoStop}%
\bibitem [{\citenamefont {Nakamura}\ \emph {et~al.}(2001)\citenamefont
  {Nakamura}, \citenamefont {Sato}, \citenamefont {Gudkov},\ and\ \citenamefont
  {Kubodera}}]{Nakamura01}%
  \BibitemOpen
  \bibfield  {author} {\bibinfo {author} {\bibfnamefont {S.}~\bibnamefont
  {Nakamura}}, \bibinfo {author} {\bibfnamefont {T.}~\bibnamefont {Sato}},
  \bibinfo {author} {\bibfnamefont {Vladimir~P.}\ \bibnamefont {Gudkov}}, \
  and\ \bibinfo {author} {\bibfnamefont {K.}~\bibnamefont {Kubodera}},\
  }\bibfield  {title} {\enquote {\bibinfo {title} {{Neutrino reactions on the
  deuteron}},}\ }\href {\doibase 10.1103/PhysRevC.63.034617} {\bibfield
  {journal} {\bibinfo  {journal} {Phys. Rev. C}\ }\textbf {\bibinfo {volume}
  {63}},\ \bibinfo {pages} {034617} (\bibinfo {year} {2001})}\BibitemShut
  {NoStop}%
\bibitem [{\citenamefont {Gazit}\ \emph {et~al.}(2009)\citenamefont {Gazit},
  \citenamefont {Quaglioni},\ and\ \citenamefont {Navr{\'a}til}}]{Gazit09}%
  \BibitemOpen
  \bibfield  {author} {\bibinfo {author} {\bibfnamefont {Doron}\ \bibnamefont
  {Gazit}}, \bibinfo {author} {\bibfnamefont {Sofia}\ \bibnamefont
  {Quaglioni}}, \ and\ \bibinfo {author} {\bibfnamefont {Petr}\ \bibnamefont
  {Navr{\'a}til}},\ }\bibfield  {title} {\enquote {\bibinfo {title}
  {{Three-Nucleon Low-Energy Constants from the Consistency of Interactions and
  Currents in Chiral Effective Field Theory}},}\ }\href {\doibase
  10.1103/PhysRevLett.103.102502} {\bibfield  {journal} {\bibinfo  {journal}
  {Phys. Rev. Lett.}\ }\textbf {\bibinfo {volume} {103}},\ \bibinfo {pages}
  {102502} (\bibinfo {year} {2009})}\BibitemShut {NoStop}%
\bibitem [{\citenamefont {Baroni}\ \emph {et~al.}(2016)\citenamefont {Baroni},
  \citenamefont {Girlanda}, \citenamefont {Kievsky}, \citenamefont {Marcucci},
  \citenamefont {Schiavilla},\ and\ \citenamefont {Viviani}}]{Baroni16}%
  \BibitemOpen
  \bibfield  {author} {\bibinfo {author} {\bibfnamefont {A.}~\bibnamefont
  {Baroni}}, \bibinfo {author} {\bibfnamefont {L.}~\bibnamefont {Girlanda}},
  \bibinfo {author} {\bibfnamefont {A.}~\bibnamefont {Kievsky}}, \bibinfo
  {author} {\bibfnamefont {L.~E.}\ \bibnamefont {Marcucci}}, \bibinfo {author}
  {\bibfnamefont {R.}~\bibnamefont {Schiavilla}}, \ and\ \bibinfo {author}
  {\bibfnamefont {M.}~\bibnamefont {Viviani}},\ }\bibfield  {title} {\enquote
  {\bibinfo {title} {{Tritium $\beta$-decay in chiral effective field
  theory}},}\ }\href {\doibase 10.1103/PhysRevC.94.024003} {\bibfield
  {journal} {\bibinfo  {journal} {Phys. Rev. C}\ }\textbf {\bibinfo {volume}
  {94}},\ \bibinfo {pages} {024003} (\bibinfo {year} {2016})}\BibitemShut
  {NoStop}%
\bibitem [{\citenamefont {Bacca}\ and\ \citenamefont
  {Pastore}(2014)}]{Bacca14}%
  \BibitemOpen
  \bibfield  {author} {\bibinfo {author} {\bibfnamefont {Sonia}\ \bibnamefont
  {Bacca}}\ and\ \bibinfo {author} {\bibfnamefont {Saori}\ \bibnamefont
  {Pastore}},\ }\bibfield  {title} {\enquote {\bibinfo {title}
  {{Electromagnetic reactions on light nuclei}},}\ }\href {\doibase
  10.1088/0954-3899/41/12/123002} {\bibfield  {journal} {\bibinfo  {journal}
  {J. Phys. G: Nucl. Part. Phys.}\ }\textbf {\bibinfo {volume} {41}},\ \bibinfo
  {pages} {123002} (\bibinfo {year} {2014})}\BibitemShut {NoStop}%
\bibitem [{\citenamefont {Marcucci}\ \emph {et~al.}(2016)\citenamefont
  {Marcucci}, \citenamefont {Gross}, \citenamefont {Pena}, \citenamefont
  {Piarulli}, \citenamefont {Schiavilla}, \citenamefont {Sick}, \citenamefont
  {Stadler}, \citenamefont {Van~Orden},\ and\ \citenamefont
  {Viviani}}]{Marcucci16}%
  \BibitemOpen
  \bibfield  {author} {\bibinfo {author} {\bibfnamefont {L.~E.}\ \bibnamefont
  {Marcucci}}, \bibinfo {author} {\bibfnamefont {F.}~\bibnamefont {Gross}},
  \bibinfo {author} {\bibfnamefont {M.~T.}\ \bibnamefont {Pena}}, \bibinfo
  {author} {\bibfnamefont {M.}~\bibnamefont {Piarulli}}, \bibinfo {author}
  {\bibfnamefont {R.}~\bibnamefont {Schiavilla}}, \bibinfo {author}
  {\bibfnamefont {I.}~\bibnamefont {Sick}}, \bibinfo {author} {\bibfnamefont
  {A.}~\bibnamefont {Stadler}}, \bibinfo {author} {\bibfnamefont {J.~W.}\
  \bibnamefont {Van~Orden}}, \ and\ \bibinfo {author} {\bibfnamefont
  {M.}~\bibnamefont {Viviani}},\ }\bibfield  {title} {\enquote {\bibinfo
  {title} {{Electromagnetic Structure of Few-Nucleon Ground States}},}\ }\href
  {\doibase 10.1088/0954-3899/43/2/023002} {\bibfield  {journal} {\bibinfo
  {journal} {J. Phys. G: Nucl. Part. Phys.}\ }\textbf {\bibinfo {volume}
  {43}},\ \bibinfo {pages} {023002} (\bibinfo {year} {2016})}\BibitemShut
  {NoStop}%
\bibitem [{\citenamefont {Klos}\ \emph {et~al.}(2013)\citenamefont {Klos},
  \citenamefont {Men{\'e}ndez}, \citenamefont {Gazit},\ and\ \citenamefont
  {Schwenk}}]{Klos14}%
  \BibitemOpen
  \bibfield  {author} {\bibinfo {author} {\bibfnamefont {P.}~\bibnamefont
  {Klos}}, \bibinfo {author} {\bibfnamefont {J.}~\bibnamefont {Men{\'e}ndez}},
  \bibinfo {author} {\bibfnamefont {D.}~\bibnamefont {Gazit}}, \ and\ \bibinfo
  {author} {\bibfnamefont {A.}~\bibnamefont {Schwenk}},\ }\bibfield  {title}
  {\enquote {\bibinfo {title} {{Large-scale nuclear structure calculations for
  spin-dependent WIMP scattering with chiral effective field theory
  currents}},}\ }\href {\doibase 10.1103/PhysRevD.89.029901,
  10.1103/PhysRevD.88.083516} {\bibfield  {journal} {\bibinfo  {journal} {Phys.
  Rev. D}\ }\textbf {\bibinfo {volume} {88}},\ \bibinfo {pages} {083516}
  (\bibinfo {year} {2013})},\ \bibinfo {note} {[Erratum: Phys. Rev. D 89,
  029901 (2014)]}\BibitemShut {NoStop}%
\bibitem [{\citenamefont {Men{\'e}ndez}()}]{men16a}%
  \BibitemOpen
  \bibfield  {author} {\bibinfo {author} {\bibfnamefont {J.}~\bibnamefont
  {Men{\'e}ndez}},\ }\bibfield  {title} {\enquote {\bibinfo {title} {{What do
  we know about neutrinoless double-beta decay nuclear matrix elements?}}}\
  }\href@noop {} {\ }\Eprint {http://arxiv.org/abs/1605.05059}
  {arXiv:1605.05059} \BibitemShut {NoStop}%
\bibitem [{\citenamefont {Zinner}\ \emph {et~al.}(2006)\citenamefont {Zinner},
  \citenamefont {Langanke},\ and\ \citenamefont {Vogel}}]{Zinner06}%
  \BibitemOpen
  \bibfield  {author} {\bibinfo {author} {\bibfnamefont {Nikolaj~Thomas}\
  \bibnamefont {Zinner}}, \bibinfo {author} {\bibfnamefont {Karlheinz}\
  \bibnamefont {Langanke}}, \ and\ \bibinfo {author} {\bibfnamefont {Petr}\
  \bibnamefont {Vogel}},\ }\bibfield  {title} {\enquote {\bibinfo {title}
  {{Muon capture on nuclei: Random phase approximation evaluation versus data
  for $6 \leq Z \leq 94$ nuclei}},}\ }\href {\doibase
  10.1103/PhysRevC.74.024326} {\bibfield  {journal} {\bibinfo  {journal} {Phys.
  Rev. C}\ }\textbf {\bibinfo {volume} {74}},\ \bibinfo {pages} {024326}
  (\bibinfo {year} {2006})}\BibitemShut {NoStop}%
\bibitem [{\citenamefont {Hayes}\ and\ \citenamefont {Towner}(2000)}]{Hayes00}%
  \BibitemOpen
  \bibfield  {author} {\bibinfo {author} {\bibfnamefont {A.~C.}\ \bibnamefont
  {Hayes}}\ and\ \bibinfo {author} {\bibfnamefont {I.~S.}\ \bibnamefont
  {Towner}},\ }\bibfield  {title} {\enquote {\bibinfo {title} {{Shell model
  calculations of neutrino scattering from C-12}},}\ }\href {\doibase
  10.1103/PhysRevC.61.044603} {\bibfield  {journal} {\bibinfo  {journal} {Phys.
  Rev. C}\ }\textbf {\bibinfo {volume} {61}},\ \bibinfo {pages} {044603}
  (\bibinfo {year} {2000})}\BibitemShut {NoStop}%
\bibitem [{\citenamefont {Volpe}\ \emph {et~al.}(2000)\citenamefont {Volpe},
  \citenamefont {Auerbach}, \citenamefont {Colo}, \citenamefont {Suzuki},\ and\
  \citenamefont {Van~Giai}}]{Volpe00}%
  \BibitemOpen
  \bibfield  {author} {\bibinfo {author} {\bibfnamefont {C.}~\bibnamefont
  {Volpe}}, \bibinfo {author} {\bibfnamefont {N.}~\bibnamefont {Auerbach}},
  \bibinfo {author} {\bibfnamefont {G.}~\bibnamefont {Colo}}, \bibinfo {author}
  {\bibfnamefont {T.}~\bibnamefont {Suzuki}}, \ and\ \bibinfo {author}
  {\bibfnamefont {N.}~\bibnamefont {Van~Giai}},\ }\bibfield  {title} {\enquote
  {\bibinfo {title} {{Microscopic theories of neutrino C-12 reactions}},}\
  }\href {\doibase 10.1103/PhysRevC.62.015501} {\bibfield  {journal} {\bibinfo
  {journal} {Phys. Rev. C}\ }\textbf {\bibinfo {volume} {62}},\ \bibinfo
  {pages} {015501} (\bibinfo {year} {2000})}\BibitemShut {NoStop}%
\bibitem [{\citenamefont {Suzuki}\ \emph {et~al.}(2006)\citenamefont {Suzuki},
  \citenamefont {Chiba}, \citenamefont {Yoshida}, \citenamefont {Kajino},\ and\
  \citenamefont {Otsuka}}]{Suzuki06}%
  \BibitemOpen
  \bibfield  {author} {\bibinfo {author} {\bibfnamefont {Toshio}\ \bibnamefont
  {Suzuki}}, \bibinfo {author} {\bibfnamefont {Satoshi}\ \bibnamefont {Chiba}},
  \bibinfo {author} {\bibfnamefont {Takashi}\ \bibnamefont {Yoshida}}, \bibinfo
  {author} {\bibfnamefont {Toshitaka}\ \bibnamefont {Kajino}}, \ and\ \bibinfo
  {author} {\bibfnamefont {Takaharu}\ \bibnamefont {Otsuka}},\ }\bibfield
  {title} {\enquote {\bibinfo {title} {{Neutrino nucleus reactions based on new
  shell model Hamiltonians}},}\ }\href {\doibase 10.1103/PhysRevC.74.034307}
  {\bibfield  {journal} {\bibinfo  {journal} {Phys. Rev. C}\ }\textbf {\bibinfo
  {volume} {74}},\ \bibinfo {pages} {034307} (\bibinfo {year}
  {2006})}\BibitemShut {NoStop}%
\bibitem [{\citenamefont {Hayes}\ \emph {et~al.}(2003)\citenamefont {Hayes},
  \citenamefont {Navr{\'a}til},\ and\ \citenamefont {Vary}}]{Hayes03}%
  \BibitemOpen
  \bibfield  {author} {\bibinfo {author} {\bibfnamefont {A.~C.}\ \bibnamefont
  {Hayes}}, \bibinfo {author} {\bibfnamefont {P.}~\bibnamefont {Navr{\'a}til}},
  \ and\ \bibinfo {author} {\bibfnamefont {J.~P.}\ \bibnamefont {Vary}},\
  }\bibfield  {title} {\enquote {\bibinfo {title} {{Neutrino C-12 scattering in
  the ab initio shell model with a realistic three body interaction}},}\ }\href
  {\doibase 10.1103/PhysRevLett.91.012502} {\bibfield  {journal} {\bibinfo
  {journal} {Phys. Rev. Lett.}\ }\textbf {\bibinfo {volume} {91}},\ \bibinfo
  {pages} {012502} (\bibinfo {year} {2003})}\BibitemShut {NoStop}%
\bibitem [{\citenamefont {Lovato}\ \emph {et~al.}(2016)\citenamefont {Lovato},
  \citenamefont {Gandolfi}, \citenamefont {Carlson}, \citenamefont {Pieper},\
  and\ \citenamefont {Schiavilla}}]{Lovato16}%
  \BibitemOpen
  \bibfield  {author} {\bibinfo {author} {\bibfnamefont {A.}~\bibnamefont
  {Lovato}}, \bibinfo {author} {\bibfnamefont {S.}~\bibnamefont {Gandolfi}},
  \bibinfo {author} {\bibfnamefont {J.}~\bibnamefont {Carlson}}, \bibinfo
  {author} {\bibfnamefont {Steven~C.}\ \bibnamefont {Pieper}}, \ and\ \bibinfo
  {author} {\bibfnamefont {R.}~\bibnamefont {Schiavilla}},\ }\bibfield  {title}
  {\enquote {\bibinfo {title} {{Electromagnetic response of $^{12}$C: A
  first-principles calculation}},}\ }\href {\doibase
  10.1103/PhysRevLett.117.082501} {\bibfield  {journal} {\bibinfo  {journal}
  {Phys. Rev. Lett.}\ }\textbf {\bibinfo {volume} {117}},\ \bibinfo {pages}
  {082501} (\bibinfo {year} {2016})}\BibitemShut {NoStop}%
\bibitem [{\citenamefont {Ekstr{\"o}m}\ \emph {et~al.}(2014)\citenamefont
  {Ekstr{\"o}m}, \citenamefont {Jansen}, \citenamefont {Wendt}, \citenamefont
  {Hagen}, \citenamefont {Papenbrock}, \citenamefont {Bacca}, \citenamefont
  {Carlsson},\ and\ \citenamefont {Gazit}}]{Ekstrom:2014iya}%
  \BibitemOpen
  \bibfield  {author} {\bibinfo {author} {\bibfnamefont {A.}~\bibnamefont
  {Ekstr{\"o}m}}, \bibinfo {author} {\bibfnamefont {G.~R.}\ \bibnamefont
  {Jansen}}, \bibinfo {author} {\bibfnamefont {K.~A.}\ \bibnamefont {Wendt}},
  \bibinfo {author} {\bibfnamefont {G.}~\bibnamefont {Hagen}}, \bibinfo
  {author} {\bibfnamefont {T.}~\bibnamefont {Papenbrock}}, \bibinfo {author}
  {\bibfnamefont {S.}~\bibnamefont {Bacca}}, \bibinfo {author} {\bibfnamefont
  {B.}~\bibnamefont {Carlsson}}, \ and\ \bibinfo {author} {\bibfnamefont
  {D.}~\bibnamefont {Gazit}},\ }\bibfield  {title} {\enquote {\bibinfo {title}
  {{Effects of three-nucleon forces and two-body currents on Gamow-Teller
  strengths}},}\ }\href {\doibase 10.1103/PhysRevLett.113.262504} {\bibfield
  {journal} {\bibinfo  {journal} {Phys. Rev. Lett.}\ }\textbf {\bibinfo
  {volume} {113}},\ \bibinfo {pages} {262504} (\bibinfo {year}
  {2014})}\BibitemShut {NoStop}%
\bibitem [{\citenamefont {Shimizu}\ \emph {et~al.}(2012)\citenamefont
  {Shimizu}, \citenamefont {Abe}, \citenamefont {Tsunoda}, \citenamefont
  {Utsuno}, \citenamefont {Yoshida}, \citenamefont {Mizusaki}, \citenamefont
  {Honma},\ and\ \citenamefont {Otsuka}}]{Shimizu12}%
  \BibitemOpen
  \bibfield  {author} {\bibinfo {author} {\bibfnamefont {Noritaka}\
  \bibnamefont {Shimizu}}, \bibinfo {author} {\bibfnamefont {Takashi}\
  \bibnamefont {Abe}}, \bibinfo {author} {\bibfnamefont {Yusuke}\ \bibnamefont
  {Tsunoda}}, \bibinfo {author} {\bibfnamefont {Yutaka}\ \bibnamefont
  {Utsuno}}, \bibinfo {author} {\bibfnamefont {Tooru}\ \bibnamefont {Yoshida}},
  \bibinfo {author} {\bibfnamefont {Takahiro}\ \bibnamefont {Mizusaki}},
  \bibinfo {author} {\bibfnamefont {Michio}\ \bibnamefont {Honma}}, \ and\
  \bibinfo {author} {\bibfnamefont {Takaharu}\ \bibnamefont {Otsuka}},\
  }\bibfield  {title} {\enquote {\bibinfo {title} {{New-generation Monte Carlo
  shell model for the K computer era}},}\ }\href {\doibase 10.1093/ptep/pts012}
  {\bibfield  {journal} {\bibinfo  {journal} {Prog. Theor. Exp. Phys.}\
  }\textbf {\bibinfo {volume} {2012}},\ \bibinfo {pages} {01A205} (\bibinfo
  {year} {2012})}\BibitemShut {NoStop}%
\bibitem [{\citenamefont {Stumpf}\ \emph {et~al.}(2016)\citenamefont {Stumpf},
  \citenamefont {Braun},\ and\ \citenamefont {Roth}}]{Stumpf16}%
  \BibitemOpen
  \bibfield  {author} {\bibinfo {author} {\bibfnamefont {Christina}\
  \bibnamefont {Stumpf}}, \bibinfo {author} {\bibfnamefont {Jonas}\
  \bibnamefont {Braun}}, \ and\ \bibinfo {author} {\bibfnamefont {Robert}\
  \bibnamefont {Roth}},\ }\bibfield  {title} {\enquote {\bibinfo {title}
  {{Importance-Truncated Large-Scale Shell Model}},}\ }\href {\doibase
  10.1103/PhysRevC.93.021301} {\bibfield  {journal} {\bibinfo  {journal} {Phys.
  Rev. C}\ }\textbf {\bibinfo {volume} {93}},\ \bibinfo {pages} {021301}
  (\bibinfo {year} {2016})}\BibitemShut {NoStop}%
\bibitem [{\citenamefont {Legeza}\ \emph {et~al.}(2015)\citenamefont {Legeza},
  \citenamefont {Veis}, \citenamefont {Poves},\ and\ \citenamefont
  {Dukelsky}}]{Legeza15}%
  \BibitemOpen
  \bibfield  {author} {\bibinfo {author} {\bibfnamefont {{\"O}.}~\bibnamefont
  {Legeza}}, \bibinfo {author} {\bibfnamefont {L.}~\bibnamefont {Veis}},
  \bibinfo {author} {\bibfnamefont {A.}~\bibnamefont {Poves}}, \ and\ \bibinfo
  {author} {\bibfnamefont {J.}~\bibnamefont {Dukelsky}},\ }\bibfield  {title}
  {\enquote {\bibinfo {title} {{Advanced density matrix renormalization group
  method for nuclear structure calculations}},}\ }\href {\doibase
  10.1103/PhysRevC.92.051303} {\bibfield  {journal} {\bibinfo  {journal} {Phys.
  Rev. C}\ }\textbf {\bibinfo {volume} {92}},\ \bibinfo {pages} {051303}
  (\bibinfo {year} {2015})}\BibitemShut {NoStop}%
\bibitem [{\citenamefont {Togashi}\ \emph {et~al.}(2016)\citenamefont
  {Togashi}, \citenamefont {Tsunoda}, \citenamefont {Otsuka},\ and\
  \citenamefont {Shimizu}}]{Togashi16}%
  \BibitemOpen
  \bibfield  {author} {\bibinfo {author} {\bibfnamefont {Tomoaki}\ \bibnamefont
  {Togashi}}, \bibinfo {author} {\bibfnamefont {Yusuke}\ \bibnamefont
  {Tsunoda}}, \bibinfo {author} {\bibfnamefont {Takaharu}\ \bibnamefont
  {Otsuka}}, \ and\ \bibinfo {author} {\bibfnamefont {Noritaka}\ \bibnamefont
  {Shimizu}},\ }\bibfield  {title} {\enquote {\bibinfo {title} {{Quantum Phase
  Transition in the Shape of Zr isotopes}},}\ }\href {\doibase
  10.1103/PhysRevLett.117.172502} {\bibfield  {journal} {\bibinfo  {journal}
  {Phys. Rev. Lett.}\ }\textbf {\bibinfo {volume} {117}},\ \bibinfo {pages}
  {172502} (\bibinfo {year} {2016})}\BibitemShut {NoStop}%
\bibitem [{\citenamefont {{\v S}imkovic}\ \emph {et~al.}(2009)\citenamefont
  {{\v S}imkovic}, \citenamefont {Faessler},\ and\ \citenamefont
  {Vogel}}]{sim09a}%
  \BibitemOpen
  \bibfield  {author} {\bibinfo {author} {\bibfnamefont {Fedor}\ \bibnamefont
  {{\v S}imkovic}}, \bibinfo {author} {\bibfnamefont {Amand}\ \bibnamefont
  {Faessler}}, \ and\ \bibinfo {author} {\bibfnamefont {Petr}\ \bibnamefont
  {Vogel}},\ }\bibfield  {title} {\enquote {\bibinfo {title} {{$0\nu\beta\beta$
  nuclear matrix elements and the occupancy of individual orbits}},}\ }\href
  {\doibase 10.1103/PhysRevC.79.015502} {\bibfield  {journal} {\bibinfo
  {journal} {Phys. Rev. C}\ }\textbf {\bibinfo {volume} {79}},\ \bibinfo
  {pages} {015502} (\bibinfo {year} {2009})}\BibitemShut {NoStop}%
\bibitem [{\citenamefont {Rodr{\'i}guez}\ \emph {et~al.}(2016)\citenamefont
  {Rodr{\'i}guez}, \citenamefont {Poves},\ and\ \citenamefont
  {Nowacki}}]{Rodriguez16}%
  \BibitemOpen
  \bibfield  {author} {\bibinfo {author} {\bibfnamefont {Tom{\'a}s~R.}\
  \bibnamefont {Rodr{\'i}guez}}, \bibinfo {author} {\bibfnamefont {Alfredo}\
  \bibnamefont {Poves}}, \ and\ \bibinfo {author} {\bibfnamefont {Frédéric}\
  \bibnamefont {Nowacki}},\ }\bibfield  {title} {\enquote {\bibinfo {title}
  {{Occupation numbers of spherical orbits in self-consistent beyond-mean-field
  methods}},}\ }\href {\doibase 10.1103/PhysRevC.93.054316} {\bibfield
  {journal} {\bibinfo  {journal} {Phys. Rev. C}\ }\textbf {\bibinfo {volume}
  {93}},\ \bibinfo {pages} {054316} (\bibinfo {year} {2016})}\BibitemShut
  {NoStop}%
\bibitem [{\citenamefont {Abe}\ \emph {et~al.}(2012)\citenamefont {Abe},
  \citenamefont {Maris}, \citenamefont {Otsuka}, \citenamefont {Shimizu},
  \citenamefont {Utsuno},\ and\ \citenamefont {Vary}}]{Abe12}%
  \BibitemOpen
  \bibfield  {author} {\bibinfo {author} {\bibfnamefont {T.}~\bibnamefont
  {Abe}}, \bibinfo {author} {\bibfnamefont {P.}~\bibnamefont {Maris}}, \bibinfo
  {author} {\bibfnamefont {T.}~\bibnamefont {Otsuka}}, \bibinfo {author}
  {\bibfnamefont {N.}~\bibnamefont {Shimizu}}, \bibinfo {author} {\bibfnamefont
  {Y.}~\bibnamefont {Utsuno}}, \ and\ \bibinfo {author} {\bibfnamefont {J.~P.}\
  \bibnamefont {Vary}},\ }\bibfield  {title} {\enquote {\bibinfo {title}
  {{Benchmarks of the full configuration interaction, Monte Carlo shell model,
  and no-core full configuration methods}},}\ }\href {\doibase
  10.1103/PhysRevC.86.054301} {\bibfield  {journal} {\bibinfo  {journal} {Phys.
  Rev. C}\ }\textbf {\bibinfo {volume} {86}},\ \bibinfo {pages} {054301}
  (\bibinfo {year} {2012})}\BibitemShut {NoStop}%
\bibitem [{\citenamefont {Navr{\'a}til}\ \emph {et~al.}(2016)\citenamefont
  {Navr{\'a}til}, \citenamefont {Quaglioni}, \citenamefont {Hupin},
  \citenamefont {Romero-Redondo},\ and\ \citenamefont {Calci}}]{Navratil16}%
  \BibitemOpen
  \bibfield  {author} {\bibinfo {author} {\bibfnamefont {Petr}\ \bibnamefont
  {Navr{\'a}til}}, \bibinfo {author} {\bibfnamefont {Sofia}\ \bibnamefont
  {Quaglioni}}, \bibinfo {author} {\bibfnamefont {Guillaume}\ \bibnamefont
  {Hupin}}, \bibinfo {author} {\bibfnamefont {Carolina}\ \bibnamefont
  {Romero-Redondo}}, \ and\ \bibinfo {author} {\bibfnamefont {Angelo}\
  \bibnamefont {Calci}},\ }\bibfield  {title} {\enquote {\bibinfo {title}
  {{Unified ab initio approaches to nuclear structure and reactions}},}\ }\href
  {\doibase 10.1088/0031-8949/91/5/053002} {\bibfield  {journal} {\bibinfo
  {journal} {Phys. Scr.}\ }\textbf {\bibinfo {volume} {91}},\ \bibinfo {pages}
  {053002} (\bibinfo {year} {2016})}\BibitemShut {NoStop}%
\bibitem [{\citenamefont {Barrett}\ \emph {et~al.}(2013)\citenamefont
  {Barrett}, \citenamefont {Navr{\'a}til},\ and\ \citenamefont
  {Vary}}]{Barrett13}%
  \BibitemOpen
  \bibfield  {author} {\bibinfo {author} {\bibfnamefont {Bruce~R.}\
  \bibnamefont {Barrett}}, \bibinfo {author} {\bibfnamefont {Petr}\
  \bibnamefont {Navr{\'a}til}}, \ and\ \bibinfo {author} {\bibfnamefont
  {James~P.}\ \bibnamefont {Vary}},\ }\bibfield  {title} {\enquote {\bibinfo
  {title} {{Ab initio no core shell model}},}\ }\href {\doibase
  10.1016/j.ppnp.2012.10.003} {\bibfield  {journal} {\bibinfo  {journal} {Prog.
  Part. Nucl. Phys.}\ }\textbf {\bibinfo {volume} {69}},\ \bibinfo {pages}
  {131--181} (\bibinfo {year} {2013})}\BibitemShut {NoStop}%
\bibitem [{\citenamefont {Zuker}(2003)}]{Zuker03}%
  \BibitemOpen
  \bibfield  {author} {\bibinfo {author} {\bibfnamefont {Andr{\'e}s~P.}\
  \bibnamefont {Zuker}},\ }\bibfield  {title} {\enquote {\bibinfo {title}
  {{Three body monopole corrections to the realistic interactions}},}\ }\href
  {\doibase 10.1103/PhysRevLett.90.042502} {\bibfield  {journal} {\bibinfo
  {journal} {Phys. Rev. Lett.}\ }\textbf {\bibinfo {volume} {90}},\ \bibinfo
  {pages} {042502} (\bibinfo {year} {2003})}\BibitemShut {NoStop}%
\bibitem [{\citenamefont {Otsuka}\ \emph {et~al.}(2010)\citenamefont {Otsuka},
  \citenamefont {Suzuki}, \citenamefont {Holt}, \citenamefont {Schwenk},\ and\
  \citenamefont {Akaishi}}]{ots10}%
  \BibitemOpen
  \bibfield  {author} {\bibinfo {author} {\bibfnamefont {T.}~\bibnamefont
  {Otsuka}}, \bibinfo {author} {\bibfnamefont {T.}~\bibnamefont {Suzuki}},
  \bibinfo {author} {\bibfnamefont {J.D.}\ \bibnamefont {Holt}}, \bibinfo
  {author} {\bibfnamefont {A.}~\bibnamefont {Schwenk}}, \ and\ \bibinfo
  {author} {\bibfnamefont {Y.}~\bibnamefont {Akaishi}},\ }\bibfield  {title}
  {\enquote {\bibinfo {title} {Three-body forces and the limit of oxygen
  isotopes},}\ }\href {\doibase 10.1103/PhysRevLett.105.032501} {\bibfield
  {journal} {\bibinfo  {journal} {Phys. Rev. Lett}\ }\textbf {\bibinfo {volume}
  {105}},\ \bibinfo {pages} {032501} (\bibinfo {year} {2010})}\BibitemShut
  {NoStop}%
\bibitem [{\citenamefont {Jansen}\ \emph {et~al.}(2014)\citenamefont {Jansen},
  \citenamefont {Engel}, \citenamefont {Hagen}, \citenamefont {Navr{\'a}til},\
  and\ \citenamefont {Signoracci}}]{Jansen14}%
  \BibitemOpen
  \bibfield  {author} {\bibinfo {author} {\bibfnamefont {G.~R.}\ \bibnamefont
  {Jansen}}, \bibinfo {author} {\bibfnamefont {J.}~\bibnamefont {Engel}},
  \bibinfo {author} {\bibfnamefont {G.}~\bibnamefont {Hagen}}, \bibinfo
  {author} {\bibfnamefont {P.}~\bibnamefont {Navr{\'a}til}}, \ and\ \bibinfo
  {author} {\bibfnamefont {A.}~\bibnamefont {Signoracci}},\ }\bibfield  {title}
  {\enquote {\bibinfo {title} {Ab-initio coupled-cluster effective interactions
  for the shell model: Application to neutron-rich oxygen and carbon
  isotopes},}\ }\href {\doibase 10.1103/PhysRevLett.113.142502} {\bibfield
  {journal} {\bibinfo  {journal} {Phys. Rev. Lett.}\ }\textbf {\bibinfo
  {volume} {113}},\ \bibinfo {pages} {142502} (\bibinfo {year}
  {2014})}\BibitemShut {NoStop}%
\bibitem [{\citenamefont {Bogner}\ \emph {et~al.}(2014)\citenamefont {Bogner},
  \citenamefont {Hergert}, \citenamefont {Holt}, \citenamefont {Schwenk},
  \citenamefont {Binder}, \citenamefont {Calci}, \citenamefont {Langhammer},\
  and\ \citenamefont {Roth}}]{bog14}%
  \BibitemOpen
  \bibfield  {author} {\bibinfo {author} {\bibfnamefont {S.~K.}\ \bibnamefont
  {Bogner}}, \bibinfo {author} {\bibfnamefont {H.}~\bibnamefont {Hergert}},
  \bibinfo {author} {\bibfnamefont {J.~D.}\ \bibnamefont {Holt}}, \bibinfo
  {author} {\bibfnamefont {A.}~\bibnamefont {Schwenk}}, \bibinfo {author}
  {\bibfnamefont {S.}~\bibnamefont {Binder}}, \bibinfo {author} {\bibfnamefont
  {A.}~\bibnamefont {Calci}}, \bibinfo {author} {\bibfnamefont
  {J.}~\bibnamefont {Langhammer}}, \ and\ \bibinfo {author} {\bibfnamefont
  {R.}~\bibnamefont {Roth}},\ }\bibfield  {title} {\enquote {\bibinfo {title}
  {Nonperturbative shell-model interactions from the in-medium similarity
  renormalization group},}\ }\href {\doibase 10.1103/PhysRevLett.113.142501}
  {\bibfield  {journal} {\bibinfo  {journal} {Phys. Rev. Lett.}\ }\textbf
  {\bibinfo {volume} {113}},\ \bibinfo {pages} {142501} (\bibinfo {year}
  {2014})}\BibitemShut {NoStop}%
\bibitem [{\citenamefont {Dikmen}\ \emph {et~al.}(2015)\citenamefont {Dikmen},
  \citenamefont {Lisetski}, \citenamefont {Barrett}, \citenamefont {Maris},
  \citenamefont {Shirokov},\ and\ \citenamefont {Vary}}]{Dikmen15}%
  \BibitemOpen
  \bibfield  {author} {\bibinfo {author} {\bibfnamefont {E.}~\bibnamefont
  {Dikmen}}, \bibinfo {author} {\bibfnamefont {A.~F.}\ \bibnamefont
  {Lisetski}}, \bibinfo {author} {\bibfnamefont {B.~R.}\ \bibnamefont
  {Barrett}}, \bibinfo {author} {\bibfnamefont {P.}~\bibnamefont {Maris}},
  \bibinfo {author} {\bibfnamefont {A.~M.}\ \bibnamefont {Shirokov}}, \ and\
  \bibinfo {author} {\bibfnamefont {J.~P.}\ \bibnamefont {Vary}},\ }\bibfield
  {title} {\enquote {\bibinfo {title} {{Ab initio effective interactions for
  sd-shell valence nucleons}},}\ }\href {\doibase 10.1103/PhysRevC.91.064301}
  {\bibfield  {journal} {\bibinfo  {journal} {Phys. Rev. C}\ }\textbf {\bibinfo
  {volume} {91}},\ \bibinfo {pages} {064301} (\bibinfo {year}
  {2015})}\BibitemShut {NoStop}%
\bibitem [{\citenamefont {Tsunoda}\ \emph {et~al.}(2017)\citenamefont
  {Tsunoda}, \citenamefont {Otsuka}, \citenamefont {Shimizu}, \citenamefont
  {Hjorth-Jensen}, \citenamefont {Takayanagi},\ and\ \citenamefont
  {Suzuki}}]{Tsunoda16}%
  \BibitemOpen
  \bibfield  {author} {\bibinfo {author} {\bibfnamefont {Naofumi}\ \bibnamefont
  {Tsunoda}}, \bibinfo {author} {\bibfnamefont {Takaharu}\ \bibnamefont
  {Otsuka}}, \bibinfo {author} {\bibfnamefont {Noritaka}\ \bibnamefont
  {Shimizu}}, \bibinfo {author} {\bibfnamefont {Morten}\ \bibnamefont
  {Hjorth-Jensen}}, \bibinfo {author} {\bibfnamefont {Kazuo}\ \bibnamefont
  {Takayanagi}}, \ and\ \bibinfo {author} {\bibfnamefont {Toshio}\ \bibnamefont
  {Suzuki}},\ }\bibfield  {title} {\enquote {\bibinfo {title} {{Exotic
  neutron-rich medium-mass nuclei with realistic nuclear forces}},}\ }\href
  {\doibase 10.1103/PhysRevC.95.021304} {\bibfield  {journal} {\bibinfo
  {journal} {Phys. Rev. C}\ }\textbf {\bibinfo {volume} {95}},\ \bibinfo
  {pages} {021304} (\bibinfo {year} {2017})}\BibitemShut {NoStop}%
\bibitem [{\citenamefont {Sheikh}\ \emph {et~al.}(2014)\citenamefont {Sheikh},
  \citenamefont {Hinohara}, \citenamefont {Dobaczewski}, \citenamefont
  {Nakatsukasa}, \citenamefont {Nazarewicz},\ and\ \citenamefont
  {Sato}}]{sheikh14}%
  \BibitemOpen
  \bibfield  {author} {\bibinfo {author} {\bibfnamefont {J.~A.}\ \bibnamefont
  {Sheikh}}, \bibinfo {author} {\bibfnamefont {N.}~\bibnamefont {Hinohara}},
  \bibinfo {author} {\bibfnamefont {J.}~\bibnamefont {Dobaczewski}}, \bibinfo
  {author} {\bibfnamefont {T.}~\bibnamefont {Nakatsukasa}}, \bibinfo {author}
  {\bibfnamefont {W.}~\bibnamefont {Nazarewicz}}, \ and\ \bibinfo {author}
  {\bibfnamefont {K.}~\bibnamefont {Sato}},\ }\bibfield  {title} {\enquote
  {\bibinfo {title} {{Isospin-invariant Skyrme energy-density-functional
  approach with axial symmetry}},}\ }\href {\doibase
  10.1103/PhysRevC.89.054317} {\bibfield  {journal} {\bibinfo  {journal} {Phys.
  Rev. C}\ }\textbf {\bibinfo {volume} {89}},\ \bibinfo {pages} {054317}
  (\bibinfo {year} {2014})}\BibitemShut {NoStop}%
\bibitem [{\citenamefont {van Isacker}\ \emph {et~al.}()\citenamefont {van
  Isacker}, \citenamefont {Engel},\ and\ \citenamefont {Nomura}}]{isa16}%
  \BibitemOpen
  \bibfield  {author} {\bibinfo {author} {\bibfnamefont {P.}~\bibnamefont {van
  Isacker}}, \bibinfo {author} {\bibfnamefont {J.}~\bibnamefont {Engel}}, \
  and\ \bibinfo {author} {\bibfnamefont {K.}~\bibnamefont {Nomura}},\
  }\href@noop {} {}\bibinfo {note} {In preparation}\BibitemShut {NoStop}%
\bibitem [{\citenamefont {Raduta}\ \emph {et~al.}(1991)\citenamefont {Raduta},
  \citenamefont {Faessler},\ and\ \citenamefont {Stoica}}]{rad91}%
  \BibitemOpen
  \bibfield  {author} {\bibinfo {author} {\bibfnamefont {A.~A.}\ \bibnamefont
  {Raduta}}, \bibinfo {author} {\bibfnamefont {A.}~\bibnamefont {Faessler}}, \
  and\ \bibinfo {author} {\bibfnamefont {S.}~\bibnamefont {Stoica}},\
  }\bibfield  {title} {\enquote {\bibinfo {title} {The $2\nu\beta\beta$ decay
  rate within a boson expansion formalism},}\ }\href {\doibase
  10.1016/0375-9474(91)90561-J} {\bibfield  {journal} {\bibinfo  {journal}
  {Nucl.\ Phys. A}\ }\textbf {\bibinfo {volume} {534}},\ \bibinfo {pages} {149}
  (\bibinfo {year} {1991})}\BibitemShut {NoStop}%
\bibitem [{\citenamefont {Stoica}\ and\ \citenamefont
  {Klapdor-Kleingrothaus}(2001)}]{sto01}%
  \BibitemOpen
  \bibfield  {author} {\bibinfo {author} {\bibfnamefont {S.}~\bibnamefont
  {Stoica}}\ and\ \bibinfo {author} {\bibfnamefont {H.~V.}\ \bibnamefont
  {Klapdor-Kleingrothaus}},\ }\bibfield  {title} {\enquote {\bibinfo {title}
  {Critical view on double-beta decay matrix elements within quasi random phase
  approximation-based methods},}\ }\href {\doibase
  10.1016/S0375-9474(01)00988-5} {\bibfield  {journal} {\bibinfo  {journal}
  {Nucl.\ Phys. A}\ }\textbf {\bibinfo {volume} {694}},\ \bibinfo {pages} {269}
  (\bibinfo {year} {2001})}\BibitemShut {NoStop}%
\bibitem [{\citenamefont {Papakonstantinou}\ and\ \citenamefont
  {Roth}(2010)}]{Papakonstantinou09}%
  \BibitemOpen
  \bibfield  {author} {\bibinfo {author} {\bibfnamefont {P.}~\bibnamefont
  {Papakonstantinou}}\ and\ \bibinfo {author} {\bibfnamefont {R.}~\bibnamefont
  {Roth}},\ }\bibfield  {title} {\enquote {\bibinfo {title} {{Large-scale
  second RPA calculations with finite-range interactions}},}\ }\href {\doibase
  10.1103/PhysRevC.81.024317} {\bibfield  {journal} {\bibinfo  {journal} {Phys.
  Rev. C}\ }\textbf {\bibinfo {volume} {81}},\ \bibinfo {pages} {024317}
  (\bibinfo {year} {2010})}\BibitemShut {NoStop}%
\bibitem [{\citenamefont {Gambacurta}\ \emph {et~al.}(2010)\citenamefont
  {Gambacurta}, \citenamefont {Grasso},\ and\ \citenamefont {Catara}}]{gam10}%
  \BibitemOpen
  \bibfield  {author} {\bibinfo {author} {\bibfnamefont {D.}~\bibnamefont
  {Gambacurta}}, \bibinfo {author} {\bibfnamefont {M.}~\bibnamefont {Grasso}},
  \ and\ \bibinfo {author} {\bibfnamefont {F.}~\bibnamefont {Catara}},\
  }\bibfield  {title} {\enquote {\bibinfo {title} {Collective nuclear
  excitations with \protect{S}kyrme-second \protect{RPA}},}\ }\href {\doibase
  10.1103/PhysRevC.81.054312} {\bibfield  {journal} {\bibinfo  {journal}
  {Phys.\ Rev.\ C}\ }\textbf {\bibinfo {volume} {81}},\ \bibinfo {pages}
  {054312} (\bibinfo {year} {2010})}\BibitemShut {NoStop}%
\bibitem [{\citenamefont {Gambacurta}\ \emph {et~al.}(2015)\citenamefont
  {Gambacurta}, \citenamefont {Grasso},\ and\ \citenamefont {Engel}}]{gam15}%
  \BibitemOpen
  \bibfield  {author} {\bibinfo {author} {\bibfnamefont {D.}~\bibnamefont
  {Gambacurta}}, \bibinfo {author} {\bibfnamefont {M.}~\bibnamefont {Grasso}},
  \ and\ \bibinfo {author} {\bibfnamefont {J.}~\bibnamefont {Engel}},\
  }\bibfield  {title} {\enquote {\bibinfo {title} {{Subtraction method in the
  second random--phase approximation: first applications with a Skyrme energy
  functional}},}\ }\href {\doibase 10.1103/PhysRevC.92.034303} {\bibfield
  {journal} {\bibinfo  {journal} {Phys. Rev. C}\ }\textbf {\bibinfo {volume}
  {92}},\ \bibinfo {pages} {034303} (\bibinfo {year} {2015})}\BibitemShut
  {NoStop}%
\bibitem [{\citenamefont {Litvinova}\ \emph {et~al.}(2008)\citenamefont
  {Litvinova}, \citenamefont {Ring},\ and\ \citenamefont
  {Tselyaev}}]{Litvinova08}%
  \BibitemOpen
  \bibfield  {author} {\bibinfo {author} {\bibfnamefont {E.}~\bibnamefont
  {Litvinova}}, \bibinfo {author} {\bibfnamefont {P.}~\bibnamefont {Ring}}, \
  and\ \bibinfo {author} {\bibfnamefont {V.}~\bibnamefont {Tselyaev}},\
  }\bibfield  {title} {\enquote {\bibinfo {title} {{Relativistic quasiparticle
  time blocking approximation. Dipole response of open-shell nuclei}},}\ }\href
  {\doibase 10.1103/PhysRevC.78.014312} {\bibfield  {journal} {\bibinfo
  {journal} {Phys. Rev. C}\ }\textbf {\bibinfo {volume} {78}},\ \bibinfo
  {pages} {014312} (\bibinfo {year} {2008})}\BibitemShut {NoStop}%
\bibitem [{\citenamefont {Robin}\ and\ \citenamefont
  {Litvinova}(2016)}]{Robin16}%
  \BibitemOpen
  \bibfield  {author} {\bibinfo {author} {\bibfnamefont {Caroline}\
  \bibnamefont {Robin}}\ and\ \bibinfo {author} {\bibfnamefont {Elena}\
  \bibnamefont {Litvinova}},\ }\bibfield  {title} {\enquote {\bibinfo {title}
  {{Nuclear response theory for spin-isospin excitations in a relativistic
  quasiparticle-phonon coupling framework}},}\ }\href {\doibase
  10.1140/epja/i2016-16205-0} {\bibfield  {journal} {\bibinfo  {journal} {Eur.
  Phys. J. A}\ }\textbf {\bibinfo {volume} {52}},\ \bibinfo {pages} {205}
  (\bibinfo {year} {2016})}\BibitemShut {NoStop}%
\bibitem [{\citenamefont {Terasaki}(2012)}]{Terasaki12}%
  \BibitemOpen
  \bibfield  {author} {\bibinfo {author} {\bibfnamefont {J.}~\bibnamefont
  {Terasaki}},\ }\bibfield  {title} {\enquote {\bibinfo {title} {{Overlap of
  quasiparticle random-phase approximation states for nuclear matrix elements
  of the neutrino-less double-$beta$ decay}},}\ }\href {\doibase
  10.1103/PhysRevC.86.021301} {\bibfield  {journal} {\bibinfo  {journal} {Phys.
  Rev. C}\ }\textbf {\bibinfo {volume} {86}},\ \bibinfo {pages} {021301}
  (\bibinfo {year} {2012})}\BibitemShut {NoStop}%
\bibitem [{\citenamefont {Terasaki}(2013)}]{Terasaki13}%
  \BibitemOpen
  \bibfield  {author} {\bibinfo {author} {\bibfnamefont {J.}~\bibnamefont
  {Terasaki}},\ }\bibfield  {title} {\enquote {\bibinfo {title} {{Overlap of
  quasiparticle random-phase approximation states based on ground states of
  different nuclei: Mathematical properties and test calculations}},}\ }\href
  {\doibase 10.1103/PhysRevC.87.024316} {\bibfield  {journal} {\bibinfo
  {journal} {Phys. Rev. C}\ }\textbf {\bibinfo {volume} {87}},\ \bibinfo
  {pages} {024316} (\bibinfo {year} {2013})}\BibitemShut {NoStop}%
\bibitem [{\citenamefont {Beane}\ \emph {et~al.}(2011)\citenamefont {Beane},
  \citenamefont {Detmold}, \citenamefont {Orginos},\ and\ \citenamefont
  {Savage}}]{Beane11}%
  \BibitemOpen
  \bibfield  {author} {\bibinfo {author} {\bibfnamefont {S.~R.}\ \bibnamefont
  {Beane}}, \bibinfo {author} {\bibfnamefont {W.}~\bibnamefont {Detmold}},
  \bibinfo {author} {\bibfnamefont {K.}~\bibnamefont {Orginos}}, \ and\
  \bibinfo {author} {\bibfnamefont {M.~J.}\ \bibnamefont {Savage}},\ }\bibfield
   {title} {\enquote {\bibinfo {title} {{Nuclear Physics from Lattice QCD}},}\
  }\href {\doibase 10.1016/j.ppnp.2010.08.002} {\bibfield  {journal} {\bibinfo
  {journal} {Prog. Part. Nucl. Phys.}\ }\textbf {\bibinfo {volume} {66}},\
  \bibinfo {pages} {1--40} (\bibinfo {year} {2011})}\BibitemShut {NoStop}%
\bibitem [{\citenamefont {Aoki}\ \emph {et~al.}(2012)\citenamefont {Aoki},
  \citenamefont {Doi}, \citenamefont {Hatsuda}, \citenamefont {Ikeda},
  \citenamefont {Inoue}, \citenamefont {Ishii}, \citenamefont {Murano},
  \citenamefont {Nemura},\ and\ \citenamefont {Sasaki}}]{Aoki12}%
  \BibitemOpen
  \bibfield  {author} {\bibinfo {author} {\bibfnamefont {Sinya}\ \bibnamefont
  {Aoki}}, \bibinfo {author} {\bibfnamefont {Takumi}\ \bibnamefont {Doi}},
  \bibinfo {author} {\bibfnamefont {Tetsuo}\ \bibnamefont {Hatsuda}}, \bibinfo
  {author} {\bibfnamefont {Yoichi}\ \bibnamefont {Ikeda}}, \bibinfo {author}
  {\bibfnamefont {Takashi}\ \bibnamefont {Inoue}}, \bibinfo {author}
  {\bibfnamefont {Noriyoshi}\ \bibnamefont {Ishii}}, \bibinfo {author}
  {\bibfnamefont {Keiko}\ \bibnamefont {Murano}}, \bibinfo {author}
  {\bibfnamefont {Hidekatsu}\ \bibnamefont {Nemura}}, \ and\ \bibinfo {author}
  {\bibfnamefont {Kenji}\ \bibnamefont {Sasaki}} (\bibinfo {collaboration} {HAL
  QCD Collaboration}),\ }\bibfield  {title} {\enquote {\bibinfo {title}
  {{Lattice QCD approach to Nuclear Physics}},}\ }\href {\doibase
  10.1093/ptep/pts010} {\bibfield  {journal} {\bibinfo  {journal} {PTEP}\
  }\textbf {\bibinfo {volume} {2012}},\ \bibinfo {pages} {01A105} (\bibinfo
  {year} {2012})}\BibitemShut {NoStop}%
\bibitem [{\citenamefont {Brice{\~n}o}\ \emph {et~al.}(2015)\citenamefont
  {Brice{\~n}o}, \citenamefont {Davoudi},\ and\ \citenamefont
  {Luu}}]{Briceno15}%
  \BibitemOpen
  \bibfield  {author} {\bibinfo {author} {\bibfnamefont {Ra{\'u}l~A.}\
  \bibnamefont {Brice{\~n}o}}, \bibinfo {author} {\bibfnamefont {Zohreh}\
  \bibnamefont {Davoudi}}, \ and\ \bibinfo {author} {\bibfnamefont {Thomas~C.}\
  \bibnamefont {Luu}},\ }\bibfield  {title} {\enquote {\bibinfo {title}
  {{Nuclear Reactions from Lattice QCD}},}\ }\href {\doibase
  10.1088/0954-3899/42/2/023101} {\bibfield  {journal} {\bibinfo  {journal} {J.
  Phys. G: Nucl. Part. Phys.}\ }\textbf {\bibinfo {volume} {42}},\ \bibinfo
  {pages} {023101} (\bibinfo {year} {2015})}\BibitemShut {NoStop}%
\bibitem [{\citenamefont {Abdel-Rehim}\ \emph {et~al.}(2015)\citenamefont
  {Abdel-Rehim} \emph {et~al.}}]{Abdel-Rehim15}%
  \BibitemOpen
  \bibfield  {author} {\bibinfo {author} {\bibfnamefont {A.}~\bibnamefont
  {Abdel-Rehim}} \emph {et~al.},\ }\bibfield  {title} {\enquote {\bibinfo
  {title} {{Nucleon and pion structure with lattice QCD simulations at physical
  value of the pion mass}},}\ }\href {\doibase 10.1103/PhysRevD.92.114513,
  10.1103/PhysRevD.93.039904} {\bibfield  {journal} {\bibinfo  {journal} {Phys.
  Rev. D}\ }\textbf {\bibinfo {volume} {92}},\ \bibinfo {pages} {114513}
  (\bibinfo {year} {2015})},\ \bibinfo {note} {[Erratum: Phys.
  Rev.D93,no.3,039904(2016)]}\BibitemShut {NoStop}%
\bibitem [{\citenamefont {Wiringa}\ \emph {et~al.}(1995)\citenamefont
  {Wiringa}, \citenamefont {Stoks},\ and\ \citenamefont {Schiavilla}}]{Wir95}%
  \BibitemOpen
  \bibfield  {author} {\bibinfo {author} {\bibfnamefont {R.~B.}\ \bibnamefont
  {Wiringa}}, \bibinfo {author} {\bibfnamefont {V.~G.~J.}\ \bibnamefont
  {Stoks}}, \ and\ \bibinfo {author} {\bibfnamefont {R.}~\bibnamefont
  {Schiavilla}},\ }\bibfield  {title} {\enquote {\bibinfo {title} {Accurate
  nucleon-nucleon potential with charge-independence breaking},}\ }\href
  {\doibase 10.1103/PhysRevC.51.38} {\bibfield  {journal} {\bibinfo  {journal}
  {Phys.\ Rev.\ C}\ }\textbf {\bibinfo {volume} {51}},\ \bibinfo {pages} {38}
  (\bibinfo {year} {1995})}\BibitemShut {NoStop}%
\bibitem [{\citenamefont {Machleidt}\ \emph {et~al.}(1987)\citenamefont
  {Machleidt}, \citenamefont {Holinde},\ and\ \citenamefont {Elster}}]{cdbonn}%
  \BibitemOpen
  \bibfield  {author} {\bibinfo {author} {\bibfnamefont {R.}~\bibnamefont
  {Machleidt}}, \bibinfo {author} {\bibfnamefont {K.}~\bibnamefont {Holinde}},
  \ and\ \bibinfo {author} {\bibfnamefont {C.}~\bibnamefont {Elster}},\
  }\bibfield  {title} {\enquote {\bibinfo {title} {The \protect{B}onn meson
  exchange model for the nucleon nucleon interaction},}\ }\href {\doibase
  10.1016/S0370-1573(87)80002-9} {\bibfield  {journal} {\bibinfo  {journal}
  {Phys. Rep.}\ }\textbf {\bibinfo {volume} {149}},\ \bibinfo {pages} {1--89}
  (\bibinfo {year} {1987})}\BibitemShut {NoStop}%
\bibitem [{\citenamefont {Stoks}\ \emph {et~al.}(1994)\citenamefont {Stoks},
  \citenamefont {Klomp}, \citenamefont {Terheggen},\ and\ \citenamefont
  {de~Swart}}]{Stoks:1994wp}%
  \BibitemOpen
  \bibfield  {author} {\bibinfo {author} {\bibfnamefont {V.~G.~J.}\
  \bibnamefont {Stoks}}, \bibinfo {author} {\bibfnamefont {R.~A.~M.}\
  \bibnamefont {Klomp}}, \bibinfo {author} {\bibfnamefont {C.~P.~F.}\
  \bibnamefont {Terheggen}}, \ and\ \bibinfo {author} {\bibfnamefont {J.~J.}\
  \bibnamefont {de~Swart}},\ }\bibfield  {title} {\enquote {\bibinfo {title}
  {{Construction of High Quality \protect{NN} Potential Models}},}\ }\href
  {\doibase 10.1103/PhysRevC.49.2950} {\bibfield  {journal} {\bibinfo
  {journal} {Phys.\ Rev.\ C}\ }\textbf {\bibinfo {volume} {49}},\ \bibinfo
  {pages} {2950--2962} (\bibinfo {year} {1994})}\BibitemShut {NoStop}%
\bibitem [{\citenamefont {Gandolfi}\ \emph {et~al.}(2014)\citenamefont
  {Gandolfi}, \citenamefont {Lovato}, \citenamefont {Carlson},\ and\
  \citenamefont {Schmidt}}]{gan14}%
  \BibitemOpen
  \bibfield  {author} {\bibinfo {author} {\bibfnamefont {S.}~\bibnamefont
  {Gandolfi}}, \bibinfo {author} {\bibfnamefont {A.}~\bibnamefont {Lovato}},
  \bibinfo {author} {\bibfnamefont {J.}~\bibnamefont {Carlson}}, \ and\
  \bibinfo {author} {\bibfnamefont {Kevin~E.}\ \bibnamefont {Schmidt}},\
  }\bibfield  {title} {\enquote {\bibinfo {title} {{From the lightest nuclei to
  the equation of state of asymmetric nuclear matter with realistic nuclear
  interactions}},}\ }\href {\doibase 10.1103/PhysRevC.90.061306} {\bibfield
  {journal} {\bibinfo  {journal} {Phys. Rev.}\ }\textbf {\bibinfo {volume}
  {C90}},\ \bibinfo {pages} {061306} (\bibinfo {year} {2014})}\BibitemShut
  {NoStop}%
\bibitem [{\citenamefont {Binder}\ \emph {et~al.}(2014)\citenamefont {Binder},
  \citenamefont {Langhammer}, \citenamefont {Calci},\ and\ \citenamefont
  {Roth}}]{Binder14}%
  \BibitemOpen
  \bibfield  {author} {\bibinfo {author} {\bibfnamefont {Sven}\ \bibnamefont
  {Binder}}, \bibinfo {author} {\bibfnamefont {Joachim}\ \bibnamefont
  {Langhammer}}, \bibinfo {author} {\bibfnamefont {Angelo}\ \bibnamefont
  {Calci}}, \ and\ \bibinfo {author} {\bibfnamefont {Robert}\ \bibnamefont
  {Roth}},\ }\bibfield  {title} {\enquote {\bibinfo {title} {{Ab Initio Path to
  Heavy Nuclei}},}\ }\href {\doibase 10.1016/j.physletb.2014.07.010} {\bibfield
   {journal} {\bibinfo  {journal} {Phys. Lett. B}\ }\textbf {\bibinfo {volume}
  {736}},\ \bibinfo {pages} {119--123} (\bibinfo {year} {2014})}\BibitemShut
  {NoStop}%
\bibitem [{\citenamefont {Ekstr\"{o}m}\ \emph {et~al.}(2015)\citenamefont
  {Ekstr\"{o}m}, \citenamefont {Jansen}, \citenamefont {Wendt}, \citenamefont
  {Hagen}, \citenamefont {Papenbrock}, \citenamefont {Carlsson}, \citenamefont
  {Forss{\'e}n}, \citenamefont {Hjorth-Jensen}, \citenamefont {Navr{\'a}til},\
  and\ \citenamefont {Nazarewicz}}]{Ekstrom:2015rta}%
  \BibitemOpen
  \bibfield  {author} {\bibinfo {author} {\bibfnamefont {A.}~\bibnamefont
  {Ekstr\"{o}m}}, \bibinfo {author} {\bibfnamefont {G.~R.}\ \bibnamefont
  {Jansen}}, \bibinfo {author} {\bibfnamefont {K.~A.}\ \bibnamefont {Wendt}},
  \bibinfo {author} {\bibfnamefont {G.}~\bibnamefont {Hagen}}, \bibinfo
  {author} {\bibfnamefont {T.}~\bibnamefont {Papenbrock}}, \bibinfo {author}
  {\bibfnamefont {B.~D.}\ \bibnamefont {Carlsson}}, \bibinfo {author}
  {\bibfnamefont {C.}~\bibnamefont {Forss{\'e}n}}, \bibinfo {author}
  {\bibfnamefont {M.}~\bibnamefont {Hjorth-Jensen}}, \bibinfo {author}
  {\bibfnamefont {P.}~\bibnamefont {Navr{\'a}til}}, \ and\ \bibinfo {author}
  {\bibfnamefont {W.}~\bibnamefont {Nazarewicz}},\ }\bibfield  {title}
  {\enquote {\bibinfo {title} {{Accurate nuclear radii and binding energies
  from a chiral interaction}},}\ }\href {\doibase 10.1103/PhysRevC.91.051301}
  {\bibfield  {journal} {\bibinfo  {journal} {Phys. Rev. C}\ }\textbf {\bibinfo
  {volume} {91}},\ \bibinfo {pages} {051301} (\bibinfo {year}
  {2015})}\BibitemShut {NoStop}%
\bibitem [{\citenamefont {Hebeler}\ \emph {et~al.}(2011)\citenamefont
  {Hebeler}, \citenamefont {Bogner}, \citenamefont {Furnstahl}, \citenamefont
  {Nogga},\ and\ \citenamefont {Schwenk}}]{Hebeler11}%
  \BibitemOpen
  \bibfield  {author} {\bibinfo {author} {\bibfnamefont {K.}~\bibnamefont
  {Hebeler}}, \bibinfo {author} {\bibfnamefont {S.~K.}\ \bibnamefont {Bogner}},
  \bibinfo {author} {\bibfnamefont {R.~J.}\ \bibnamefont {Furnstahl}}, \bibinfo
  {author} {\bibfnamefont {A.}~\bibnamefont {Nogga}}, \ and\ \bibinfo {author}
  {\bibfnamefont {A.}~\bibnamefont {Schwenk}},\ }\bibfield  {title} {\enquote
  {\bibinfo {title} {{Improved nuclear matter calculations from chiral
  low-momentum interactions}},}\ }\href {\doibase 10.1103/PhysRevC.83.031301}
  {\bibfield  {journal} {\bibinfo  {journal} {Phys. Rev. C}\ }\textbf {\bibinfo
  {volume} {83}},\ \bibinfo {pages} {031301} (\bibinfo {year}
  {2011})}\BibitemShut {NoStop}%
\bibitem [{\citenamefont {Garcia~Ruiz}\ \emph {et~al.}(2016)\citenamefont
  {Garcia~Ruiz} \emph {et~al.}}]{GarciaRuiz16}%
  \BibitemOpen
  \bibfield  {author} {\bibinfo {author} {\bibfnamefont {R.~F.}\ \bibnamefont
  {Garcia~Ruiz}} \emph {et~al.},\ }\bibfield  {title} {\enquote {\bibinfo
  {title} {{Unexpectedly large charge radii of neutron-rich calcium
  isotopes}},}\ }\href {\doibase 10.1038/nphys3645} {\bibfield  {journal}
  {\bibinfo  {journal} {Nature Phys.}\ }\textbf {\bibinfo {volume} {12}},\
  \bibinfo {pages} {594} (\bibinfo {year} {2016})}\BibitemShut {NoStop}%
\bibitem [{\citenamefont {Hebeler}\ \emph
  {et~al.}(2015{\natexlab{b}})\citenamefont {Hebeler}, \citenamefont {Krebs},
  \citenamefont {Epelbaum}, \citenamefont {Golak},\ and\ \citenamefont
  {Skibinski}}]{Hebeler15a}%
  \BibitemOpen
  \bibfield  {author} {\bibinfo {author} {\bibfnamefont {K.}~\bibnamefont
  {Hebeler}}, \bibinfo {author} {\bibfnamefont {H.}~\bibnamefont {Krebs}},
  \bibinfo {author} {\bibfnamefont {E.}~\bibnamefont {Epelbaum}}, \bibinfo
  {author} {\bibfnamefont {J.}~\bibnamefont {Golak}}, \ and\ \bibinfo {author}
  {\bibfnamefont {R.}~\bibnamefont {Skibinski}},\ }\bibfield  {title} {\enquote
  {\bibinfo {title} {{Efficient calculation of chiral three-nucleon forces up
  to N$^3$LO for ab initio studies}},}\ }\href {\doibase
  10.1103/PhysRevC.91.044001} {\bibfield  {journal} {\bibinfo  {journal} {Phys.
  Rev. C}\ }\textbf {\bibinfo {volume} {91}},\ \bibinfo {pages} {044001}
  (\bibinfo {year} {2015}{\natexlab{b}})}\BibitemShut {NoStop}%
\bibitem [{\citenamefont {Drischler}\ \emph {et~al.}(2016)\citenamefont
  {Drischler}, \citenamefont {Carbone}, \citenamefont {Hebeler},\ and\
  \citenamefont {Schwenk}}]{Drischler16}%
  \BibitemOpen
  \bibfield  {author} {\bibinfo {author} {\bibfnamefont {C.}~\bibnamefont
  {Drischler}}, \bibinfo {author} {\bibfnamefont {A.}~\bibnamefont {Carbone}},
  \bibinfo {author} {\bibfnamefont {K.}~\bibnamefont {Hebeler}}, \ and\
  \bibinfo {author} {\bibfnamefont {A.}~\bibnamefont {Schwenk}},\ }\bibfield
  {title} {\enquote {\bibinfo {title} {{Neutron matter from chiral two- and
  three-nucleon calculations up to N$^3$LO}},}\ }\href {\doibase
  10.1103/PhysRevC.94.054307} {\bibfield  {journal} {\bibinfo  {journal} {Phys.
  Rev. C}\ }\textbf {\bibinfo {volume} {94}},\ \bibinfo {pages} {054307}
  (\bibinfo {year} {2016})}\BibitemShut {NoStop}%
\bibitem [{\citenamefont {Kummel}\ \emph {et~al.}(1978)\citenamefont {Kummel},
  \citenamefont {Luhrmann},\ and\ \citenamefont {Zabolitzky}}]{Kummel78}%
  \BibitemOpen
  \bibfield  {author} {\bibinfo {author} {\bibfnamefont {H.}~\bibnamefont
  {Kummel}}, \bibinfo {author} {\bibfnamefont {K.~H.}\ \bibnamefont
  {Luhrmann}}, \ and\ \bibinfo {author} {\bibfnamefont {J.~G.}\ \bibnamefont
  {Zabolitzky}},\ }\bibfield  {title} {\enquote {\bibinfo {title}
  {{Many-Fermion theory in expS- (or coupled cluster) form}},}\ }\href
  {\doibase 10.1016/0370-1573(78)90081-9} {\bibfield  {journal} {\bibinfo
  {journal} {Phys. Rept.}\ }\textbf {\bibinfo {volume} {36}},\ \bibinfo {pages}
  {1--36} (\bibinfo {year} {1978})}\BibitemShut {NoStop}%
\bibitem [{\citenamefont {Mihaila}\ and\ \citenamefont
  {Heisenberg}(2000)}]{PhysRevC.61.054309}%
  \BibitemOpen
  \bibfield  {author} {\bibinfo {author} {\bibfnamefont {Bogdan}\ \bibnamefont
  {Mihaila}}\ and\ \bibinfo {author} {\bibfnamefont {Jochen~H.}\ \bibnamefont
  {Heisenberg}},\ }\bibfield  {title} {\enquote {\bibinfo {title} {{Ground
  state correlations and mean field in ${}^{16}\mathrm{O}.$ \protect{II}.
  Effects of a three-nucleon interaction}},}\ }\href {\doibase
  10.1103/PhysRevC.61.054309} {\bibfield  {journal} {\bibinfo  {journal} {Phys.
  Rev. C}\ }\textbf {\bibinfo {volume} {61}},\ \bibinfo {pages} {054309}
  (\bibinfo {year} {2000})}\BibitemShut {NoStop}%
\bibitem [{\citenamefont {Hagen}\ \emph
  {et~al.}(2007{\natexlab{a}})\citenamefont {Hagen}, \citenamefont {Dean},
  \citenamefont {Hjorth-Jensen}, \citenamefont {Papenbrock},\ and\
  \citenamefont {Schwenk}}]{PhysRevC.76.044305}%
  \BibitemOpen
  \bibfield  {author} {\bibinfo {author} {\bibfnamefont {G.}~\bibnamefont
  {Hagen}}, \bibinfo {author} {\bibfnamefont {D.~J.}\ \bibnamefont {Dean}},
  \bibinfo {author} {\bibfnamefont {M.}~\bibnamefont {Hjorth-Jensen}}, \bibinfo
  {author} {\bibfnamefont {T.}~\bibnamefont {Papenbrock}}, \ and\ \bibinfo
  {author} {\bibfnamefont {A.}~\bibnamefont {Schwenk}},\ }\bibfield  {title}
  {\enquote {\bibinfo {title} {Benchmark calculations for $^{3}\mathrm{H}$,
  $^{4}\mathrm{He}$, $^{16}\mathrm{O}$, and $^{40}\mathrm{Ca}$ with \textit{ab
  initio} coupled-cluster theory},}\ }\href {\doibase
  10.1103/PhysRevC.76.044305} {\bibfield  {journal} {\bibinfo  {journal} {Phys.
  Rev. C}\ }\textbf {\bibinfo {volume} {76}},\ \bibinfo {pages} {044305}
  (\bibinfo {year} {2007}{\natexlab{a}})}\BibitemShut {NoStop}%
\bibitem [{\citenamefont {Hagen}\ \emph
  {et~al.}(2007{\natexlab{b}})\citenamefont {Hagen}, \citenamefont
  {Papenbrock}, \citenamefont {Dean}, \citenamefont {Schwenk}, \citenamefont
  {Nogga}, \citenamefont {W\l{}och},\ and\ \citenamefont
  {Piecuch}}]{PhysRevC.76.034302}%
  \BibitemOpen
  \bibfield  {author} {\bibinfo {author} {\bibfnamefont {G.}~\bibnamefont
  {Hagen}}, \bibinfo {author} {\bibfnamefont {T.}~\bibnamefont {Papenbrock}},
  \bibinfo {author} {\bibfnamefont {D.~J.}\ \bibnamefont {Dean}}, \bibinfo
  {author} {\bibfnamefont {A.}~\bibnamefont {Schwenk}}, \bibinfo {author}
  {\bibfnamefont {A.}~\bibnamefont {Nogga}}, \bibinfo {author} {\bibfnamefont
  {M.}~\bibnamefont {W\l{}och}}, \ and\ \bibinfo {author} {\bibfnamefont
  {P.}~\bibnamefont {Piecuch}},\ }\bibfield  {title} {\enquote {\bibinfo
  {title} {Coupled-cluster theory for three-body hamiltonians},}\ }\href
  {\doibase 10.1103/PhysRevC.76.034302} {\bibfield  {journal} {\bibinfo
  {journal} {Phys. Rev. C}\ }\textbf {\bibinfo {volume} {76}},\ \bibinfo
  {pages} {034302} (\bibinfo {year} {2007}{\natexlab{b}})}\BibitemShut
  {NoStop}%
\bibitem [{\citenamefont {Lee}(2009)}]{Lee09}%
  \BibitemOpen
  \bibfield  {author} {\bibinfo {author} {\bibfnamefont {Dean}\ \bibnamefont
  {Lee}},\ }\bibfield  {title} {\enquote {\bibinfo {title} {{Lattice
  simulations for few- and many-body systems}},}\ }\href {\doibase
  10.1016/j.ppnp.2008.12.001} {\bibfield  {journal} {\bibinfo  {journal} {Prog.
  Part. Nucl. Phys.}\ }\textbf {\bibinfo {volume} {63}},\ \bibinfo {pages}
  {117--154} (\bibinfo {year} {2009})}\BibitemShut {NoStop}%
\bibitem [{\citenamefont {Pudliner}\ \emph {et~al.}(1997)\citenamefont
  {Pudliner}, \citenamefont {Pandharipande}, \citenamefont {Carlson},
  \citenamefont {Pieper},\ and\ \citenamefont {Wiringa}}]{pudliner97}%
  \BibitemOpen
  \bibfield  {author} {\bibinfo {author} {\bibfnamefont {B.~S.}\ \bibnamefont
  {Pudliner}}, \bibinfo {author} {\bibfnamefont {V.~R.}\ \bibnamefont
  {Pandharipande}}, \bibinfo {author} {\bibfnamefont {J.}~\bibnamefont
  {Carlson}}, \bibinfo {author} {\bibfnamefont {Steven~C.}\ \bibnamefont
  {Pieper}}, \ and\ \bibinfo {author} {\bibfnamefont {R.~B.}\ \bibnamefont
  {Wiringa}},\ }\bibfield  {title} {\enquote {\bibinfo {title} {{Quantum Monte
  Carlo calculations of nuclei with $A\lesssim 7$}},}\ }\href {\doibase
  10.1103/PhysRevC.56.1720} {\bibfield  {journal} {\bibinfo  {journal} {Phys.
  Rev. C}\ }\textbf {\bibinfo {volume} {56}},\ \bibinfo {pages} {1720--1750}
  (\bibinfo {year} {1997})}\BibitemShut {NoStop}%
\bibitem [{\citenamefont {Lynn}\ \emph {et~al.}(2016)\citenamefont {Lynn},
  \citenamefont {Tews}, \citenamefont {Carlson}, \citenamefont {Gandolfi},
  \citenamefont {Gezerlis}, \citenamefont {Schmidt},\ and\ \citenamefont
  {Schwenk}}]{Lynn15}%
  \BibitemOpen
  \bibfield  {author} {\bibinfo {author} {\bibfnamefont {J.~E.}\ \bibnamefont
  {Lynn}}, \bibinfo {author} {\bibfnamefont {I.}~\bibnamefont {Tews}}, \bibinfo
  {author} {\bibfnamefont {J.}~\bibnamefont {Carlson}}, \bibinfo {author}
  {\bibfnamefont {S.}~\bibnamefont {Gandolfi}}, \bibinfo {author}
  {\bibfnamefont {A.}~\bibnamefont {Gezerlis}}, \bibinfo {author}
  {\bibfnamefont {K.~E.}\ \bibnamefont {Schmidt}}, \ and\ \bibinfo {author}
  {\bibfnamefont {A.}~\bibnamefont {Schwenk}},\ }\bibfield  {title} {\enquote
  {\bibinfo {title} {{Chiral Three-Nucleon Interactions in Light Nuclei,
  Neutron-$\alpha$ Scattering, and Neutron Matter}},}\ }\href {\doibase
  10.1103/PhysRevLett.116.062501} {\bibfield  {journal} {\bibinfo  {journal}
  {Phys. Rev. Lett.}\ }\textbf {\bibinfo {volume} {116}},\ \bibinfo {pages}
  {062501} (\bibinfo {year} {2016})}\BibitemShut {NoStop}%
\bibitem [{\citenamefont {Roth}(2009)}]{Roth09}%
  \BibitemOpen
  \bibfield  {author} {\bibinfo {author} {\bibfnamefont {Robert}\ \bibnamefont
  {Roth}},\ }\bibfield  {title} {\enquote {\bibinfo {title} {{Importance
  Truncation for Large-Scale Configuration Interaction Approaches}},}\ }\href
  {\doibase 10.1103/PhysRevC.79.064324} {\bibfield  {journal} {\bibinfo
  {journal} {Phys. Rev. C}\ }\textbf {\bibinfo {volume} {79}},\ \bibinfo
  {pages} {064324} (\bibinfo {year} {2009})}\BibitemShut {NoStop}%
\bibitem [{\citenamefont {Maris}\ \emph {et~al.}(2011)\citenamefont {Maris},
  \citenamefont {Vary}, \citenamefont {Navr{\'a}til}, \citenamefont {Ormand},
  \citenamefont {Nam},\ and\ \citenamefont {Dean}}]{Maris11}%
  \BibitemOpen
  \bibfield  {author} {\bibinfo {author} {\bibfnamefont {P.}~\bibnamefont
  {Maris}}, \bibinfo {author} {\bibfnamefont {J.~P.}\ \bibnamefont {Vary}},
  \bibinfo {author} {\bibfnamefont {P.}~\bibnamefont {Navr{\'a}til}}, \bibinfo
  {author} {\bibfnamefont {W.~E.}\ \bibnamefont {Ormand}}, \bibinfo {author}
  {\bibfnamefont {H.}~\bibnamefont {Nam}}, \ and\ \bibinfo {author}
  {\bibfnamefont {D.~J.}\ \bibnamefont {Dean}},\ }\bibfield  {title} {\enquote
  {\bibinfo {title} {{Origin of the anomalous long lifetime of $^{14}$C}},}\
  }\href {\doibase 10.1103/PhysRevLett.106.202502} {\bibfield  {journal}
  {\bibinfo  {journal} {Phys. Rev. Lett.}\ }\textbf {\bibinfo {volume} {106}},\
  \bibinfo {pages} {202502} (\bibinfo {year} {2011})}\BibitemShut {NoStop}%
\bibitem [{\citenamefont {Hagen}\ \emph {et~al.}(2015)\citenamefont {Hagen}
  \emph {et~al.}}]{Hagen15}%
  \BibitemOpen
  \bibfield  {author} {\bibinfo {author} {\bibfnamefont {G.}~\bibnamefont
  {Hagen}} \emph {et~al.},\ }\bibfield  {title} {\enquote {\bibinfo {title}
  {{Neutron and weak-charge distributions of the $^{48}$Ca nucleus}},}\ }\href
  {\doibase 10.1038/nphys3529} {\bibfield  {journal} {\bibinfo  {journal}
  {Nature Phys.}\ }\textbf {\bibinfo {volume} {12}},\ \bibinfo {pages}
  {186--190} (\bibinfo {year} {2015})}\BibitemShut {NoStop}%
\bibitem [{\citenamefont {Morris}\ \emph {et~al.}(2015)\citenamefont {Morris},
  \citenamefont {Parzuchowski},\ and\ \citenamefont {Bogner}}]{Morris15}%
  \BibitemOpen
  \bibfield  {author} {\bibinfo {author} {\bibfnamefont {T.~D.}\ \bibnamefont
  {Morris}}, \bibinfo {author} {\bibfnamefont {N.}~\bibnamefont
  {Parzuchowski}}, \ and\ \bibinfo {author} {\bibfnamefont {S.~K.}\
  \bibnamefont {Bogner}},\ }\bibfield  {title} {\enquote {\bibinfo {title}
  {{Magnus expansion and in-medium similarity renormalization group}},}\ }\href
  {\doibase 10.1103/PhysRevC.92.034331} {\bibfield  {journal} {\bibinfo
  {journal} {Phys. Rev. C}\ }\textbf {\bibinfo {volume} {92}},\ \bibinfo
  {pages} {034331} (\bibinfo {year} {2015})}\BibitemShut {NoStop}%
\bibitem [{\citenamefont {Lapoux}\ \emph {et~al.}(2016)\citenamefont {Lapoux},
  \citenamefont {Som{\`a}}, \citenamefont {Barbieri}, \citenamefont {Hergert},
  \citenamefont {Holt},\ and\ \citenamefont {Stroberg}}]{Lapoux16}%
  \BibitemOpen
  \bibfield  {author} {\bibinfo {author} {\bibfnamefont {V.}~\bibnamefont
  {Lapoux}}, \bibinfo {author} {\bibfnamefont {V.}~\bibnamefont {Som{\`a}}},
  \bibinfo {author} {\bibfnamefont {C.}~\bibnamefont {Barbieri}}, \bibinfo
  {author} {\bibfnamefont {H.}~\bibnamefont {Hergert}}, \bibinfo {author}
  {\bibfnamefont {J.~D.}\ \bibnamefont {Holt}}, \ and\ \bibinfo {author}
  {\bibfnamefont {S. R.}\ \bibnamefont {Stroberg}},\ }\bibfield  {title}
  {\enquote {\bibinfo {title} {{Radii and Binding Energies in Oxygen Isotopes:
  A Challenge for Nuclear Forces}},}\ }\href {\doibase
  10.1103/PhysRevLett.117.052501} {\bibfield  {journal} {\bibinfo  {journal}
  {Phys. Rev. Lett.}\ }\textbf {\bibinfo {volume} {117}},\ \bibinfo {pages}
  {052501} (\bibinfo {year} {2016})}\BibitemShut {NoStop}%
\bibitem [{\citenamefont {Okubo}(1954)}]{oku54}%
  \BibitemOpen
  \bibfield  {author} {\bibinfo {author} {\bibfnamefont {S.}~\bibnamefont
  {Okubo}},\ }\bibfield  {title} {\enquote {\bibinfo {title} {Diagonalization
  of hamiltonian and tamm-dancoff equation},}\ }\href {\doibase
  10.1143/PTP.12.603} {\bibfield  {journal} {\bibinfo  {journal} {Prog. Theor.
  Phys}\ }\textbf {\bibinfo {volume} {12}},\ \bibinfo {pages} {603} (\bibinfo
  {year} {1954})}\BibitemShut {NoStop}%
\bibitem [{\citenamefont {Lee}\ and\ \citenamefont {Suzuki}(1980)}]{lee80}%
  \BibitemOpen
  \bibfield  {author} {\bibinfo {author} {\bibfnamefont {S.Y.}\ \bibnamefont
  {Lee}}\ and\ \bibinfo {author} {\bibfnamefont {K.}~\bibnamefont {Suzuki}},\
  }\bibfield  {title} {\enquote {\bibinfo {title} {The effective interaction of
  two nucleons in the s-d shell},}\ }\href {\doibase
  10.1016/0370-2693(80)90423-2} {\bibfield  {journal} {\bibinfo  {journal}
  {Phys.\ Lett.\ B}\ }\textbf {\bibinfo {volume} {91}},\ \bibinfo {pages} {173}
  (\bibinfo {year} {1980})}\BibitemShut {NoStop}%
\bibitem [{\citenamefont {Suzuki}\ and\ \citenamefont {Lee}(1980)}]{suzuki80}%
  \BibitemOpen
  \bibfield  {author} {\bibinfo {author} {\bibfnamefont {Kenji}\ \bibnamefont
  {Suzuki}}\ and\ \bibinfo {author} {\bibfnamefont {Shyh~Yuan}\ \bibnamefont
  {Lee}},\ }\bibfield  {title} {\enquote {\bibinfo {title} {Convergent theory
  for effective interaction in nuclei},}\ }\href {\doibase 10.1143/PTP.64.2091}
  {\bibfield  {journal} {\bibinfo  {journal} {Prog.\ Theor.\ Phys.}\ }\textbf
  {\bibinfo {volume} {64}},\ \bibinfo {pages} {2091} (\bibinfo {year}
  {1980})}\BibitemShut {NoStop}%
\bibitem [{\citenamefont {Lisetskiy}\ \emph {et~al.}(2008)\citenamefont
  {Lisetskiy}, \citenamefont {Barrett}, \citenamefont {Kruse}, \citenamefont
  {Navr{\'a}til}, \citenamefont {Stetcu},\ and\ \citenamefont {Vary}}]{lis08}%
  \BibitemOpen
  \bibfield  {author} {\bibinfo {author} {\bibfnamefont {A.~F.}\ \bibnamefont
  {Lisetskiy}}, \bibinfo {author} {\bibfnamefont {B.~R.}\ \bibnamefont
  {Barrett}}, \bibinfo {author} {\bibfnamefont {M.~K.~G.}\ \bibnamefont
  {Kruse}}, \bibinfo {author} {\bibfnamefont {P.}~\bibnamefont {Navr{\'a}til}},
  \bibinfo {author} {\bibfnamefont {I.}~\bibnamefont {Stetcu}}, \ and\ \bibinfo
  {author} {\bibfnamefont {J.~P.}\ \bibnamefont {Vary}},\ }\bibfield  {title}
  {\enquote {\bibinfo {title} {Ab initio shell model with a core},}\ }\href
  {\doibase 10.1103/PhysRevC.78.044302} {\bibfield  {journal} {\bibinfo
  {journal} {Phys. Rev. C}\ }\textbf {\bibinfo {volume} {78}},\ \bibinfo
  {pages} {044302} (\bibinfo {year} {2008})}\BibitemShut {NoStop}%
\bibitem [{\citenamefont {Jansen}\ \emph {et~al.}(2016)\citenamefont {Jansen},
  \citenamefont {Signoracci}, \citenamefont {Hagen},\ and\ \citenamefont
  {Navr{\'a}til}}]{Jansen16}%
  \BibitemOpen
  \bibfield  {author} {\bibinfo {author} {\bibfnamefont {G.~R.}\ \bibnamefont
  {Jansen}}, \bibinfo {author} {\bibfnamefont {A.}~\bibnamefont {Signoracci}},
  \bibinfo {author} {\bibfnamefont {G.}~\bibnamefont {Hagen}}, \ and\ \bibinfo
  {author} {\bibfnamefont {P.}~\bibnamefont {Navr{\'a}til}},\ }\bibfield
  {title} {\enquote {\bibinfo {title} {{Open $sd$-shell nuclei from first
  principles}},}\ }\href {\doibase 10.1103/PhysRevC.94.011301} {\bibfield
  {journal} {\bibinfo  {journal} {Phys. Rev. C}\ }\textbf {\bibinfo {volume}
  {94}},\ \bibinfo {pages} {011301} (\bibinfo {year} {2016})}\BibitemShut
  {NoStop}%
\bibitem [{\citenamefont {Stroberg}\ \emph {et~al.}(2017)\citenamefont
  {Stroberg}, \citenamefont {Calci}, \citenamefont {Hergert}, \citenamefont
  {Holt}, \citenamefont {Bogner}, \citenamefont {Roth},\ and\ \citenamefont
  {Schwenk}}]{Stroberg16a}%
  \BibitemOpen
  \bibfield  {author} {\bibinfo {author} {\bibfnamefont {S.~R.}\ \bibnamefont
  {Stroberg}}, \bibinfo {author} {\bibfnamefont {A.}~\bibnamefont {Calci}},
  \bibinfo {author} {\bibfnamefont {H.}~\bibnamefont {Hergert}}, \bibinfo
  {author} {\bibfnamefont {J.~D.}\ \bibnamefont {Holt}}, \bibinfo {author}
  {\bibfnamefont {S.~K.}\ \bibnamefont {Bogner}}, \bibinfo {author}
  {\bibfnamefont {R.}~\bibnamefont {Roth}}, \ and\ \bibinfo {author}
  {\bibfnamefont {A.}~\bibnamefont {Schwenk}},\ }\bibfield  {title} {\enquote
  {\bibinfo {title} {{A nucleus-dependent valence-space approach to nuclear
  structure}},}\ }\href {\doibase 10.1103/PhysRevLett.118.032502} {\bibfield
  {journal} {\bibinfo  {journal} {Phys. Rev. Lett.}\ }\textbf {\bibinfo
  {volume} {118}},\ \bibinfo {pages} {032502} (\bibinfo {year}
  {2017})}\BibitemShut {NoStop}%
\bibitem [{\citenamefont {Stroberg}\ \emph {et~al.}(2016)\citenamefont
  {Stroberg}, \citenamefont {Hergert}, \citenamefont {Holt}, \citenamefont
  {Bogner},\ and\ \citenamefont {Schwenk}}]{Stroberg16}%
  \BibitemOpen
  \bibfield  {author} {\bibinfo {author} {\bibfnamefont {S.~R.}\ \bibnamefont
  {Stroberg}}, \bibinfo {author} {\bibfnamefont {H.}~\bibnamefont {Hergert}},
  \bibinfo {author} {\bibfnamefont {J.~D.}\ \bibnamefont {Holt}}, \bibinfo
  {author} {\bibfnamefont {S.~K.}\ \bibnamefont {Bogner}}, \ and\ \bibinfo
  {author} {\bibfnamefont {A.}~\bibnamefont {Schwenk}},\ }\bibfield  {title}
  {\enquote {\bibinfo {title} {{Ground and excited states of doubly open-shell
  nuclei from ab initio valence-space Hamiltonians}},}\ }\href {\doibase
  10.1103/PhysRevC.93.051301} {\bibfield  {journal} {\bibinfo  {journal} {Phys.
  Rev. C}\ }\textbf {\bibinfo {volume} {93}},\ \bibinfo {pages} {051301}
  (\bibinfo {year} {2016})}\BibitemShut {NoStop}%
\bibitem [{\citenamefont {Brown}\ and\ \citenamefont
  {Richter}(2006)}]{brown06}%
  \BibitemOpen
  \bibfield  {author} {\bibinfo {author} {\bibfnamefont {B.~Alex}\ \bibnamefont
  {Brown}}\ and\ \bibinfo {author} {\bibfnamefont {W.~A.}\ \bibnamefont
  {Richter}},\ }\bibfield  {title} {\enquote {\bibinfo {title} {{New 'USD'
  Hamiltonians for the sd shell}},}\ }\href {\doibase
  10.1103/PhysRevC.74.034315} {\bibfield  {journal} {\bibinfo  {journal} {Phys.
  Rev. C}\ }\textbf {\bibinfo {volume} {74}},\ \bibinfo {pages} {034315}
  (\bibinfo {year} {2006})}\BibitemShut {NoStop}%
\bibitem [{\citenamefont {Hergert}(2017)}]{Hergert16a}%
  \BibitemOpen
  \bibfield  {author} {\bibinfo {author} {\bibfnamefont {H.}~\bibnamefont
  {Hergert}},\ }\bibfield  {title} {\enquote {\bibinfo {title} {{In-Medium
  Similarity Renormalization Group for Closed and Open-Shell Nuclei}},}\ }\href
  {\doibase 10.1088/1402-4896/92/2/023002} {\bibfield  {journal} {\bibinfo
  {journal} {Phys. Scr.}\ }\textbf {\bibinfo {volume} {92}},\ \bibinfo {pages}
  {023002} (\bibinfo {year} {2017})}\BibitemShut {NoStop}%
\bibitem [{\citenamefont {Kutzelnigg}\ and\ \citenamefont
  {Mukherjee}(1997)}]{Kutzelnigg97}%
  \BibitemOpen
  \bibfield  {author} {\bibinfo {author} {\bibfnamefont {Werner}\ \bibnamefont
  {Kutzelnigg}}\ and\ \bibinfo {author} {\bibfnamefont {Debashis}\ \bibnamefont
  {Mukherjee}},\ }\bibfield  {title} {\enquote {\bibinfo {title} {Normal order
  and extended wick theorem for a multiconfiguration reference wave
  function},}\ }\href {\doibase 10.1063/1.474405} {\bibfield  {journal}
  {\bibinfo  {journal} {J. Chem. Phys.}\ }\textbf {\bibinfo {volume} {107}},\
  \bibinfo {pages} {432--449} (\bibinfo {year} {1997})}\BibitemShut {NoStop}%
\bibitem [{\citenamefont {Mukherjee}(1997)}]{Mukherjee97}%
  \BibitemOpen
  \bibfield  {author} {\bibinfo {author} {\bibfnamefont {Debashis}\
  \bibnamefont {Mukherjee}},\ }\bibfield  {title} {\enquote {\bibinfo {title}
  {Normal ordering and a wick-like reduction theorem for fermions with respect
  to a multi-determinantal reference state},}\ }\href {\doibase
  10.1016/S0009-2614(97)00714-8} {\bibfield  {journal} {\bibinfo  {journal}
  {Chem. Phys. Lett.}\ }\textbf {\bibinfo {volume} {274}},\ \bibinfo {pages}
  {561 -- 566} (\bibinfo {year} {1997})}\BibitemShut {NoStop}%
\bibitem [{\citenamefont {Epelbaum}\ \emph {et~al.}(2015)\citenamefont
  {Epelbaum}, \citenamefont {Krebs},\ and\ \citenamefont
  {Mei{\ss}ner}}]{Epelbaum15}%
  \BibitemOpen
  \bibfield  {author} {\bibinfo {author} {\bibfnamefont {E.}~\bibnamefont
  {Epelbaum}}, \bibinfo {author} {\bibfnamefont {H.}~\bibnamefont {Krebs}}, \
  and\ \bibinfo {author} {\bibfnamefont {U.~G.}\ \bibnamefont {Mei{\ss}ner}},\
  }\bibfield  {title} {\enquote {\bibinfo {title} {{Improved chiral
  nucleon-nucleon potential up to next-to-next-to-next-to-leading order}},}\
  }\href {\doibase 10.1140/epja/i2015-15053-8} {\bibfield  {journal} {\bibinfo
  {journal} {Eur. Phys. J. A}\ }\textbf {\bibinfo {volume} {51}},\ \bibinfo
  {pages} {53} (\bibinfo {year} {2015})}\BibitemShut {NoStop}%
\bibitem [{\citenamefont {Furnstahl}\ \emph {et~al.}(2015)\citenamefont
  {Furnstahl}, \citenamefont {Klco}, \citenamefont {Phillips},\ and\
  \citenamefont {Wesolowski}}]{Furnstahl:2015rha}%
  \BibitemOpen
  \bibfield  {author} {\bibinfo {author} {\bibfnamefont {R.~J.}\ \bibnamefont
  {Furnstahl}}, \bibinfo {author} {\bibfnamefont {N.}~\bibnamefont {Klco}},
  \bibinfo {author} {\bibfnamefont {D.~R.}\ \bibnamefont {Phillips}}, \ and\
  \bibinfo {author} {\bibfnamefont {S.}~\bibnamefont {Wesolowski}},\ }\bibfield
   {title} {\enquote {\bibinfo {title} {{Quantifying truncation errors in
  effective field theory}},}\ }\href {\doibase 10.1103/PhysRevC.92.024005}
  {\bibfield  {journal} {\bibinfo  {journal} {Phys. Rev. C}\ }\textbf {\bibinfo
  {volume} {92}},\ \bibinfo {pages} {024005} (\bibinfo {year}
  {2015})}\BibitemShut {NoStop}%
\bibitem [{\citenamefont {Carlsson}\ \emph {et~al.}(2016)\citenamefont
  {Carlsson}, \citenamefont {Ekstr{\"o}m}, \citenamefont {Forss{\'e}n},
  \citenamefont {Str{\"o}mberg}, \citenamefont {Jansen}, \citenamefont {Lilja},
  \citenamefont {Lindby}, \citenamefont {Mattsson},\ and\ \citenamefont
  {Wendt}}]{Carlsson16}%
  \BibitemOpen
  \bibfield  {author} {\bibinfo {author} {\bibfnamefont {B.~D.}\ \bibnamefont
  {Carlsson}}, \bibinfo {author} {\bibfnamefont {A.}~\bibnamefont
  {Ekstr{\"o}m}}, \bibinfo {author} {\bibfnamefont {C.}~\bibnamefont
  {Forss{\'e}n}}, \bibinfo {author} {\bibfnamefont {D.~Fahlin}\ \bibnamefont
  {Str{\"o}mberg}}, \bibinfo {author} {\bibfnamefont {G.~R.}\ \bibnamefont
  {Jansen}}, \bibinfo {author} {\bibfnamefont {O.}~\bibnamefont {Lilja}},
  \bibinfo {author} {\bibfnamefont {M.}~\bibnamefont {Lindby}}, \bibinfo
  {author} {\bibfnamefont {B.~A.}\ \bibnamefont {Mattsson}}, \ and\ \bibinfo
  {author} {\bibfnamefont {K.~A.}\ \bibnamefont {Wendt}},\ }\bibfield  {title}
  {\enquote {\bibinfo {title} {{Uncertainty analysis and order-by-order
  optimization of chiral nuclear interactions}},}\ }\href {\doibase
  10.1103/PhysRevX.6.011019} {\bibfield  {journal} {\bibinfo  {journal} {Phys.
  Rev. X}\ }\textbf {\bibinfo {volume} {6}},\ \bibinfo {pages} {011019}
  (\bibinfo {year} {2016})}\BibitemShut {NoStop}%
\bibitem [{\citenamefont {Simonis}\ \emph {et~al.}(2016)\citenamefont
  {Simonis}, \citenamefont {Hebeler}, \citenamefont {Holt}, \citenamefont
  {Men{\'e}ndez},\ and\ \citenamefont {Schwenk}}]{Simonis16}%
  \BibitemOpen
  \bibfield  {author} {\bibinfo {author} {\bibfnamefont {J.}~\bibnamefont
  {Simonis}}, \bibinfo {author} {\bibfnamefont {K.}~\bibnamefont {Hebeler}},
  \bibinfo {author} {\bibfnamefont {J.~D.}\ \bibnamefont {Holt}}, \bibinfo
  {author} {\bibfnamefont {J.}~\bibnamefont {Men{\'e}ndez}}, \ and\ \bibinfo
  {author} {\bibfnamefont {A.}~\bibnamefont {Schwenk}},\ }\bibfield  {title}
  {\enquote {\bibinfo {title} {{Exploring sd-shell nuclei from two- and
  three-nucleon interactions with realistic saturation properties}},}\ }\href
  {\doibase 10.1103/PhysRevC.93.011302} {\bibfield  {journal} {\bibinfo
  {journal} {Phys. Rev. C}\ }\textbf {\bibinfo {volume} {93}},\ \bibinfo
  {pages} {011302} (\bibinfo {year} {2016})}\BibitemShut {NoStop}%
\bibitem [{\citenamefont {Freeman}\ and\ \citenamefont
  {Schiffer}(2012)}]{Freeman12}%
  \BibitemOpen
  \bibfield  {author} {\bibinfo {author} {\bibfnamefont {S.~J.}\ \bibnamefont
  {Freeman}}\ and\ \bibinfo {author} {\bibfnamefont {J.~P.}\ \bibnamefont
  {Schiffer}},\ }\bibfield  {title} {\enquote {\bibinfo {title} {{Constraining
  the $0{\nu}2{\beta}$ matrix elements by nuclear structure observables}},}\
  }\href {\doibase 10.1088/0954-3899/39/12/124004} {\bibfield  {journal}
  {\bibinfo  {journal} {J. Phys. G: Nucl. Part. Phys.}\ }\textbf {\bibinfo
  {volume} {39}},\ \bibinfo {pages} {124004} (\bibinfo {year}
  {2012})}\BibitemShut {NoStop}%
\bibitem [{\citenamefont {Brown}\ \emph {et~al.}(2014)\citenamefont {Brown},
  \citenamefont {Horoi},\ and\ \citenamefont {Sen'kov}}]{Brown14}%
  \BibitemOpen
  \bibfield  {author} {\bibinfo {author} {\bibfnamefont {B.~A.}\ \bibnamefont
  {Brown}}, \bibinfo {author} {\bibfnamefont {M.}~\bibnamefont {Horoi}}, \ and\
  \bibinfo {author} {\bibfnamefont {R.~A.}\ \bibnamefont {Sen'kov}},\
  }\bibfield  {title} {\enquote {\bibinfo {title} {{Nuclear Structure Aspects
  of Neutrinoless Double-$\beta$ Decay}},}\ }\href {\doibase
  10.1103/PhysRevLett.113.262501} {\bibfield  {journal} {\bibinfo  {journal}
  {Phys. Rev. Lett.}\ }\textbf {\bibinfo {volume} {113}},\ \bibinfo {pages}
  {262501} (\bibinfo {year} {2014})}\BibitemShut {NoStop}%
\bibitem [{\citenamefont {Faessler}\ \emph {et~al.}(2009)\citenamefont
  {Faessler}, \citenamefont {Fogli}, \citenamefont {Lisi}, \citenamefont
  {Rodin}, \citenamefont {Rotunno},\ and\ \citenamefont
  {\v{S}imkovic}}]{fae09}%
  \BibitemOpen
  \bibfield  {author} {\bibinfo {author} {\bibfnamefont {Amand}\ \bibnamefont
  {Faessler}}, \bibinfo {author} {\bibfnamefont {G.~L.}\ \bibnamefont {Fogli}},
  \bibinfo {author} {\bibfnamefont {E.}~\bibnamefont {Lisi}}, \bibinfo {author}
  {\bibfnamefont {V.}~\bibnamefont {Rodin}}, \bibinfo {author} {\bibfnamefont
  {A.~M.}\ \bibnamefont {Rotunno}}, \ and\ \bibinfo {author} {\bibfnamefont
  {F.}~\bibnamefont {\v{S}imkovic}},\ }\bibfield  {title} {\enquote {\bibinfo
  {title} {\protect{QRPA} uncertainties and their correlations in the analysis
  of $0\nu\beta\beta$ decay},}\ }\href {\doibase 10.1103/PhysRevD.79.053001}
  {\bibfield  {journal} {\bibinfo  {journal} {Phys.\ Rev.\ D}\ }\textbf
  {\bibinfo {volume} {79}},\ \bibinfo {pages} {053001} (\bibinfo {year}
  {2009})}\BibitemShut {NoStop}%
\bibitem [{\citenamefont {Faessler}\ \emph
  {et~al.}(2008{\natexlab{b}})\citenamefont {Faessler}, \citenamefont {Fogli},
  \citenamefont {Lisi}, \citenamefont {Rodin}, \citenamefont {Rotunno},\ and\
  \citenamefont {\v{S}imkovic}}]{fae08}%
  \BibitemOpen
  \bibfield  {author} {\bibinfo {author} {\bibfnamefont {Amand}\ \bibnamefont
  {Faessler}}, \bibinfo {author} {\bibfnamefont {G.~L.}\ \bibnamefont {Fogli}},
  \bibinfo {author} {\bibfnamefont {E.}~\bibnamefont {Lisi}}, \bibinfo {author}
  {\bibfnamefont {V.}~\bibnamefont {Rodin}}, \bibinfo {author} {\bibfnamefont
  {A.M.}\ \bibnamefont {Rotunno}}, \ and\ \bibinfo {author} {\bibfnamefont
  {F.}~\bibnamefont {\v{S}imkovic}},\ }\bibfield  {title} {\enquote {\bibinfo
  {title} {{Overconstrained Estimates of Neutrinoless Double Beta Decay Within
  the QRPA}},}\ }\href {\doibase 10.1088/0954-3899/35/7/075104} {\bibfield
  {journal} {\bibinfo  {journal} {J.\ Phys.\ G}\ }\textbf {\bibinfo {volume}
  {35}},\ \bibinfo {pages} {075104} (\bibinfo {year}
  {2008}{\natexlab{b}})}\BibitemShut {NoStop}%
\bibitem [{\citenamefont {Deppisch}\ and\ \citenamefont
  {Suhonen}(2016)}]{Deppisch16}%
  \BibitemOpen
  \bibfield  {author} {\bibinfo {author} {\bibfnamefont {Frank~F.}\
  \bibnamefont {Deppisch}}\ and\ \bibinfo {author} {\bibfnamefont {Jouni}\
  \bibnamefont {Suhonen}},\ }\bibfield  {title} {\enquote {\bibinfo {title}
  {{Statistical analysis of beta decays and the effective value of $g_A$ in the
  proton-neutron quasiparticle random-phase approximation framework}},}\ }\href
  {\doibase 10.1103/PhysRevC.94.055501} {\bibfield  {journal} {\bibinfo
  {journal} {Phys. Rev. C}\ }\textbf {\bibinfo {volume} {94}},\ \bibinfo
  {pages} {055501} (\bibinfo {year} {2016})}\BibitemShut {NoStop}%
\bibitem [{\citenamefont {Dobaczewski}\ \emph {et~al.}(2014)\citenamefont
  {Dobaczewski}, \citenamefont {Nazarewicz},\ and\ \citenamefont
  {Reinhard}}]{dobaczewski14}%
  \BibitemOpen
  \bibfield  {author} {\bibinfo {author} {\bibfnamefont {J.}~\bibnamefont
  {Dobaczewski}}, \bibinfo {author} {\bibfnamefont {W.}~\bibnamefont
  {Nazarewicz}}, \ and\ \bibinfo {author} {\bibfnamefont {P.~G.}\ \bibnamefont
  {Reinhard}},\ }\bibfield  {title} {\enquote {\bibinfo {title} {Error
  estimates of theoretical models: a guide},}\ }\href {\doibase
  10.1088/0954-3899/41/7/074001} {\bibfield  {journal} {\bibinfo  {journal} {J.
  Phys. G: Nucl. Part. Phys.}\ }\textbf {\bibinfo {volume} {41}},\ \bibinfo
  {pages} {074001} (\bibinfo {year} {2014})}\BibitemShut {NoStop}%
\end{thebibliography}%

\end{document}